\documentclass[
reprint,onecolumn,
superscriptaddress, 12pt, 
nofootinbib,
 amsmath,amssymb,
floatfix
]{revtex4-2}

\usepackage{physics} 
\usepackage{dcolumn}
\usepackage{bm}
\usepackage{cancel}
\usepackage{xcolor}
\usepackage{tcolorbox}
\newcommand{\beqa}{\begin{eqnarray}}
\newcommand{\eeqa}{\end{eqnarray}}
\newcommand{\nn}{\nonumber}

\usepackage{subcaption}
\usepackage{booktabs}
\usepackage{graphicx}
\usepackage{dcolumn}
\usepackage{bm}

\usepackage{mathtools}
\usepackage{graphicx}

\newcommand{\Acoef}{\;A\;}                 
\newcommand{\Bcoef}{\;B\;}                 
\newcommand{\Ccoef}{\;C\;}                 
\newcommand{\Delt}{\Delta}                 
\newcommand{\SQ}{\sqrt{\Delt}}             

\begin{document}
\title{Force Dipole Interactions in Membranes with Odd Viscosity}
\author{Sneha Krishnan}
\affiliation{Birla Institute of Technology and Science, Pilani, Hyderabad Campus, Telangana 500078, India}
\author{Udaya Maurya}
\affiliation{Institute for Plasma Research, Bhat, Gandhinagar, India}
\author{Rickmoy Samanta}
\affiliation{Birla Institute of Technology and Science, Pilani, Hyderabad Campus, Telangana 500078, India}

\begin{abstract}
We develop a hydrodynamic framework for the interactions and collective
dynamics of force dipoles embedded in a compressible fluid membrane
supported by a shallow viscous subphase. Starting from the generalized
two-dimensional Stokes equations with shear, dilatational, and odd
(Hall) viscosities, we derive an exact real-space Green tensor using
Hankel transforms. The resulting tensor is characterized by three
hydrodynamic screening scales associated with shear, compressional,
and odd-viscous modes, and smoothly reduces to the standard limiting
cases of incompressible membranes and compressible parity-symmetric
membranes, while also capturing the chiral response generated by odd
viscosity. Using this Green tensor we obtain the velocity and vorticity
fields generated by a force dipole and formulate the dynamical system
governing interacting dipoles. The analysis reveals several distinct
dynamical regimes and identifies observables that isolate the
antisymmetric odd-viscous contribution to dipole interactions,
including transverse drift and chiral relative motion.
\end{abstract}

\maketitle
\tableofcontents
\section{Introduction}
\label{intro}

At microscopic length scales viscous stresses dominate inertial
effects, so the motion of swimmers and membrane--bound motor
assemblies is governed by Stokesian low--Reynolds--number hydrodynamics
\cite{purcell,lauga2009}. In this regime the linearity of the Stokes
equations allows flow fields to be represented as superpositions of
fundamental singularity solutions. Because freely swimming bodies
exert no net force or torque on the surrounding fluid, the leading
far--field disturbance is not a Stokeslet but a force dipole, or
\emph{stresslet}, which distinguishes extensile (pusher) from
contractile (puller) activity. Membrane--bound motor proteins,
enzymatic complexes, and cytoskeletal assemblies naturally act as
force dipoles, converting chemical energy into mechanical stresses on
the surrounding membrane fluid and cytoskeleton
\cite{gov2004,camley2013,mik,aggexp1,aggexp2}. Force dipoles therefore
provide a minimal hydrodynamic description of interactions between
active inclusions.

In unbounded momentum--conserving suspensions, hydrodynamic
interactions between such dipoles give rise to a wide range of
collective phenomena. Extensile (pusher) dipoles destabilize
orientationally ordered states, leading to large--scale hydrodynamic
instabilities and chaotic low--Reynolds--number turbulent flows
\cite{SimhaRamaswamy2002,SaintillanShelley2008,SaintillanShelley2009,
Ramaswamy2010,Marchetti2013,KochSubramanian2011,Saintillan2018}.
Contractile (puller) dipoles exhibit qualitatively different nonlinear
behavior and may instead promote clustering or aggregation depending
on parameters and dimensionality
\cite{BaskaranMarchetti2009,BaskaranMarchetti2010,IshikawaPedley2008,
AlarconPagonabarraga2013,Theers2018,ZottlStark2016}. When swimmers or
active inclusions are confined to interfaces or membranes these
interactions are strongly modified by the effectively two--dimensional
hydrodynamics of the membrane environment. In such systems momentum is
transported within the membrane but also dissipates into the
surrounding three--dimensional bulk fluid, leading to characteristic
screening lengths and mobility kernels first captured in the
Saffman--Delbrück theory and its many extensions
\cite{saff1,saff2,hughes,evans,lg96,staj,fischer}.

Recent theoretical and experimental work has shown that pair
interactions of inclusions in membranes depend sensitively on membrane
rheology, curvature, and coupling to the surrounding fluid
\cite{Levine2004,OppenheimerDiamant2009,OppenheimerDiamant2010,OppenheimerDiamant2011, SamantaOppenheimer2021, manikantan2020, bagaria2022,jain2023, Shoham2023, krishnan2025}. In particular, supported
membranes with finite subphase depth exhibit screened hydrodynamic
interactions governed by characteristic shear and compressional
screening lengths. Despite these advances, a systematic hydrodynamic
description of force--dipole interactions in compressible membranes
remains relatively unexplored. Membrane compressibility introduces
additional longitudinal modes that modify the structure of the
mobility kernel and can influence dipole coupling and collective
dynamics.

Beyond the conventional parity-symmetric viscous response, two-dimensional fluids can also support an antisymmetric contribution to the stress tensor proportional to the strain rate rotated by the Levi-Civita tensor, characterized by the odd (Hall) viscosity. This term, commonly referred to as \emph{odd viscosity}
or Hall viscosity, breaks parity and time--reversal symmetry and
generates transverse momentum transport
\cite{avron1998,banerjee2017,Soni2019,alizzi2023}. Odd viscosity has
been predicted or observed in a variety of chiral active and quantum
fluids and is known to produce circulating flows and chiral responses
near boundaries and defects. However, its implications for
hydrodynamic interactions between active inclusions in membranes
remain only partially understood.

In this work we develop a hydrodynamic framework for
force--dipole interactions in supported membranes that incorporates
both membrane compressibility and odd viscosity. Starting from the
generalized two--dimensional Stokes equation with momentum leakage
into the surrounding fluid, we derive the real--space Green tensor of
the membrane mobility. The resulting response tensor takes the form
\[
G_{ij}(\mathbf r)
=
A_0(r)\delta_{ij}
+
A_1(r)\hat r_i \hat r_j
+
A_2(r)\varepsilon_{ij},
\]
where the symmetric kernels $A_0$ and $A_1$ describe the conventional
dissipative hydrodynamic response while the antisymmetric kernel
$A_2$ encodes the chiral contribution associated with odd viscosity.
Observables that isolate this antisymmetric sector therefore provide
a natural way to characterize odd--viscous effects in dipolar
interactions. The radial kernels are governed by screened hydrodynamic
modes determined by the interplay between shear stresses, membrane
compressibility, and momentum leakage into the bulk subphase.

Using this Green tensor we derive explicit expressions for the
velocity and vorticity fields generated by a force dipole in a
compressible membrane. These fields form the basic building blocks for
describing interactions between active inclusions. Several limiting
regimes of the theory are analyzed, including the parity--symmetric
case without odd viscosity, the incompressible membrane limit, and a
degenerate regime in which the hydrodynamic screening lengths
coincide. In this latter case the Green tensor simplifies and allows
compact analytic expressions for the dipolar flow fields. The
near--field and far--field structures of the resulting flows are
examined in detail in order to characterize how the odd sector of the
response modifies the transverse velocity field and the vorticity
distribution around active dipoles.

Building on the single--dipole solution, we formulate the dynamical
equations governing interacting dipoles in the membrane. The resulting
many--body system couples dipole positions and orientations through
hydrodynamic interactions mediated by the Green tensor. This
formulation yields explicit evolution equations for pair motion as
well as for global observables such as center--of--mass drift and the
polarization of the dipole ensemble. Specializing to the two--dipole
case leads to a closed dynamical system describing the relative
separation and orientation dynamics of dipolar pairs. Within this
framework odd viscosity generates chiral relative motion and spiral
trajectories that do not arise in the parity--symmetric membrane case.

More broadly, the framework developed here provides a systematic
continuum description of active dipoles in compressible chiral
membrane fluids. The theory connects microscopic dipolar activity with
hydrodynamic observables and establishes a basis for studying
collective phenomena in active membrane systems. In forthcoming work
\cite{krishnan2026} we will present  more systematic multidipole simulations in
compressible membranes with odd viscosity based on the dynamical
equations derived in this paper.

Finally, we note that the present work builds on and complements a
number of previous studies on membrane hydrodynamics and odd viscosity,
including Refs.~\cite{fischer,staj,Levine2004,OppenheimerDiamant2009,OppenheimerDiamant2010,OppenheimerDiamant2011, manikantan2020,bagaria2022,jain2023, Samanta2025, krishnan2025,
Soni2019,HosakaKomuraAndelman2021,Hosaka2023}. While this manuscript
was in preparation a related preprint appeared that also analyzes the
hydrodynamic Green's functions and singularity flows in a compressible
supported fluid layer with odd viscosity \cite{hosaka2026}. The two
approaches share a similar physical setup but differ somewhat in
formulation. In particular, the present work organizes the hydrodynamic
response in terms of three characteristic screening scales associated
with shear stresses, compressional modes, and the odd-viscous sector.
When the antisymmetric bulk coupling vanishes and the substrate
interaction reduces to the conventional Brinkman friction describing
momentum leakage into the supporting fluid, the present formulation
reduces to the setup in Ref.~\cite{hosaka2026},
and the Green tensor obtained here recovers the corresponding solution
derived in that work. In addition, beyond the real-space Green tensor,
the present work develops the associated force-dipole velocity and
vorticity fields, analyzes several limiting regimes including the incompressible case, parity symmetric compressible case and a
degenerate odd viscous limit, and formulates the resulting interacting
two-dipole and many-dipole dynamical systems.
\paragraph{Organization of the paper.}
The remainder of the paper is organized as follows.
In Sec.~\ref{gveq} we formulate the continuum description of a
compressible supported membrane with shear, dilatational, and odd
viscosities and obtain the real--space Green tensor of the generalized
Stokes operator using a Hankel transform solution. From this tensor we
derive the velocity and vorticity fields generated by a force dipole
and introduce the three characteristic hydrodynamic screening lengths
associated with shear, compressional, and odd-sector couplings.

In Sec.~\ref{sec:limits} we examine several useful limits of the Green
tensor and the dipolar flow fields, including the parity--symmetric
compressible case, the incompressible supported membrane limit, and
the identical--screening degenerate regime in which the hydrodynamic
poles coalesce.

Subsequently we use the dipolar solution to construct the dynamical
equations governing interacting active dipoles embedded in the
membrane. We derive the general many--dipole dynamical system, analyze
global observables such as center--of--mass motion and polarization,
and then specialize to the two--dipole problem. Analytical results are
presented for both near--field and far--field regimes, including the
degenerate screening limit where the pair dynamics admits a compact
description and exhibits chiral spiral trajectories.

Finally, we discuss the structure of the odd--viscosity contribution to
pair motion and identify observables that isolate the antisymmetric
hydrodynamic response. Technical details of the Hankel transform derivation of the real--space Green tensor, the pole structure of the generalized Stokes operator and detailed simulation figures in all regimes are provided in the Appendix.

\begin{figure}[t]
    \centering
    \includegraphics[width=0.7\linewidth]{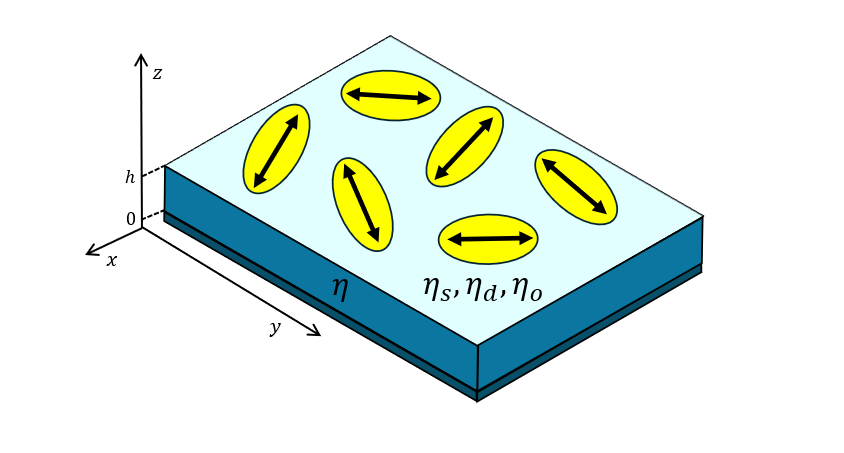}
    \caption{Schematic illustration of interacting force dipole motors
    confined to a fluid interface. Each yellow ellipse represents a particle,
    with the double-headed arrow indicating its orientation. The light blue layer
    corresponds to a two-dimensional membrane with viscosity 
    $\eta_s, \eta_d, \eta_o$, supported by a fluid subphase of viscosity $\eta$
    above a rigid substrate.}
    \label{schm3d}
\end{figure}
\section{Basic setup and analytic solution using Hankel transform}
\label{gveq}
Our explorations are based on the setup presented in Fig.~\ref{schm3d} in which we model the membrane as a  two-dimensional fluid layer with shear viscosity ($\eta_s$) and dilatation viscosity ($\eta_d$) in the presence of odd viscosity ($\eta_o$), on top of a subphase of height $h$ and viscosity $\eta$.
At the length and time scales relevant for membrane-embedded inclusions,
inertial terms are negligible and the membrane velocity is determined by
instantaneous balance between viscous stresses and applied forces
\cite{purcell,lauga2009}.
We model the membrane as a two-dimensional isotropic fluid layer that supports
both shear and dilatational stresses, characterized by a shear viscosity
$\eta_s$ and a dilatation (compressional) viscosity $\eta_d$
\cite{saff1,saff2,camley2013}.
When microscopic activity breaks time-reversal and
parity, the general linear, isotropic constitutive relation admits
a nondissipative, antisymmetric contribution known as odd (Hall) viscosity
\cite{avron1998,banerjee2017,Soni2019,alizzi2023,HosakaKomuraAndelman2021,Hosaka2023}.
Following Refs.~\cite{HosakaKomuraAndelman2021,Hosaka2023}, we treat $\eta_o$ as an
effective coarse-grained parameter that captures the macroscopic response of a
homogeneously distributed chiral active medium, such as rotor- or
torque-generating inclusions embedded in the membrane. In a strictly incompressible membrane ($\eta_d\!\to\!\infty$), the velocity field
and Green’s function are independent of the odd viscosity, rendering odd-viscous
effects hydrodynamically invisible, except via edge modes. Allowing finite compressibility activates
longitudinal modes that couple directly to antisymmetric stresses, thereby
restoring odd-viscosity contributions to the compressible membrane mobility tensor. The resulting in-plane membrane dynamics are governed by a generalized
two-dimensional Stokes equation. In the overdamped regime, instantaneous force
balance requires that viscous stresses within the membrane, pressure gradients
arising from coupling to the surrounding fluid, and externally applied force
densities balance. The membrane velocity field $\mathbf v(\mathbf r)$ thus
satisfies
\begin{equation}
\eta_s \nabla^2 \mathbf v
+ \eta_d \nabla(\nabla\!\cdot\!\mathbf v)
+ \eta_o \nabla^2 \mathbf v^{*}
- \frac{h}{2}\nabla p
+ \mathbf f^{\mathrm{3D}}
+ \mathbf F
= \mathbf 0 ,
\label{eqgen_intro}
\end{equation}
where $\mathbf v^{*}_i=\varepsilon_{ij}v_j$ denotes the velocity rotated by
$\pi/2$, $p$ is the pressure field in the underlying three-dimensional fluid,
$\mathbf f^{\mathrm{3D}}$ represents momentum exchange between the membrane and
its environment, and $\mathbf F$ is the in-plane force density generated by
embedded inclusions, such as that generated by
membrane-anchored force dipoles. The first two terms describe dissipative shear and
dilatational stresses, while the third term, proportional to the odd viscosity
$\eta_o$, is antisymmetric and nondissipative, encoding a transverse stress
response that breaks parity and time-reversal symmetry. Coupling to the bulk
fluid leads to momentum leakage, which we model as \cite{evans}
Brinkman-like friction coefficients $\zeta_{\parallel} $ and $\zeta_{\perp}
$ 
\begin{equation}
f^{3D}_i = -\big( \zeta_{\parallel}\,\delta_{ij} + \zeta_{\perp}\,\epsilon_{ij} \big)\, v_j.
\end{equation}
In the presence of a \textit{shallow} subphase, the longitudinal coupling can be well approximated \cite{staj}  as
\begin{equation}
\zeta_{\parallel}\sim\frac{\eta}{h},
\label{brnkfric}
\end{equation}
 For completeness, we note that in the  limit of an unsupported (free) membrane, the longitudinal friction coefficient $\zeta_{\parallel}(q)$ is in general \emph{linear} in the in-plane wavenumber $q$, reflecting the well-known Saffman--Delbrück momentum leakage into a semi-infinite fluid. Since this introduces qualitatively distinct hydrodynamic behavior, we do not consider the unsupported-membrane limit in this work and defer a detailed analysis to a future communication.\\ The force balance equation is closed by a compressibility
relation obtained under the lubrication approximation \cite{HosakaKomuraAndelman2021,Hosaka2023},
\begin{equation}
\nabla \cdot \mathbf{v}
= \frac{h^2}{6\eta}\,\nabla^2 p.
\label{eq:divv_intro}
\end{equation}
where $\eta$ is the bulk fluid
shear viscosity and $h$ is the subphase depth. In the lubrication limit, appropriate for a shallow subphase where $h$ is small
compared to lateral hydrodynamic length scales, the three-dimensional Stokes
equation in the subphase can be solved explicitly with no-slip boundary
conditions at the substrate and velocity continuity at the membrane interface
\cite{Hosaka2023,Barentin1999}.
Taking the divergence of the bulk flow and integrating across the subphase
thickness $0\le z\le h$ yields the closure relation Eq.~(\ref{eq:divv_intro}) linking the in-plane
compressibility of the membrane flow to pressure variations in the subphase. More complex situations involving surfactant-mediated interfacial stresses and
additional hidden dynamical variables can further modify this coupling mechanism \cite{Manikantansquires}; however, for analytical tractability we restrict our
attention here to the lubrication regime. Together, Eqs.~\eqref{eqgen_intro} and \eqref{eq:divv_intro} provide a minimal
continuum description of a compressible supported membrane with shear and odd
viscosities, capturing conventional dissipative in-plane momentum transport,
screening due to bulk momentum leakage, and nondissipative chiral transport
arising from the breaking of time-reversal and parity symmetries
\cite{HosakaKomuraAndelman2021,Hosaka2023}.\\
 Because the Stokes operator defined by Eqs.~\eqref{eqgen_intro} is linear and translationally invariant in the membrane plane, the velocity generated by any localized forcing can be written in terms of the associated Green’s tensor $G_{ij}(\mathbf r)$. For force-free active inclusions, the monopole (Stokeslet) contribution vanishes, and the leading term in the multipole expansion is the force dipole (stresslet). Differentiating the Stokeslet  along the dipole axis then yields the dipolar flow, taking the form,
\begin{equation}
v_i(\mathbf{r}) = \sigma \, \hat{d}_k \, \partial_k G_{ij}(\mathbf{r}) \, \hat{d}_j,
\label{vdipolegen}
\end{equation}
where $\hat{\mathbf d}$ denotes the dipole orientation and
$\sigma \equiv \mathrm{FL}\,$ is the dipole strength, with $ \mathrm{L}$ the dipole size
($\sigma>0$ for pushers and $\sigma<0$ for pullers).
The full Green’s function for a Stokeslet point force $\mathbf F$ in this system
is given by (see Appendix~\ref{app:greens_real_space} for a detailed derivation
using Hankel transforms),
\beqa
G_{ij}(\mathbf r)=A_0(r)\,\delta_{ij}+A_1(r)\,\hat r_i\hat r_j + A_2(r)\,\varepsilon_{ij},
\qquad r=|\mathbf r|,\ \hat r_i = r_i/r,
\label{greengen}
\eeqa
where
\beqa
&A_0(r)=\frac{1}{2\pi}\sum_{i=1}^2 R^{(\beta)}_i K_0(m_i r)
+\frac{C_0}{2\pi}\frac{1}{r^2}
+\frac{1}{2\pi}\sum_{i=1}^2 C_i\frac{m_i K_1(m_i r)}{r} \nn\\
&A_1(r)=-\frac{1}{2\pi}\sum_{i=1}^2 C_i m_i^2 K_0(m_i r)
-\frac{1}{\pi}\sum_{i=1}^2 C_i\frac{m_i K_1(m_i r)}{r}
-\frac{C_0}{\pi}\frac{1}{r^2}\nn\\
&A_2(r)=\frac{1}{2\pi}\sum_{i=1}^2 R^{(\gamma)}_i\,K_0(m_i r)
\label{adefs}
\eeqa
where $K_0,K_1$ are modified Bessel functions of the second kind. The method to compute the coefficients $m_1,m_2, C_i$ and the residues $R^{(\beta)},R^{(\gamma)},$ is provided explicitly below. We first define
\[
\begin{aligned}
\Acoef &= \eta_s(\eta_s+\eta_d)+\eta_o^2,\\[6pt]
\Bcoef &= \eta_s(\eta_s+\eta_d)(\kappa^2+\lambda^2)+2\eta_o^2\nu^2,\\[6pt]
\Ccoef &= \eta_s(\eta_s+\eta_d)\kappa^2\lambda^2+\eta_o^2\nu^4.
\end{aligned}
\]
where the hydrodynamic screening lengths are defined as
\begin{equation}
\kappa^{-1}=\sqrt{\frac{\eta_s}{\zeta_{\parallel}}},\qquad
\lambda^{-1}=\sqrt{\frac{h(\eta_s+\eta_d)}{3\eta + h\,\zeta_{\parallel}}},\qquad
\nu^{-1}=\sqrt{\frac{\eta_o}{\zeta_{\perp}}}\,.
\label{scrn}
\end{equation}
This helps us construct $\Delt$ 

\[
\Delt \;=\; B^2 - 4AC ,
\]
and importantly, the two \emph{screening masses}, named in analogy with Yukawa interactions,
\[
 \;
m_1 \;=\; \sqrt{\dfrac{\Bcoef - \SQ}{2\Acoef}}\;,\qquad
m_2 \;=\; \sqrt{\dfrac{\Bcoef +\SQ}{2\Acoef}}\;.
\]

The $R^{(\beta)}$ and $R^{(\gamma)}$ residues are given by
\[
\begin{aligned}
R^{(\beta)}_1 &= \frac{(\eta_s+\eta_d)}{\SQ}\left(\frac{-\Bcoef+\SQ}{2\Acoef}+\lambda^2\right)
= \frac{\eta_s+\eta_d}{2\Acoef\,\SQ}\big(-\Bcoef+\SQ + 2\Acoef\lambda^2\big), \\[8pt]
R^{(\beta)}_2 &= -\frac{(\eta_s+\eta_d)}{\SQ}\left(\frac{-\Bcoef-\SQ}{2\Acoef}+\lambda^2\right)
= -\frac{\eta_s+\eta_d}{2\Acoef\,\SQ}\big(-\Bcoef-\SQ + 2\Acoef\lambda^2\big),
\end{aligned}
\]

\[
\begin{aligned}
R^{(\gamma)}_1 &= -\frac{\eta_o}{\SQ}\left(\frac{-\Bcoef+\SQ}{2\Acoef}+\nu^2\right)
= -\frac{\eta_o}{2\Acoef\,\SQ}\big(-\Bcoef+\SQ + 2\Acoef\nu^2\big),\\[8pt]
R^{(\gamma)}_2 &= \frac{\eta_o}{\SQ}\left(\frac{-\Bcoef-\SQ}{2\Acoef}+\nu^2\right)
= \frac{\eta_o}{2\Acoef\,\SQ}\big(-\Bcoef-\SQ + 2\Acoef\nu^2\big).
\end{aligned}
\]

The coefficients $C_0,C_1,C_2$ are given by
\[
\; C_0  \;=\; \frac{\eta_s\kappa^2 - (\eta_s+\eta_d)\lambda^2}{\Ccoef} \; . \;
\]
\[
\begin{aligned}
C_1= \frac{ -\eta_d(-\Bcoef+\SQ) + 2\Acoef N_0 }{ (-\Bcoef+\SQ)\SQ }, \\[8pt]
C_2= \frac{ \eta_d(\Bcoef+\SQ) + 2\Acoef N_0 }{ (\Bcoef+\SQ)\SQ } .
\end{aligned}
\]
where  \(N_0=\eta_s\kappa^2-(\eta_s+\eta_d)\lambda^2\)).
The velocity field generated by a force dipole
of strength $\sigma$ and orientation $\hat{\mathbf d}$ takes the form
\begin{equation}
\mathbf v(\mathbf r)
=
\sigma
\left[
\alpha\!\left(A_0'(r)+\frac{A_1(r)}{r}\right)\hat{\mathbf d}
+
\left(
\frac{A_1(r)}{r}
+
\alpha^2\!\left(A_1'(r)-\frac{2A_1(r)}{r}\right)
\right)\hat{\mathbf r}
+
\alpha A_2'(r)(\varepsilon\!\cdot\!\hat{\mathbf d})
\right],
\label{vgen}
\end{equation}
while the associated vorticity field is
\begin{equation}
\omega(\mathbf r)
=
\sigma
\left[
\alpha\beta
\left(
A_0''(r)-\frac{A_0'(r)}{r}-\frac{A_1'(r)}{r}+\frac{2A_1(r)}{r^2}
\right)
-
\left(
\frac{A_2'(r)}{r}\beta^2+\alpha^2A_2''(r)
\right)
\right].
\label{vorgen}
\end{equation}
These expressions provide the complete hydrodynamic velocity and
vorticity fields generated by a force dipole in the membrane once the
radial Green tensor functions $A_0(r)$, $A_1(r)$ and $A_2(r)$ are
specified. The geometry of the flow relative to the dipole orientation
is conveniently described using the  scalars
\begin{equation}
\alpha = d_i \hat r_i ,
\label{alphadef}
\end{equation}
and
\begin{equation}
\beta = \varepsilon_{ij} d_j \hat r_i ,
\label{betadef}
\end{equation}
which represent the projections of the dipole orientation along the
radial and transverse directions, respectively.  Because the dipole
orientation vector $\mathbf d$ is normalized, these scalars satisfy
\begin{equation}
\alpha^2+\beta^2=1.
\end{equation}
Together, Eqs.~(\ref{alphadef})-(\ref{betadef}) provide a compact
parameterization of the relative orientation between the dipole axis
and the observation direction $\hat{\mathbf r}$, allowing the dipolar
flow and vorticity fields to be expressed entirely in terms of the
scalar functions $A_0(r)$, $A_1(r)$ and $A_2(r)$ and their radial
derivatives.

\subsection{Comments on screening lengths}

The three screening lengths in Eq.~(\ref{scrn}) set the spatial range of
distinct hydrodynamic modes in a supported membrane and delimit crossovers
between qualitatively different flow regimes.
The shear screening length
$\kappa^{-1}=\sqrt{\eta_s/\zeta_{\parallel}}$ controls shear modes: for $r\ll\kappa^{-1}$ the flow resembles unscreened 2D Stokes
hydrodynamics, while for $r\gg\kappa^{-1}$ shear disturbances are suppressed by
momentum leakage into the subphase.
The compressional length
$\lambda^{-1}=\sqrt{h(\eta_s+\eta_d)/(3\eta+h\zeta_{\parallel})}$ governs
longitudinal (dilation/compression) modes, reflecting the competition between
membrane compressibility and pressure-driven drainage in the underlying 3D
fluid.
The odd-viscous length
$\nu^{-1}=\sqrt{\eta_o/\zeta_{\perp}}$ sets the range of chiral, Hall-like
momentum transport that appears only when both parity and time-reversal
symmetries are broken; this channel is nondissipative and produces transverse,
lift-like responses without entropy production.

In the shallow-subphase (Brinkman) approximation one has
$\zeta_{\parallel}\sim\eta/h$ and often $\zeta_{\perp}$ is approximated as \cite{HosakaKomuraAndelman2021,Hosaka2023}
\begin{equation}
\zeta_{\perp}\sim\left(\frac{\eta_{o}}{\eta_{s}}\right)\zeta_{\parallel}\equiv \mu~\zeta_{\parallel}.
\label{brnkfricperp}
\end{equation}
so that $\mu=\eta_o/\eta_s$ is a dimensionless measure of odd-to-shear response.
In this limit $\mu$  links
$\nu^{-1}$ to $\kappa^{-1}$; relaxing this proportionality makes $\nu^{-1}$ an
independent crossover scale and allows odd-viscous transport to dominate at
intermediate distances even when conventional shear is strongly screened. We will not use the approximation of Eq.~(\ref{brnkfricperp}) except while specializing to a degenerate limit in Sec.~\ref{deg}.

\section{Useful limits of the membrane Green tensor, fluid velocity and vorticity}
\label{sec:limits}

We now record several limiting cases of the general Green tensor derived
above. These limits provide useful checks for the full Green tensor presented here and
clarify how the general solution reduces to familiar parity-symmetric
membrane hydrodynamics when the odd viscosity vanishes, as well as how
it simplifies in the identical-screening degenerate regime. 

\subsection{Compressible limit with vanishing odd viscosity}

It is useful to verify that the general Green tensor derived above reduces to the
standard compressible membrane mobility when odd viscosity is switched off.
This provides a nontrivial consistency check of the analytic solution obtained
from the Hankel transform.

We therefore consider the limit
\[
\eta_o \rightarrow 0 .
\]

In this case the antisymmetric stresses vanish and the hydrodynamics becomes
parity symmetric. The characteristic coefficients introduced in the general
solution simplify considerably. From the definitions
\[
\Acoef=\eta_s(\eta_s+\eta_d)+\eta_o^2, \qquad
\Bcoef=\eta_s(\eta_s+\eta_d)(\kappa^2+\lambda^2)+2\eta_o^2\nu^2, \qquad
\Ccoef=\eta_s(\eta_s+\eta_d)\kappa^2\lambda^2+\eta_o^2\nu^4,
\]
we obtain
\[
\Acoef=\eta_s(\eta_s+\eta_d),
\]
\[
\Bcoef=\eta_s(\eta_s+\eta_d)(\kappa^2+\lambda^2),
\]
\[
\Ccoef=\eta_s(\eta_s+\eta_d)\kappa^2\lambda^2 .
\]

The discriminant entering the pole structure becomes
\[
\Delta = \Bcoef^2-4\Acoef\Ccoef
       =[\eta_s(\eta_s+\eta_d)]^2(\kappa^2-\lambda^2)^2 .
\]

Consequently
\[
\sqrt{\Delta}=\eta_s(\eta_s+\eta_d)\,|\kappa^2-\lambda^2|.
\]

Substituting these expressions into the definitions of the screening masses
\[
m_{1,2}^2=\frac{\Bcoef\mp\sqrt{\Delta}}{2\Acoef},
\]
we find that the two poles reduce to the physical shear and compressional
screening scales,
\[
m_1^2=\lambda^2, \qquad m_2^2=\kappa^2 ,
\]
up to a trivial relabeling of the indices depending on the relative magnitude
of $\kappa$ and $\lambda$.

The residues appearing in the Green tensor also simplify.
The antisymmetric residues are proportional to $\eta_o$ and therefore vanish
identically,
\[
R^{(\gamma)}_1=R^{(\gamma)}_2=0 .
\]

As a result the antisymmetric kernel disappears,
\[
A_2(r)=0 .
\]

The symmetric residues become
\[
R^{(\beta)}_1 =0,
\qquad
R^{(\beta)}_2=\frac{1}{\eta_s}.
\]

The coefficients entering the algebraic part of the Green tensor are
\[
C_0=\frac{\eta_s\kappa^2-(\eta_s+\eta_d)\lambda^2}
{\eta_s(\eta_s+\eta_d)\kappa^2\lambda^2}
=\frac{1}{(\eta_s+\eta_d)\lambda^2}-\frac{1}{\eta_s\kappa^2},
\]
\[
C_1=-\frac{1}{(\eta_s+\eta_d)\lambda^2},
\qquad
C_2=\frac{1}{\eta_s\kappa^2}.
\]

Substituting these simplified coefficients into the general expressions for the
radial Green tensor functions yields
\[
A_0(r)
=
\frac{1}{2\pi}
\left[
\frac{1}{\eta_s}
\left(
K_0(\kappa r)+\frac{K_1(\kappa r)}{\kappa r}-\frac{1}{(\kappa r)^2}
\right)
-
\frac{1}{\eta_s+\eta_d}
\left(
\frac{K_1(\lambda r)}{\lambda r}-\frac{1}{(\lambda r)^2}
\right)
\right],
\]

\[
A_1(r)
=
\frac{1}{2\pi}
\left[
\frac{1}{\eta_s+\eta_d}
\left(
K_0(\lambda r)+\frac{2K_1(\lambda r)}{\lambda r}-\frac{2}{(\lambda r)^2}
\right)
-
\frac{1}{\eta_s}
\left(
K_0(\kappa r)+\frac{2K_1(\kappa r)}{\kappa r}-\frac{2}{(\kappa r)^2}
\right)
\right].
\]

The antisymmetric coefficient vanishes,
\[
A_2(r)=0 .
\]

The real-space Green tensor therefore reduces to
\[
G_{ij}(\mathbf r)
=
A_0(r)\delta_{ij}
+
A_1(r)\hat r_i\hat r_j .
\]

This expression is the Green function for a compressible supported membrane
without odd viscosity. The hydrodynamic response separates into two screened
channels corresponding to transverse shear and longitudinal compression.
The shear sector is screened over the length scale $\kappa^{-1}$, while the
compressional sector is screened over the length scale $\lambda^{-1}$.
In contrast to the general case with odd viscosity, no antisymmetric component
proportional to $\varepsilon_{ij}$ survives, reflecting the restoration of
parity symmetry in the membrane dynamics.  In
this case it is useful to write the Green tensor in the Helmholtz form
\begin{equation}
G_{ij}(\mathbf r)
=
G_L(r)\,\hat r_i\hat r_j
+
G_T(r)\,\big(\delta_{ij}-\hat r_i\hat r_j\big),
\end{equation}
with
\begin{align}
G_L(r)
&=
\frac{1}{2\pi}
\Bigg\{
\frac{1}{\eta_s+\eta_d}
\left[
K_0(\lambda r)
+\frac{K_1(\lambda r)}{\lambda r}
-\frac{1}{\lambda^2 r^2}
\right]
+
\frac{1}{\eta_s}
\left[
\frac{1}{\kappa^2 r^2}
-\frac{K_1(\kappa r)}{\kappa r}
\right]
\Bigg\},
\\
G_T(r)
&=
\frac{1}{2\pi}
\Bigg\{
\frac{1}{\eta_s+\eta_d}
\left[
\frac{1}{\lambda^2 r^2}
-\frac{K_1(\lambda r)}{\lambda r}
\right]
+
\frac{1}{\eta_s}
\left[
K_0(\kappa r)
+\frac{K_1(\kappa r)}{\kappa r}
-\frac{1}{\kappa^2 r^2}
\right]
\Bigg\}.
\end{align}
The corresponding force-dipole velocity and vorticity fields then
follow directly from the general formulas derived in the previous
section. The incompressible result is recovered smoothly in the limit
\(\eta_d\to\infty\), for which the longitudinal mode is suppressed.
In the compressible parity--symmetric case with vanishing odd viscosity
($\eta_o=0$), the antisymmetric Green--tensor component vanishes,
$A_2(r)=0$, and the dipole velocity field follows from
Eq.~(\ref{vgen}) after substituting the compressible Green--tensor
coefficients $A_0(r)$ and $A_1(r)$.  The resulting flow can be written
in the compact form
\begin{equation}
\mathbf v(\mathbf r)
=
\sigma
\left[
\alpha\,C_d(r)\,\hat{\mathbf d}
+
C_r(r)\,\hat{\mathbf r}
\right],
\end{equation}
where $\alpha=\hat{\mathbf d}\!\cdot\!\hat{\mathbf r}$.  The radial
coefficients receive independent contributions from the shear
($\kappa$) and compressional ($\lambda$) hydrodynamic modes and are
given by

\begin{equation}
C_d(r)=
\frac{1}{2\pi}
\left[
\frac{1}{\eta_s}
\left(
-\kappa K_1(\kappa r)
-\frac{2K_2(\kappa r)}{r}
+\frac{4}{\kappa^2 r^3}
\right)
+
\frac{1}{\eta_s+\eta_d}
\left(
\frac{2K_2(\lambda r)}{r}
-\frac{4}{\lambda^2 r^3}
\right)
\right].
\end{equation}

\begin{equation}
\begin{aligned}
C_r(r)=
\frac{1}{2\pi}\Bigg[
&\frac{1}{\eta_s}
\left(
\alpha^2\kappa K_1(\kappa r)
+
\frac{(4\alpha^2-1)K_2(\kappa r)}{r}
+
\frac{2-8\alpha^2}{\kappa^2 r^3}
\right)
\\[6pt]
&-
\frac{1}{\eta_s+\eta_d}
\left(
\alpha^2\lambda K_1(\lambda r)
+
\frac{(4\alpha^2-1)K_2(\lambda r)}{r}
+
\frac{2-8\alpha^2}{\lambda^2 r^3}
\right)
\Bigg].
\end{aligned}
\end{equation}
Here $\kappa^{-1}=\sqrt{\eta_s/\zeta_\parallel}$ is the shear screening
length, and $\lambda^{-1}=\sqrt{(\eta_s+\eta_d)/\zeta_\parallel}$ is the
compressional screening length.  In the incompressible limit
$\eta_d\to\infty$, the compressional contribution vanishes and the
velocity reduces smoothly to the incompressible dipole flow obtained in
the previous section. The vorticity field follows from Eq.~(\ref{vorgen}). 
For the parity--symmetric case with vanishing odd viscosity
($\eta_o=0$) the antisymmetric Green tensor component vanishes,
$A_2(r)=0$, and the vorticity reduces to

\begin{equation}
\omega(\mathbf r)
=
\sigma\,\alpha\beta
\left(
A_0''(r)-\frac{A_0'(r)}{r}-\frac{A_1'(r)}{r}
+\frac{2A_1(r)}{r^2}
\right).
\end{equation}

Substituting the compressible Green tensor coefficients
$A_0(r)$ and $A_1(r)$ and simplifying yields the remarkably
simple result

\begin{equation}
\omega(\mathbf r)
=
\frac{\sigma}{2\pi\eta_s}\,
\alpha\beta\,\kappa^2 K_2(\kappa r).
\end{equation}

Thus the compressional hydrodynamic mode does not contribute
to the dipolar vorticity; it produces only irrotational
(longitudinal) flow.  The vorticity is therefore identical to
the incompressible supported--membrane result.
\subsection{Incompressible parity--symmetric limit of the Green tensor}

It is useful to examine the limit in which both odd viscosity and
membrane compressibility are removed.  This corresponds to taking
\[
\eta_o \rightarrow 0,
\qquad
\eta_d \rightarrow \infty .
\]
In this regime the antisymmetric stresses vanish and the membrane flow
becomes incompressible.  The resulting Green tensor therefore provides a
useful consistency check of the general solution derived above.

We begin from the definitions of the coefficients appearing in the
general solution,
\[
A=\eta_s(\eta_s+\eta_d)+\eta_o^2,
\qquad
B=\eta_s(\eta_s+\eta_d)(\kappa^2+\lambda^2)+2\eta_o^2\nu^2,
\qquad
C=\eta_s(\eta_s+\eta_d)\kappa^2\lambda^2+\eta_o^2\nu^4 .
\]

Taking the parity--symmetric limit $\eta_o\to0$ gives
\[
A=\eta_s(\eta_s+\eta_d),
\qquad
B=\eta_s(\eta_s+\eta_d)(\kappa^2+\lambda^2),
\qquad
C=\eta_s(\eta_s+\eta_d)\kappa^2\lambda^2 .
\]

The hydrodynamic screening lengths are defined by
\[
\kappa^{-1}=\sqrt{\frac{\eta_s}{\zeta_\parallel}},
\qquad
\lambda^{-1}=\sqrt{\frac{h(\eta_s+\eta_d)}{3\eta+h\zeta_\parallel}} .
\]

In the incompressible limit $\eta_d\rightarrow\infty$ the longitudinal
screening length diverges,
\[
\lambda^{-1}\rightarrow\infty,
\qquad
\lambda\rightarrow0 .
\]

The discriminant entering the pole structure of the Green tensor is
\[
\Delta=B^2-4AC .
\]
Substituting the simplified coefficients gives
\[
\Delta=[\eta_s(\eta_s+\eta_d)]^2\kappa^4 .
\]

Consequently
\[
\sqrt{\Delta}=\eta_s(\eta_s+\eta_d)\kappa^2 .
\]

The screening masses defined by
\[
m_{1,2}^2=\frac{B\mp\sqrt{\Delta}}{2A}
\]
then reduce to
\[
m_1^2=0,
\qquad
m_2^2=\kappa^2 .
\]

Thus the hydrodynamics contains a massless incompressible mode together
with a single screened shear mode characterized by the screening length
$\kappa^{-1}$.

The residues appearing in the antisymmetric sector are proportional to
$\eta_o$ and therefore vanish,
\[
R^{(\gamma)}_1=R^{(\gamma)}_2=0 .
\]
As a result the antisymmetric Green tensor component disappears,
\[
A_2(r)=0 .
\]

The symmetric residues become
\[
R^{(\beta)}_1=0,
\qquad
R^{(\beta)}_2=\frac{1}{\eta_s}.
\]

Substituting these results into the general expressions for the radial
functions yields
\[
A_0(r)
=
\frac{1}{2\pi\eta_s}
\left[
K_0(\kappa r)
+
\frac{K_1(\kappa r)}{\kappa r}
-
\frac{1}{(\kappa r)^2}
\right],
\]

\[
A_1(r)
=
-\frac{1}{2\pi\eta_s}
\left[
K_0(\kappa r)
+
\frac{2K_1(\kappa r)}{\kappa r}
-
\frac{2}{(\kappa r)^2}
\right].
\]

The antisymmetric coefficient vanishes,
\[
A_2(r)=0 .
\]

The Green tensor therefore reduces to the purely symmetric form
\[
G_{ij}(\mathbf r)
=
A_0(r)\delta_{ij}
+
A_1(r)\hat r_i\hat r_j ,
\qquad r=|\mathbf r| .
\]

This expression represents the velocity response of an incompressible
supported membrane with momentum leakage to the surrounding fluid.  The
hydrodynamic flow is screened over the characteristic length scale
$\kappa^{-1}$ and contains only transverse shear modes, reflecting both
the incompressibility of the membrane and the restoration of parity
symmetry when odd viscosity is removed. Substituting the incompressible Green–tensor coefficients into the
general dipolar velocity expression yields the force–dipole flow in the
compact form
\begin{equation}
\mathbf v(\mathbf r)
=
\sigma
\left[
\alpha\,C_d(r)\,\hat{\mathbf d}
+
C_r(r)\,\hat{\mathbf r}
\right],
\end{equation}
where $\alpha=\hat{\mathbf d}\!\cdot\!\hat{\mathbf r}$ and the radial
coefficients are
\begin{equation}
C_d(r)=
\frac{1}{2\pi\eta_s}
\left[
-\kappa K_1(\kappa r)
-\frac{2K_2(\kappa r)}{r}
+\frac{4}{\kappa^2 r^3}
\right],
\end{equation}
\begin{equation}
C_r(r)=
\frac{1}{2\pi\eta_s}
\left[
\alpha^2\kappa K_1(\kappa r)
+\frac{(4\alpha^2-1)K_2(\kappa r)}{r}
+\frac{2-8\alpha^2}{\kappa^2 r^3}
\right].
\end{equation}
The incompressible dipole velocity therefore depends only
on the screened kernels $K_1$ and $K_2$, while the transverse
odd--viscous contribution vanishes identically in this limit.
In polar  it reads
\begin{align}
v_r(\mathbf r)
&=
\frac{\sigma}{2\pi\eta_s}
(2\alpha^2-1)
\left[
\frac{K_2(\kappa r)}{r}
-
\frac{2}{\kappa^2 r^3}
\right], \\
v_\theta(\mathbf r)
&=
\frac{\sigma\,\alpha\beta}{2\pi\eta_s}
\left[
-\kappa K_1(\kappa r)
-\frac{2K_2(\kappa r)}{r}
+\frac{4}{\kappa^2 r^3}
\right],
\end{align}
where \(r=|\mathbf r|\), \(\kappa^{-1}=\sqrt{\eta_s h/\eta}\) is the
hydrodynamic screening length,  Since $A_2=0$, the vorticity reduces to
\begin{equation}
\omega(\mathbf r)
=
\sigma\,\alpha\beta
\left(
A_0''(r)-\frac{A_0'(r)}{r}-\frac{A_1'(r)}{r}+\frac{2A_1(r)}{r^2}
\right),
\end{equation}
which simplifies to the compact form
\begin{equation}
\omega(\mathbf r)
=
\frac{\sigma}{2\pi\eta_s}\,
\alpha\beta\,\kappa^2 K_2(\kappa r).
\end{equation}
Thus, in the incompressible supported-membrane limit, the dipolar
vorticity is controlled entirely by the screened kernel \(K_2(\kappa r)\).

\subsection{Degenerate identical-screening limit}
\label{deg}
Finally, we consider the identical-screening degenerate limit of the
full compressible odd-viscous theory. Starting from the general
solution with screening masses
\begin{equation}
m_{1,2}^2=\frac{B\mp\sqrt{\Delta}}{2A},
\qquad
\Delta=B^2-4AC,\nn
\end{equation}
where
\begin{align}
A &= \eta_s(\eta_s+\eta_d)+\eta_o^2,\nn
\\
B &= \eta_s(\eta_s+\eta_d)(\kappa^2+\lambda^2)+2\eta_o^2\nu^2,\nn
\\
C &= \eta_s(\eta_s+\eta_d)\kappa^2\lambda^2+\eta_o^2\nu^4,\nn
\end{align}
we impose
\begin{equation}
\zeta_\parallel=\frac{\eta}{h},
\qquad
\eta_d=3\eta_s,
\qquad
\zeta_\perp=\frac{\eta_o}{\eta_s}\zeta_\parallel.
\end{equation}
The three screening lengths then coincide,
\begin{equation}
\kappa=\lambda=\nu=m,
\qquad
m=\frac{1}{\ell},
\qquad
\ell=\sqrt{\frac{\eta_s h}{\eta}},
\end{equation}
and, since \(\eta_s+\eta_d=4\eta_s\),
\begin{equation}
A=4\eta_s^2+\eta_o^2,
\qquad
B=2Am^2,
\qquad
C=Am^4.\nn
\end{equation}
Hence
\begin{equation}
\Delta=0,
\qquad
m_1=m_2=m,
\end{equation}
so the two screening poles coalesce into a single mode. In this limit
the Green tensor retains the form
\begin{equation}
G_{ij}(\mathbf r)
=
A_0(r)\,\delta_{ij}
+
A_1(r)\,\hat r_i\hat r_j
+
A_2(r)\,\varepsilon_{ij},\nn
\end{equation}
but the radial functions collapse to
\begin{align}
A_0(r) &=
\frac{\eta_s}{2\pi(4\eta_s^2+\eta_o^2)}
\left[
4K_0(mr)
-\frac{3}{m^2r^2}
+\frac{3}{mr}K_1(mr)
\right],\nn
\\
A_1(r) &=
\frac{\eta_s}{2\pi(4\eta_s^2+\eta_o^2)}
\left[
-3K_0(mr)
-\frac{6}{mr}K_1(mr)
+\frac{6}{m^2r^2}
\right],\nn
\\
A_2(r) &=
-\frac{\eta_o}{2\pi(4\eta_s^2+\eta_o^2)}K_0(mr).
\end{align}
Thus the general two-mode Green tensor reduces to a single-screening
form controlled by the length scale \(\ell\). The odd viscosity enters
through the antisymmetric coefficient \(A_2(r)\) and through the
overall normalization factor \(4\eta_s^2+\eta_o^2\). Substituting these expressions into the general dipole velocity formula
\begin{equation}
\mathbf v(\mathbf r)
=
\sigma
\left[
\alpha\!\left(A_0'(r)+\frac{A_1(r)}{r}\right)\hat{\mathbf d}
+
\left(
\frac{A_1(r)}{r}
+
\alpha^2\!\left(A_1'(r)-\frac{2A_1(r)}{r}\right)
\right)\hat{\mathbf r}
+
\alpha A_2'(r)(\varepsilon\!\cdot\!\hat{\mathbf d})
\right],
\end{equation}
we obtain the compact form
\begin{equation}
\mathbf v(\mathbf r)
=
\sigma
\left[
\alpha\,C_d(r)\,\hat{\mathbf d}
+
C_r(r)\,\hat{\mathbf r}
+
\alpha\,C_\perp(r)(\varepsilon\!\cdot\!\hat{\mathbf d})
\right].
\end{equation}
The radial coefficients are
\begin{align}
C_d(r)
&=
\frac{\eta_s}{\pi(4\eta_s^2+\eta_o^2)}
\left[
-\frac{3K_2(mr)}{r}
-2mK_1(mr)
+\frac{6}{m^2r^3}
\right],\nn
\\[6pt]
C_r(r) 
&=
\frac{3\eta_s}{2\pi(4\eta_s^2+\eta_o^2)}
\left[
\alpha^2 mK_1(mr)
+
\frac{(4\alpha^2-1)K_2(mr)}{r}
+
\frac{2-8\alpha^2}{m^2r^3}
\right],\nn
\\[6pt]
C_\perp(r)
&=
\frac{m\eta_o}{2\pi(4\eta_s^2+\eta_o^2)}K_1(mr).\nn
\end{align}
The transverse coefficient $C_\perp$ arises entirely from the
antisymmetric Green tensor component and therefore vanishes in the
absence of odd viscosity. The vorticity follows from the general expression
\begin{equation}
\omega(\mathbf r)
=
\sigma
\left[
\alpha\beta
\left(
A_0''-\frac{A_0'}{r}-\frac{A_1'}{r}+\frac{2A_1}{r^2}
\right)
-
\left(
\frac{A_2'}{r}\beta^2+\alpha^2A_2''
\right)
\right],\nn
\end{equation}
which yields
\begin{equation}
\omega(\mathbf r)
=
\frac{\sigma}{2\pi(4\eta_s^2+\eta_o^2)}
\left[
4\eta_s\,\alpha\beta\,m^2K_2(mr)
+
\eta_o
\left(
\alpha^2 m^2K_0(mr)
+
\frac{(2\alpha^2-1)m}{r}K_1(mr)
\right)
\right].\nn
\end{equation}
Thus the vorticity naturally separates into even-- and odd--viscosity
contributions,
\begin{equation}
\omega=\omega_{\rm even}+\omega_{\rm odd},\nn
\end{equation}
with
\begin{align}
\omega_{\rm even}(\mathbf r)
&=
\frac{4\sigma\eta_s}{2\pi(4\eta_s^2+\eta_o^2)}
\,\alpha\beta\,m^2K_2(mr),
\\
\omega_{\rm odd}(\mathbf r)\nn
&=
\frac{\sigma\eta_o}{2\pi(4\eta_s^2+\eta_o^2)}
\left[
\alpha^2 m^2K_0(mr)
+
\frac{(2\alpha^2-1)m}{r}K_1(mr)
\right].\nn
\end{align}
In the limit $\eta_o\to0$ the transverse velocity disappears and the
vorticity reduces to the purely even--viscosity dipolar form.
\subsubsection{Near Field}
We now examine the near--field behavior of the degenerate Green tensor.
In the regime
\[
mr\ll1,
\]
corresponding to distances much smaller than the common screening
length \(\ell=1/m\), the modified Bessel functions admit the
expansions
\[
K_0(mr)=-\ln\!\left(\frac{mr}{2}\right)-\gamma+\mathcal O(r^2\ln r),
\qquad
K_1(mr)=\frac{1}{mr}+\mathcal O(r\ln r).
\]

Substituting these expressions into the degenerate coefficients
\(A_0(r)\), \(A_1(r)\), and \(A_2(r)\), one finds that the explicit
\(1/r^2\) singularities appearing in \(A_0\) and \(A_1\) cancel.
The leading near--field behavior therefore reduces to
\begin{align}
A_0(r) &\simeq
-\frac{5\eta_s}{4\pi(4\eta_s^2+\eta_o^2)}\ln(mr)+\mathcal O(1),\\
A_1(r) &\simeq
\frac{3\eta_s}{4\pi(4\eta_s^2+\eta_o^2)}+\mathcal O(r^2\ln r),\\
A_2(r) &\simeq
\frac{\eta_o}{2\pi(4\eta_s^2+\eta_o^2)}\ln(mr)+\mathcal O(1).
\end{align}

Consequently the Green tensor assumes the logarithmic form
characteristic of two--dimensional hydrodynamics,
\begin{equation}
G_{ij}(\mathbf r)\simeq
\frac{1}{8\pi(4\eta_s^2+\eta_o^2)}
\left[
\eta_s\!\left(-3-10\gamma+10\ln\frac{2}{mr}\right)\delta_{ij}
+6\eta_s\,\hat r_i\hat r_j
+4\eta_o\!\left(\gamma-\ln\frac{2}{mr}\right)\varepsilon_{ij}
\right].
\end{equation}

Substituting this result into the general dipole velocity
formula~(\ref{vgen}) yields the leading near--field dipolar flow,
\begin{equation}
\mathbf v(\mathbf r)=
\frac{\sigma}{8\pi(4\eta_s^2+\eta_o^2)}\,\frac{1}{r}
\left[
-4\eta_s\,\alpha\,\mathbf d
+6\eta_s(1-2\alpha^2)\hat{\mathbf r}
+4\eta_o\,\alpha\,\mathbf d_\perp
\right],
\end{equation}
where \(\mathbf d_\perp=\varepsilon\!\cdot\!\mathbf d\).
The first two terms correspond to the usual even--viscous dipolar flow,
while the third term represents a transverse contribution generated
entirely by the odd viscosity.

Using the vorticity expression~(\ref{vorgen}), the corresponding
near--field vorticity becomes
\begin{equation}
\omega(\mathbf r)=
\frac{\sigma}{2\pi(4\eta_s^2+\eta_o^2)}
\frac{\eta_o(\alpha^2-\beta^2)+8\eta_s\alpha\beta}{r^2}.
\end{equation}

Expressed in polar coordinates, where
\(\alpha=\cos(\theta-\phi)\) and
\(\beta=\sin(\theta-\phi)\), the vorticity takes the compact angular
form
\begin{equation}
\omega(r,\theta,\phi)=
\frac{\sigma}{2\pi(4\eta_s^2+\eta_o^2)}
\frac{\eta_o\cos2(\theta-\phi)+4\eta_s\sin2(\theta-\phi)}{r^2}.
\end{equation}

Thus the degenerate near field retains the logarithmic structure of
two--dimensional hydrodynamics.  The dipolar velocity decays as
\(r^{-1}\) while the vorticity exhibits a quadrupolar \(r^{-2}\)
structure.  Odd viscosity generates a transverse component of the
dipolar flow and rotates the vorticity quadrupole, producing a
characteristic chiral distortion of the flow pattern.
\subsubsection{Far field}
We now turn to the opposite asymptotic regime in which the observation
point lies far from the dipole,
\[
mr\gg1 .
\]
In this limit the modified Bessel functions exhibit the large--argument
behavior
\[
K_\nu(z)\simeq
\sqrt{\frac{\pi}{2z}}\,e^{-z}
\left(1+\frac{4\nu^2-1}{8z}+\mathcal O(z^{-2})\right).
\]
Substituting these expansions into the degenerate coefficients
\(A_0(r)\), \(A_1(r)\), and \(A_2(r)\) yields
\begin{align}
A_0(r) &=
-\frac{3\eta_s}{2\pi(4\eta_s^2+\eta_o^2)}
\frac{1}{m^2 r^2}
+
\frac{\sqrt{2\pi}\,\eta_s\!\left(9+4mr(5+8mr)\right)}
{32\pi(4\eta_s^2+\eta_o^2)(mr)^{5/2}}
e^{-mr},
\\[6pt]
A_1(r) &=
\frac{3\eta_s}{\pi(4\eta_s^2+\eta_o^2)}
\frac{1}{m^2 r^2}
-
\frac{3\sqrt{2\pi}\,\eta_s\!\left(6+mr(15+8mr)\right)}
{32\pi(4\eta_s^2+\eta_o^2)(mr)^{5/2}}
e^{-mr},
\\[6pt]
A_2(r) &=
\frac{\eta_o}{16\sqrt{2\pi}(4\eta_s^2+\eta_o^2)}
\frac{(1-8mr)e^{-mr}}{(mr)^{3/2}} .
\end{align}
Thus the even sector of the Green tensor retains a long--range
algebraic tail proportional to \(r^{-2}\), while the odd sector is
exponentially screened by the common screening length \(\ell=1/m\).

Substituting these asymptotic forms into the general velocity
expression~(\ref{vgen}) gives the leading far--field dipolar velocity,
\begin{equation}
\mathbf v(\mathbf r)\simeq
\frac{3\sigma\eta_s}{\pi(4\eta_s^2+\eta_o^2)}
\frac{1}{m^2 r^3}
\left[
2\alpha\,\mathbf d
+(1-4\alpha^2)\hat{\mathbf r}
\right]
+\mathcal O\!\left(e^{-mr}\right).
\end{equation}

The velocity therefore decays algebraically as \(r^{-3}\), reflecting
the long--range dipolar structure of the even--viscosity component of
the Green tensor.

The vorticity may be obtained from the invariant expression
(\ref{vorgen}).  Substituting the algebraic parts of \(A_0\) and
\(A_1\) shows that the leading \(r^{-2}\) contributions cancel
identically in the vorticity operator,
\[
A_0''-\frac{A_0'}{r}-\frac{A_1'}{r}+\frac{2A_1}{r^2}=0 .
\]
Consequently the algebraic contributions cancel in the vorticity
operator,
\[
A_0''-\frac{A_0'}{r}-\frac{A_1'}{r}+\frac{2A_1}{r^2}=0 ,
\]
so that the far--field vorticity is determined entirely by the
exponentially small screened terms originating from the Bessel
functions. Expanding the resulting expression for large $r$ yields
\begin{equation}
\omega(\mathbf r)\simeq
\frac{2\,\sigma\eta_s}{\sqrt{2\pi}(4\eta_s^2+\eta_o^2)}
\,\alpha\beta\,
m^{3/2}\,
\frac{e^{-mr}}{\sqrt r}.
\end{equation}
\section{Dynamical system for interacting force dipoles on the membrane}

The velocity and vorticity fields generated by a single force dipole
derived above allow us to formulate the dynamics of a system of
interacting dipoles embedded in the membrane.

Consider a collection of $N$ force dipoles labeled by
$a=1,\dots,N$. Each dipole is characterized by its position
$R_{a i}(t)$, orientation $d_{a i}(t)$, and dipole strength $\sigma_a$.
The orientation vector is normalized,
\begin{equation}
d_{a i} d_{a i}=1 .
\end{equation}

For two dipoles $a$ and $b$ we define the relative separation
\begin{equation}
r_{ab,i}=R_{a i}-R_{b i},
\qquad
r_{ab}=|{\mathbf r}_{ab}|=\sqrt{r_{ab,i}r_{ab,i}},
\qquad
\hat r_{ab,i}=\frac{r_{ab,i}}{r_{ab}} .
\end{equation}

The rotational invariants that characterize the relative orientation of
dipole $b$ with respect to the separation direction are
\begin{equation}
\alpha_{ab}=d_{b i}\hat r_{ab,i},
\qquad
\beta_{ab}=\varepsilon_{ij}d_{b j}\hat r_{ab,i},
\end{equation}
which satisfy $\alpha_{ab}^2+\beta_{ab}^2=1$. Using the single--dipole solution derived above, the velocity induced by
dipole $b$ at the position of dipole $a$ is
\begin{align}
v_{ab,i}
=
\sigma_b
\Bigg[
&
\alpha_{ab}
\left(A_0'(r_{ab})+\frac{A_1(r_{ab})}{r_{ab}}\right)
d_{b i}
\\
&+
\left(
\frac{A_1(r_{ab})}{r_{ab}}
+
\alpha_{ab}^2
\left(
A_1'(r_{ab})-\frac{2A_1(r_{ab})}{r_{ab}}
\right)
\right)
\hat r_{ab,i}
\\
&+
\alpha_{ab}A_2'(r_{ab})\varepsilon_{ij}d_{b j}
\Bigg].
\end{align}

Each dipole is advected by the velocity generated by all other dipoles,
\begin{equation}
\dot R_{a i}
=
\sum_{b\neq a} v_{ab,i}.
\end{equation}

The vorticity produced by dipole $b$ at the position of dipole $a$
follows from the single--dipole expression,
\begin{align}
\omega_{ab}
=
\sigma_b
\Bigg[
&
\alpha_{ab}\beta_{ab}
\left(
A_0''(r_{ab})
-\frac{A_0'(r_{ab})}{r_{ab}}
-\frac{A_1'(r_{ab})}{r_{ab}}
+\frac{2A_1(r_{ab})}{r_{ab}^2}
\right)
\\
&
-
\left(
\frac{A_2'(r_{ab})}{r_{ab}}\beta_{ab}^2
+
\alpha_{ab}^2A_2''(r_{ab})
\right)
\Bigg].
\end{align}

The total vorticity acting on dipole $a$ is therefore
\begin{equation}
\omega_a=\sum_{b\neq a}\omega_{ab}.
\end{equation}

The orientation of each dipole rotates with the local vorticity of the
flow.  The angular velocity of a material vector equals one half of the
fluid vorticity, leading to
\begin{equation}
\dot d_{a i}
=
\frac{1}{2}\,\omega_a\,\varepsilon_{ij} d_{a j},
\end{equation}
which preserves the normalization $d_{a i}d_{a i}=1$.

The interacting force--dipole dynamics on the membrane is therefore
described by the coupled system
\begin{align}
\dot R_{a i} &= \sum_{b\neq a} v_{ab,i}, \\
\dot d_{a i} &= \frac12\,\omega_a\,\varepsilon_{ij} d_{a j},
\end{align}
where the pairwise velocity $v_{ab,i}$ and vorticity $\omega_{ab}$
are determined entirely by the Green tensor functions
$A_0(r)$, $A_1(r)$ and $A_2(r)$. The dynamical system described above possesses only a limited set of
exact invariants.  The orientation equation
\[
\dot d_{a i}=\tfrac12\,\omega_a\,\varepsilon_{ij}d_{a j}
\]
preserves the normalization of each dipole orientation vector,
since
\[
\frac{d}{dt}(d_{a i}d_{a i})
=
2d_{a i}\dot d_{a i}
=
\omega_a\,d_{a i}\varepsilon_{ij}d_{a j}=0,
\]
and therefore $d_{a i}d_{a i}=1$ for all time.  Apart from this trivial
constraint, the interacting $N$--dipole system does not generally admit
additional conserved quantities.  Although the equations of motion are
invariant under global translations and rotations of the membrane,
these symmetries do not produce conserved linear or angular momenta
because the dynamics is driven by active stresses rather
than inertial forces.  In particular, the center of mass
$R_{\mathrm{cm},i}=\tfrac1N\sum_a R_{a i}$ evolves according to
\[
\dot R_{\mathrm{cm},i}=\frac1N\sum_a\sum_{b\neq a}v_{ab,i},
\]
which does not vanish in general since the pairwise velocities
$v_{ab,i}$ and $v_{ba,i}$ are not antisymmetric for active dipoles.
Consequently the system lacks a conserved energy or momentum and the
many--body dynamics is generically dissipative. Additional invariants may nevertheless arise in special situations,
such as for two identical dipoles or in particular limits of the Green
tensor where the interaction kernel acquires additional symmetries.
In particular, in the far--field regime of the incompressible limit,
where the leading vorticity contribution vanishes, the dynamics reduces
to a purely positional interaction and admits an exact Hamiltonian
description, as recently demonstrated in \cite{krishnan2025}, see also \cite{Shoham2023}.
Moreover, for systems with quenched dipole orientations in the
incompressible limit, the dynamics can admit a Hamiltonian formulation
even in the near field, since the evolution then reduces to a closed
system for the dipole positions alone \cite{krishnan2025}.
\subsection{COM drift and Polarization evolution equations}
We may also derive evolution equations for global observables of the
dipole ensemble.  The center of mass of the system is defined by
\[
R_{\mathrm{cm},i}=\frac{1}{N}\sum_{a=1}^N R_{a i}.
\]
Using the positional dynamics $\dot R_{a i}=\sum_{b\neq a}v_{ab,i}$ we obtain
\[
\dot R_{\mathrm{cm},i}
=
\frac{1}{N}
\sum_a\sum_{b\neq a} v_{ab,i}.
\]
Substituting the pair velocity expression gives
\begin{align}
\dot R_{\mathrm{cm},i}
&=
\frac{1}{N}
\sum_a\sum_{b\neq a}
\sigma_b
\Bigg[
\alpha_{ab}\!\left(A_0'(r_{ab})+\frac{A_1(r_{ab})}{r_{ab}}\right)d_{b i}
+
\left(
\frac{A_1(r_{ab})}{r_{ab}}
+
\alpha_{ab}^2\!\left(A_1'(r_{ab})-\frac{2A_1(r_{ab})}{r_{ab}}\right)
\right)\hat r_{ab,i}\nn
\\
&\qquad
+
\alpha_{ab}A_2'(r_{ab})\varepsilon_{ij}d_{b j}
\Bigg].
\end{align}
Thus the center of mass motion is generated by the collective
hydrodynamic interactions among the dipoles and does not generally
vanish for active stresses.

The total polarization of the system is defined by
\[
P_i=\sum_{a=1}^N d_{a i}.
\]
Using the orientation equation
$\dot d_{a i}=\tfrac12\,\omega_a\,\varepsilon_{ij}d_{a j}$
we obtain
\[
\dot P_i
=
\frac12
\sum_a \omega_a \varepsilon_{ij} d_{a j}
=
\frac12
\sum_a\sum_{b\neq a}
\omega_{ab}\varepsilon_{ij} d_{a j}.
\]
Substituting the pair vorticity expression yields
\[
\dot P_i
=
\frac12
\sum_a\sum_{b\neq a}
\sigma_b
\Bigg[
\alpha_{ab}\beta_{ab}
\left(
A_0''-\frac{A_0'}{r_{ab}}
-\frac{A_1'}{r_{ab}}
+\frac{2A_1}{r_{ab}^2}
\right)
-
\left(
\frac{A_2'}{r_{ab}}\beta_{ab}^2
+
\alpha_{ab}^2A_2''
\right)
\Bigg]
\varepsilon_{ij} d_{a j}.
\]
These equations show that the center of mass evolves through the
collective hydrodynamic velocities while the total polarization rotates
under the vorticity field generated by the dipoles.
\subsection{Pair Geometry}

It is convenient to describe the relative configuration of two dipoles
in terms of the pair separation $r_{ab}$ and the orientation scalars
\[
\alpha_{ab}=d_{b i}\hat r_{ab,i},
\qquad
\beta_{ab}=\varepsilon_{ij}d_{b j}\hat r_{ab,i},
\]
which satisfy $\alpha_{ab}^2+\beta_{ab}^2=1$.

The relative velocity between the dipoles is
\[
\Delta V_{ab,i}:=\dot R_{a i}-\dot R_{b i}
=
v_{ab,i}-v_{ba,i}
+
\sum_{c\neq a,b}(v_{ac,i}-v_{bc,i}),
\]
where the first two terms represent the direct interaction between the
pair and the remaining sum accounts for advection by the other dipoles.

The pair geometry evolves according to
\[
\dot r_{ab}=\hat r_{ab,i}\Delta V_{ab,i},
\]
\[
\dot\alpha_{ab}
=
\frac12\,\omega_b\,\beta_{ab}
+
\frac{1}{r_{ab}}
\left(d_{b i}-\alpha_{ab}\hat r_{ab,i}\right)\Delta V_{ab,i},
\]
\[
\dot\beta_{ab}
=
-\frac12\,\omega_b\,\alpha_{ab}
+
\frac{1}{r_{ab}}
\left(\varepsilon_{ij}d_{b j}-\beta_{ab}\hat r_{ab,i}\right)\Delta V_{ab,i}.
\]

Using the explicit dipole velocity field, the radial evolution becomes
\[
\begin{aligned}
\dot r_{ab}
&=
\sigma_b
\left[
\alpha_{ab}^2\big(A_0'(r_{ab})+A_1'(r_{ab})\big)
+
\beta_{ab}^2\frac{A_1(r_{ab})}{r_{ab}}
+
\alpha_{ab}\beta_{ab}A_2'(r_{ab})
\right]
\\
&\quad
+
\sigma_a
\left[
\alpha_{ba}^2\big(A_0'(r_{ab})+A_1'(r_{ab})\big)
+
\beta_{ba}^2\frac{A_1(r_{ab})}{r_{ab}}
+
\alpha_{ba}\beta_{ba}A_2'(r_{ab})
\right]
\\
&\quad
+
\hat r_{ab,i}\sum_{c\neq a,b}(v_{ac,i}-v_{bc,i}).
\end{aligned}
\]

The orientation scalars evolve according to
\[
\begin{aligned}
\dot\alpha_{ab}
&=
\frac12\,\omega_b\,\beta_{ab}
+
\frac{\sigma_b\alpha_{ab}}{r_{ab}}
\left[
\beta_{ab}^2
\left(A_0'(r_{ab})+\frac{A_1(r_{ab})}{r_{ab}}\right)
-
\alpha_{ab}\beta_{ab}A_2'(r_{ab})
\right]
\\
&\quad
-
\frac{1}{r_{ab}}
(d_{b i}-\alpha_{ab}\hat r_{ab,i})v_{ba,i}
+
\frac{1}{r_{ab}}
(d_{b i}-\alpha_{ab}\hat r_{ab,i})
\sum_{c\neq a,b}(v_{ac,i}-v_{bc,i}),
\end{aligned}
\]

\[
\begin{aligned}
\dot\beta_{ab}
&=
-\frac12\,\omega_b\,\alpha_{ab}
+
\frac{\sigma_b}{r_{ab}}
\left[
-\alpha_{ab}^2\beta_{ab}
\left(A_0'(r_{ab})+\frac{A_1(r_{ab})}{r_{ab}}\right)
+
\alpha_{ab}^3 A_2'(r_{ab})
\right]
\\
&\quad
-
\frac{1}{r_{ab}}
(\varepsilon_{ij}d_{b j}-\beta_{ab}\hat r_{ab,i})v_{ba,i}
+
\frac{1}{r_{ab}}
(\varepsilon_{ij}d_{b j}-\beta_{ab}\hat r_{ab,i})
\sum_{c\neq a,b}(v_{ac,i}-v_{bc,i}).
\end{aligned}
\]

These equations show that the pair geometry evolves through three
distinct mechanisms: direct dipole--dipole interactions, advection by
the background flow generated by the remaining dipoles, and rotation of
the dipole orientations by the local vorticity field.
\section{Specialization to the two--dipole system}

We now specialize the general $N$--dipole dynamics to the case of two
interacting dipoles, labeled by $a=1,2$.  Their positions and
orientations are denoted by $R_{1i},R_{2i}$ and $d_{1i},d_{2i}$, with
fixed dipole strengths $\sigma_1,\sigma_2$.  It is convenient to
introduce the relative separation vector
\[
r_i:=R_{1i}-R_{2i},
\qquad
r=|\mathbf r|,
\qquad
\hat r_i=\frac{r_i}{r}.
\]
The geometry of the pair is characterized by the scalars
\[
\alpha_{12}=d_{2i}\hat r_i,
\qquad
\beta_{12}=\varepsilon_{ij}d_{2j}\hat r_i,
\]
and
\[
\alpha_{21}=d_{1i}\hat r_{21,i}=-\,d_{1i}\hat r_i,
\qquad
\beta_{21}=\varepsilon_{ij}d_{1j}\hat r_{21,i}
=-\,\varepsilon_{ij}d_{1j}\hat r_i.
\]
These satisfy
\[
\alpha_{12}^2+\beta_{12}^2=1,
\qquad
\alpha_{21}^2+\beta_{21}^2=1.
\]

Since there are only two dipoles, the translational equations reduce to
\[
\dot R_{1i}=v_{12,i},
\qquad
\dot R_{2i}=v_{21,i},
\]
where
\[
\begin{aligned}
v_{12,i}
&=
\sigma_2
\Bigg[
\alpha_{12}
\left(A_0'(r)+\frac{A_1(r)}{r}\right)d_{2i}
+
\left(
\frac{A_1(r)}{r}
+
\alpha_{12}^2
\left(
A_1'(r)-\frac{2A_1(r)}{r}
\right)
\right)\hat r_i
\\
&\qquad\qquad
+
\alpha_{12}A_2'(r)\varepsilon_{ij}d_{2j}
\Bigg],
\end{aligned}
\]
and
\[
\begin{aligned}
v_{21,i}
&=
\sigma_1
\Bigg[
\alpha_{21}
\left(A_0'(r)+\frac{A_1(r)}{r}\right)d_{1i}
-
\left(
\frac{A_1(r)}{r}
+
\alpha_{21}^2
\left(
A_1'(r)-\frac{2A_1(r)}{r}
\right)
\right)\hat r_i
\\
&\qquad\qquad
+
\alpha_{21}A_2'(r)\varepsilon_{ij}d_{1j}
\Bigg].
\end{aligned}
\]
The minus sign in the second line follows from
$\hat r_{21,i}=-\hat r_i$.

The vorticities acting on the two dipoles are simply
\[
\omega_1=\omega_{12},
\qquad
\omega_2=\omega_{21},
\]
with
\[
\omega_{12}
=
\sigma_2
\left[
\alpha_{12}\beta_{12}
\left(
A_0''(r)
-\frac{A_0'(r)}{r}
-\frac{A_1'(r)}{r}
+\frac{2A_1(r)}{r^2}
\right)
-
\left(
\frac{A_2'(r)}{r}\beta_{12}^2
+
\alpha_{12}^2A_2''(r)
\right)
\right],
\]
and
\[
\omega_{21}
=
\sigma_1
\left[
\alpha_{21}\beta_{21}
\left(
A_0''(r)
-\frac{A_0'(r)}{r}
-\frac{A_1'(r)}{r}
+\frac{2A_1(r)}{r^2}
\right)
-
\left(
\frac{A_2'(r)}{r}\beta_{21}^2
+
\alpha_{21}^2A_2''(r)
\right)
\right].
\]

The orientation dynamics therefore becomes
\[
\dot d_{1i}=\frac12\,\omega_{12}\,\varepsilon_{ij}d_{1j},
\qquad
\dot d_{2i}=\frac12\,\omega_{21}\,\varepsilon_{ij}d_{2j}.
\]

It is often preferable to work directly with the relative coordinate
$r_i$ and the center of mass
\[
R_{{\rm cm},i}=\frac12(R_{1i}+R_{2i}).
\]
Their evolution is
\[
\dot r_i=\dot R_{1i}-\dot R_{2i}=v_{12,i}-v_{21,i},
\qquad
\dot R_{{\rm cm},i}=\frac12\,(v_{12,i}+v_{21,i}).
\]
Substituting the explicit velocities yields
\[
\begin{aligned}
\dot r_i
&=
\sigma_2
\Bigg[
\alpha_{12}
\left(A_0'(r)+\frac{A_1(r)}{r}\right)d_{2i}
+
\left(
\frac{A_1(r)}{r}
+
\alpha_{12}^2
\left(
A_1'(r)-\frac{2A_1(r)}{r}
\right)
\right)\hat r_i
+
\alpha_{12}A_2'(r)\varepsilon_{ij}d_{2j}
\Bigg]
\\
&\quad
-
\sigma_1
\Bigg[
\alpha_{21}
\left(A_0'(r)+\frac{A_1(r)}{r}\right)d_{1i}
-
\left(
\frac{A_1(r)}{r}
+
\alpha_{21}^2
\left(
A_1'(r)-\frac{2A_1(r)}{r}
\right)
\right)\hat r_i
+
\alpha_{21}A_2'(r)\varepsilon_{ij}d_{1j}
\Bigg],
\end{aligned}
\]
and
\[
\begin{aligned}
\dot R_{{\rm cm},i}
&=
\frac12\,\sigma_2
\Bigg[
\alpha_{12}
\left(A_0'(r)+\frac{A_1(r)}{r}\right)d_{2i}
+
\left(
\frac{A_1(r)}{r}
+
\alpha_{12}^2
\left(
A_1'(r)-\frac{2A_1(r)}{r}
\right)
\right)\hat r_i
+
\alpha_{12}A_2'(r)\varepsilon_{ij}d_{2j}
\Bigg]
\\
&\quad
+
\frac12\,\sigma_1
\Bigg[
\alpha_{21}
\left(A_0'(r)+\frac{A_1(r)}{r}\right)d_{1i}
-
\left(
\frac{A_1(r)}{r}
+
\alpha_{21}^2
\left(
A_1'(r)-\frac{2A_1(r)}{r}
\right)
\right)\hat r_i
+
\alpha_{21}A_2'(r)\varepsilon_{ij}d_{1j}
\Bigg].
\end{aligned}
\]

Because there are no additional dipoles, the pair-geometry subsystem
closes exactly.  Defining
\[
\Delta V_i:=\dot r_i=v_{12,i}-v_{21,i},
\]
we obtain
\[
\dot r=\hat r_i\Delta V_i,
\]
\[
\dot\alpha_{12}
=
\frac12\,\omega_{21}\,\beta_{12}
+
\frac{1}{r}
\left(d_{2i}-\alpha_{12}\hat r_i\right)\Delta V_i,
\]
\[
\dot\beta_{12}
=
-\frac12\,\omega_{21}\,\alpha_{12}
+
\frac{1}{r}
\left(\varepsilon_{ij}d_{2j}-\beta_{12}\hat r_i\right)\Delta V_i.
\]
Similarly,
\[
\dot\alpha_{21}
=
\frac12\,\omega_{12}\,\beta_{21}
-
\frac{1}{r}
\left(d_{1i}+\alpha_{21}\hat r_i\right)\Delta V_i,
\]
\[
\dot\beta_{21}
=
-\frac12\,\omega_{12}\,\alpha_{21}
-
\frac{1}{r}
\left(\varepsilon_{ij}d_{1j}+\beta_{21}\hat r_i\right)\Delta V_i.
\]

The radial evolution simplifies to
\[
\begin{aligned}
\dot r
&=
\sigma_2
\left[
\alpha_{12}^2\big(A_0'(r)+A_1'(r)\big)
+\beta_{12}^2\frac{A_1(r)}{r}
+\alpha_{12}\beta_{12}A_2'(r)
\right]
\\
&\quad
+
\sigma_1
\left[
\alpha_{21}^2\big(A_0'(r)+A_1'(r)\big)
+\beta_{21}^2\frac{A_1(r)}{r}
+\alpha_{21}\beta_{21}A_2'(r)
\right].
\end{aligned}
\]

The alignment scalars satisfy
\[
\begin{aligned}
\dot\alpha_{12}
&=
\frac12\,\omega_{21}\,\beta_{12}
+
\frac{\sigma_2\alpha_{12}}{r}
\left[
\beta_{12}^2\left(A_0'(r)+\frac{A_1(r)}{r}\right)
-\alpha_{12}\beta_{12}A_2'(r)
\right]
\\
&\quad
-
\frac{1}{r}
\left(d_{2i}-\alpha_{12}\hat r_i\right)v_{21,i},
\end{aligned}
\]
and
\[
\begin{aligned}
\dot\beta_{12}
&=
-\frac12\,\omega_{21}\,\alpha_{12}
+
\frac{\sigma_2}{r}
\left[
-\alpha_{12}^2\beta_{12}
\left(A_0'(r)+\frac{A_1(r)}{r}\right)
+\alpha_{12}^3A_2'(r)
\right]
\\
&\quad
-
\frac{1}{r}
\left(\varepsilon_{ij}d_{2j}-\beta_{12}\hat r_i\right)v_{21,i}.
\end{aligned}
\]
The corresponding equations for $(\alpha_{21},\beta_{21})$ follow by
interchanging $1\leftrightarrow2$.

Thus, in the two--dipole case, the many--body problem reduces to a
closed nonlinear system for the center of mass, relative separation, and
the two dipole orientations.  In particular, the relative dynamics is
fully determined by the scalar Green-tensor kernels
$A_0(r)$, $A_1(r)$, and $A_2(r)$. Finally the polarization evolution becomes
\begin{equation}
\dot P_i
=
\frac12
\Bigg[
\sigma_2
\left(
\alpha_{12}\beta_{12}Q(r)
-
\frac{A_2'(r)}{r}\beta_{12}^2
-
\alpha_{12}^2A_2''(r)
\right)
\varepsilon_{ij} d_{1j}
+
\sigma_1
\left(
\alpha_{21}\beta_{21}Q(r)
-
\frac{A_2'(r)}{r}\beta_{21}^2
-
\alpha_{21}^2A_2''(r)
\right)
\varepsilon_{ij} d_{2j}
\Bigg],
\end{equation}
where
\[
Q(r)=
A_0''(r)-\frac{A_0'(r)}{r}-\frac{A_1'(r)}{r}+\frac{2A_1(r)}{r^2}.
\]
Equivalently, in terms of the pair vorticities $\omega_{12}$ and
$\omega_{21}$,
\begin{equation}
\dot P_i
=
\frac12
\left(
\omega_{12}\,\varepsilon_{ij} d_{1j}
+
\omega_{21}\,\varepsilon_{ij} d_{2j}
\right).
\end{equation}
\section{Two--dipole dynamics in the near--field degenerate limit}

We now specialize the interacting dipole dynamics to the near--field
regime of the degenerate Green tensor.  In this limit the screening
lengths coincide and the real--space Green tensor reduces to a
logarithmic form.  The leading behavior of the radial kernels for
$r\ll m^{-1}$ is
\begin{align}
A_0(r) &\simeq
-\frac{5\eta_s}{4\pi(4\eta_s^2+\eta_o^2)}\ln(mr),
\\
A_1(r) &\simeq
\frac{3\eta_s}{4\pi(4\eta_s^2+\eta_o^2)},
\\
A_2(r) &\simeq
\frac{\eta_o}{2\pi(4\eta_s^2+\eta_o^2)}\ln(mr).
\end{align}
For later convenience we define the denominator
\[
D = 4\eta_s^2+\eta_o^2 .
\]

The derivatives entering the velocity and vorticity fields therefore
reduce to
\begin{align}
A_0'(r) &= -\frac{5\eta_s}{4\pi D}\frac{1}{r},
&
A_1'(r) &= 0,
&
A_2'(r) &= \frac{\eta_o}{2\pi D}\frac{1}{r}, \nn
\\
A_2''(r) &= -\frac{\eta_o}{2\pi D}\frac{1}{r^2},
&
Q(r) &= A_0''-\frac{A_0'}{r}-\frac{A_1'}{r}+\frac{2A_1}{r^2}
     = \frac{4\eta_s}{\pi D}\frac{1}{r^2}.
\end{align}
Considering two dipoles with positions $R_1,R_2$, orientations
$d_1,d_2$, and strengths $\sigma_1,\sigma_2$.
We define the separation vector
\[
\mathbf r = R_1-R_2,
\qquad
r = |\mathbf r|,
\qquad
\hat{\mathbf r} = \mathbf r/r .
\]
The pair geometry is characterized by the scalars
\[
\alpha_{12}=d_2\cdot\hat{\mathbf r},
\qquad
\beta_{12}=(\varepsilon\cdot d_2)\cdot\hat{\mathbf r},
\]
\[
\alpha_{21}=-d_1\cdot\hat{\mathbf r},
\qquad
\beta_{21}=-(\varepsilon\cdot d_1)\cdot\hat{\mathbf r},
\]
together with the dipole--dipole alignment scalars
\[
\gamma = d_1\cdot d_2,
\qquad
\delta = (\varepsilon\cdot d_1)\cdot d_2 .
\]

In the near--field limit the pair velocities take the form
\begin{align}
v_{12}
&=
\frac{\sigma_2}{r}
\left[
-\frac{\eta_s}{2\pi D}\alpha_{12}d_2
+
\frac{3\eta_s}{4\pi D}(1-2\alpha_{12}^2)\hat{\mathbf r}
+
\frac{\eta_o}{2\pi D}\alpha_{12}(\varepsilon\cdot d_2)
\right],\nn
\\
v_{21}
&=
\frac{\sigma_1}{r}
\left[
-\frac{\eta_s}{2\pi D}\alpha_{21}d_1
-
\frac{3\eta_s}{4\pi D}(1-2\alpha_{21}^2)\hat{\mathbf r}
+
\frac{\eta_o}{2\pi D}\alpha_{21}(\varepsilon\cdot d_1)\nn
\right].
\end{align}
The relative separation evolves according to
\[
\dot r = \hat{\mathbf r}\cdot(v_{12}-v_{21}),
\]
which yields
\begin{equation}
\dot r
=
\frac{1}{4\pi D\,r}
\Big[
-8\eta_s(\sigma_2\alpha_{12}^2+\sigma_1\alpha_{21}^2)
+
2\eta_o(\sigma_2\alpha_{12}\beta_{12}+\sigma_1\alpha_{21}\beta_{21})
+
3\eta_s(\sigma_1+\sigma_2)
\Big].
\end{equation}
The vorticities generated by the dipoles reduce to
\begin{align}
\omega_{12}
&=
\frac{\sigma_2}{2\pi D\,r^2}
\left(
\eta_o(\alpha_{12}^2-\beta_{12}^2)
+
8\eta_s\alpha_{12}\beta_{12}
\right),\nn
\\
\omega_{21}
&=
\frac{\sigma_1}{2\pi D\,r^2}
\left(
\eta_o(\alpha_{21}^2-\beta_{21}^2)
+
8\eta_s\alpha_{21}\beta_{21}
\right).
\end{align}
The orientation vectors rotate according to
\[
\dot d_1 = \tfrac12 \omega_{12}(\varepsilon\cdot d_1),
\qquad
\dot d_2 = \tfrac12 \omega_{21}(\varepsilon\cdot d_2).
\]
It is often convenient to express the orientation dynamics in terms of
the alignment scalars.  Using
\[
\dot\alpha_{12}
=
\frac12\omega_{21}\beta_{12}
+
\frac{1}{r}(d_2-\alpha_{12}\hat r)\cdot(v_{12}-v_{21}),
\]
\[
\dot\beta_{12}
=
-\frac12\omega_{21}\alpha_{12}
+
\frac{1}{r}(\varepsilon\cdot d_2-\beta_{12}\hat r)\cdot(v_{12}-v_{21}),
\]
and evaluating the projections explicitly in the near--field limit,
one finds
\begin{align}
(d_2-\alpha_{12}\hat r)\cdot v_{21}
&=
\frac{\sigma_1\alpha_{21}}{r}
\left[
-\frac{\eta_s}{2\pi D}(\gamma+\alpha_{12}\alpha_{21})
+
\frac{\eta_o}{2\pi D}(-\delta+\alpha_{12}\beta_{21})
\right],\nn
\\
(\varepsilon\cdot d_2-\beta_{12}\hat r)\cdot v_{21}
&=
\frac{\sigma_1\alpha_{21}}{r}
\left[
-\frac{\eta_s}{2\pi D}(\delta+\beta_{12}\alpha_{21})
+
\frac{\eta_o}{2\pi D}(\gamma+\beta_{12}\beta_{21})
\right].\nn
\end{align}

The structure of these equations reveals two distinct physical
contributions.  Terms proportional to $\eta_s$ originate from the
symmetric hydrodynamic response of the membrane and depend only on
alignment combinations such as $\gamma$ and
$\alpha_{ab}\alpha_{ba}$.  In contrast, the odd viscosity $\eta_o$
enters exclusively through pseudoscalar combinations involving the
Levi--Civita tensor, producing chiral couplings proportional to
$\delta$ and $\alpha_{ab}\beta_{ba}$.  These terms break mirror
symmetry of the pair dynamics and generate intrinsically chiral
relative motion of the dipoles. Substituting these expressions into the pair vorticities yields
\begin{align}
\omega_{12}
&=
\frac{\sigma_2}{2\pi D r^2}
\left[
\eta_o(\alpha_{12}^2-\beta_{12}^2)
+
8\eta_s\alpha_{12}\beta_{12}
\right],\nn
\\
\omega_{21}
&=
\frac{\sigma_1}{2\pi D r^2}
\left[
\eta_o(\alpha_{21}^2-\beta_{21}^2)
+
8\eta_s\alpha_{21}\beta_{21}
\right].\nn
\end{align}
The polarization evolution therefore reduces to
\begin{equation}
\dot P_i
=
\frac{1}{4\pi D r^2}
\Big[
\sigma_2
\left(
\eta_o(\alpha_{12}^2-\beta_{12}^2)
+
8\eta_s\alpha_{12}\beta_{12}
\right)
\varepsilon_{ij} d_{1j}
+
\sigma_1
\left(
\eta_o(\alpha_{21}^2-\beta_{21}^2)
+
8\eta_s\alpha_{21}\beta_{21}
\right)
\varepsilon_{ij} d_{2j}
\Big].
\end{equation}
To analyze the structure of the two--dipole dynamics it is convenient to
express the system in terms of variables defined relative to the
separation vector between the dipoles.
Let
\[
\mathbf r = r\,\hat{\mathbf r},
\qquad
\hat{\mathbf r}=(\cos\theta,\sin\theta),
\]
denote the relative separation between dipoles $1$ and $2$, where
$r=|\mathbf r|$ and $\theta$ is the polar angle of the separation
vector.  The dipole orientations are described by unit vectors
\[
\mathbf d_a=(\cos\phi_a,\sin\phi_a),
\qquad a=1,2.
\]
It is convenient to introduce orientation angles measured relative to
the line of centers,
\[
\psi_1=\phi_1-\theta,
\qquad
\psi_2=\phi_2-\theta .
\]
In terms of these variables the geometric scalars that enter the
pair interaction simplify to
\[
\alpha_{12}=\mathbf d_2\cdot\hat{\mathbf r}=\cos\psi_2,
\qquad
\beta_{12}=(\varepsilon\!\cdot\!\mathbf d_2)\cdot\hat{\mathbf r}=\sin\psi_2,
\]
\[
\alpha_{21}=-\mathbf d_1\cdot\hat{\mathbf r}=-\cos\psi_1,
\qquad
\beta_{21}=-(\varepsilon\!\cdot\!\mathbf d_1)\cdot\hat{\mathbf r}
=-\sin\psi_1 .
\]
Here $\varepsilon_{ij}$ denotes the two--dimensional Levi--Civita tensor.
The pair dynamics can then be written entirely in terms of the three
variables $(r,\psi_1,\psi_2)$, while the angle $\theta$ follows from the
relative motion. In the near--field degenerate limit the evolution equations become (obtained using Mathematica)
\begin{align}
\dot r &=
\frac{
-\eta_s(\sigma_1+\sigma_2)
-4\eta_s\!\left(\sigma_1\cos2\psi_1+\sigma_2\cos2\psi_2\right)
+\eta_o\!\left(\sigma_1\sin2\psi_1+\sigma_2\sin2\psi_2\right)
}{
4\pi r\,(\eta_o^2+4\eta_s^2)
},\nn
\\[6pt]
\dot\theta &=
-\frac{
\eta_o(\sigma_1+\sigma_2)
+\eta_o\!\left(\sigma_1\cos2\psi_1+\sigma_2\cos2\psi_2\right)
+\eta_s\!\left(\sigma_1\sin2\psi_1+\sigma_2\sin2\psi_2\right)
}{
4\pi r^2(\eta_o^2+4\eta_s^2)
},\nn
\\[6pt]
\dot\psi_1 &=
\frac{
\eta_o(\sigma_1+\sigma_2)
+\eta_o\sigma_1\cos2\psi_1
+2\eta_o\sigma_2\cos2\psi_2
+\eta_s\sigma_1\sin2\psi_1
+5\eta_s\sigma_2\sin2\psi_2
}{
4\pi r^2(\eta_o^2+4\eta_s^2)
},\nn
\\[6pt]
\dot\psi_2 &=
\frac{
\eta_o(\sigma_1+\sigma_2)
+2\eta_o\sigma_1\cos2\psi_1
+\eta_o\sigma_2\cos2\psi_2
+5\eta_s\sigma_1\sin2\psi_1
+\eta_s\sigma_2\sin2\psi_2
}{
4\pi r^2(\eta_o^2+4\eta_s^2)
}.
\end{align}

The two--dipole dynamics therefore reduces to a three--dimensional
nonlinear system for $(r,\psi_1,\psi_2)$, while $\theta$ evolves
kinematically from the relative velocity.
A notable feature of this system is the separation of time scales:
the radial evolution scales as $\dot r\sim r^{-1}$, whereas the angular
variables evolve as $\dot\theta,\dot\psi_a\sim r^{-2}$.
Consequently the dipole orientations evolve faster than the separation
at small distances, which naturally produces chiral spiralling
trajectories in the near--field regime.
To further expose the symmetry structure of the two--dipole dynamics it
is useful to introduce the sum and difference orientation variables
\[
\Sigma=\psi_1+\psi_2,
\qquad
\Delta=\psi_1-\psi_2 .
\]
These variables describe, respectively, the mean orientation of the
dipole pair relative to the separation vector and the relative
orientation of the two dipoles.  In terms of $(\Sigma,\Delta)$ the
individual angles become
\[
\psi_1=\frac{\Sigma+\Delta}{2},
\qquad
\psi_2=\frac{\Sigma-\Delta}{2}.
\]
Using the trigonometric identities
\[
\cos 2\psi_1=\cos(\Sigma+\Delta),
\qquad
\cos 2\psi_2=\cos(\Sigma-\Delta),
\]
\[
\sin 2\psi_1=\sin(\Sigma+\Delta),
\qquad
\sin 2\psi_2=\sin(\Sigma-\Delta),
\]
the near--field degenerate equations can be rewritten entirely in terms
of $(r,\Sigma,\Delta)$. The radial evolution becomes (obtained using Mathematica)
\begin{align}
\dot r &=
\frac{
-\eta_o\sigma_1\sin(\Delta+\Sigma)
+\eta_o\sigma_2\sin(\Delta-\Sigma)
+4\eta_s\sigma_1\cos(\Delta+\Sigma)
+4\eta_s\sigma_2\cos(\Delta-\Sigma)
+\eta_s(\sigma_1+\sigma_2)
}{
4\pi r(\eta_o^2+4\eta_s^2)
}.
\end{align}

The polar angular velocity of the separation vector is
\begin{align}
\dot\theta &=
-\frac{
\eta_o\sigma_1\cos(\Delta+\Sigma)
+\eta_o\sigma_2\cos(\Delta-\Sigma)
+\eta_s\sigma_1\sin(\Delta+\Sigma)
-\eta_s\sigma_2\sin(\Delta-\Sigma)
+\eta_o(\sigma_1+\sigma_2)
}{
4\pi r^2(\eta_o^2+4\eta_s^2)
}.
\end{align}

The evolution of the orientation variables follows from
\[
\dot\Sigma=\dot\psi_1+\dot\psi_2,
\qquad
\dot\Delta=\dot\psi_1-\dot\psi_2 ,
\]
which yields 
\begin{align}
\dot\Sigma &=
\frac{
3\sigma_1\!\left(
\eta_o\cos(\Delta+\Sigma)
+2\eta_s\sin(\Delta+\Sigma)
\right)
+
3\sigma_2\!\left(
\eta_o\cos(\Delta-\Sigma)
-2\eta_s\sin(\Delta-\Sigma)
\right)
+
2\eta_o(\sigma_1+\sigma_2)
}{
4\pi r^2(\eta_o^2+4\eta_s^2)
},
\\[6pt]
\dot\Delta &=
-\frac{
\eta_o\sigma_1\cos(\Delta+\Sigma)
-\eta_o\sigma_2\cos(\Delta-\Sigma)
+4\eta_s\sigma_1\sin(\Delta+\Sigma)
+4\eta_s\sigma_2\sin(\Delta-\Sigma)
}{
4\pi r^2(\eta_o^2+4\eta_s^2)
}.
\end{align}
The pair dynamics therefore reduces to a closed nonlinear system for the
three variables $(r,\Sigma,\Delta)$, while the polar angle $\theta$
evolves kinematically from the instantaneous velocity.  The system
exhibits a clear separation of time scales,
\[
\dot r\sim r^{-1},
\qquad
\dot\Sigma,\dot\Delta,\dot\theta\sim r^{-2},
\]
so that the orientation variables evolve more rapidly than the radial
separation at small distances. A useful way to analyze the geometry of the relative trajectories is to
eliminate the time variable and derive an equation for the separation as
a function of the polar angle.  Using
\[
\frac{dr}{d\theta}=\frac{\dot r}{\dot\theta},
\]
one obtains  (using Mathematica)
\begin{equation}
\frac{dr}{d\theta}
=
r\,
\frac{
-\eta_o\sigma_1\sin(\Delta+\Sigma)
+\eta_o\sigma_2\sin(\Delta-\Sigma)
+4\eta_s\sigma_1\cos(\Delta+\Sigma)
+4\eta_s\sigma_2\cos(\Delta-\Sigma)
+\eta_s(\sigma_1+\sigma_2)
}{
\eta_o\sigma_1\cos(\Delta+\Sigma)
+\eta_o\sigma_2\cos(\Delta-\Sigma)
+\eta_s\sigma_1\sin(\Delta+\Sigma)
-\eta_s\sigma_2\sin(\Delta-\Sigma)
+\eta_o(\sigma_1+\sigma_2)
}.
\end{equation}
It is convenient to write this relation in logarithmic form,
\begin{equation}
\frac{d\ln r}{d\theta}=\kappa(\Sigma,\Delta),
\end{equation}
where the spiral function
\begin{equation}
\kappa(\Sigma,\Delta)=
\frac{
-\eta_o\sigma_1\sin(\Delta+\Sigma)
+\eta_o\sigma_2\sin(\Delta-\Sigma)
+4\eta_s\sigma_1\cos(\Delta+\Sigma)
+4\eta_s\sigma_2\cos(\Delta-\Sigma)
+\eta_s(\sigma_1+\sigma_2)
}{
\eta_o\sigma_1\cos(\Delta+\Sigma)
+\eta_o\sigma_2\cos(\Delta-\Sigma)
+\eta_s\sigma_1\sin(\Delta+\Sigma)
-\eta_s\sigma_2\sin(\Delta-\Sigma)
+\eta_o(\sigma_1+\sigma_2)
}\nn
\end{equation}
controls the radial evolution of the trajectory.

When $\kappa$ varies slowly along the trajectory the solution locally
approximates a logarithmic spiral,
\[
r(\theta)\approx r_0 e^{\kappa\theta}.
\]
The sign of $\kappa$ determines the character of the motion.  Negative
values of $\kappa$ correspond to spiralling collision trajectories,
where the dipoles approach one another while rotating, whereas positive
values correspond to spiralling escape trajectories.  The odd viscosity
coefficient $\eta_o$ appears in both the numerator and denominator of
$\kappa$ and therefore controls the handedness of the spiral: reversing
the sign of $\eta_o$ reverses the direction of rotation while leaving
the radial dynamics unchanged.

These results show that the near--field degenerate dipole interaction
naturally generates chiral spiral trajectories whose orientation is set
by the sign of the odd viscosity and whose radial behavior is governed
by the instantaneous values of the orientation variables $(\Sigma,\Delta)$.
\paragraph{Equal--strength dipoles and spiral dynamics.}

Additional analytical insight can be obtained in the symmetric case
where the dipole strengths are equal,
\[
\sigma_1=\sigma_2=\sigma .
\]
In this case the dynamical system simplifies considerably.  The radial
evolution becomes
\begin{align}
\dot r &=
\frac{
\eta_o \sigma \cos\Delta\,\sin\Sigma
-4\eta_s \sigma \cos\Delta\,\cos\Sigma
-\eta_s\sigma
}{
2\pi r(\eta_o^2+4\eta_s^2)
}.
\end{align}

The angular velocity of the separation vector is
\begin{align}
\dot\theta &=
-\frac{
\sigma\left[
\cos\Delta\left(\eta_o\cos\Sigma+\eta_s\sin\Sigma\right)
+\eta_o
\right]
}{
2\pi r^2(\eta_o^2+4\eta_s^2)
}.
\end{align}

The orientation variables satisfy
\begin{align}
\dot\Sigma &=
\frac{
\sigma\left[
3\cos\Delta\left(\eta_o\cos\Sigma+2\eta_s\sin\Sigma\right)
+2\eta_o
\right]
}{
2\pi r^2(\eta_o^2+4\eta_s^2)
},
\\[6pt]
\dot\Delta &=
\frac{
\sigma\sin\Delta\left(
\eta_o\sin\Sigma-4\eta_s\cos\Sigma
\right)
}{
2\pi r^2(\eta_o^2+4\eta_s^2)
}.
\end{align}

As in the general case, the separation of time scales persists,
\[
\dot r\sim r^{-1},
\qquad
\dot\theta,\dot\Sigma,\dot\Delta\sim r^{-2},
\]
so that the dipole orientations evolve more rapidly than the radial
distance at small separations.

The geometry of the relative trajectories may again be obtained by
eliminating the time variable,
\[
\frac{dr}{d\theta}=\frac{\dot r}{\dot\theta}.
\]
This yields
\begin{equation}
\frac{dr}{d\theta}
=
r\,
\frac{
-\eta_o\cos\Delta\,\sin\Sigma
+4\eta_s\cos\Delta\,\cos\Sigma
+\eta_s
}{
\eta_o\cos\Delta\,\cos\Sigma
+\eta_s\cos\Delta\,\sin\Sigma
+\eta_o
}.
\end{equation}
Equivalently,
\begin{equation}
\frac{d\ln r}{d\theta}
=
\kappa(\Sigma,\Delta),\nn
\end{equation}
where the spiral function is
\begin{equation}
\kappa(\Sigma,\Delta)
=
\frac{
-\eta_o\cos\Delta\,\sin\Sigma
+4\eta_s\cos\Delta\,\cos\Sigma
+\eta_s
}{
\eta_o\cos\Delta\,\cos\Sigma
+\eta_s\cos\Delta\,\sin\Sigma
+\eta_o
}.\nn
\end{equation}

Although the relation
\[
\frac{d\ln r}{d\theta}=\kappa(\Sigma,\Delta)
\]
has the form of a spiral equation, the near--field trajectories are not
generically logarithmic spirals because $\Sigma$ and $\Delta$ typically
evolve by $O(1)$ over the same angular timescale as $\theta$.  A
logarithmic--spiral approximation is valid only when
$\kappa(\Sigma,\Delta)$ is approximately constant along a given
trajectory.
\[
r(\theta)\approx r_0 e^{\kappa\theta}.
\]
The sign of $\kappa$ determines whether the dipoles spiral toward one
another or separate, while the odd viscosity $\eta_o$ controls the
handedness of the spiral through the pseudoscalar combinations
appearing in both the numerator and denominator.
\section{Two--dipole dynamics in the far--field degenerate limit}

We now specialize the general two--dipole dynamical system derived in the
previous section to the far--field limit of the degenerate Green tensor.
In this regime the dipole separation is large compared with the
hydrodynamic screening length,
\[
mr \gg 1 .
\]

The leading algebraic parts of the degenerate Green tensor are
\begin{equation}
A_0(r)=-\frac{C}{r^2},
\qquad
A_1(r)=\frac{2C}{r^2},
\end{equation}
with
\begin{equation}
C=\frac{3\eta_s}{2\pi(4\eta_s^2+\eta_o^2)m^2}.
\end{equation}
The antisymmetric component is exponentially screened,
\begin{equation}
A_2(r)=
D\,\frac{(1-8mr)e^{-mr}}{(mr)^{3/2}},
\qquad
D=\frac{\eta_o}{16\sqrt{2\pi}(4\eta_s^2+\eta_o^2)} .
\end{equation}

The derivatives entering the dynamical equations are
\begin{align}
A_0'(r)&=\frac{2C}{r^3},
&
A_1'(r)&=-\frac{4C}{r^3},\nn
\\[4pt]
A_0'(r)+\frac{A_1(r)}{r}&=\frac{4C}{r^3},
&
A_1'(r)-\frac{2A_1(r)}{r}&=-\frac{8C}{r^3}.\nn
\end{align}
Substituting these expressions into the general dipole velocity law
yields the far--field pair velocity
\begin{equation}
v_{ab,i}
=
\sigma_b
\left[
\frac{4C\,\alpha_{ab}}{r^3} d_{bi}
+
\frac{2C}{r^3}(1-4\alpha_{ab}^2)\hat r_i
+
\alpha_{ab}A_2'(r)\varepsilon_{ij}d_{bj}
\right].\nn
\end{equation}
The relative velocity $\dot r_i=v_{12,i}-v_{21,i}$ becomes
\begin{align}
\dot r_i
&=
\frac{4C}{r^3}
\left(
\sigma_2\alpha_{12}d_{2i}
-
\sigma_1\alpha_{21}d_{1i}
\right)
\nonumber\\
&\quad
+
\frac{2C}{r^3}
\Big[
\sigma_2(1-4\alpha_{12}^2)
+
\sigma_1(1-4\alpha_{21}^2)
\Big]\hat r_i
\nonumber\\
&\quad
+
\sigma_2\alpha_{12}A_2'(r)\varepsilon_{ij}d_{2j}
-
\sigma_1\alpha_{21}A_2'(r)\varepsilon_{ij}d_{1j}.
\end{align}
Projecting along the separation direction $\hat{\mathbf r}$ gives the
radial evolution
\begin{equation}
\dot r
=
\frac{2C}{r^3}
\Big[
(-\alpha_{21}^2+\beta_{21}^2)\sigma_1
+
(-\alpha_{12}^2+\beta_{12}^2)\sigma_2
\Big]
+
A_2'(r)
(\alpha_{21}\beta_{21}\sigma_1+\alpha_{12}\beta_{12}\sigma_2).
\end{equation}
It is therefore natural to decompose the radial dynamics into parity--even
and parity--odd sectors,
$
\dot r = \dot r_{\rm even} + \dot r_{\rm odd}.\nn
$
The symmetric contribution reads
\begin{equation}
\dot r_{\rm even}
=
\frac{2C}{r^3}
\Big[
(-\alpha_{21}^2+\beta_{21}^2)\sigma_1
+
(-\alpha_{12}^2+\beta_{12}^2)\sigma_2
\Big],
\end{equation}
while the odd--viscosity contribution is
\begin{equation}
\dot r_{\rm odd}
=
\frac{D e^{-mr} m(-3+2mr(3+8mr))}
{2(mr)^{5/2}}
(\alpha_{21}\beta_{21}\sigma_1+\alpha_{12}\beta_{12}\sigma_2).
\end{equation}

The vorticity acting on each dipole is determined by
\[
\omega_{ab}
=
\sigma_b
\left[
\alpha_{ab}\beta_{ab}Q(r)
-
\left(
\frac{A_2'(r)}{r}\beta_{ab}^2
+
\alpha_{ab}^2A_2''(r)
\right)
\right],
\]
where
\[
Q(r)=
A_0''-\frac{A_0'}{r}-\frac{A_1'}{r}+\frac{2A_1}{r^2}.
\]

Using only the leading algebraic far--field pieces
\[
A_0(r)=-\frac{C}{r^2}, \qquad A_1(r)=\frac{2C}{r^2},
\]
the vorticity combination reduces to
\begin{equation}
Q(r)=A_0''-\frac{A_0'}{r}-\frac{A_1'}{r}+\frac{2A_1}{r^2}=0 .
\end{equation}
Thus the symmetric sector does not contribute to the dipole rotation
at algebraic order in the far field. However, the full far--field expressions for $A_0$ and $A_1$
contain exponentially small corrections proportional to $e^{-mr}$.
Retaining these terms yields a nonvanishing contribution to $Q(r)$.
Using the complete expressions for $A_0$ and $A_1$ one finds
\begin{equation}
Q(r)=
\frac{\eta_s e^{-mr}}
{64\sqrt{2\pi}\,(4\eta_s^2+\eta_o^2)}
\,
\frac{
81+2mr\!\left(-33+8mr\!\left(6+mr(15+8mr)\right)\right)
}{r^2 (mr)^{5/2}} .
\end{equation}
For large separations $mr\gg1$ the leading behavior becomes
\begin{equation}
Q(r)\sim
\frac{\sqrt{2}}{\sqrt{\pi}}
\frac{\eta_s m^{3/2}}{4\eta_s^2+\eta_o^2}
\frac{e^{-mr}}{\sqrt r}.
\end{equation}
This term is exponentially suppressed but nonzero, and it has the
same asymptotic scaling as the odd--viscosity contribution arising
from $A_2''(r)$,
\[
A_2''(r)\sim
-8D m^{3/2}\frac{e^{-mr}}{\sqrt r}.
\]
Consequently the far--field rotational dynamics receives
comparable exponentially small contributions from both the
symmetric and antisymmetric sectors of the Green tensor, even
though the symmetric contribution vanishes at leading algebraic
order.
The pair vorticities follow from the general relation
\[
\omega_{ab}
=
\sigma_b
\left[
\alpha_{ab}\beta_{ab}Q(r)
-
\left(
\frac{A_2'(r)}{r}\beta_{ab}^2
+
\alpha_{ab}^2A_2''(r)
\right)
\right],
\]
where
\[
Q(r)=
A_0''-\frac{A_0'}{r}-\frac{A_1'}{r}+\frac{2A_1}{r^2}.
\]
Using the full far--field expressions of the degenerate Green tensor,
Mathematica yields the explicit result
\begin{equation}
Q(r)=
\frac{\eta_s e^{-mr}
\left(81+2mr(-33+8mr(6+mr(15+8mr)))\right)}
{64\sqrt{2\pi}\,r^2(mr)^{5/2}(\eta_o^2+4\eta_s^2)} .
\end{equation}
The derivatives of the antisymmetric kernel are
\begin{align}
A_2'(r)
&=
\frac{D e^{-mr} m(-3+2mr(3+8mr))}
{2(mr)^{5/2}},
\\[6pt]
A_2''(r)
&=
\frac{D e^{-mr}(15-4mr(3+mr(7+8mr)))}
{4r^2(mr)^{3/2}} .
\end{align}
Substituting these expressions gives the full pair vorticity
\begin{align}
\omega_{ab}
&=
\frac{\sigma_b e^{-mr}}
{64\sqrt{2\pi}r^2(mr)^{5/2}(\eta_o^2+4\eta_s^2)}
\nonumber\\
&\quad\times
\Big[
\eta_o
\left(
mr(-15+4mr(3+mr(7+8mr)))\alpha_{ab}^2
-
2(-3+2mr(3+8mr))\beta_{ab}^2
\right)
\nonumber\\
&\qquad
+
\eta_s
\left(
81+2mr(-33+8mr(6+mr(15+8mr)))
\right)
\alpha_{ab}\beta_{ab}
\Big].
\end{align}
For large separations $mr\gg1$ the dominant asymptotic behavior becomes
\begin{equation}
\omega_{ab}
=
\frac{m^{3/2}e^{-mr}}
{2\sqrt{2\pi}(\eta_o^2+4\eta_s^2)}
\left(
\alpha_{ab}^2\eta_o
+
4\alpha_{ab}\beta_{ab}\eta_s
\right)
\frac{\sigma_b}{\sqrt r}
+
O\!\left(\frac{e^{-mr}}{r^{3/2}}\right).
\end{equation}
It is useful to express the angular dependence in terms of the
orientation of dipole $b$ relative to the separation direction.
Let
\[
\psi_b=\phi_b-\theta
\]
denote the angle between the dipole orientation and the line of
centers.  The geometric scalars entering the pair interaction then
reduce to
\[
\alpha_{ab}=\cos\psi_b,
\qquad
\beta_{ab}=\sin\psi_b .
\]
Substituting these relations into the asymptotic vorticity yields
\begin{equation}
\omega_{ab}
=
\frac{m^{3/2}e^{-mr}\sigma_b}
{2\sqrt{2\pi}(\eta_o^2+4\eta_s^2)\sqrt r}
\left[
\frac{\eta_o}{2}(1+\cos2\psi_b)
+
2\eta_s\sin2\psi_b
\right]
+
O\!\left(\frac{e^{-mr}}{r^{3/2}}\right).
\end{equation}
This representation makes the angular structure of the far--field
rotational coupling transparent.  The odd viscosity contributes
through the even harmonic $(1+\cos2\psi_b)$, while the symmetric
viscosity generates the chiral component proportional to
$\sin2\psi_b$.  Both contributions are exponentially screened with
distance, reflecting the short--range character of the rotational
interaction in the degenerate far--field regime.
The orientation dynamics therefore obeys
\[
\dot d_{ai}
=
\frac12\,\omega_a\,\varepsilon_{ij}d_{aj},
\qquad
\omega_a=\omega_{ab}.
\]
Thus the far--field degenerate two--dipole system exhibits the scalings
\begin{align}
\dot r_i &\sim r^{-3},\nn\\
\dot r &\sim r^{-3},\nn\\
\omega_{ab} &\sim e^{-mr}r^{-1/2}\nn.
\end{align}
The translational interaction therefore decays algebraically as $r^{-3}$,
while the rotational coupling is exponentially suppressed.  Although the
algebraic symmetric contribution cancels exactly in the far field, the
exponentially small corrections to $A_0$ and $A_1$ generate a nonvanishing
term proportional to $\eta_s$ in $\omega_{ab}$, so that both the symmetric
and antisymmetric sectors contribute at the same asymptotic order.
The translational interaction therefore decays algebraically as
$r^{-3}$, while the rotational coupling is exponentially screened and
arises entirely from the odd--viscosity sector of the Green tensor.
\paragraph{Equal--strength dipoles }
We now reconsider the far--field dynamics of a pair of identical dipoles
in the degenerate regime, retaining the full far--field expressions of
the Green tensor rather than truncating to the leading algebraic terms
only.  We therefore set
\[
\sigma_1=\sigma_2=\sigma .
\]
Let the dipole orientations be described by the laboratory angles
$\phi_1$ and $\phi_2$, and define the collective variables
\[
\Sigma=\phi_1+\phi_2,
\qquad
\Delta=\phi_1-\phi_2 .
\]
The separation direction is
\[
\hat{\mathbf r}=(\cos\theta,\sin\theta),
\]
and we work throughout in the regime
\[
mr\gg1 .
\]
For convenience, we introduce
\[
D=4\eta_s^2+\eta_o^2,
\qquad
z=mr ,
\]
together with the polynomials
\begin{align}
U(z)&=81+2z\bigl(-33+8z(6+z(15+8z))\bigr),\nn\\
P_1(z)&=-3+2z(3+8z),\nn\\
P_2(z)&=-21+4z(6+z(15+8z)),\nn\\
R(z)&=81+8z(21+z(15+8z)),\nn\\
S(z)&=9+z\bigl(3-2z(3+8z)\bigr),\nn\\
T(z)&=81+z\bigl(183+2z(45-8z)\bigr).
\end{align}
The full far--field degenerate kernels are (obtained using Mathematica)
\begin{align}
A_0(r)
&=
-\frac{3\eta_s}{2\pi D}\frac{1}{m^2r^2}
+
\frac{\sqrt{2\pi}\,\eta_s\left(9+4mr(5+8mr)\right)}
{32\pi D (mr)^{5/2}}\,e^{-mr},
\nn\\[4pt]
A_1(r)
&=
\frac{3\eta_s}{\pi D}\frac{1}{m^2r^2}
-
\frac{3\sqrt{2\pi}\,\eta_s\left(6+mr(15+8mr)\right)}
{32\pi D (mr)^{5/2}}\,e^{-mr},
\nn\\[4pt]
A_2(r)
&=
\frac{\eta_o}{16\sqrt{2\pi}D}
\frac{(1-8mr)e^{-mr}}{(mr)^{3/2}} . \nn
\end{align}
The combination entering the vorticity formula in the symmetric sector is
\begin{equation}
Q(r)=A_0''-\frac{A_0'}{r}-\frac{A_1'}{r}+\frac{2A_1}{r^2}.\nn
\end{equation}
Using the full far--field kernels, one finds
\begin{equation}
Q(r)=
\frac{\eta_s e^{-z}}{64\sqrt{2\pi}\,r^2 z^{5/2}D}\,U(z).\nn
\end{equation}
Thus the symmetric sector cancels at algebraic order but reappears
through exponentially small corrections.  The large--distance expansion
is
\begin{equation}
Q(r)\sim
\frac{\sqrt{2/\pi}\,\eta_s m^{3/2}}{D}\,
\frac{e^{-mr}}{\sqrt r}.
\end{equation}
The pair vorticities are obtained from
\[
\omega_{ab}
=
\sigma
\left[
\alpha_{ab}\beta_{ab}Q(r)
-
\left(
\frac{A_2'(r)}{r}\beta_{ab}^2
+\alpha_{ab}^2A_2''(r)
\right)
\right].
\]
For the full far--field kernels this yields the asymptotic form
\begin{equation}
\omega_{ab}
=
\frac{m^{3/2}e^{-mr}\sigma}{2\sqrt{2\pi}D}
\left(
\alpha_{ab}^2\eta_o+4\alpha_{ab}\beta_{ab}\eta_s
\right)\frac{1}{\sqrt r}
+
O\!\left(\frac{e^{-mr}}{r^{3/2}}\right).
\end{equation}
If $\psi_b=\phi_b-\theta$ denotes the orientation of dipole $b$
relative to the line of centers, so that
\[
\alpha_{ab}=\cos\psi_b,
\qquad
\beta_{ab}=\sin\psi_b,
\]
then the far--field vorticity may equivalently be written as
\begin{equation}
\omega_{ab}
=
\frac{m^{3/2}e^{-mr}\sigma}{2\sqrt{2\pi}D\sqrt r}
\left[
\frac{\eta_o}{2}(1+\cos2\psi_b)
+
2\eta_s\sin2\psi_b
\right]
+
O\!\left(\frac{e^{-mr}}{r^{3/2}}\right).
\end{equation}
Hence the rotational coupling is exponentially suppressed, with
comparable contributions from the antisymmetric and symmetric sectors at
the first nonvanishing order. Substituting the full kernels into the two--dipole velocity law and
specializing to equal strengths gives the relative radial and orbital
velocities.  Writing the result in terms of $(r,\Sigma,\Delta,\theta)$,
one finds
\begin{align}
\dot r
&=
\frac{\sigma m e^{-z}}{64\pi z^{7/2}D}
\Big[
-384 e^z\sqrt z\,\eta_s \cos\Delta \cos(\Sigma-2\theta)
\nonumber\\
&\qquad
+
\sqrt{2\pi}\,\eta_s
\bigl(S(z)+T(z)\cos\Delta\cos(\Sigma-2\theta)\bigr)
\nonumber\\
&\qquad
-
\sqrt{2\pi}\,zP_1(z)\eta_o
\cos\Delta\sin(\Sigma-2\theta)
\Big],
\end{align}
and
\begin{align}
\dot\theta
&=
\frac{\sigma e^{-z}}{64\pi r^2 z^{5/2}D}
\Big[
384 e^z\sqrt z\,\eta_s \cos\Delta \sin(\Sigma-2\theta)
\nonumber\\
&\qquad
+
\sqrt{2\pi}
\Big(
zP_1(z)\eta_o\bigl(1+\cos\Delta\cos(\Sigma-2\theta)\bigr)
-
R(z)\eta_s\cos\Delta\sin(\Sigma-2\theta)
\Big)
\Big].
\end{align}

The orientation dynamics follows from the pair vorticities.  In terms of
\[
\dot\Sigma=\frac{\omega_{12}+\omega_{21}}{2},
\qquad
\dot\Delta=\frac{\omega_{12}-\omega_{21}}{2},
\]
the full far--field equal--strength system becomes
\begin{align}
\dot\Sigma
&=
\frac{\sigma e^{-z}}{128\sqrt{2\pi}\,r^2 z^{5/2}D}
\Big[
z\eta_o\bigl(-9+4z^2(-1+8z)+P_2(z)\cos\Delta\cos(\Sigma-2\theta)\bigr)
\nonumber\\
&\qquad
+
U(z)\eta_s\cos\Delta\sin(\Sigma-2\theta)
\Big],
\\[6pt]
\dot\Delta
&=
\frac{\sigma e^{-z}\sin\Delta}{128\sqrt{2\pi}\,r^2 z^{5/2}D}
\Big[
-U(z)\eta_s\cos(\Sigma-2\theta)
+
zP_2(z)\eta_o\sin(\Sigma-2\theta)
\Big].
\end{align}

These formulas exhibit the main structural features of the corrected
far--field dynamics.  First, the branch
\[
\Delta=0
\qquad\text{or}\qquad
\Delta=\pi
\]
remains invariant because $\dot\Delta$ carries an overall factor
$\sin\Delta$.  Second, the translational sector is algebraic, whereas
the orientational sector is exponentially weak:
\[
\dot r_i\sim r^{-3},
\qquad
\dot r\sim r^{-3},
\qquad
\omega_{ab}\sim e^{-mr}r^{-1/2},
\qquad
\dot\Sigma,\dot\Delta\sim e^{-mr}r^{-1/2}.
\]
Thus the far--field motion is dominated by algebraic translation, while
rotation and orientational relaxation are exponentially slow.

On the invariant branch $\Delta=0$, the equations reduce to
\begin{align}
\dot r\big|_{\Delta=0}
&=
\frac{\sigma m e^{-z}}{64\pi z^{7/2}D}
\Big[
\bigl(-384 e^z\sqrt z+\sqrt{2\pi}\,T(z)\bigr)\eta_s\cos(\Sigma-2\theta)
\nonumber\\
&\qquad
+
\sqrt{2\pi}\,S(z)\eta_s
-
\sqrt{2\pi}\,zP_1(z)\eta_o\sin(\Sigma-2\theta)
\Big],
\\[6pt]
\dot\theta\big|_{\Delta=0}
&=
\frac{\sigma e^{-z}}{64\pi r^2 z^{5/2}D}
\Big[
384 e^z\sqrt z\,\eta_s\sin(\Sigma-2\theta)
\nonumber\\
&\qquad
+
\sqrt{2\pi}
\Big(
zP_1(z)\eta_o\bigl(1+\cos(\Sigma-2\theta)\bigr)
-
R(z)\eta_s\sin(\Sigma-2\theta)
\Big)
\Big],
\\[6pt]
\dot\Sigma\big|_{\Delta=0}
&=
\frac{\sigma e^{-z}}{128\sqrt{2\pi}\,r^2 z^{5/2}D}
\Big[
z\eta_o\bigl(-9+4z^2(-1+8z)+P_2(z)\cos(\Sigma-2\theta)\bigr)
\nonumber\\
&\qquad
+
U(z)\eta_s\sin(\Sigma-2\theta)
\Big],
\\[6pt]
\dot\Delta\big|_{\Delta=0}&=0 .
\end{align}
Eliminating time between the radial and orbital equations gives the
orbital equation
$
\frac{dr}{d\theta}=\frac{\dot r}{\dot\theta}.
$
On the branch $\Delta=0$ this becomes
\begin{equation}
\frac{dr}{d\theta}
=
r\,
\frac{
\bigl(-384 e^z\sqrt z+\sqrt{2\pi}T(z)\bigr)\eta_s\cos\Psi
+\sqrt{2\pi}S(z)\eta_s
-\sqrt{2\pi}zP_1(z)\eta_o\sin\Psi
}{
384 e^z\sqrt z\,\eta_s\sin\Psi
+\sqrt{2\pi}\bigl(zP_1(z)\eta_o(1+\cos\Psi)-R(z)\eta_s\sin\Psi\bigr)
},
\end{equation}
where
\[
\Psi=\Sigma-2\theta .
\]
Thus the orbital law has the generic spiral form
\[
\frac{dr}{d\theta}=r\,F(r,\Psi).
\]
At leading large distance the algebraic terms dominate and this reduces
to
\[
\frac{dr}{d\theta}\sim -\,r\cot\Psi .
\]
Hence, when $\Psi$ varies slowly, the orbit is asymptotically
logarithmic.

The phase dynamics is obtained from
\[
\dot\Psi=\dot\Sigma-2\dot\theta .
\]
Using the above branch equations, one finds that $\dot\Psi$ is again
exponentially small.  Consequently the corrected full far--field
picture is the following: algebraic translation persists, while the
relative orientation variables evolve only through exponentially
suppressed corrections.  This slow phase dynamics allows approximate
orientation locking and supports asymptotic spiral trajectories, but the
locking mechanism is generated by both the symmetric and antisymmetric
sectors at the first nonzero order.

Finally, in the convenient gauge $\theta=0$, the corrected equal--strength
system becomes
\begin{align}
\dot r
&=
\frac{\sigma m e^{-z}}{64\pi z^{7/2}D}
\Big[
-384 e^z\sqrt z\,\eta_s\cos\Delta\cos\Sigma
+\sqrt{2\pi}\eta_s S(z)
\nonumber\\
&\qquad
+
\sqrt{2\pi}\cos\Delta
\bigl(T(z)\eta_s\cos\Sigma-zP_1(z)\eta_o\sin\Sigma\bigr)
\Big],
\\[6pt]
\dot\Sigma
&=
\frac{\sigma e^{-z}}{128\sqrt{2\pi}\,r^2 z^{5/2}D}
\Big[
z\eta_o\bigl(-9+4z^2(-1+8z)+P_2(z)\cos\Delta\cos\Sigma\bigr)
\nonumber\\
&\qquad
+
U(z)\eta_s\cos\Delta\sin\Sigma
\Big],
\\[6pt]
\dot\Delta
&=
\frac{\sigma e^{-z}\sin\Delta}{128\sqrt{2\pi}\,r^2 z^{5/2}D}
\Big[
-U(z)\eta_s\cos\Sigma
+
zP_2(z)\eta_o\sin\Sigma
\Big].
\end{align}
These equations provide a compact explicit description of the corrected
full far--field equal--strength dynamics in the degenerate regime.

\section{Odd contribution to relative pair motion and identification of odd-sector observables}

We now examine how the antisymmetric part of the hydrodynamic response
influences the dynamics of interacting dipoles.  Starting from the
general pairwise velocity and vorticity laws derived earlier, we analyze
the relative motion of a dipole pair and identify the contributions that
arise specifically from the antisymmetric Green--tensor coefficient
$A_2(r)$.  The analysis is carried out for the full general solution,
so that both symmetric and antisymmetric sectors of the response are
retained throughout.  By separating the terms proportional to $A_2(r)$
and its derivatives, we construct diagnostic quantities that isolate
the odd-sector contribution to pair translation, orbital motion, and
orientational dynamics. We consider two dipoles labeled $1$ and $2$ with positions $R_{ai}$ and
orientations $d_{ai}$.  Define the relative separation
\[
r_i=R_{1i}-R_{2i},
\qquad
r=|\mathbf r|,
\qquad
\hat r_i=\frac{r_i}{r}.
\]
The relative velocity is
\begin{equation}
\dot r_i=v_{12,i}-v_{21,i}.\nn
\end{equation}
For a dipole pair the geometric scalars
\begin{equation}
\alpha_{ab}=d_{bi}\hat r_{ab,i},
\qquad
\beta_{ab}=\varepsilon_{ij}d_{bj}\hat r_{ab,i}\nn
\end{equation}
describe the orientation of dipole $b$ relative to the separation
direction. From the general pairwise velocity law, the part generated by the
antisymmetric Green tensor coefficient $A_2(r)$ is
\begin{equation}
v_{ab,i}^{(\mathrm{odd})}
=
\sigma_b\,\alpha_{ab}\,A_2'(r_{ab})\,\varepsilon_{ij}d_{bj}.\nn
\end{equation}
A convenient observable characterizing the relative motion is the
transverse velocity
\begin{equation}
V_\perp=\varepsilon_{ij}\hat r_i \dot r_j .\nn
\end{equation}
Using the pairwise velocity law one finds
\begin{equation}
V_\perp
=
-\Big(\sigma_2\alpha_{12}\beta_{12}+\sigma_1\alpha_{21}\beta_{21}\Big)
\left(A_0'(r)+\frac{A_1(r)}{r}\right)
-
\Big(\sigma_2\alpha_{12}^2+\sigma_1\alpha_{21}^2\Big)A_2'(r).
\end{equation}
The first term originates from the symmetric Green tensor coefficients
$A_0$ and $A_1$, while the second term arises entirely from the
antisymmetric coefficient $A_2$.  The purely odd contribution to the
transverse pair motion is therefore
\begin{equation}
V_\perp^{(\mathrm{odd})}
=
-\Big(\sigma_2\alpha_{12}^2+\sigma_1\alpha_{21}^2\Big)A_2'(r).
\end{equation}

A related observable is the orbital angular velocity of the pair,
\begin{equation}
\Omega
=
\frac{1}{r^2}\varepsilon_{ij}r_i\dot r_j
=
\frac{1}{r}V_\perp ,
\end{equation}
which gives
\begin{equation}
\Omega
=
-\frac{1}{r}
\Big(\sigma_2\alpha_{12}\beta_{12}+\sigma_1\alpha_{21}\beta_{21}\Big)
\left(A_0'(r)+\frac{A_1(r)}{r}\right)
-
\frac{1}{r}
\Big(\sigma_2\alpha_{12}^2+\sigma_1\alpha_{21}^2\Big)A_2'(r).
\end{equation}
The odd contribution to the orbital rotation is therefore
\begin{equation}
\Omega^{(\mathrm{odd})}
=
-\frac{1}{r}
\Big(\sigma_2\alpha_{12}^2+\sigma_1\alpha_{21}^2\Big)A_2'(r).
\end{equation}
Odd viscosity also influences the orientational dynamics through the
vorticity field generated by the dipoles.  The orientation of dipole
$a$ evolves according to
\begin{equation}
\dot d_{ai}
=
\frac12\,\omega_a\,\varepsilon_{ij}d_{aj},
\qquad
\omega_a=\sum_{b\neq a}\omega_{ab}.\nn
\end{equation}
The antisymmetric contribution to the pairwise vorticity is
\begin{equation}
\omega_{ab}^{(\mathrm{odd})}
=
-\sigma_b
\left[
\frac{A_2'(r_{ab})}{r_{ab}}\beta_{ab}^2
+
\alpha_{ab}^2A_2''(r_{ab})
\right].
\end{equation}
A convenient scalar built out of the odd sector is obtained from the difference of the two orientation rates,
\begin{equation}
\Delta\dot\phi^{(\mathrm{odd})}
=
\frac12\left(\omega_{12}^{(\mathrm{odd})}-\omega_{21}^{(\mathrm{odd})}\right),
\end{equation}
which depends only on the radial kernels $A_2'(r)$ and $A_2''(r)$.
The full observables $V_\perp$, $\Omega$, and $\dot\phi_a$ generally
contain both symmetric and antisymmetric contributions.  The quantities
$V_\perp^{(\mathrm{odd})}$, $\Omega^{(\mathrm{odd})}$, and
$\omega_{ab}^{(\mathrm{odd})}$ isolate the part of the dynamics
generated by the antisymmetric Green--tensor coefficient $A_2(r)$, but
they represent extracted components of the full observables rather
than standalone single--run measurements.

In practice the antisymmetric contribution may nevertheless be
identified experimentally in several ways.  First, since the odd
viscosity coefficient $\eta_o$ changes sign under reversal of the
medium chirality, comparison of the pair dynamics in systems with
opposite chirality isolates the part of the motion proportional to
$A_2(r)$ to leading order.  Second, in the degenerate model the symmetric
and antisymmetric responses exhibit distinct radial dependences: the
symmetric contributions arise from the algebraic kernels $A_0(r)$ and
$A_1(r)$, whereas the antisymmetric contribution is controlled by the
screened kernel $A_2(r)$ and its derivatives.  Measurements of the
distance dependence of the pair motion can therefore help separate the
two responses.  Finally, the theoretical pair dynamics derived above
may be fitted to experimentally measured dipole trajectories, allowing
the antisymmetric kernel $A_2(r)$ to be inferred together with the
symmetric response functions.
\section{Conclusions}

We have developed a continuum hydrodynamic framework for the dynamics
of active force dipoles embedded in compressible supported membranes
with odd (Hall) viscosity. Starting from a generalized two--dimensional
Stokes equation with momentum leakage to the surrounding fluid, we
derived the real--space Green tensor describing the membrane mobility
using a Hankel transform solution. The resulting response tensor has
the structure
\[
G_{ij}(\mathbf r)
=
A_0(r)\,\delta_{ij}
+
A_1(r)\,\hat r_i \hat r_j
+
A_2(r)\,\varepsilon_{ij},
\]
where the symmetric kernels $A_0$ and $A_1$ describe the conventional
dissipative hydrodynamic response while the antisymmetric kernel
$A_2$ encodes the chiral contribution associated with odd viscosity.
The radial functions are governed by screened hydrodynamic modes
determined by the interplay between membrane viscosity,
compressibility, and momentum leakage into the surrounding fluid.

Using this Green tensor we obtained explicit expressions for the
velocity and vorticity fields generated by an active force dipole in
a compressible membrane. The dipolar flow separates naturally into
longitudinal, radial, and transverse components, with the transverse
component receiving contributions from the odd-viscous sector. The
vorticity field likewise decomposes into even and odd parts with
distinct angular structures.

Several useful limits of the theory were analyzed. In the
parity--symmetric limit $\eta_o \to 0$, the antisymmetric kernel
vanishes and the Green tensor reduces to the standard compressible
membrane mobility with shear and compressional screening lengths.
In the incompressible limit the longitudinal mode is suppressed and
the flow reduces to the screened two--dimensional Stokes response.
A particularly transparent regime arises in the degenerate case where
the hydrodynamic screening lengths coincide, allowing compact
expressions for the Green tensor and the dipolar flow fields.

In the near field the Green tensor retains the logarithmic structure
characteristic of two--dimensional hydrodynamics. The dipolar velocity
decays as $r^{-1}$ and the vorticity exhibits a quadrupolar
$r^{-2}$ structure, while odd viscosity contributes a transverse flow
component and a chiral distortion of the vorticity pattern. In the
far field the symmetric sector generates algebraically decaying
translational interactions whereas the antisymmetric contribution is
exponentially screened. As a consequence, dipole translation decays
as $r^{-3}$ while rotational couplings become exponentially small at
large separations.

Building on the single--dipole solution, we formulated the dynamical
equations governing interacting dipoles on the membrane. The resulting
many--body system couples dipole positions and orientations through
hydrodynamic interactions determined by the Green tensor kernels.
The framework also yields evolution equations for global observables
such as the center--of--mass motion and the polarization of the dipole
ensemble. Specializing to two dipoles produces a closed nonlinear
system whose near--field dynamics exhibits chiral spiral trajectories,
while in the far field translational motion dominates and orientational
relaxation becomes exponentially slow.

Finally, we examined how odd viscosity enters observable features of
dipole pair dynamics. Quantities such as transverse pair motion and
orbital rotation generally contain contributions from both symmetric
and antisymmetric sectors of the Green tensor, whereas terms involving
derivatives of the kernel $A_2(r)$ isolate the purely odd-viscous
response. These results therefore provide a systematic framework for
analyzing and interpreting odd-viscous contributions in simulations
and experiments involving active inclusions in membrane systems.

The dynamical formulation developed here also provides a natural basis
for studying collective phenomena in large assemblies of active
dipoles. In forthcoming work \cite{krishnan2026} we will present more
multidipole simulations in compressible and odd-viscous membranes
based on the framework developed in this paper, exploring the emergent
dynamics of interacting active inclusions in chiral membrane fluids.
\section{Acknowledgments}
We are very thankful to Sarthak Bagaria, Michael D Graham, Naomi Oppenheimer, Haim Diamant and Mark Henle. S.K. is supported by an Institute fellowship from Birla Institute of Technology and Science, Pilani (Hyderabad Campus). R.S is supported by DST INSPIRE Faculty fellowship, India (Grant No.IFA19-PH231), NFSG and OPERA Research Grant from Birla Institute of Technology and Science, Pilani (Hyderabad Campus).
\section{Data Availability} The data that supports the findings of this study are available within the article.

\appendix
\section{Detailed Simulation Figures}
\label{appsim}
\begin{figure*}[!t]
  \centering
  \begin{tabular}{cccc}
    \multicolumn{4}{c}{\textbf{Incompressible membrane – Near-zone – Dynamical orientations}} \\[4pt]
  \multicolumn{2}{c}{\textbf{Pusher}} & \multicolumn{2}{c}{\textbf{Puller}} \\
    traj & $\langle d_{ij}\rangle(t)$ & traj & $\langle d_{ij}\rangle(t)$ \\[6pt]

    \includegraphics[width=0.18\linewidth]{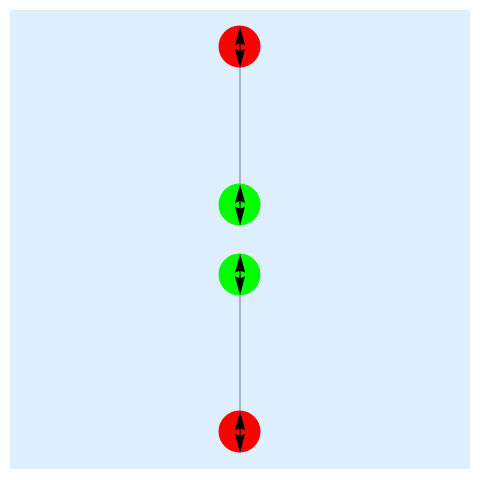} &
    \includegraphics[width=0.18\linewidth]{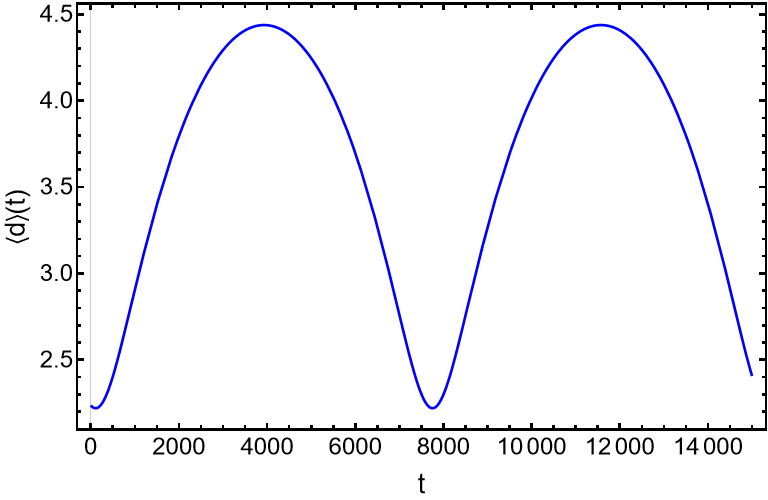} &
    \includegraphics[width=0.18\linewidth]{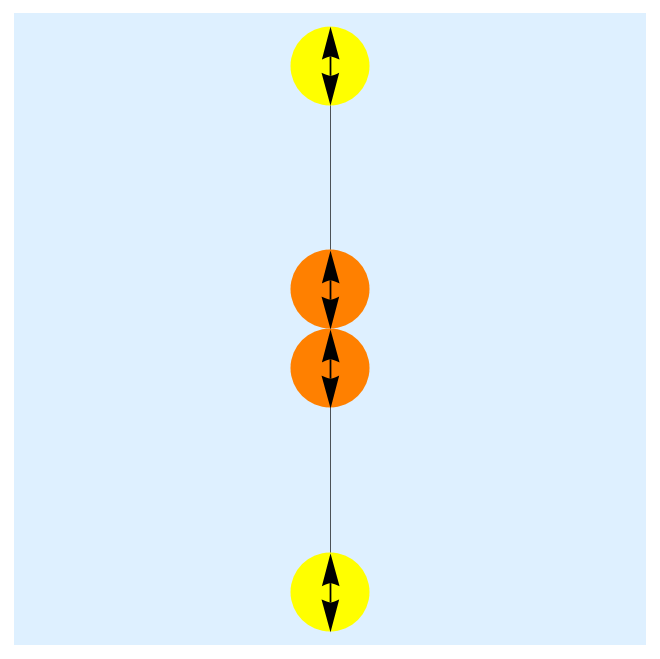} &
    \includegraphics[width=0.18\linewidth]{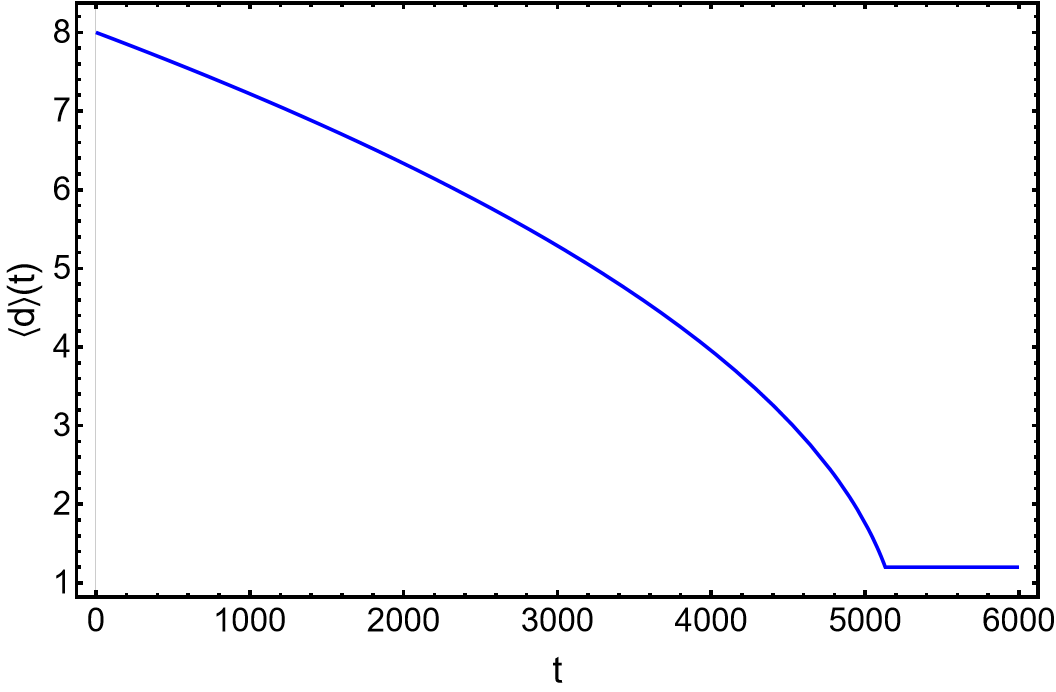} \\[4pt]

    \includegraphics[width=0.18\linewidth]{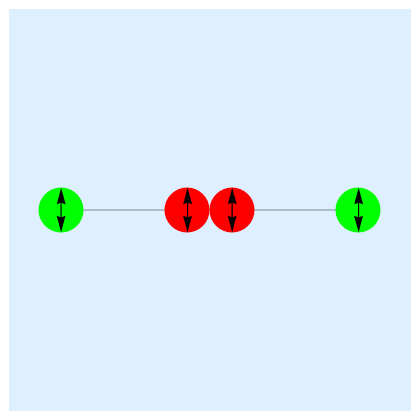} &
    \includegraphics[width=0.18\linewidth]{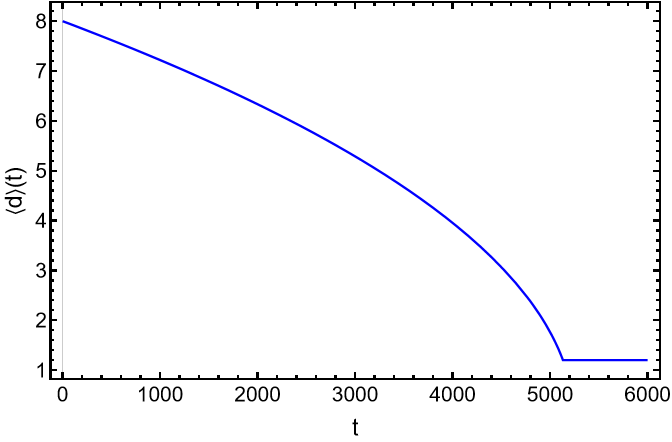} &
    \includegraphics[width=0.18\linewidth]{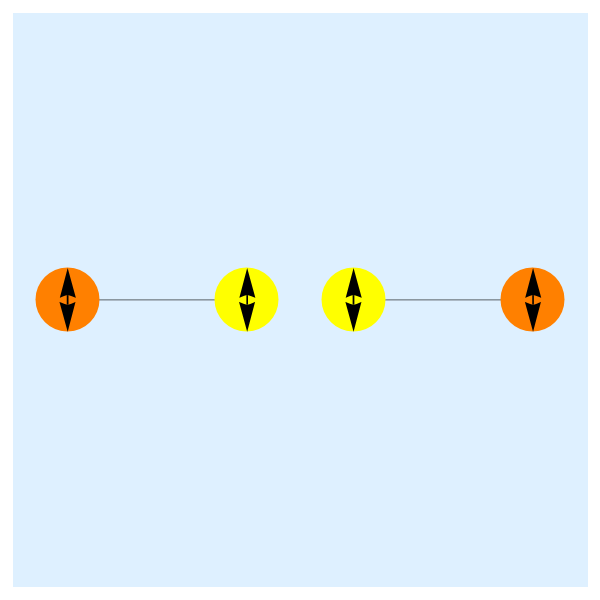} &
    \includegraphics[width=0.18\linewidth]{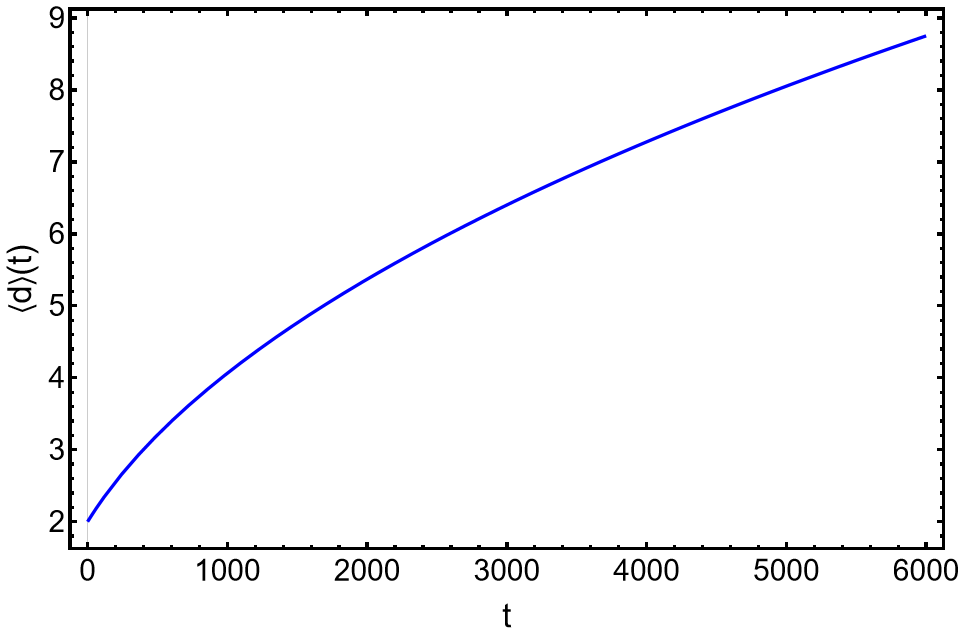} \\[4pt]

    \includegraphics[width=0.18\linewidth]{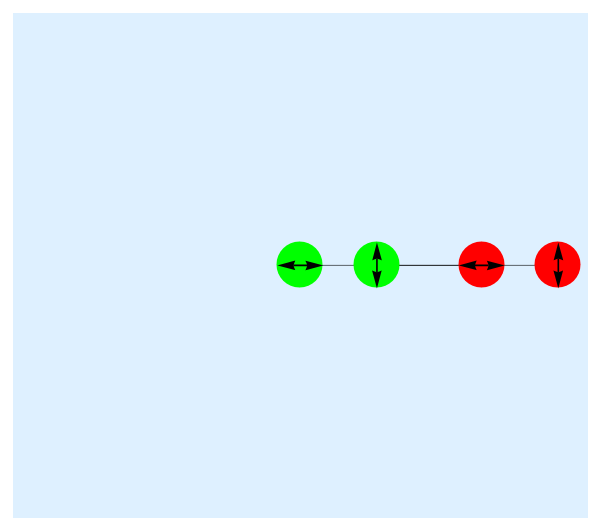} &
    \includegraphics[width=0.18\linewidth]{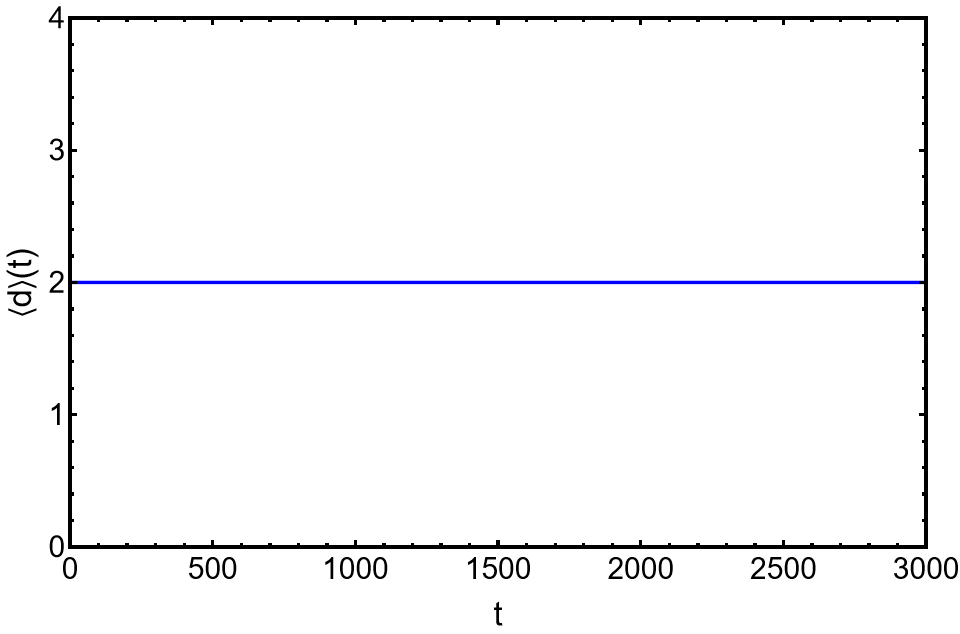} &
    \includegraphics[width=0.18\linewidth]{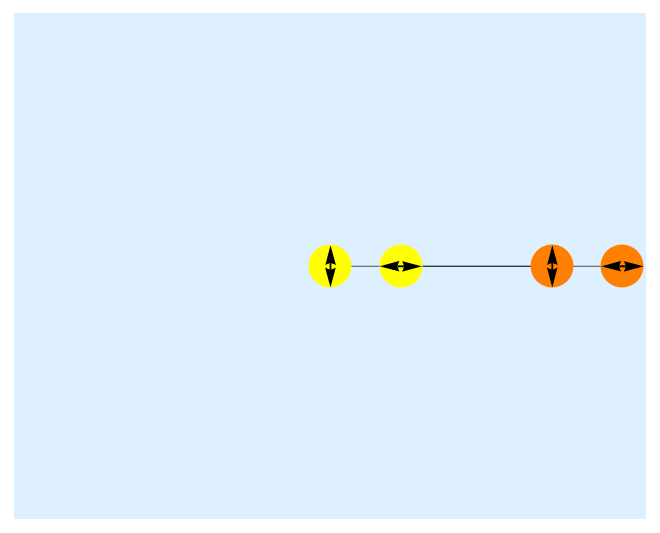} &
    \includegraphics[width=0.18\linewidth]{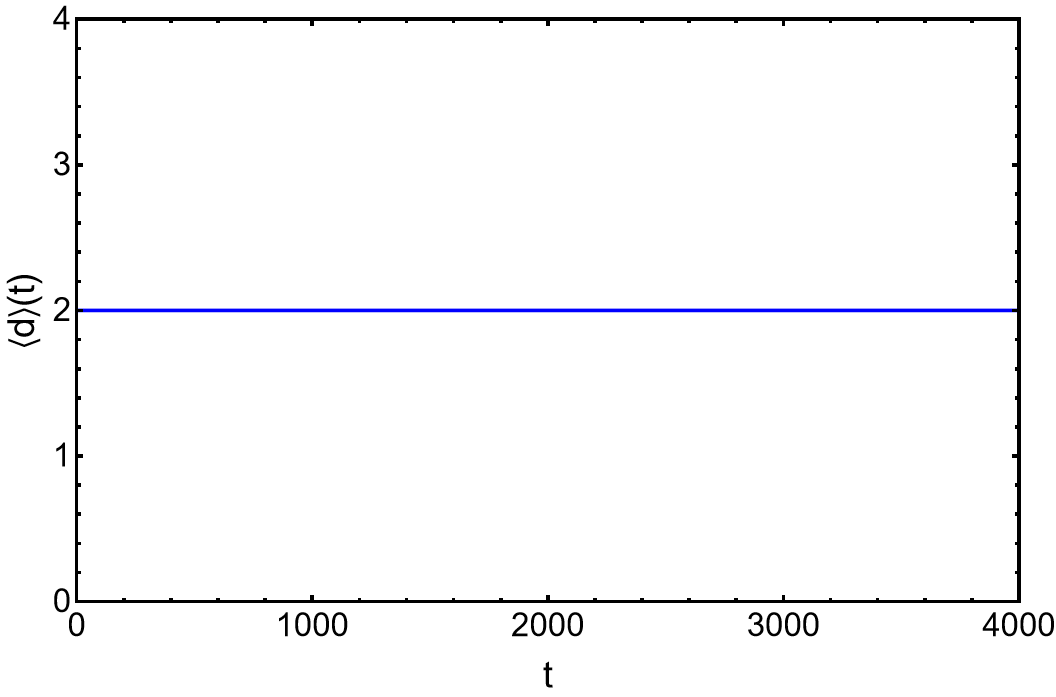} \\[4pt]

    \includegraphics[width=0.18\linewidth]{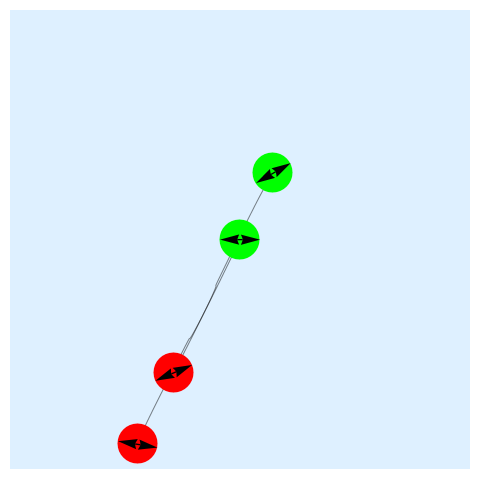} &
    \includegraphics[width=0.18\linewidth]{images/Fincmpnqnzpusherarbdist.png} &
    \includegraphics[width=0.18\linewidth]{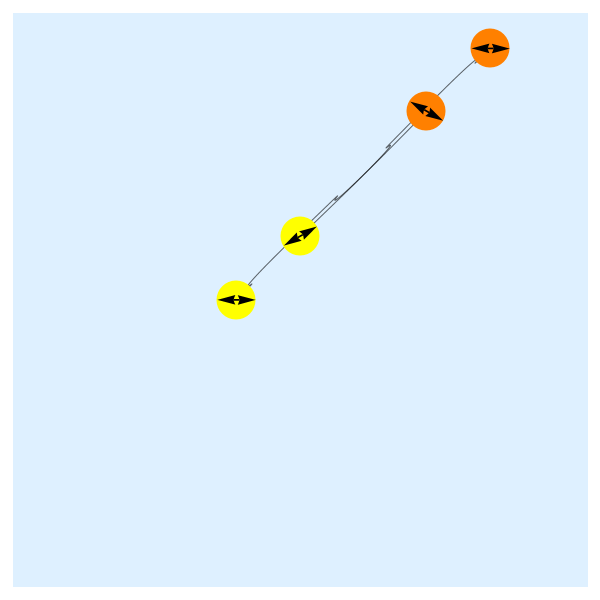} &
    \includegraphics[width=0.18\linewidth]{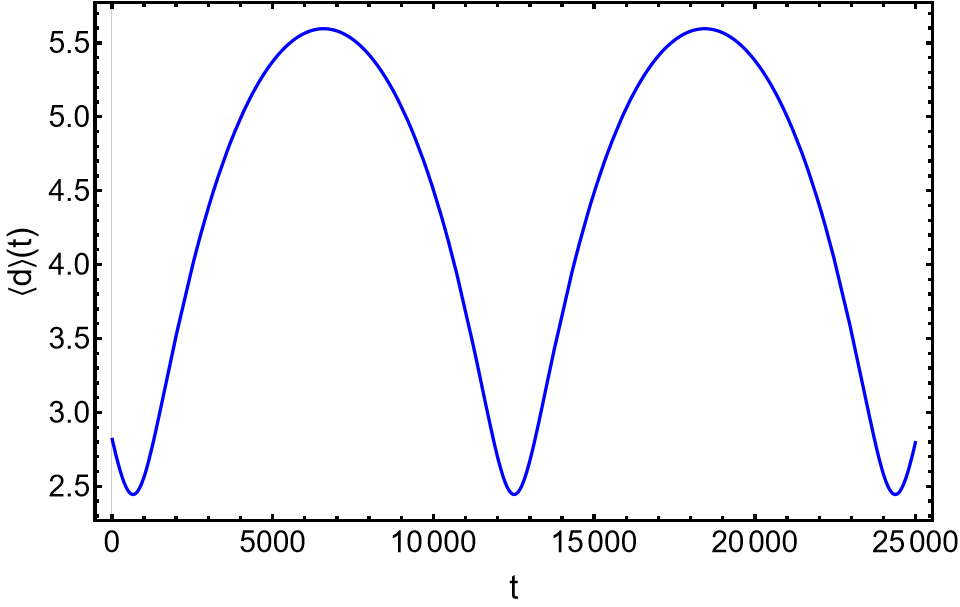} \\[4pt]

    \includegraphics[width=0.18\linewidth]{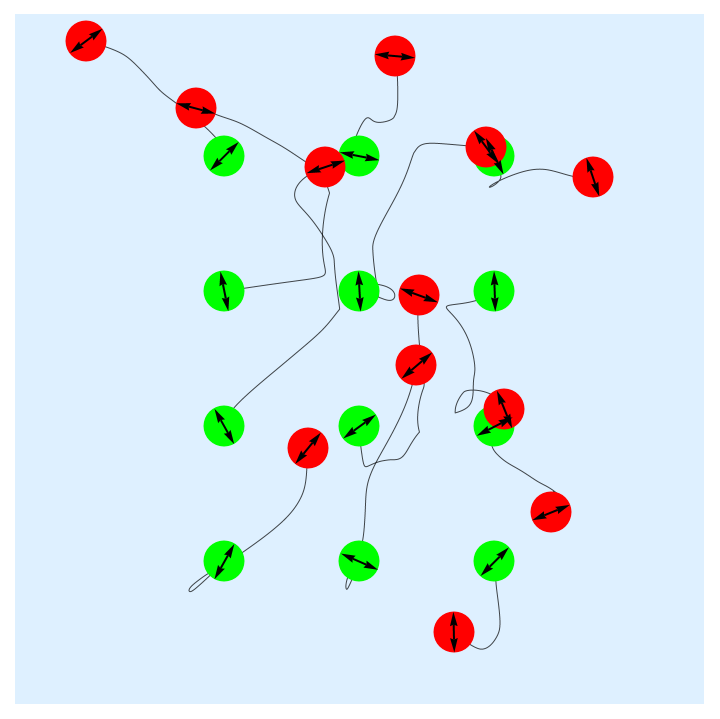} &
    \includegraphics[width=0.18\linewidth]{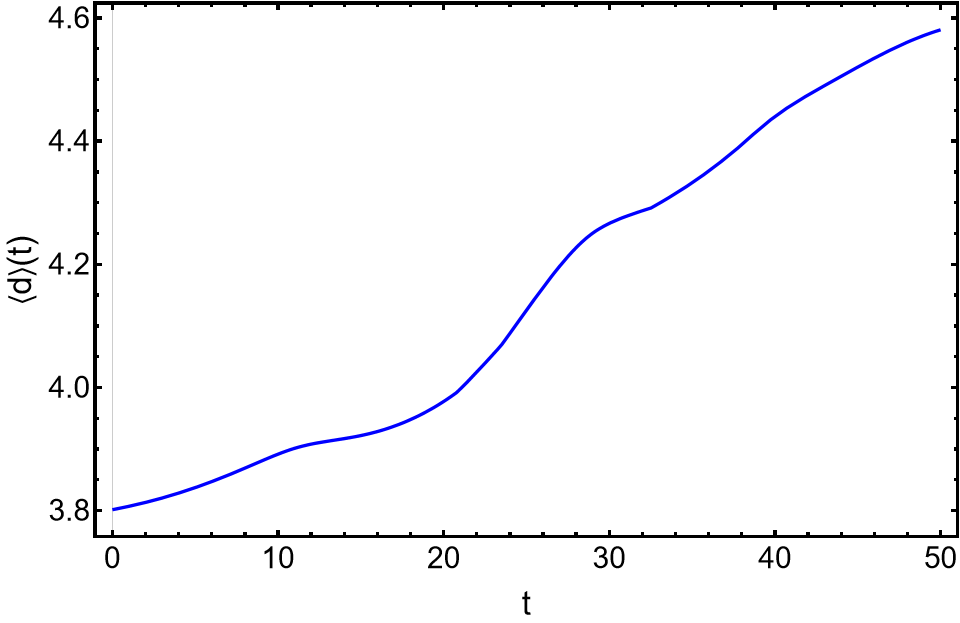} &
    \includegraphics[width=0.18\linewidth]{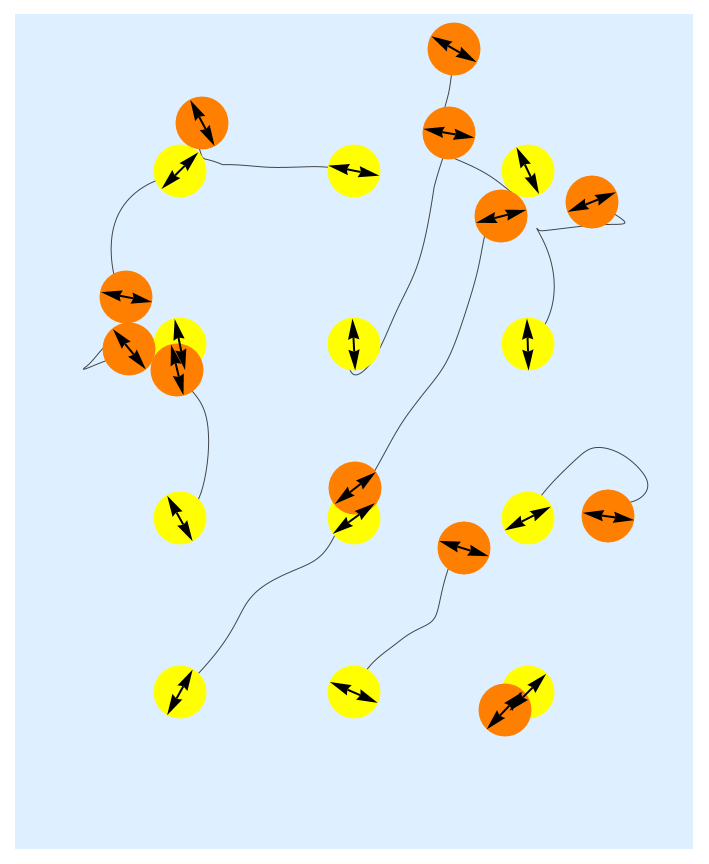} &
    \includegraphics[width=0.18\linewidth]{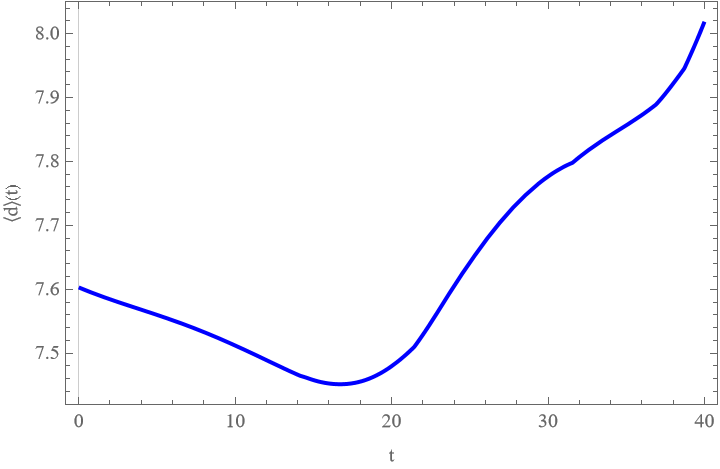} \\
  \end{tabular}

  \caption{
  Near–zone dynamics of pusher and puller motors in an incompressible supported membrane 
  with dynamical orientations.  
  Rows: axial, side-by-side, perpendicular, random pair, 12-dipole cluster.  
  Columns: trajectories and mean pair separation for pusher (left) and puller (right) dipoles.
  }
  \label{fig:incmp_nz_dyn_all}
\end{figure*}

\begin{figure*}[t]
  \centering
  \begin{tabular}{cccc}
    \multicolumn{4}{c}{\textbf{Incompressible membrane – Far-zone – Dynamical orientations}} \\[6pt]
  \multicolumn{2}{c}{\textbf{Pusher}} & \multicolumn{2}{c}{\textbf{Puller}} \\
    traj & $\langle d_{ij}\rangle(t)$ & traj & $\langle d_{ij}\rangle(t)$ \\[6pt]

    \includegraphics[width=0.18\linewidth]{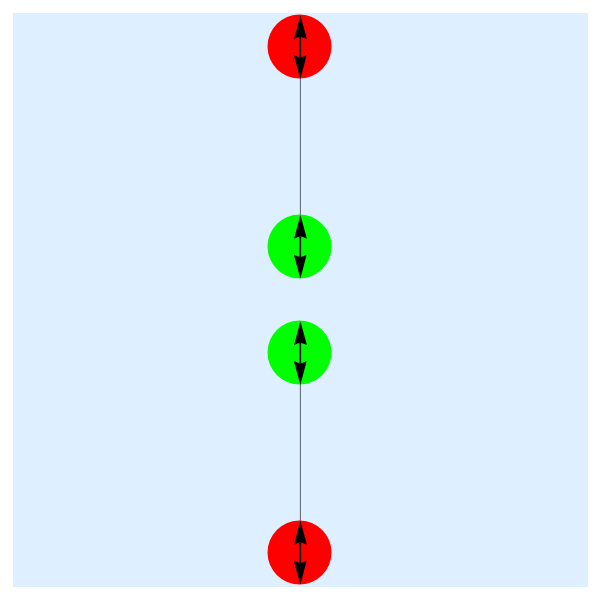} &
    \includegraphics[width=0.18\linewidth]{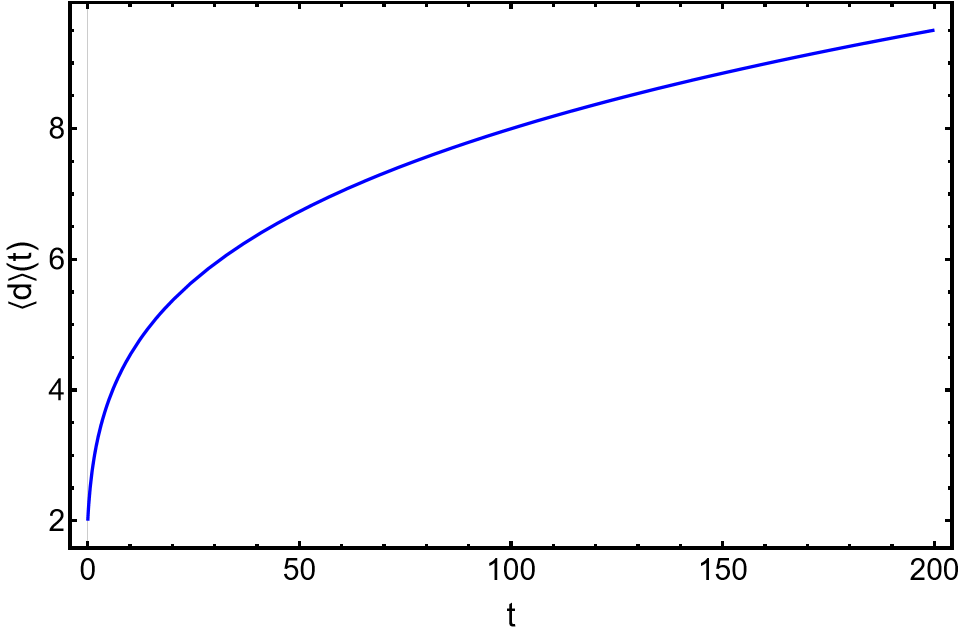} &
    \includegraphics[width=0.18\linewidth]{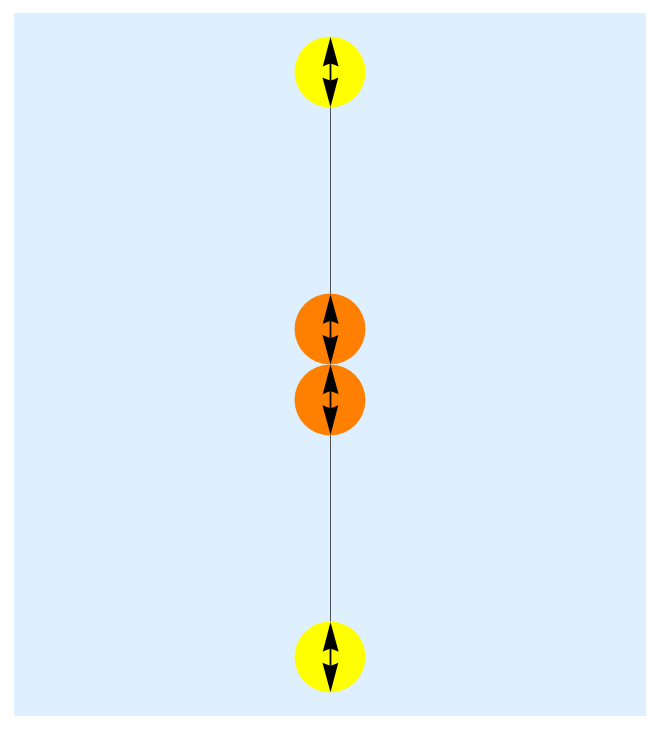} &
    \includegraphics[width=0.18\linewidth]{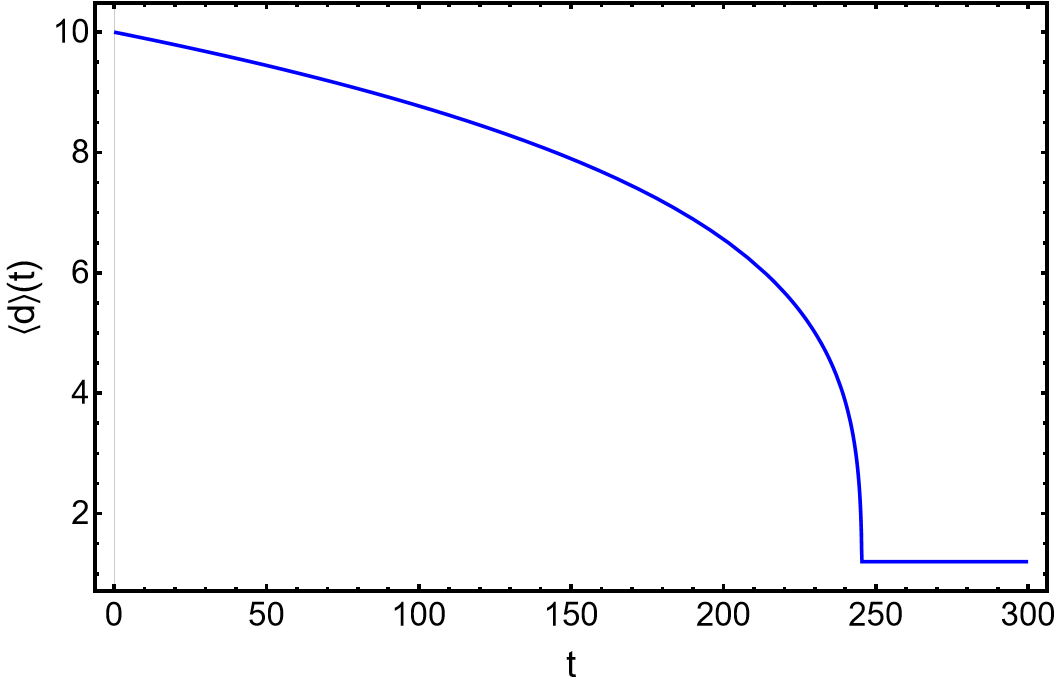} \\[4pt]

    \includegraphics[width=0.18\linewidth]{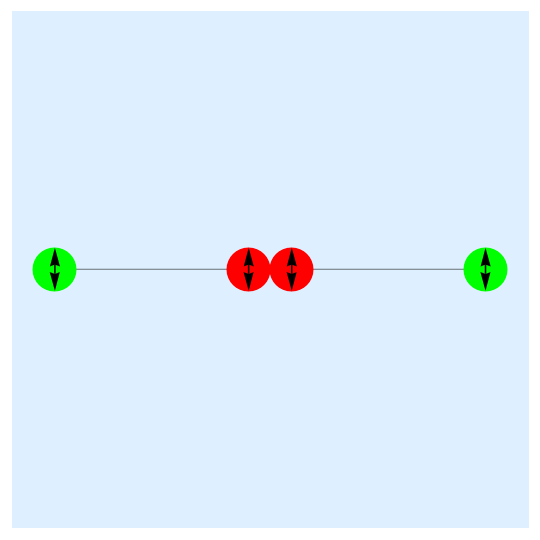} &
    \includegraphics[width=0.18\linewidth]{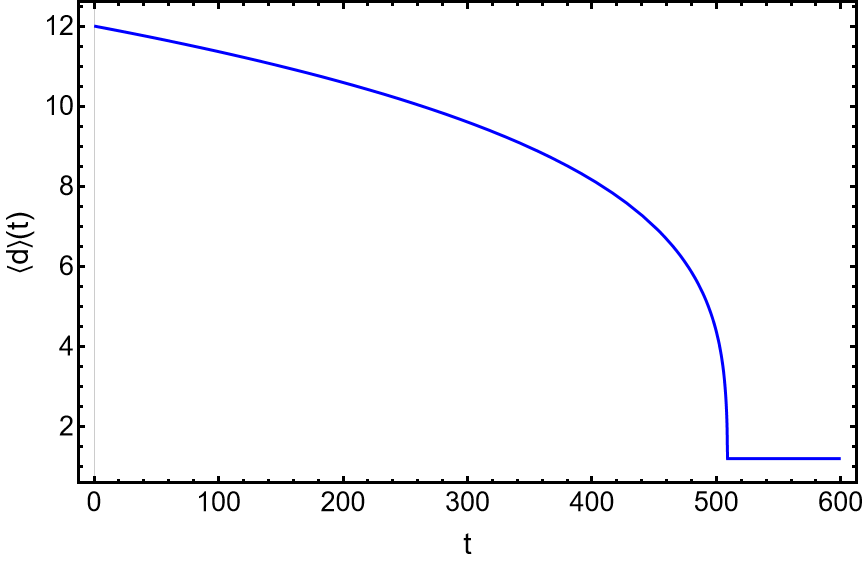} &
    \includegraphics[width=0.18\linewidth]{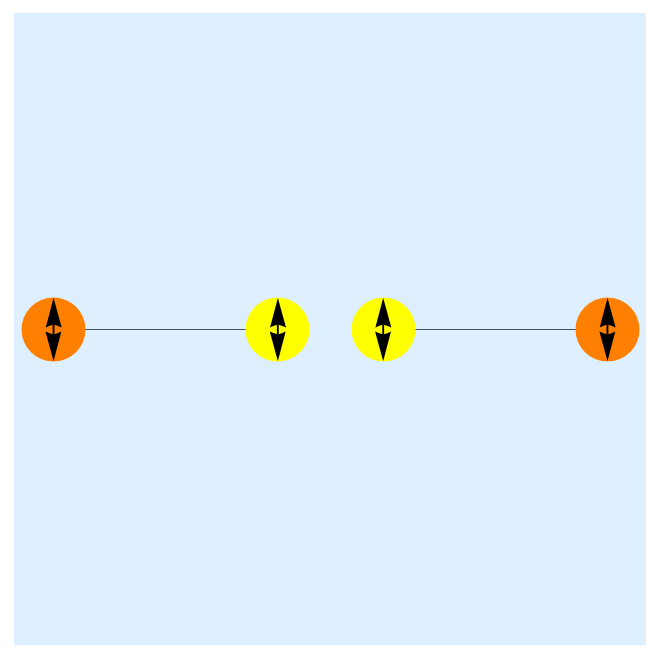} &
    \includegraphics[width=0.18\linewidth]{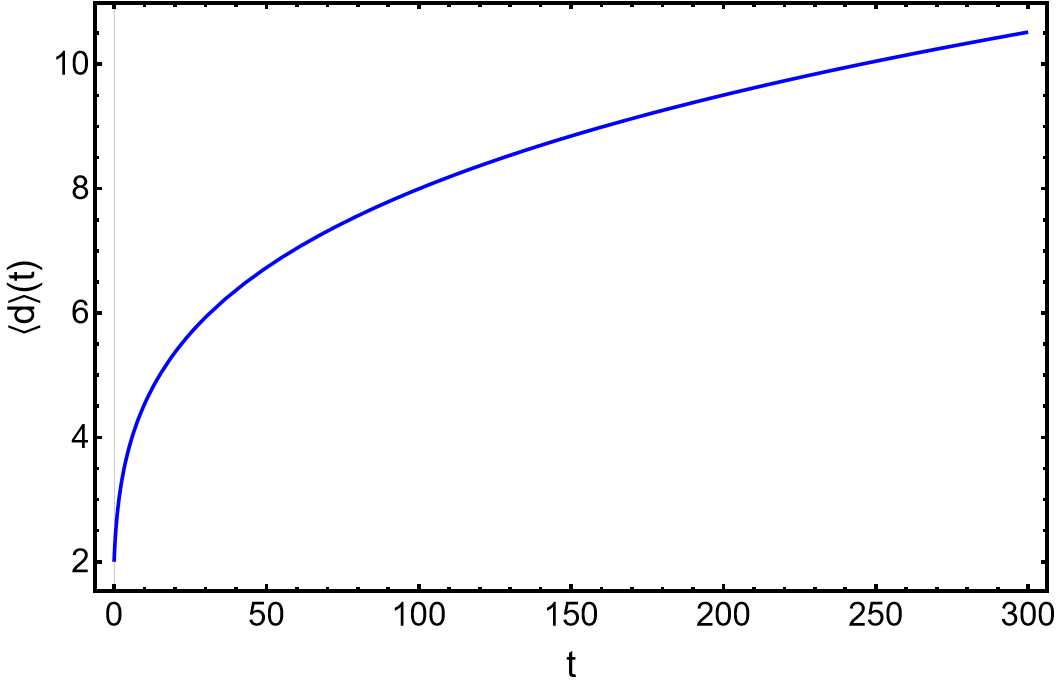} \\[4pt]

    \includegraphics[width=0.18\linewidth]{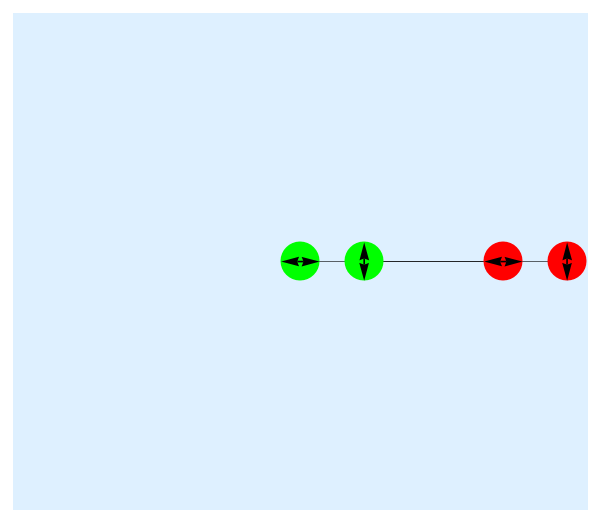} &
    \includegraphics[width=0.18\linewidth]{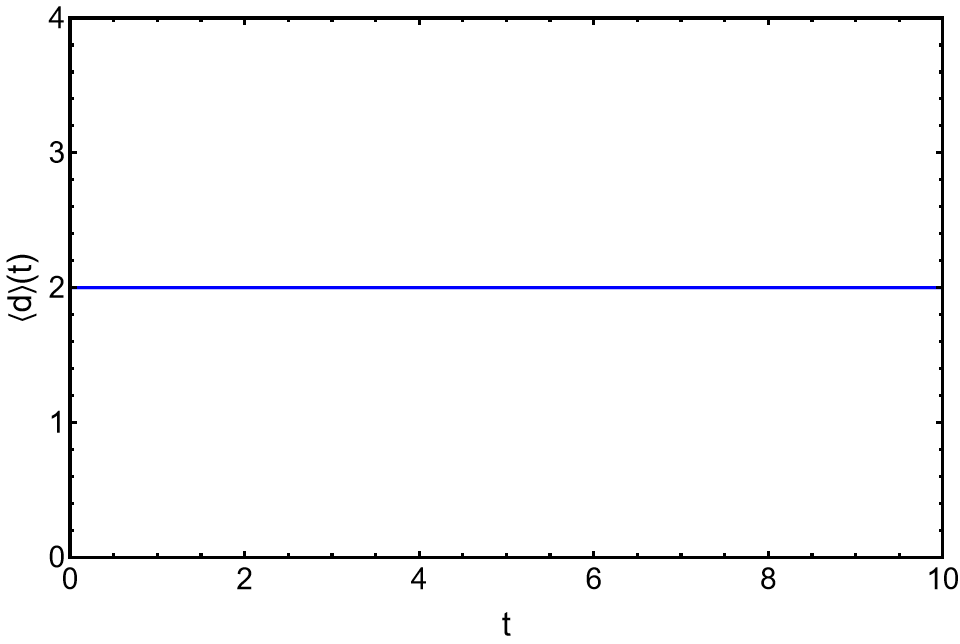} &
    \includegraphics[width=0.18\linewidth]{images/Fincmpnqfzpusherperp.png} &
    \includegraphics[width=0.18\linewidth]{images/Fincmpnqfzpusherperpdist.png} \\[4pt]

    \includegraphics[width=0.18\linewidth]{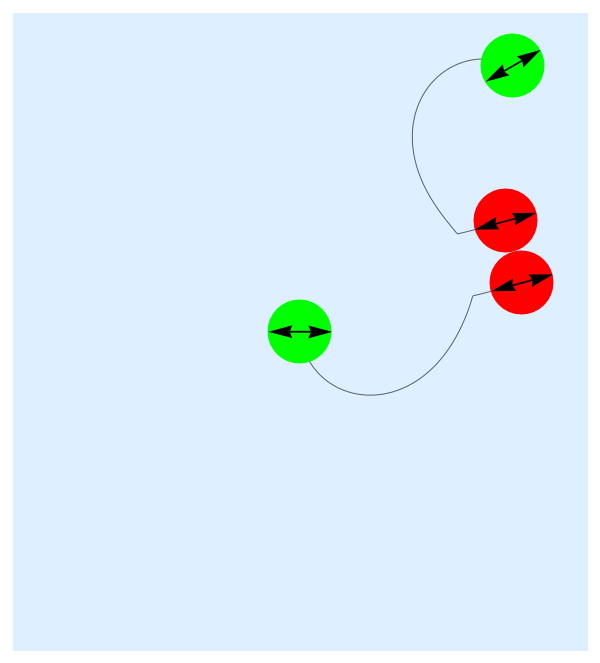} &
    \includegraphics[width=0.18\linewidth]{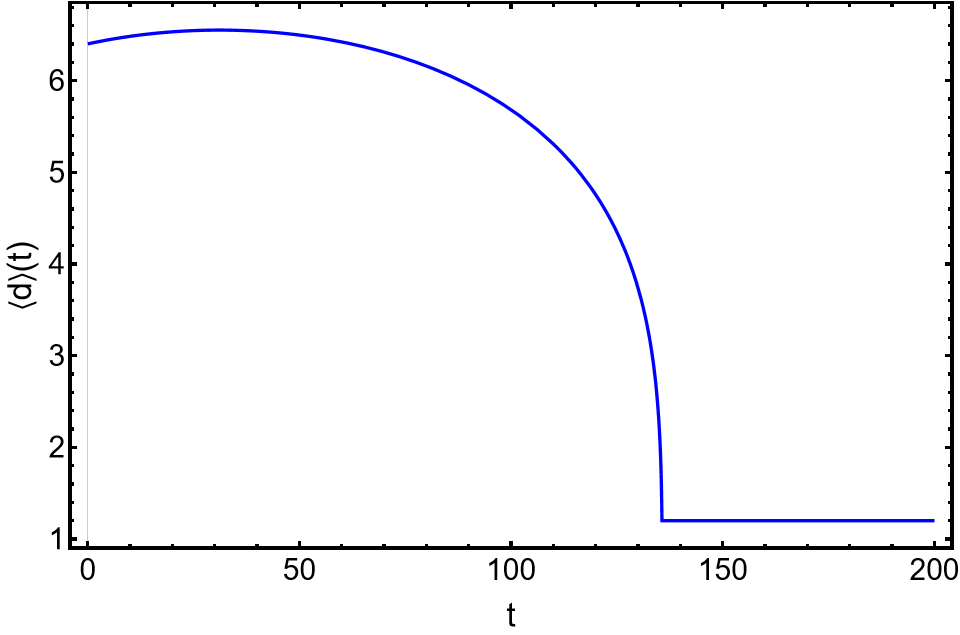} &
    \includegraphics[width=0.18\linewidth]{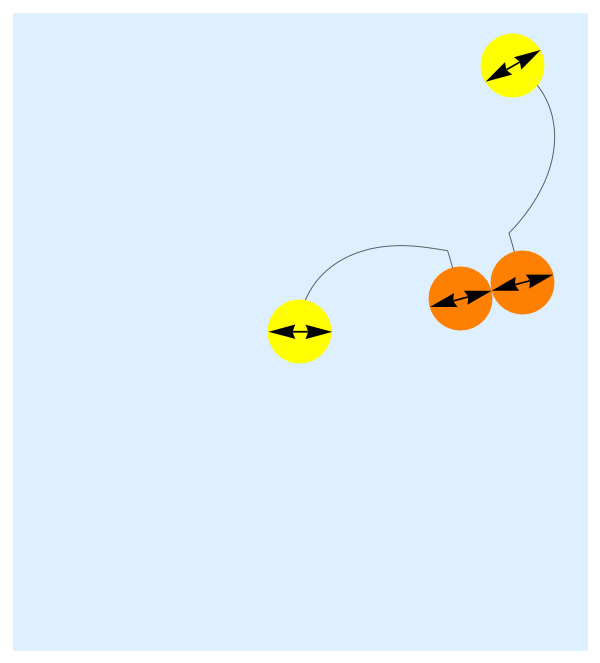} &
    \includegraphics[width=0.18\linewidth]{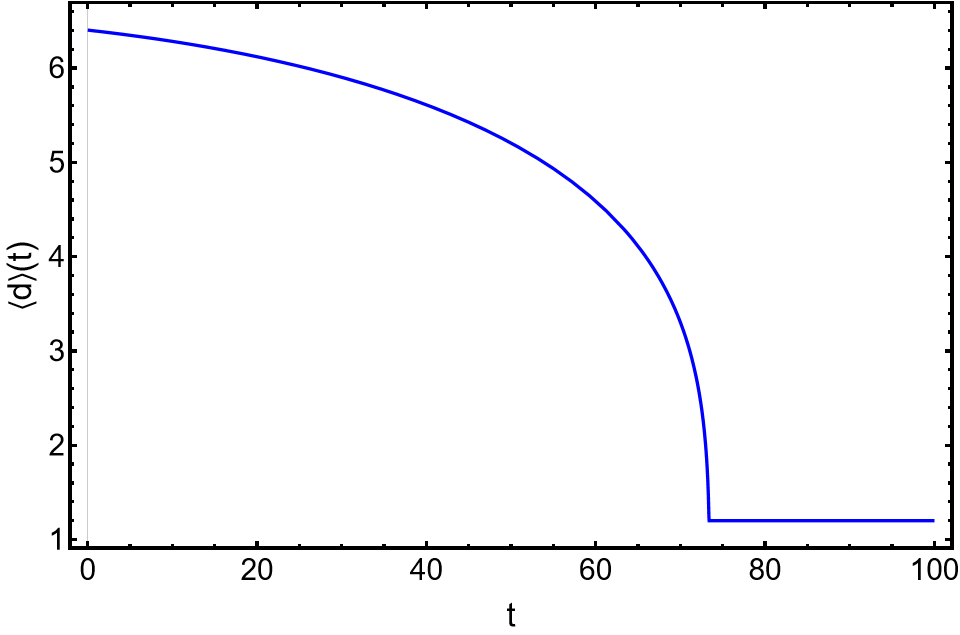} \\[4pt]

    \includegraphics[width=0.18\linewidth]{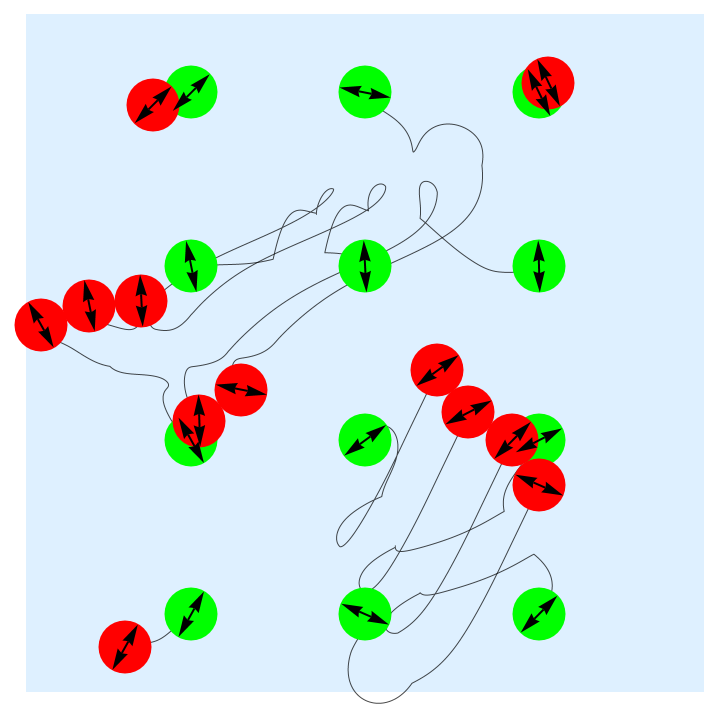} &
    \includegraphics[width=0.18\linewidth]{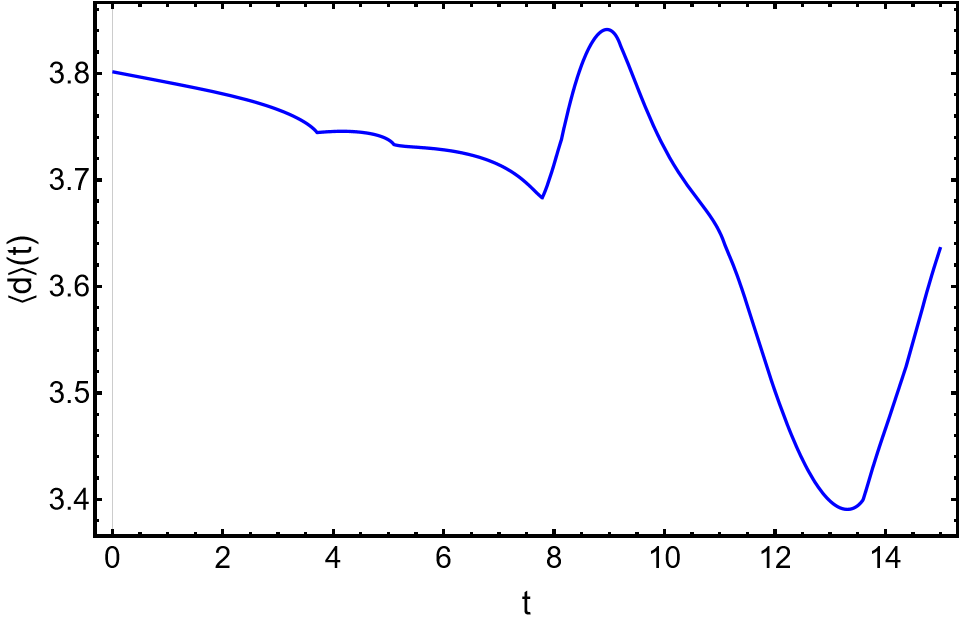} &
    \includegraphics[width=0.18\linewidth]{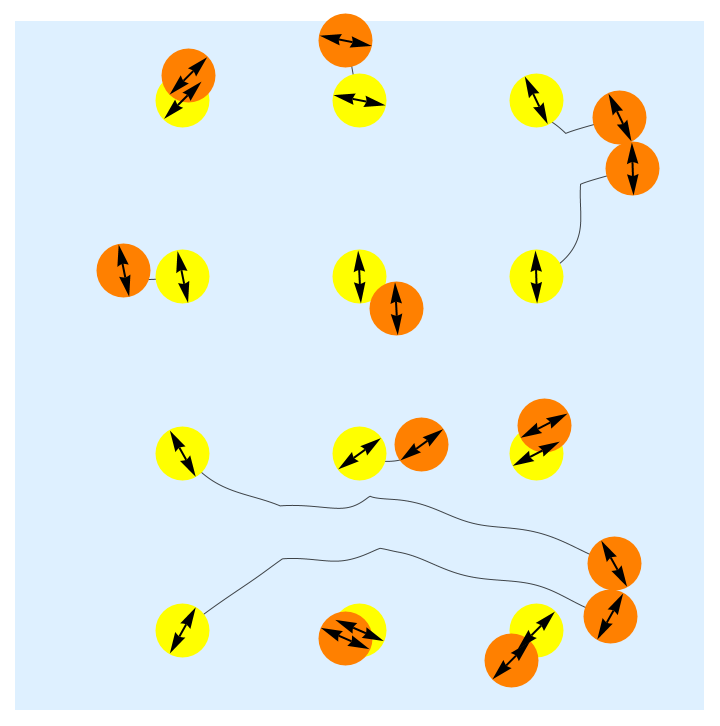} &
    \includegraphics[width=0.18\linewidth]{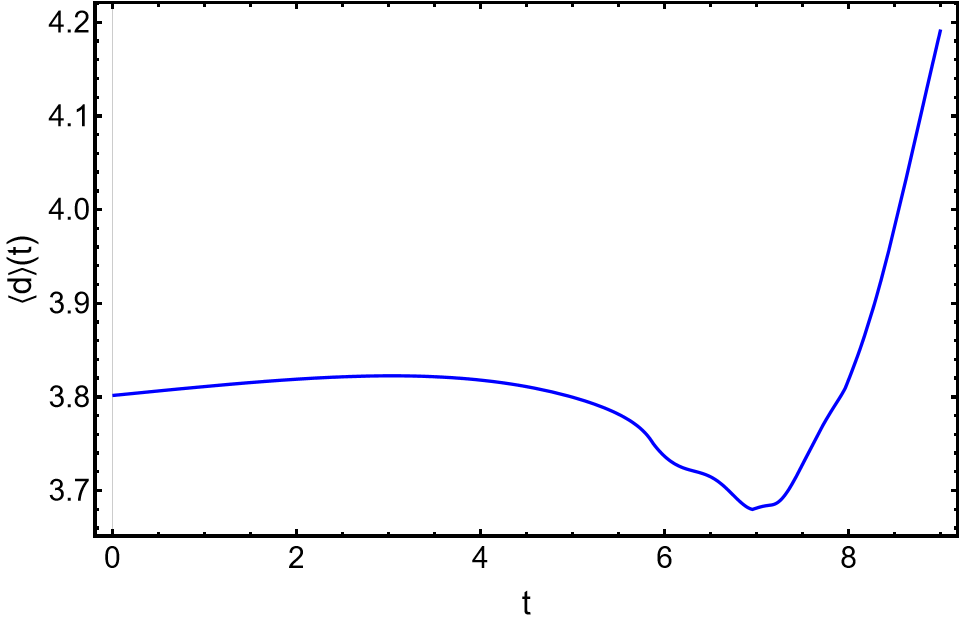} \\
  \end{tabular}

  \caption{
  Far–zone dynamics of pusher and puller dipoles in an incompressible supported membrane with
  \emph{dynamical} orientations.  
  Rows correspond to axial, side-by-side, perpendicular, random, and cluster configurations.  
  Columns show trajectories and mean pair separations for pushers (left) and pullers (right).
  }
  \label{fig:incmp_fz_dyn_all}
\end{figure*}
\begin{figure*}[t]
  \centering
  \begin{tabular}{cccc}
    \multicolumn{4}{c}{\textbf{Incompressible membrane – Near-zone – Quenched orientations}} \\[6pt]
    \multicolumn{2}{c}{\textbf{Pusher}} & \multicolumn{2}{c}{\textbf{Puller}} \\
    traj & $\langle d_{ij}\rangle(t)$ & traj & $\langle d_{ij}\rangle(t)$ \\[6pt]

    \includegraphics[width=0.18\linewidth]{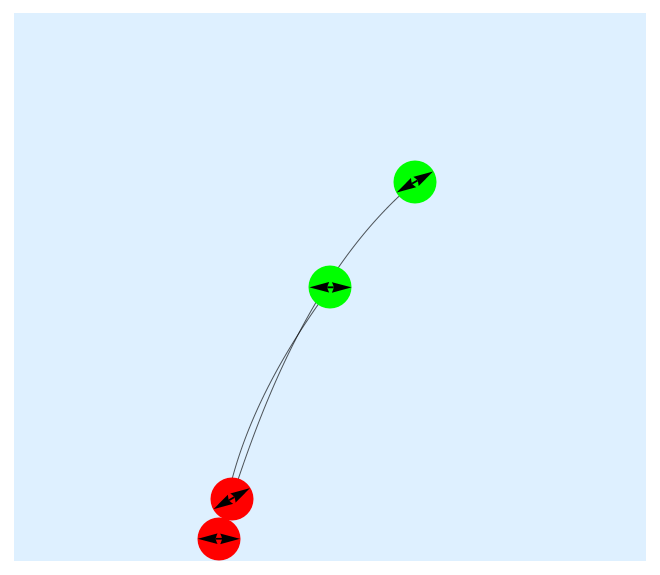} &
    \includegraphics[width=0.18\linewidth]{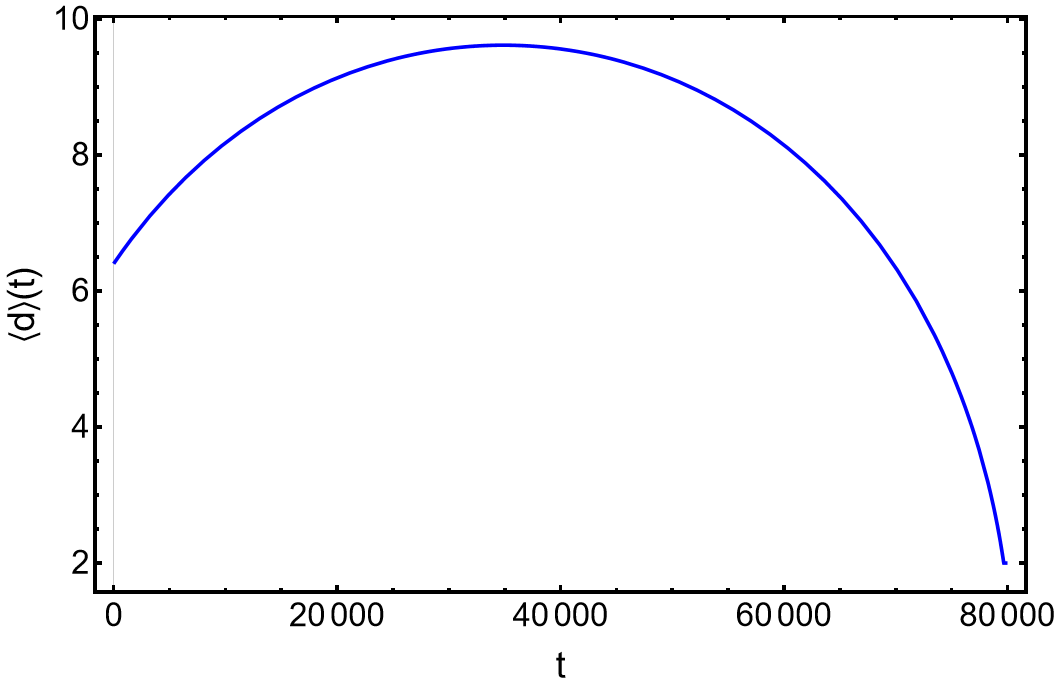} &
    \includegraphics[width=0.18\linewidth]{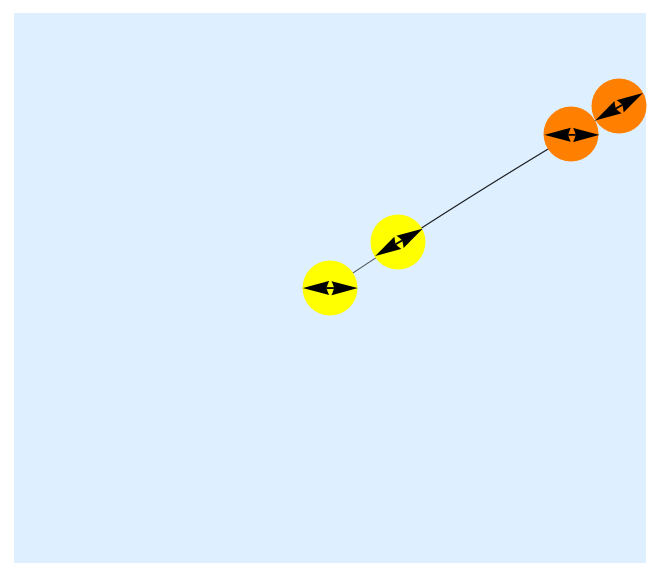} &
    \includegraphics[width=0.18\linewidth]{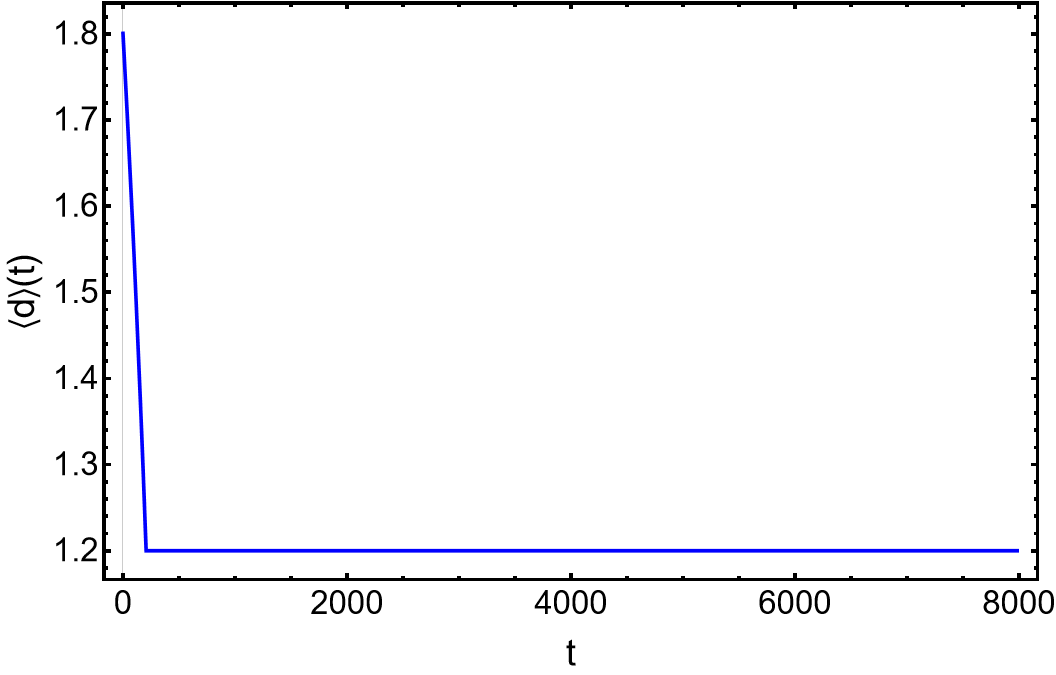} \\[4pt]

    \includegraphics[width=0.18\linewidth]{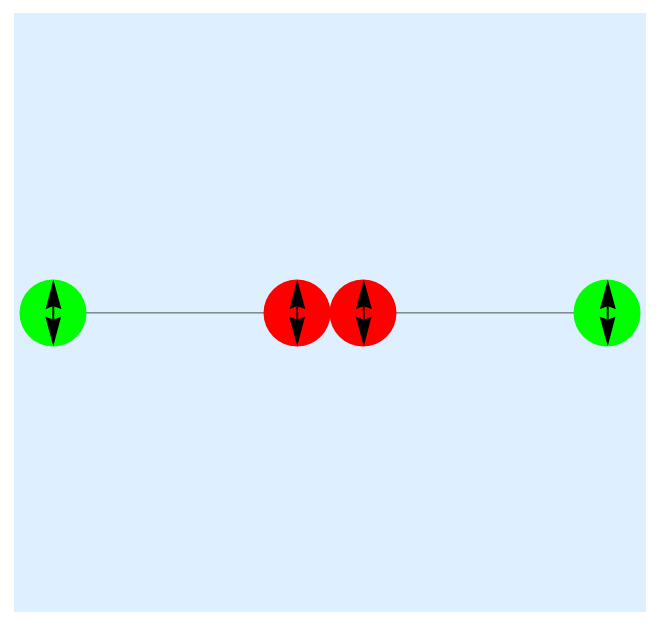} &
    \includegraphics[width=0.18\linewidth]{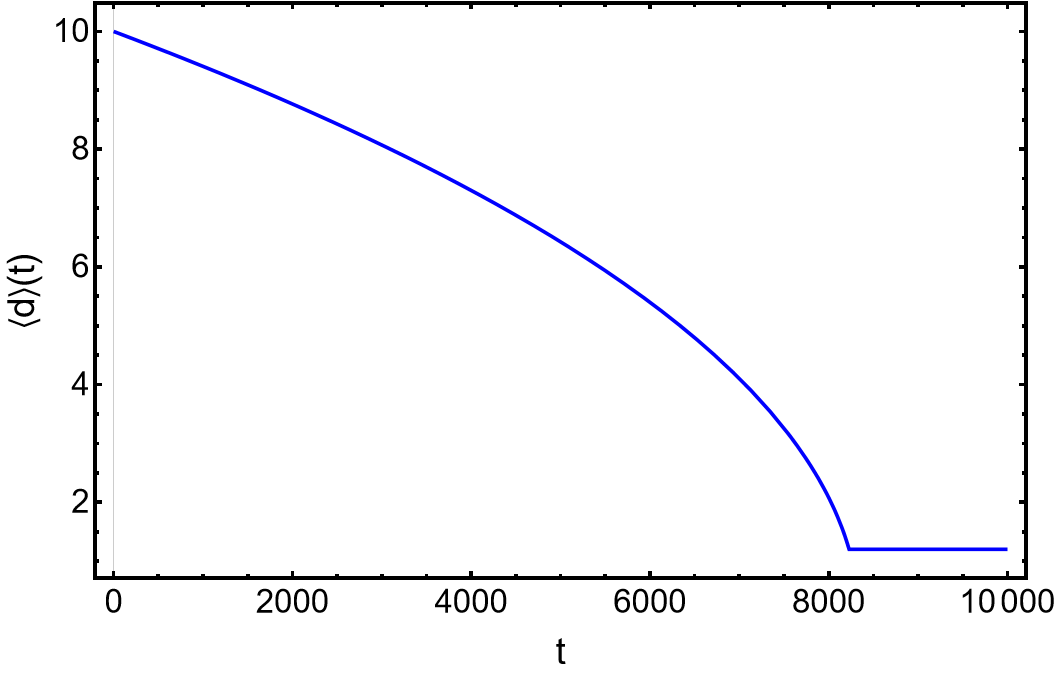} &
    \includegraphics[width=0.18\linewidth]{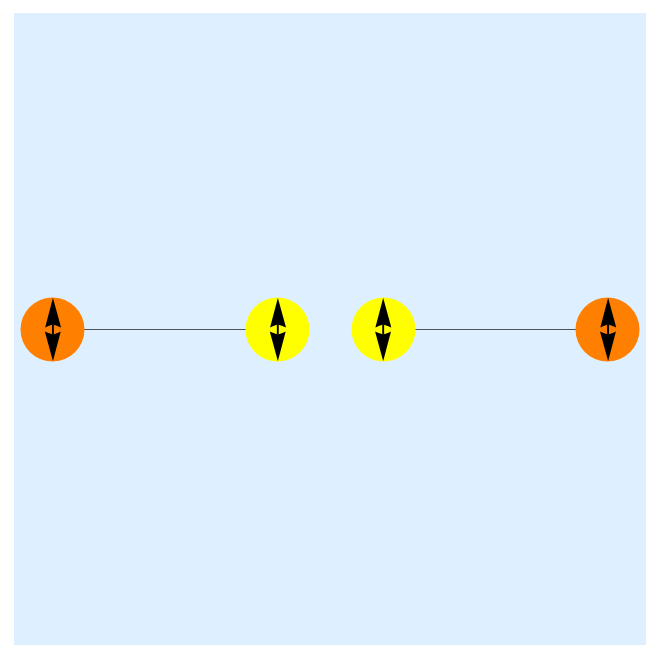} &
    \includegraphics[width=0.18\linewidth]{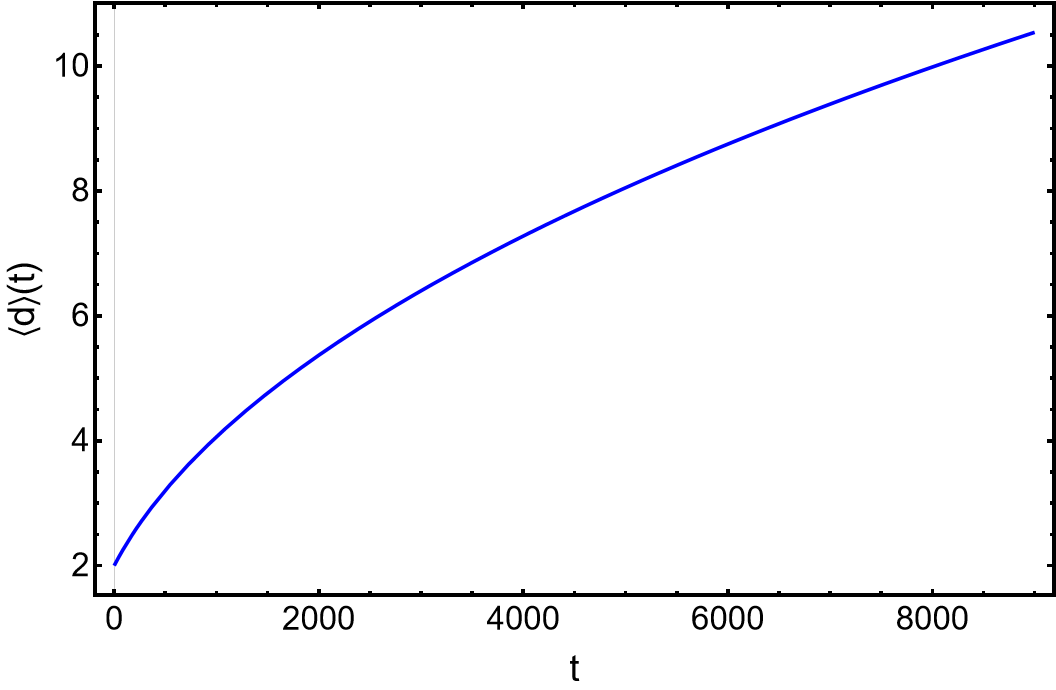} \\[4pt]

    \includegraphics[width=0.18\linewidth]{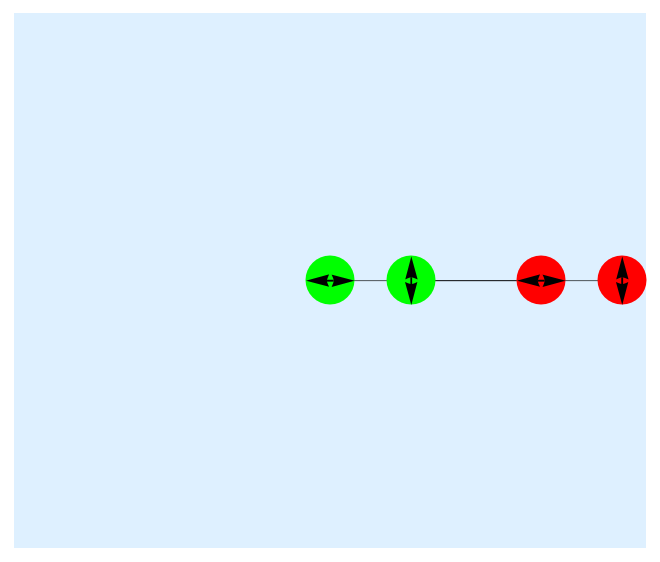} &
    \includegraphics[width=0.18\linewidth]{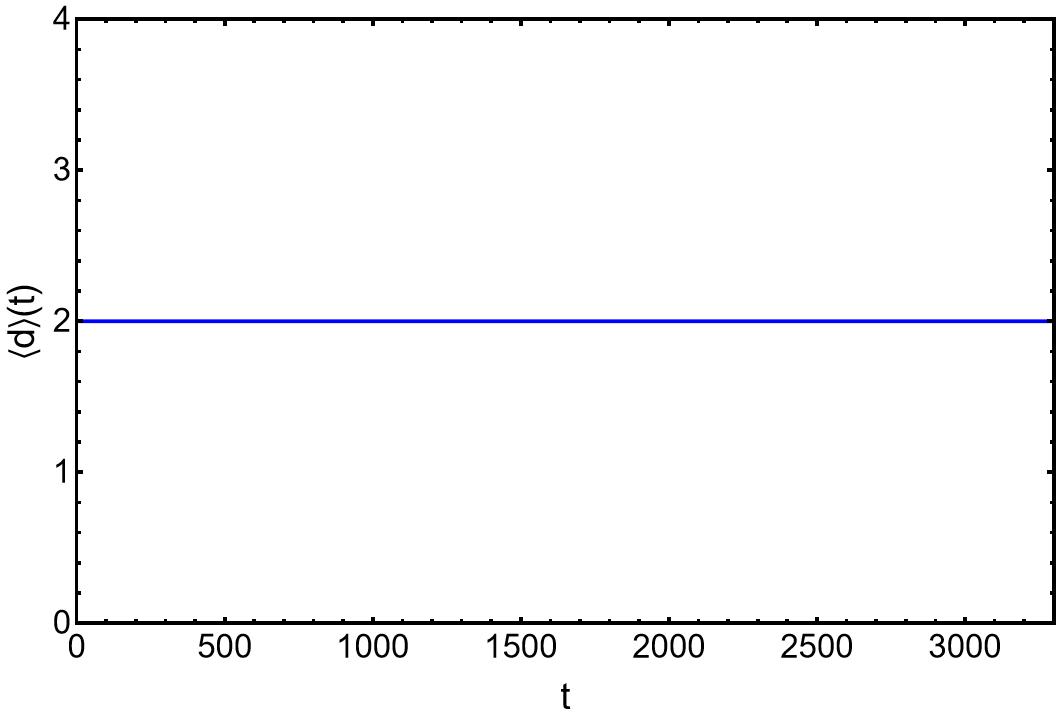} &
    \includegraphics[width=0.18\linewidth]{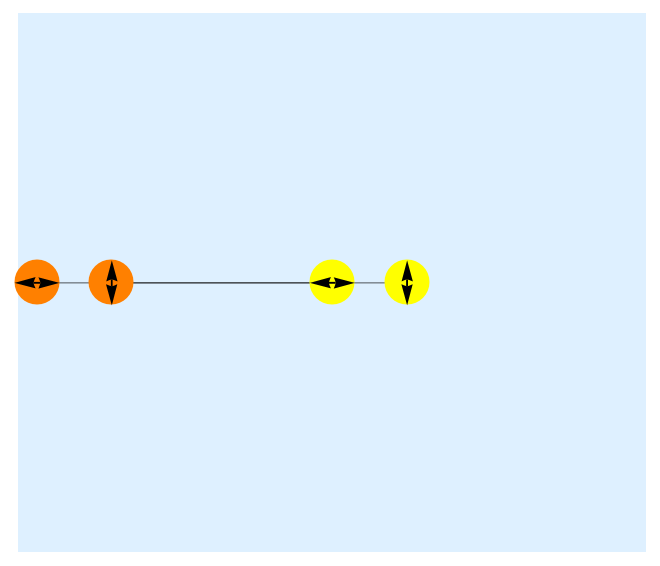} &
    \includegraphics[width=0.18\linewidth]{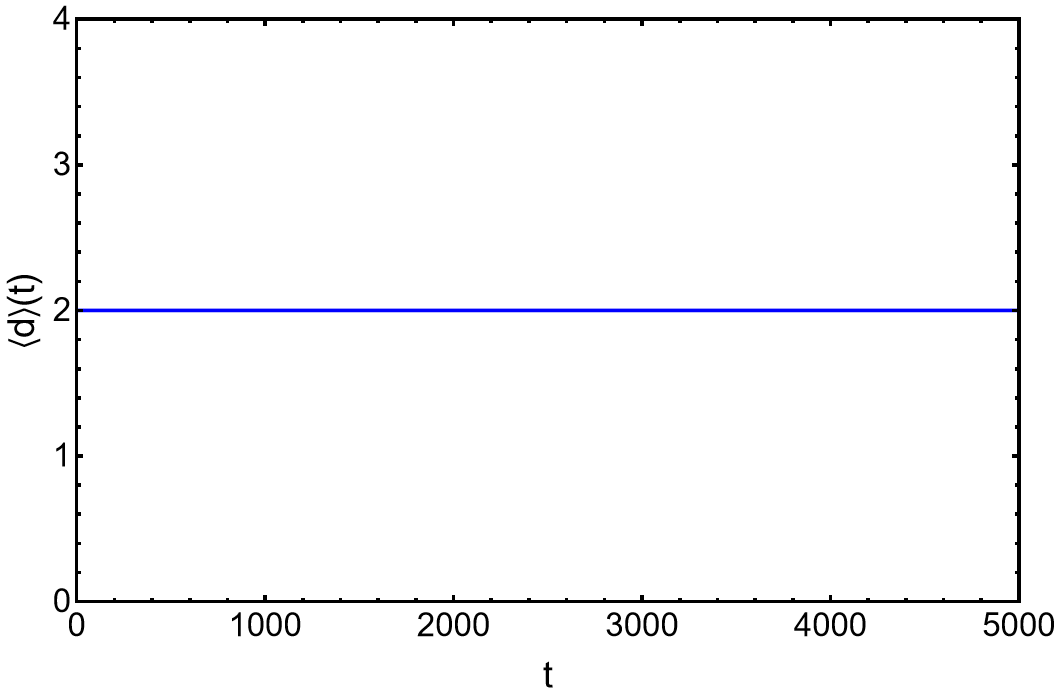} \\[4pt]

    \includegraphics[width=0.18\linewidth]{images/Fincmpqnzpusherarb.png} &
    \includegraphics[width=0.18\linewidth]{images/Fincmpqnzpusherarbdist.png} &
    \includegraphics[width=0.18\linewidth]{images/Fincmpqnzpullerarb.png} &
    \includegraphics[width=0.18\linewidth]{images/Fincmpqnzpullerarbdist.png} \\[4pt]

    \includegraphics[width=0.18\linewidth]{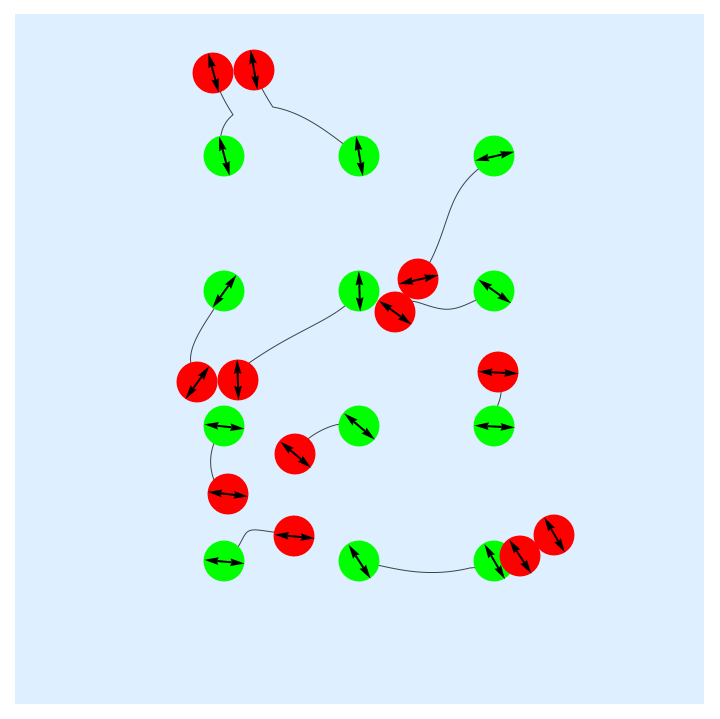} &
    \includegraphics[width=0.18\linewidth]{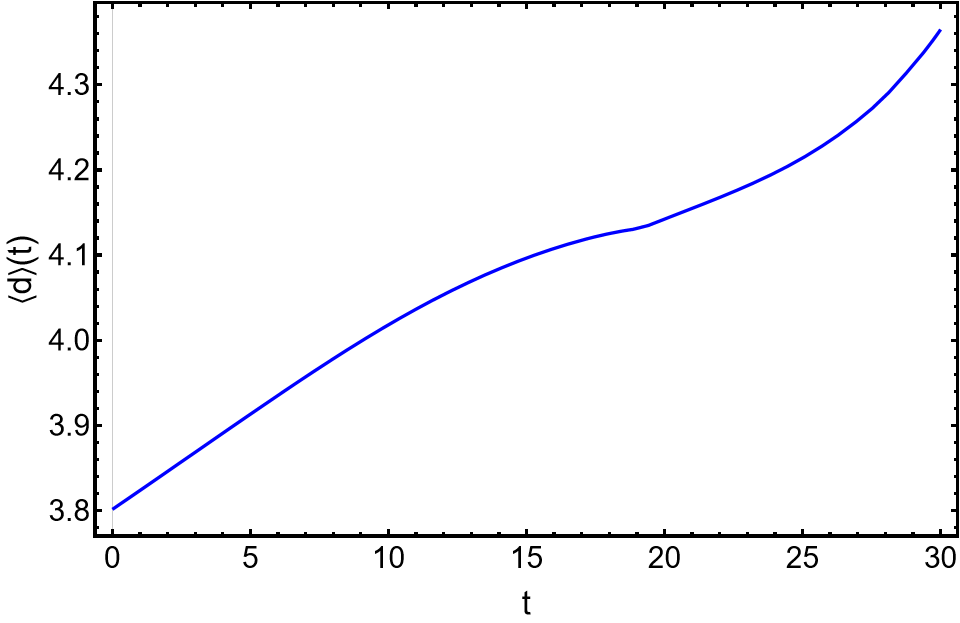} &
    \includegraphics[width=0.18\linewidth]{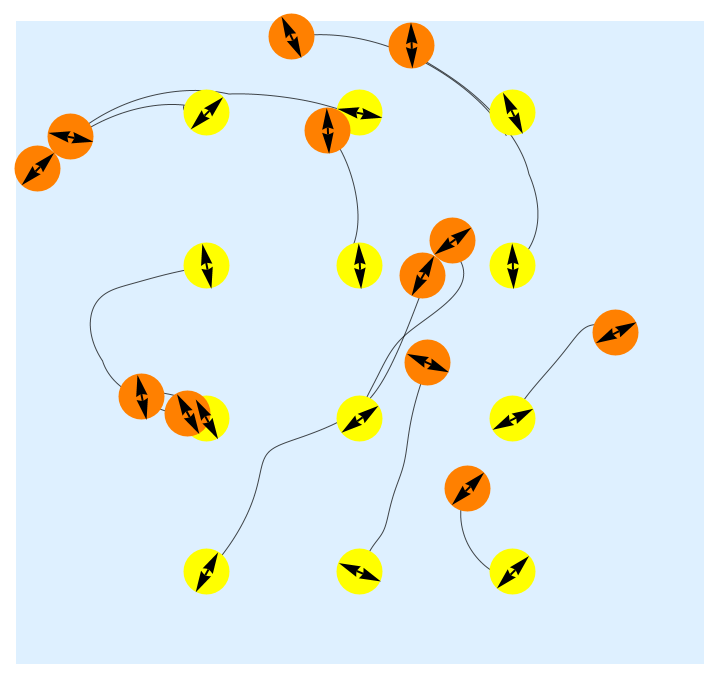} &
    \includegraphics[width=0.18\linewidth]{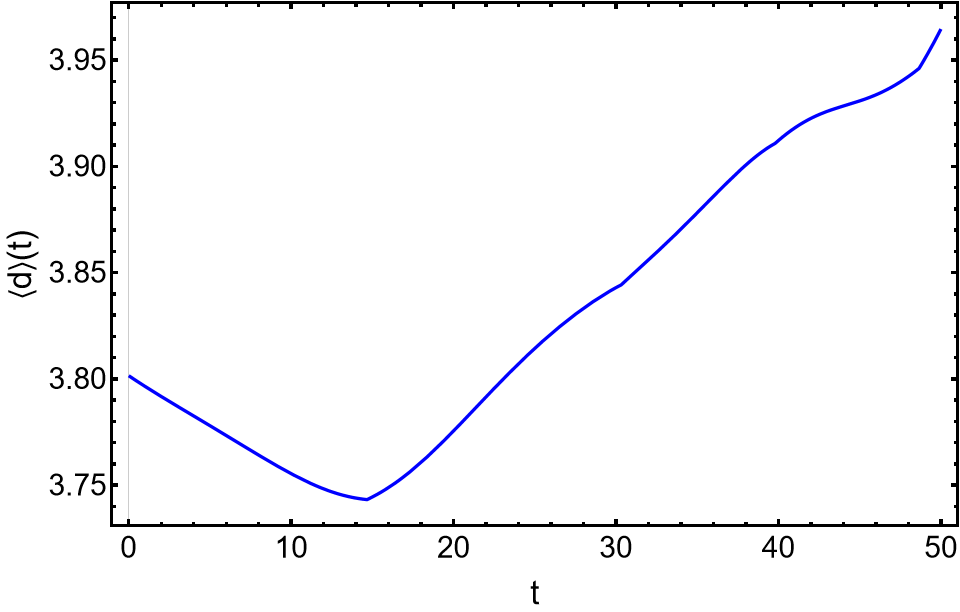} \\
  \end{tabular}

  \caption{
  Near–zone dynamics of pusher and puller dipoles in an incompressible supported membrane with
  \emph{quenched} orientations.  
  Rows correspond to axial, side-by-side, perpendicular, random, and cluster configurations.  
  Columns show trajectories and mean separations for pushers (left) and pullers (right).
  }
  \label{fig:incmp_nz_quenched_all}
\end{figure*}
\begin{figure*}[t]
  \centering
  \begin{tabular}{cccc}
    \multicolumn{4}{c}{\textbf{Incompressible membrane – Far-zone – Quenched orientations}} \\[6pt]
    \multicolumn{2}{c}{\textbf{Pusher}} & \multicolumn{2}{c}{\textbf{Puller}} \\
    traj & $\langle d_{ij}\rangle(t)$ & traj & $\langle d_{ij}\rangle(t)$ \\[6pt]

    \includegraphics[width=0.18\linewidth]{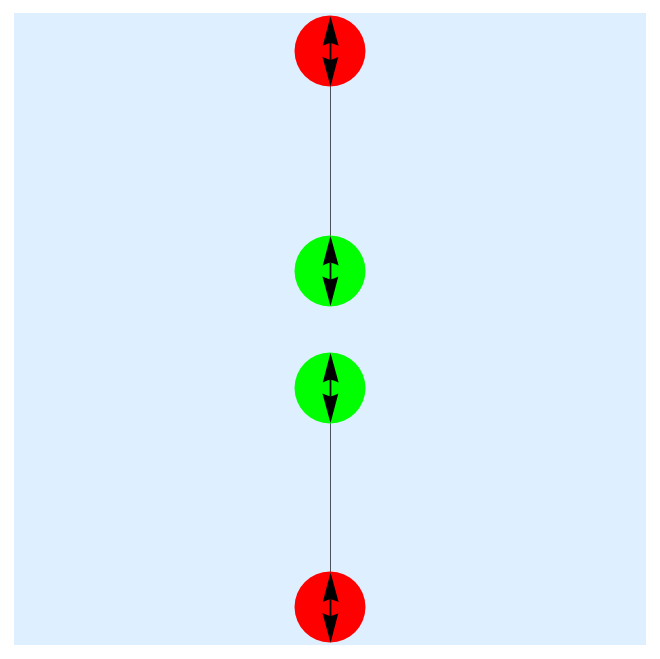} &
    \includegraphics[width=0.18\linewidth]{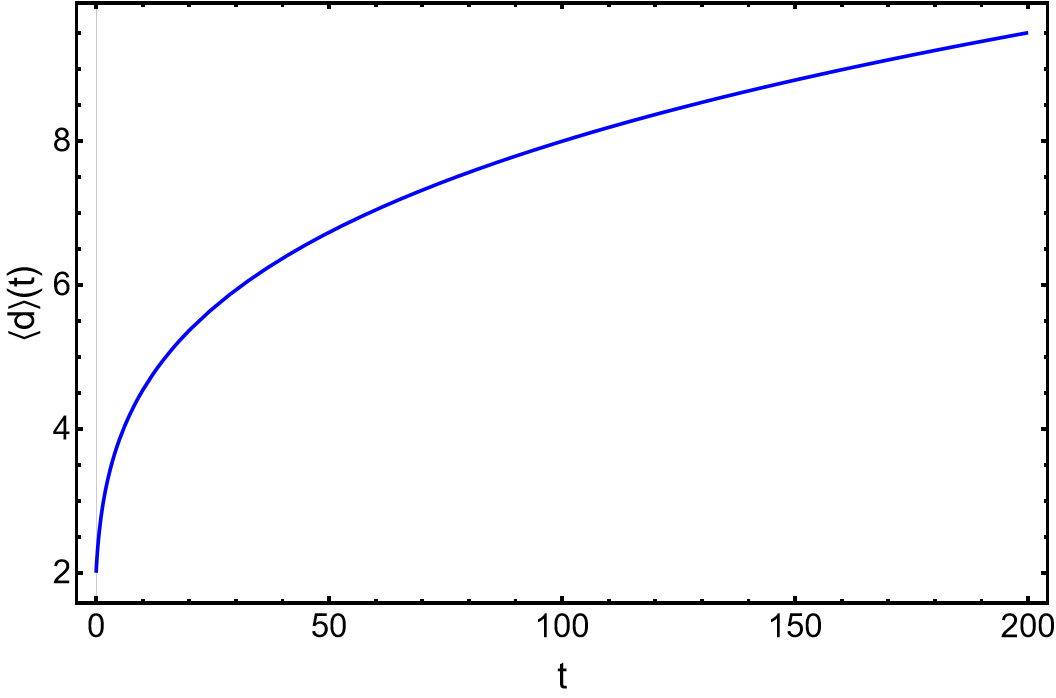} &
    \includegraphics[width=0.18\linewidth]{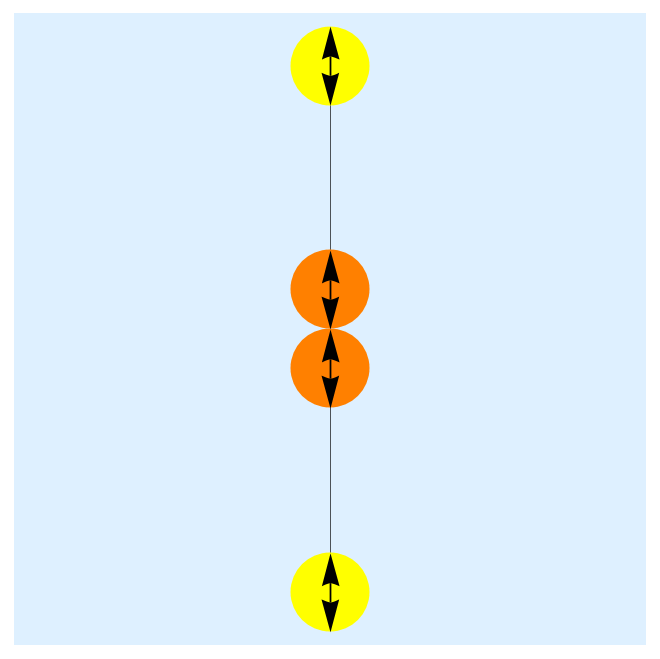} &
    \includegraphics[width=0.18\linewidth]{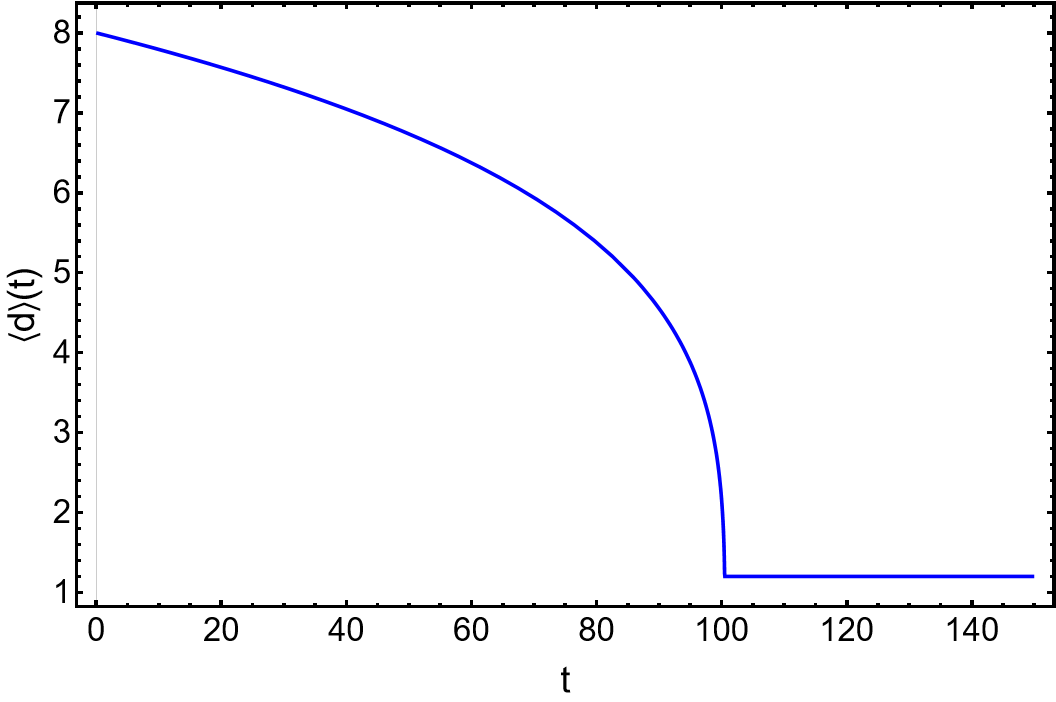} \\[4pt]

    \includegraphics[width=0.18\linewidth]{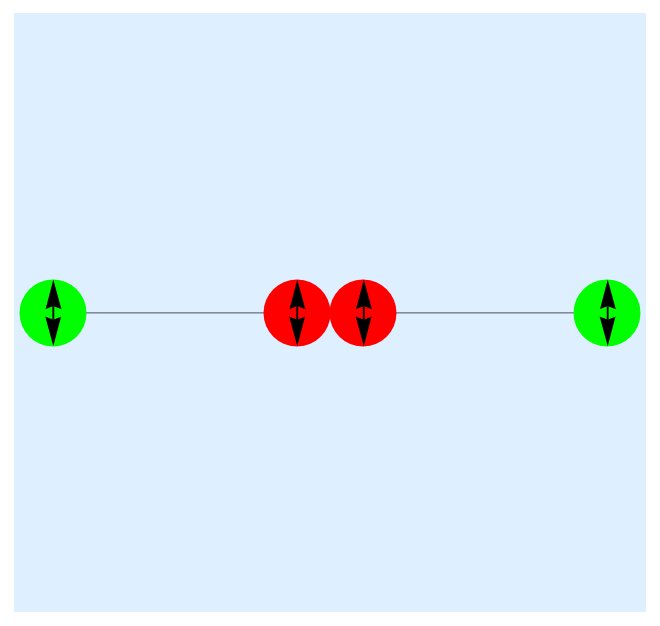} &
    \includegraphics[width=0.18\linewidth]{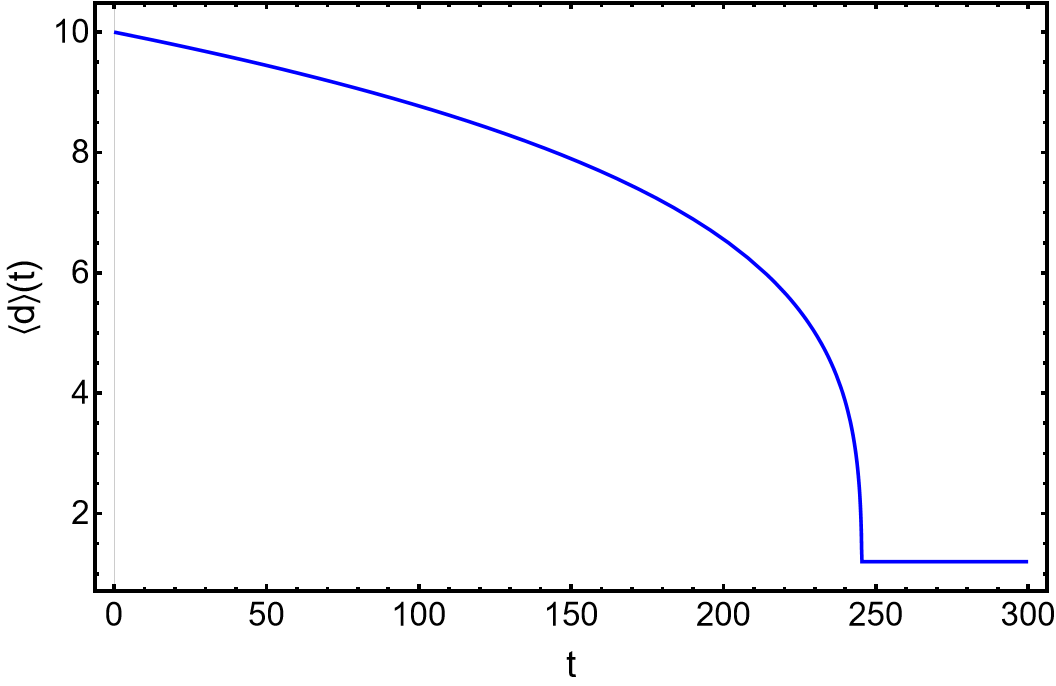} &
    \includegraphics[width=0.18\linewidth]{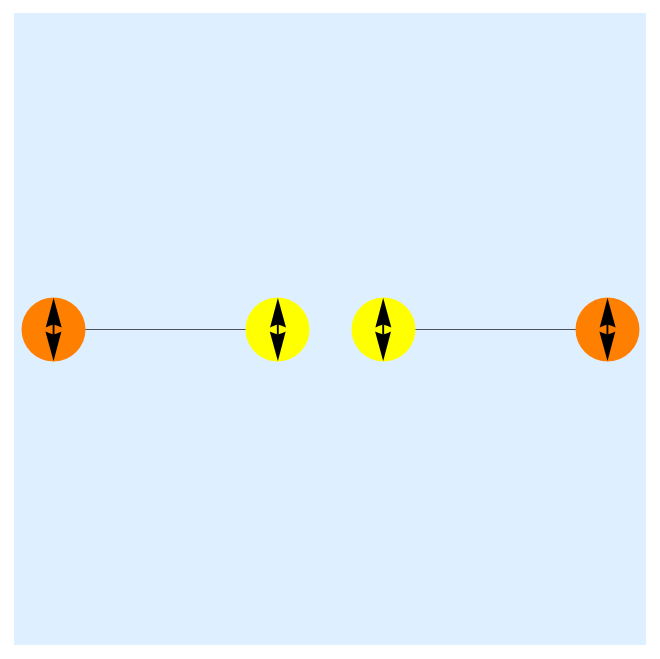} &
    \includegraphics[width=0.18\linewidth]{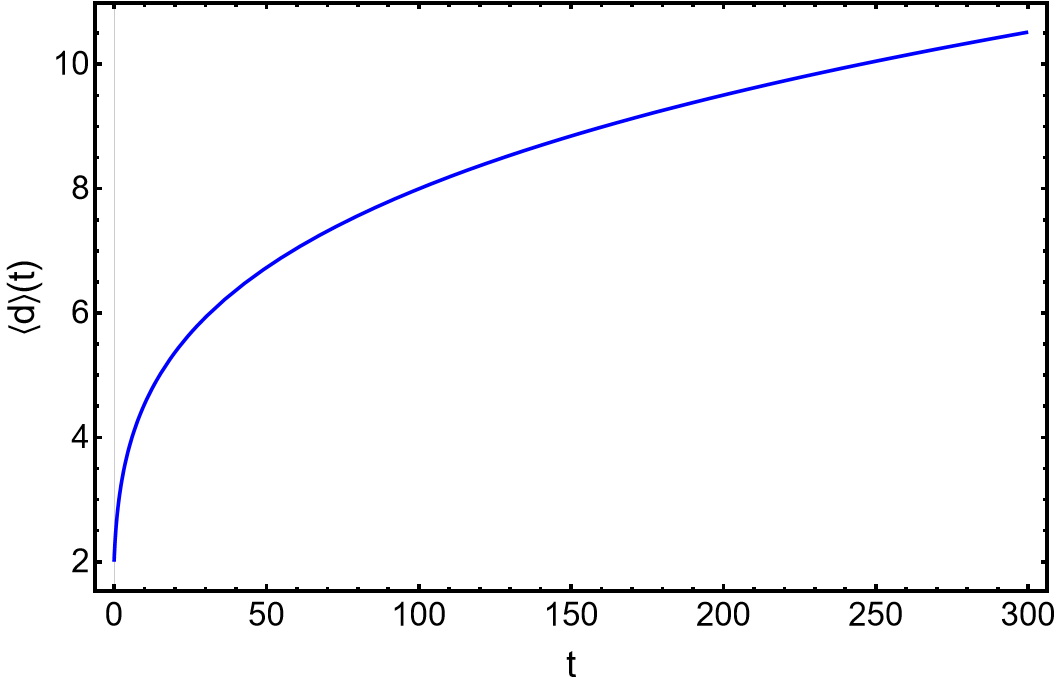} \\[4pt]

    \includegraphics[width=0.18\linewidth]{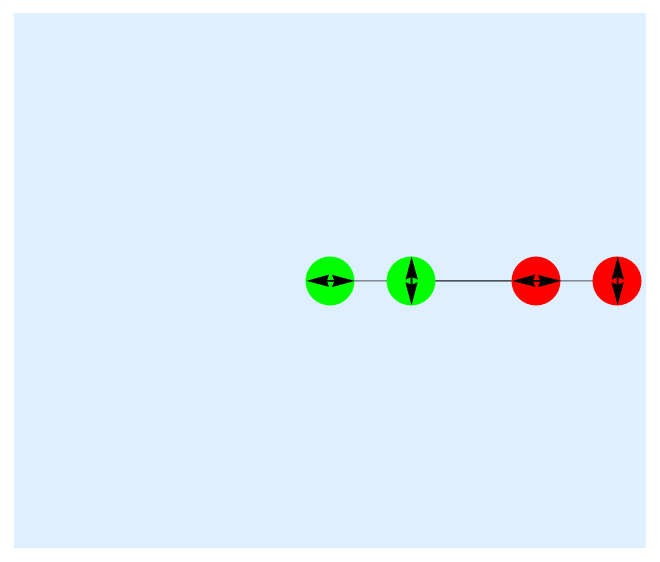} &
    \includegraphics[width=0.18\linewidth]{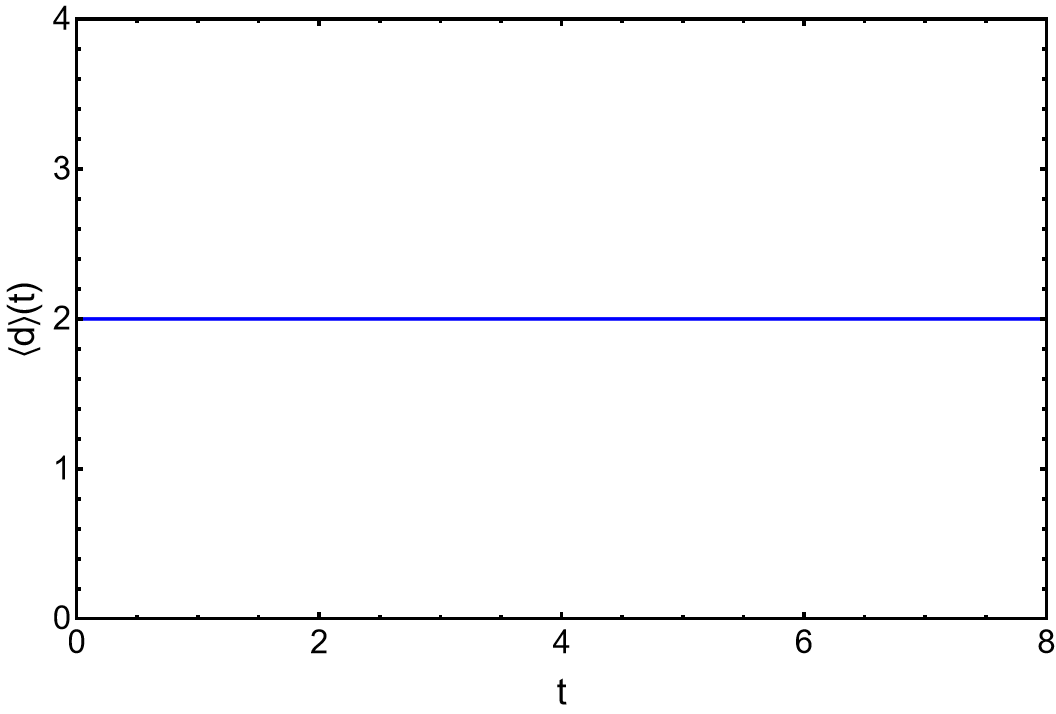} &
    \includegraphics[width=0.18\linewidth]{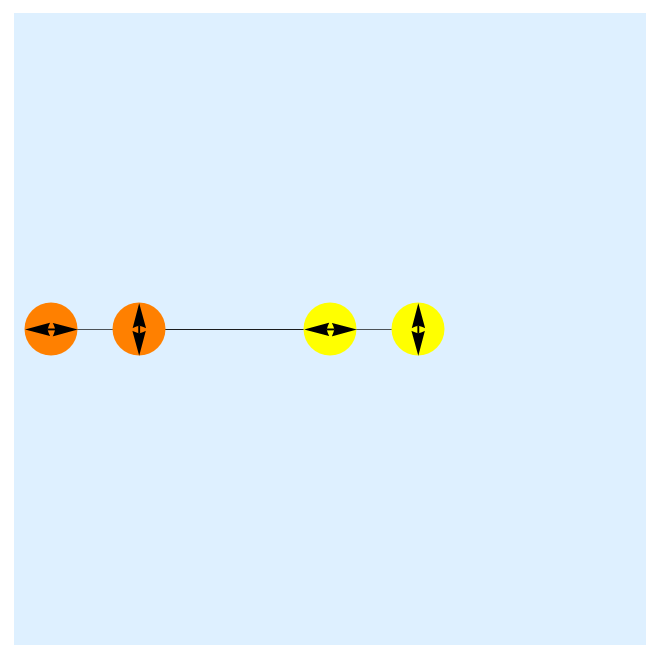} &
    \includegraphics[width=0.18\linewidth]{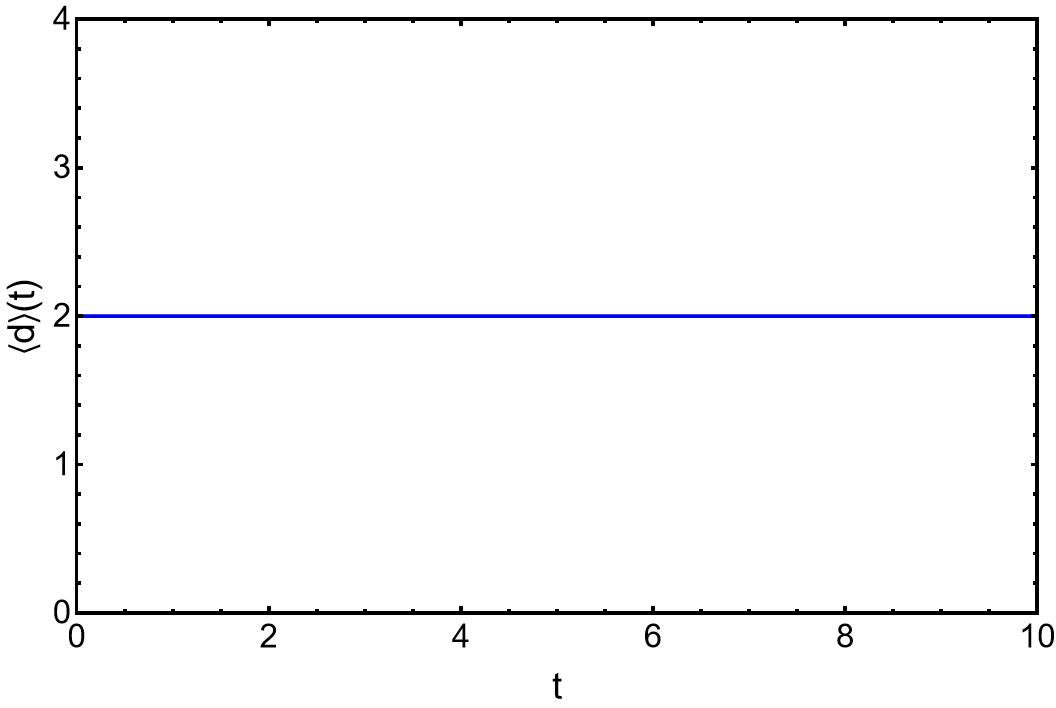} \\[4pt]

    \includegraphics[width=0.18\linewidth]{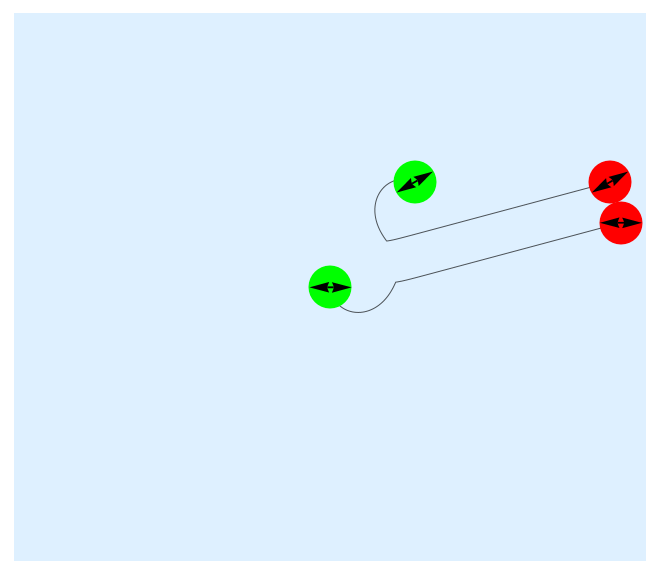} &
    \includegraphics[width=0.18\linewidth]{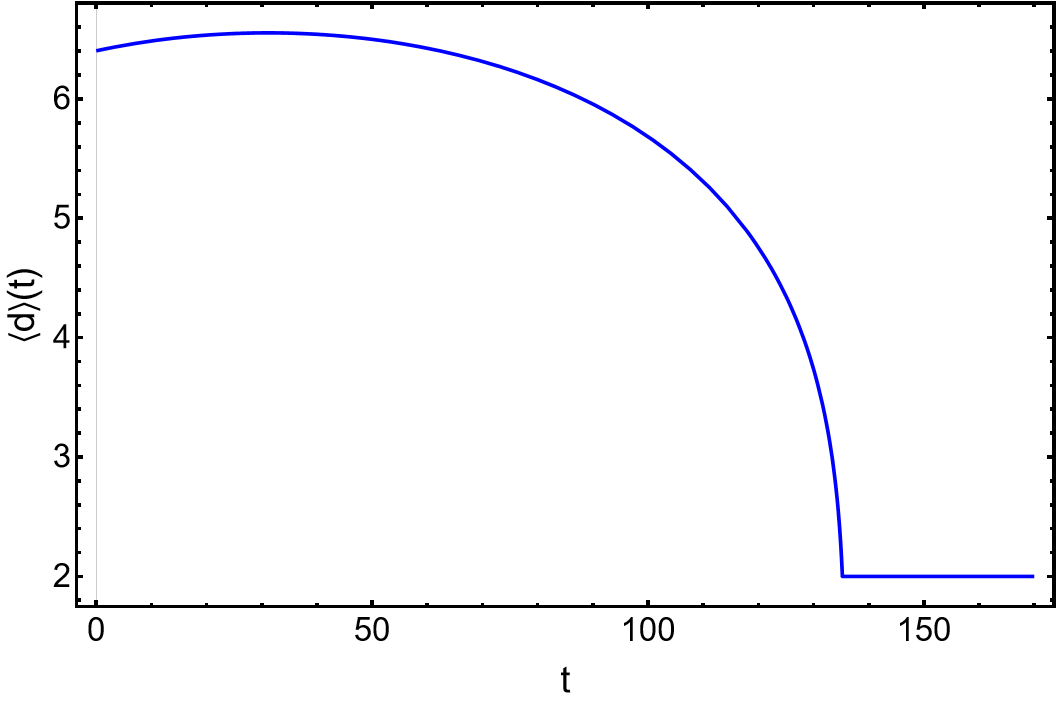} &
    \includegraphics[width=0.18\linewidth]{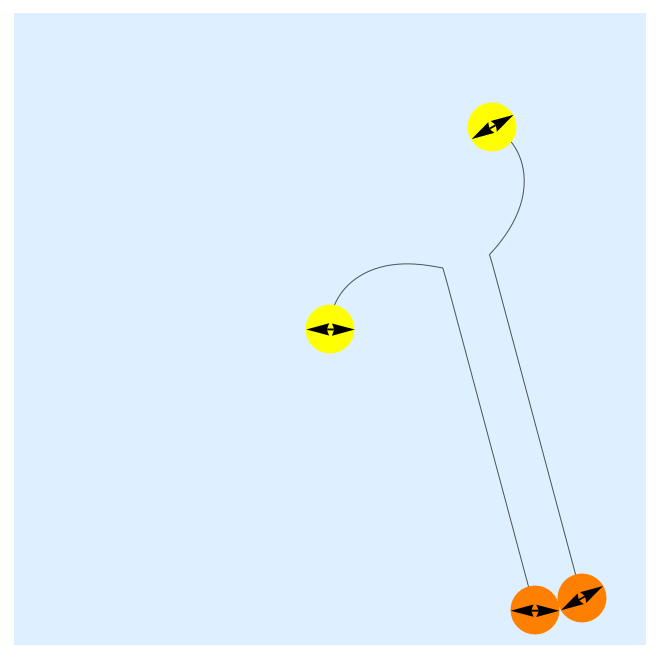} &
    \includegraphics[width=0.18\linewidth]{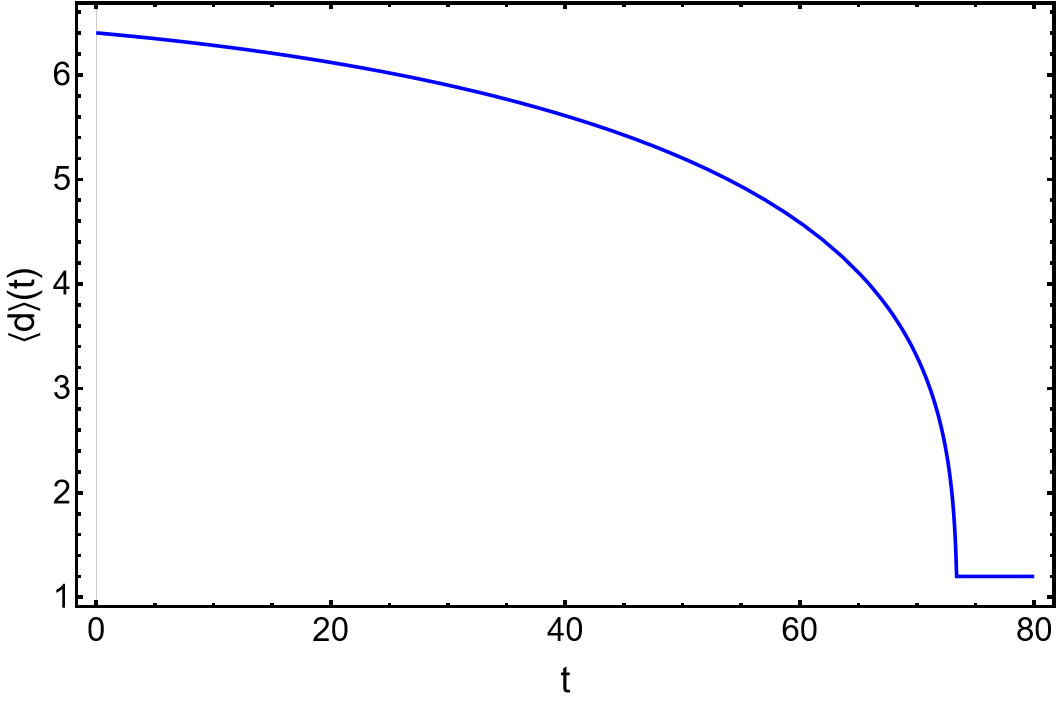} \\[4pt]

    \includegraphics[width=0.18\linewidth]{images/incmp_nq_fz_pusher_cluster} &
    \includegraphics[width=0.18\linewidth]{images/incmp_nq_fz_pusher_cluster_dist} &
    \includegraphics[width=0.18\linewidth]{images/incmp_nq_fz_puller_cluster} &
    \includegraphics[width=0.18\linewidth]{images/incmp_nq_fz_puller_cluster_dist} \\
  \end{tabular}

  \caption{
  Far–zone dynamics of pusher and puller dipoles in an incompressible supported membrane with
  \emph{quenched} orientations.  
  Rows correspond to axial, side-by-side, perpendicular, random, and cluster configurations.  
  Columns show trajectories and mean separations for pushers (left) and pullers (right).
  }
  \label{fig:incmp_fz_quenched_all}
\end{figure*}

\begin{figure*}[t]
  \centering
  \begin{tabular}{cccc}
    \multicolumn{4}{c}{\textbf{Compressible membrane (zero odd viscosity) – Near zone – Dynamical orientations}} \\[6pt]
    \multicolumn{2}{c}{\textbf{Pusher}} & \multicolumn{2}{c}{\textbf{Puller}} \\
    traj & $\langle d_{ij}\rangle(t)$ & traj & $\langle d_{ij}\rangle(t)$ \\[6pt]

    \includegraphics[width=0.18\linewidth]{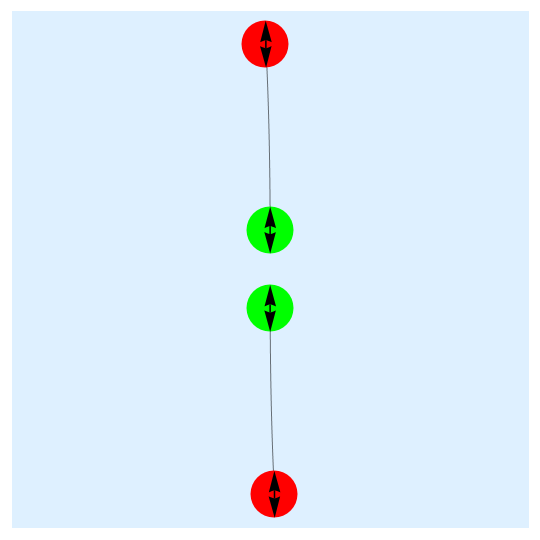} &
    \includegraphics[width=0.18\linewidth]{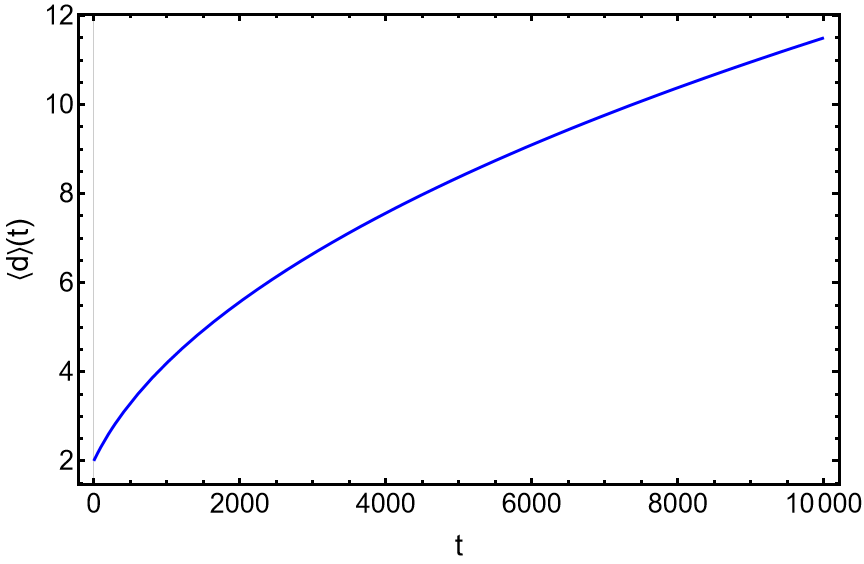} &
    \includegraphics[width=0.18\linewidth]{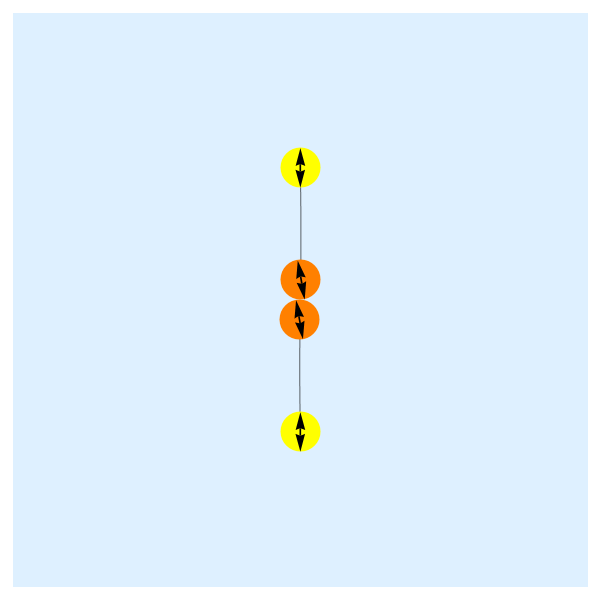} &
    \includegraphics[width=0.18\linewidth]{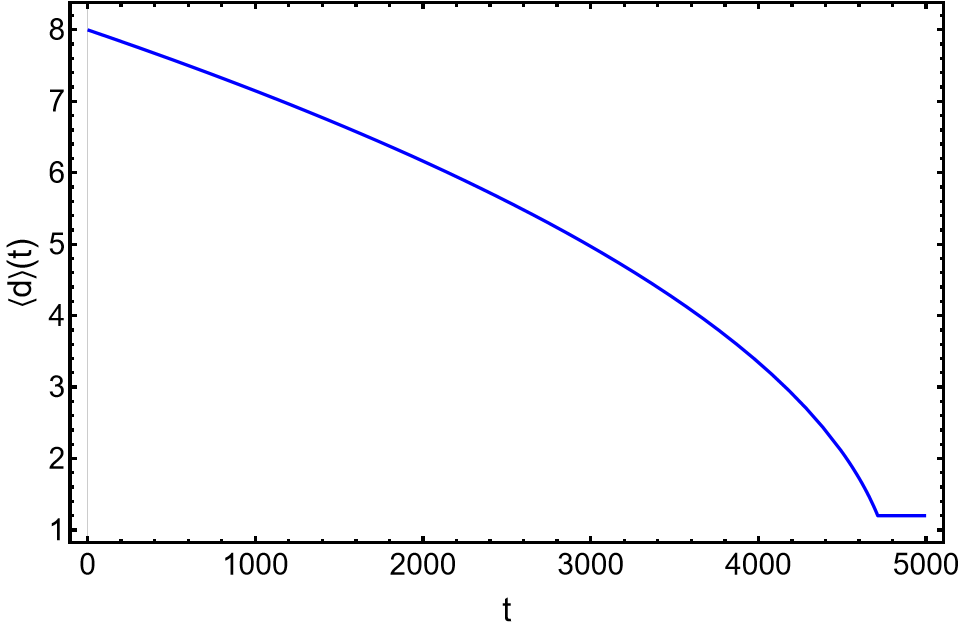} \\[4pt]

    \includegraphics[width=0.18\linewidth]{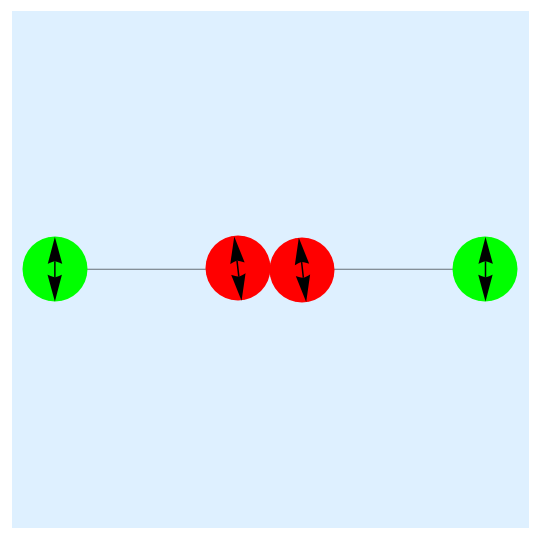} &
    \includegraphics[width=0.18\linewidth]{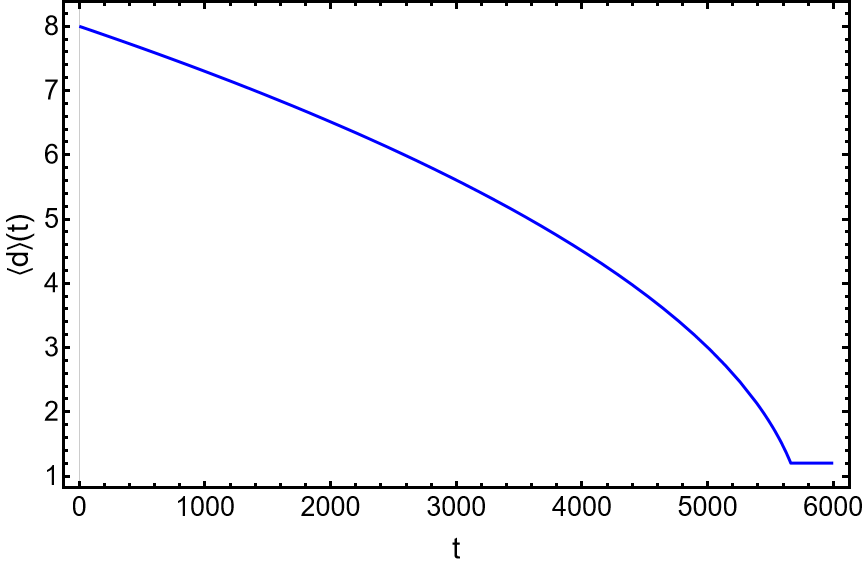} &
    \includegraphics[width=0.18\linewidth]{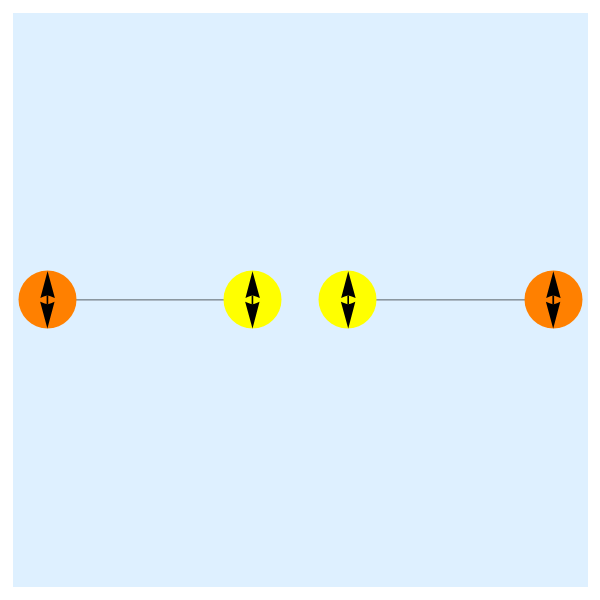} &
    \includegraphics[width=0.18\linewidth]{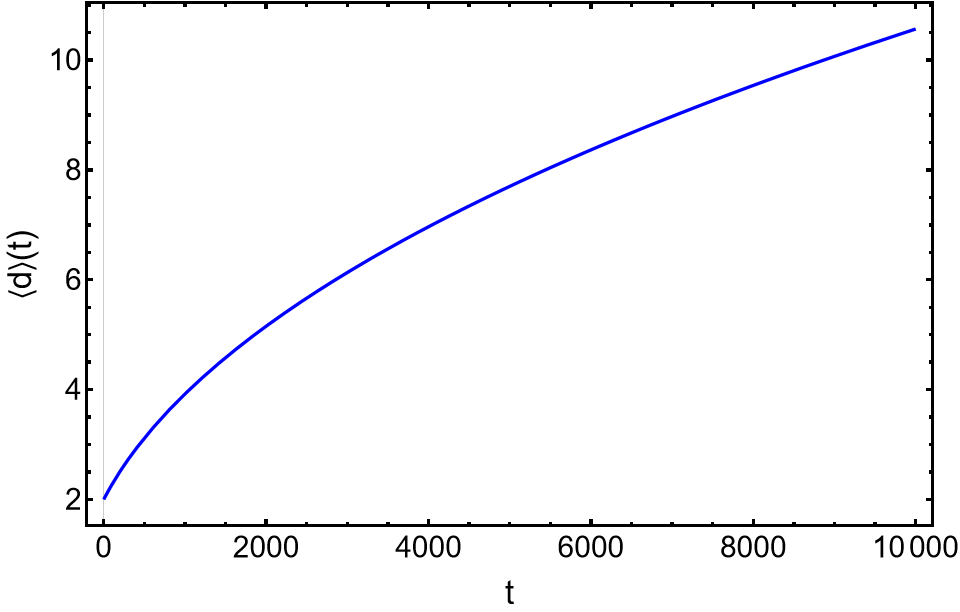} \\[4pt]

    \includegraphics[width=0.18\linewidth]{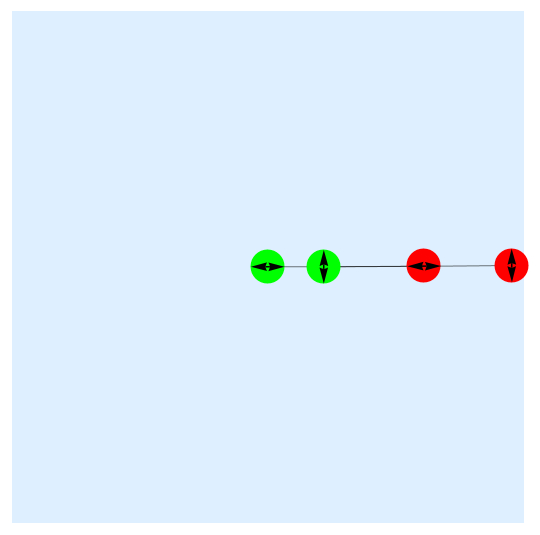} &
    \includegraphics[width=0.18\linewidth]{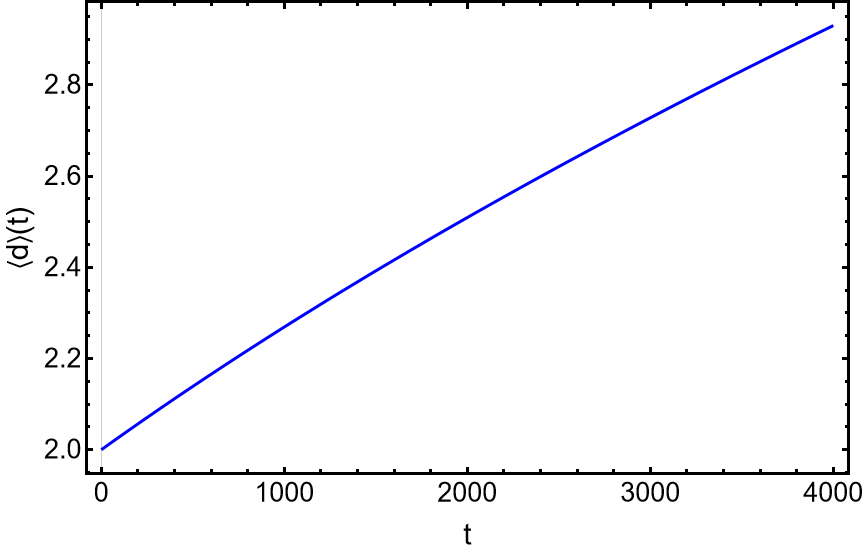} &
    \includegraphics[width=0.18\linewidth]{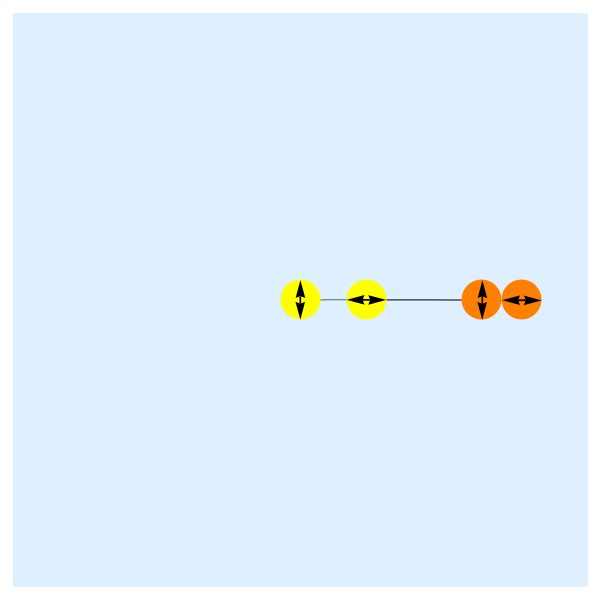} &
    \includegraphics[width=0.18\linewidth]{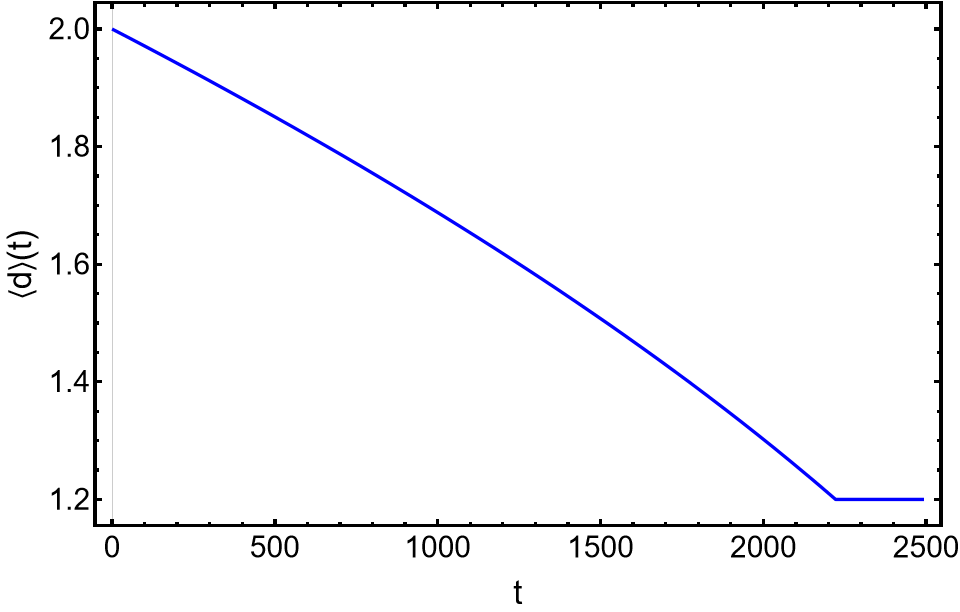} \\[4pt]

    \includegraphics[width=0.18\linewidth]{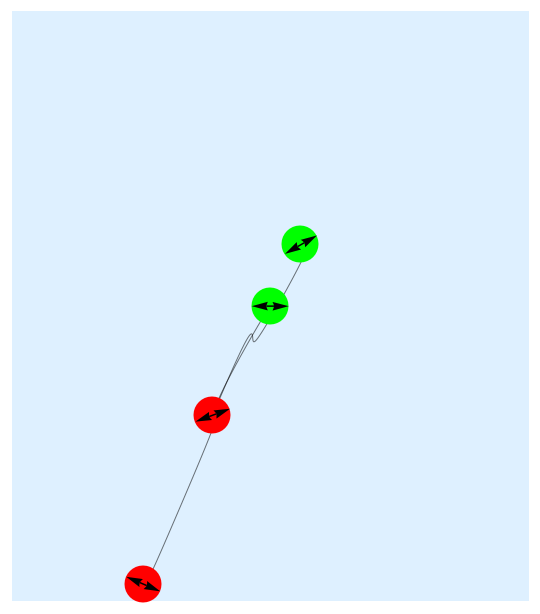} &
    \includegraphics[width=0.18\linewidth]{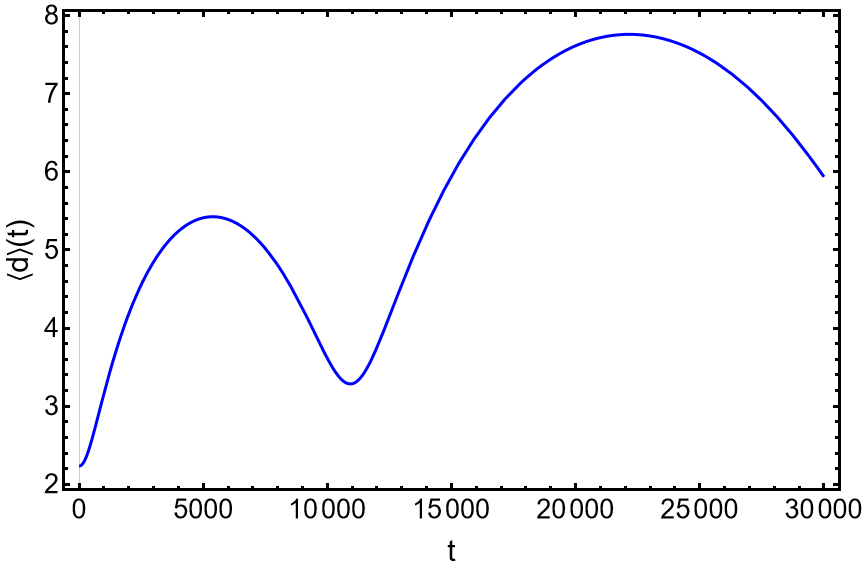} &
    \includegraphics[width=0.18\linewidth]{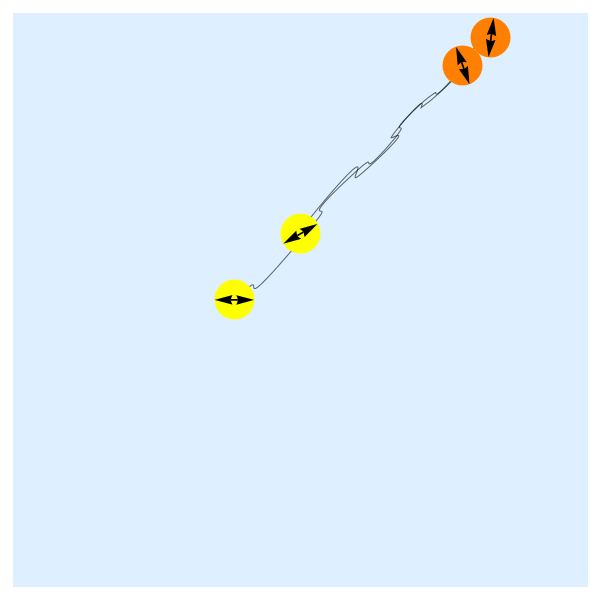}&
    \includegraphics[width=0.18\linewidth]{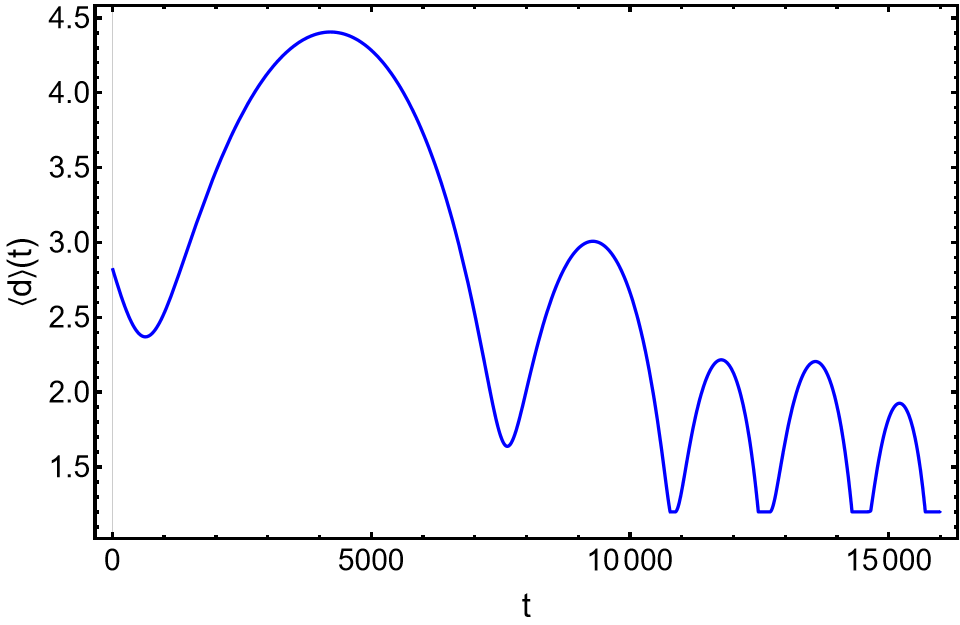}\\[4pt]

    \includegraphics[width=0.18\linewidth]{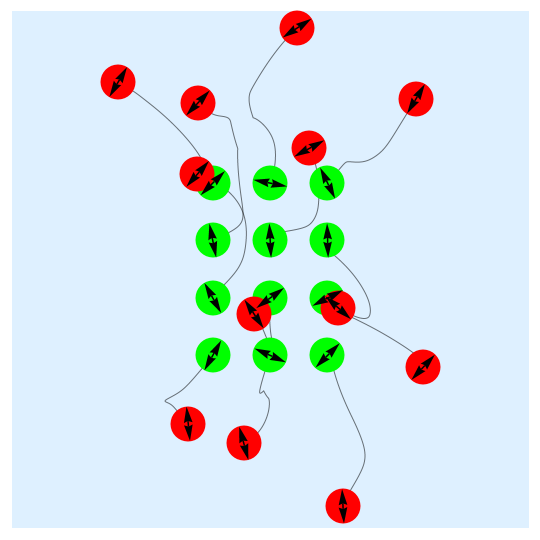} &
    \includegraphics[width=0.18\linewidth]{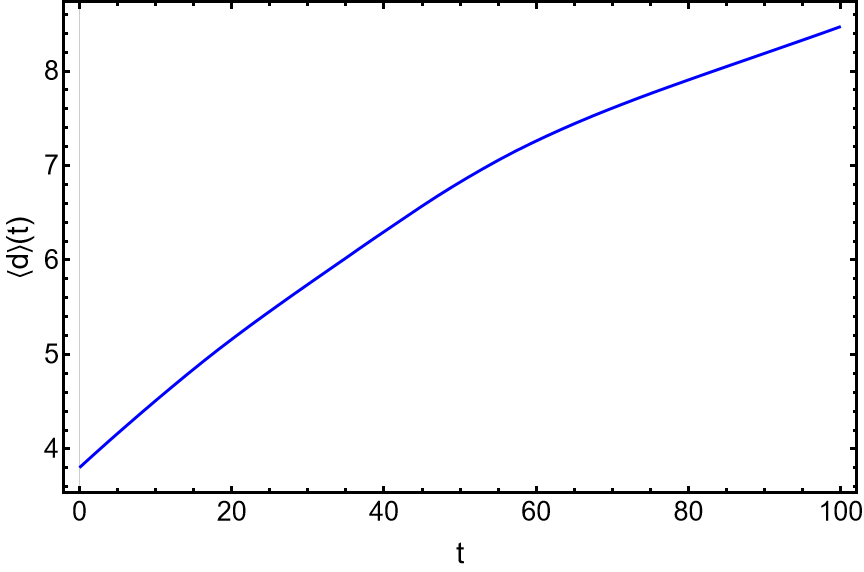} &
    \includegraphics[width=0.18\linewidth]{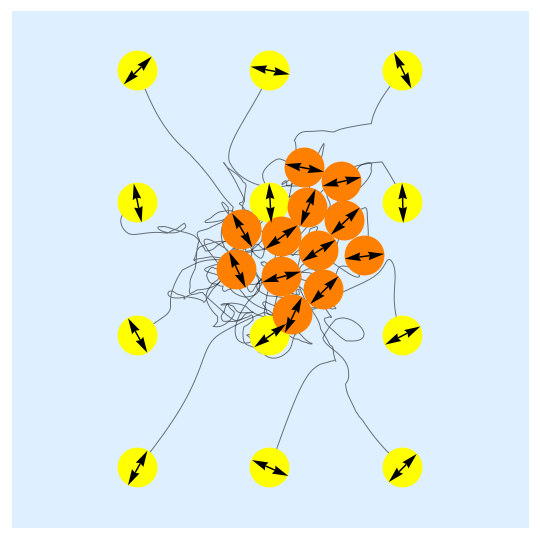} &
    \includegraphics[width=0.18\linewidth]{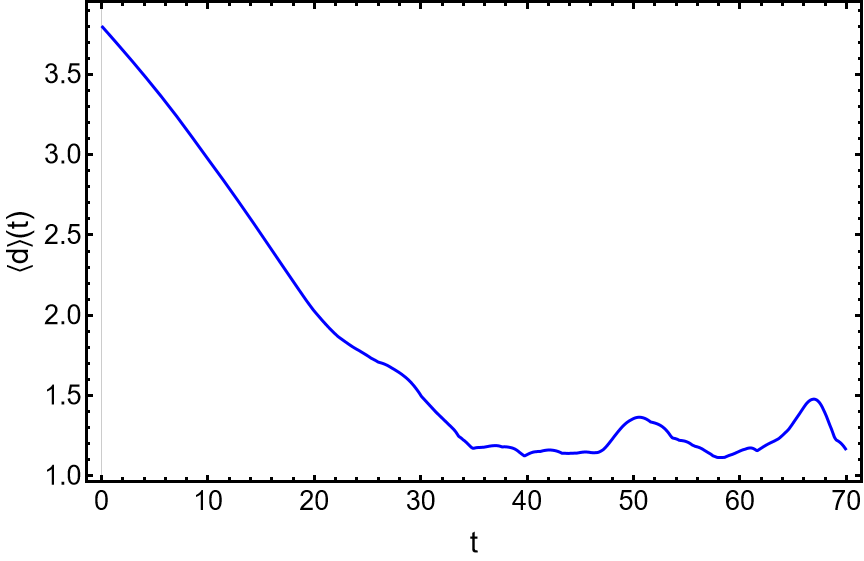} \\
  \end{tabular}

  \caption{
  Near–zone dynamics of pusher and puller dipoles in a compressible supported membrane
  with vanishing odd viscosity and \emph{dynamical} orientations.  
  Rows: axial, side-by-side, perpendicular, random pair, and 12-dipole cluster.  
  Columns: trajectories and mean pair separation for pushers (left) and pullers (right).
  }
  \label{fig:cmp_nz_dyn_all}
\end{figure*}
\begin{figure*}[t]
  \centering
  \begin{tabular}{cccc}
    \multicolumn{4}{c}{\textbf{Compressible membrane (zero odd viscosity) – Far zone – Dynamical orientations}} \\[6pt]
    \multicolumn{2}{c}{\textbf{Pusher}} & \multicolumn{2}{c}{\textbf{Puller}} \\
    traj & $\langle d_{ij}\rangle(t)$ & traj & $\langle d_{ij}\rangle(t)$ \\[6pt]

    \includegraphics[width=0.18\linewidth]{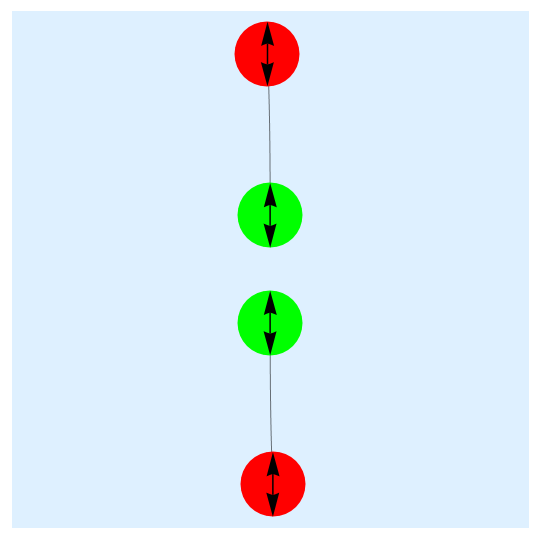} &
    \includegraphics[width=0.18\linewidth]{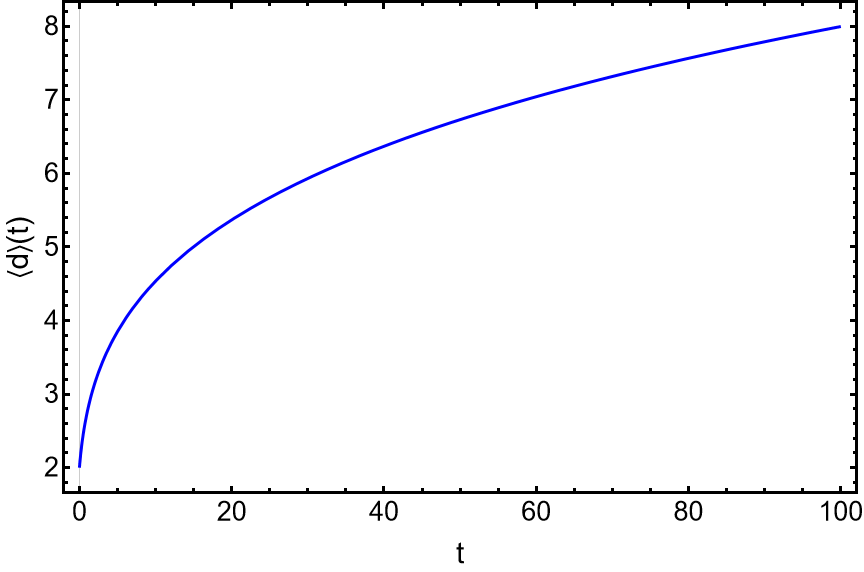} &
    \includegraphics[width=0.18\linewidth]{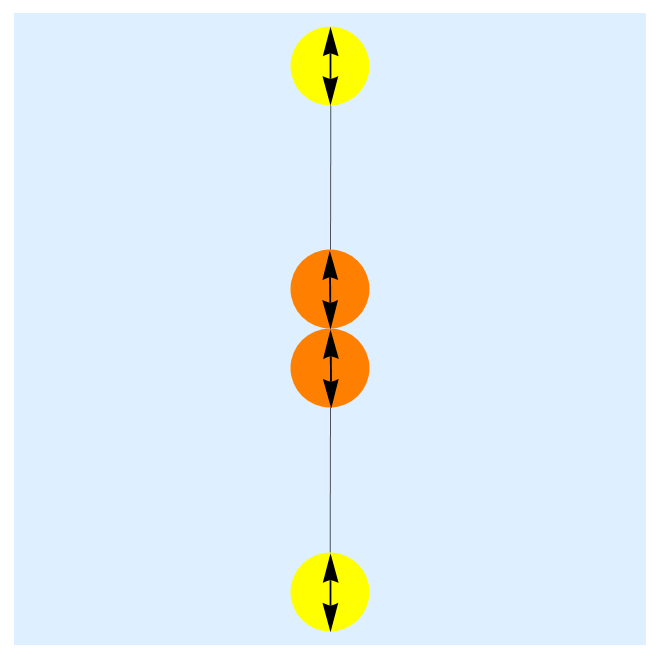} &
    \includegraphics[width=0.18\linewidth]{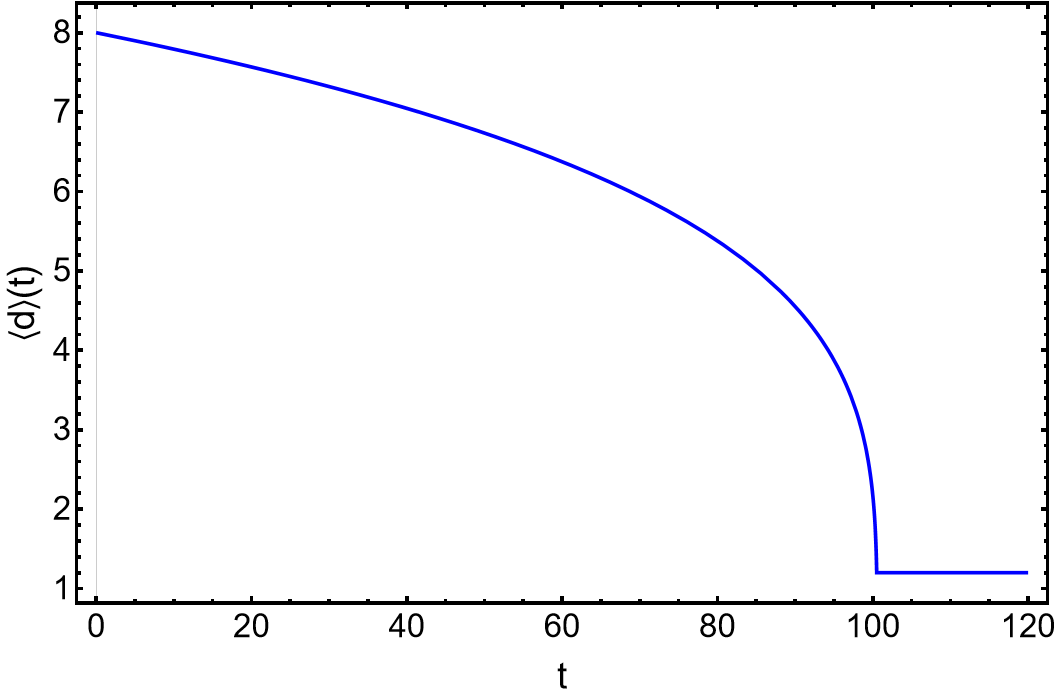} \\[4pt]

    \includegraphics[width=0.18\linewidth]{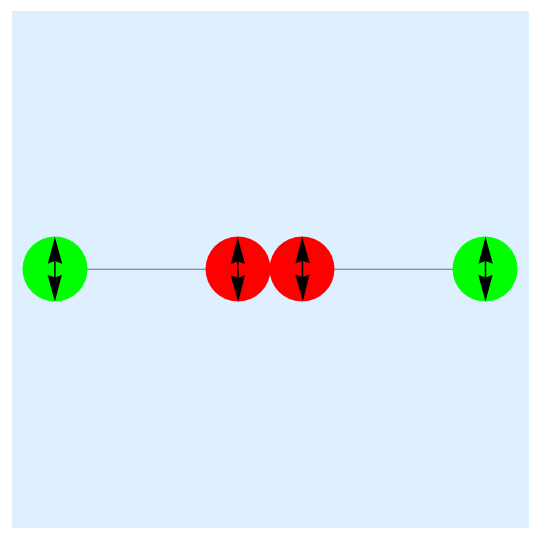} &
    \includegraphics[width=0.18\linewidth]{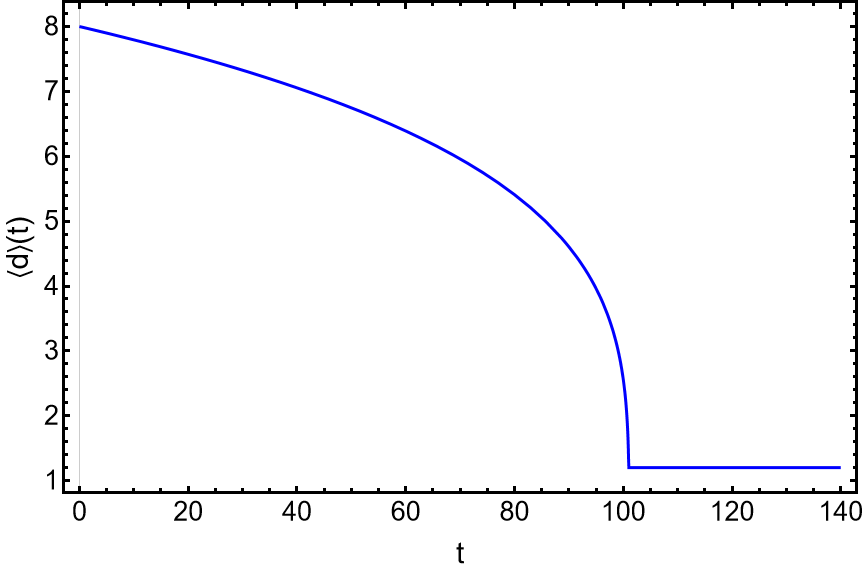} &
    \includegraphics[width=0.18\linewidth]{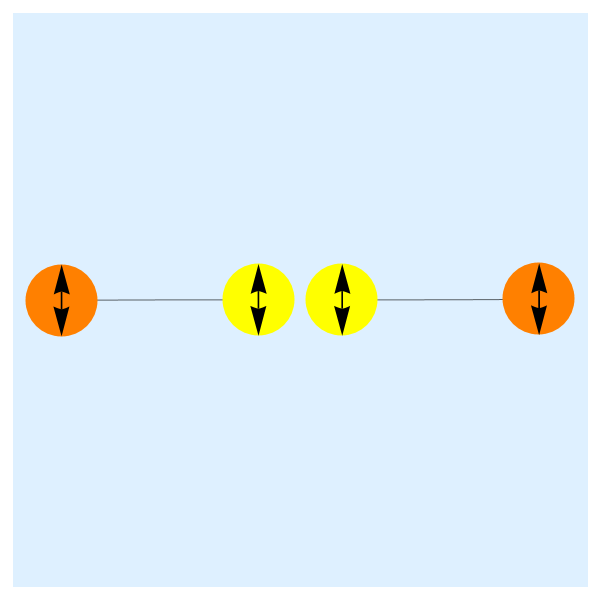} &
    \includegraphics[width=0.18\linewidth]{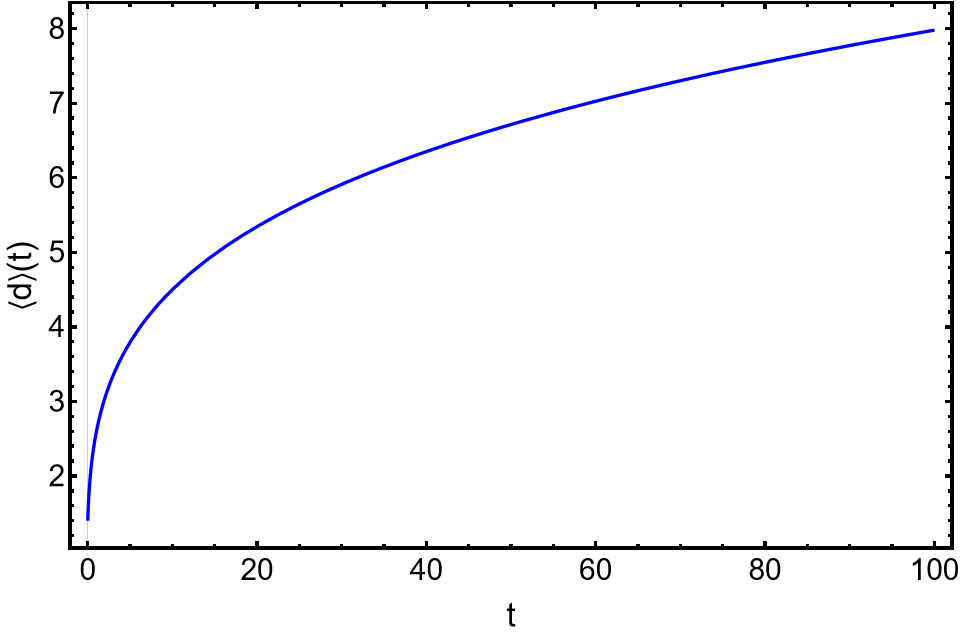} \\[4pt]

    \includegraphics[width=0.18\linewidth]{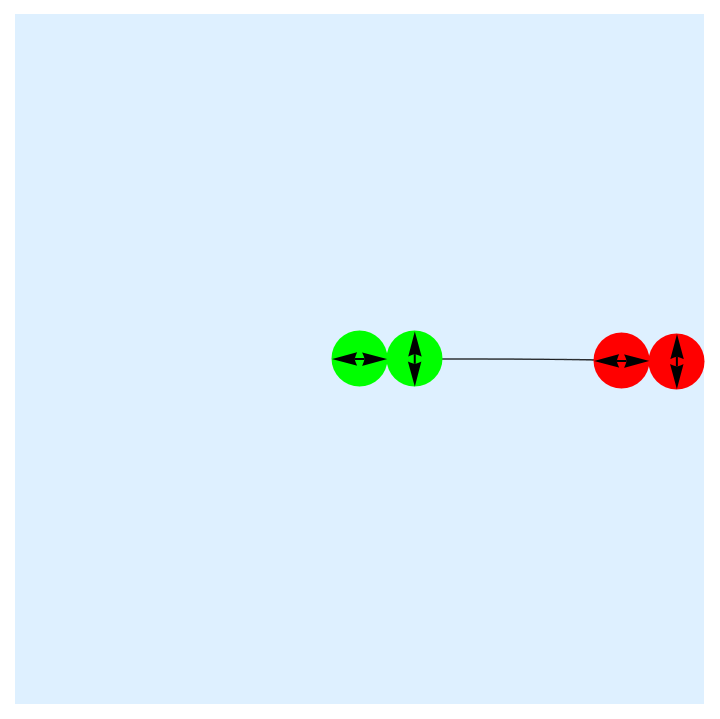} &
    \includegraphics[width=0.18\linewidth]{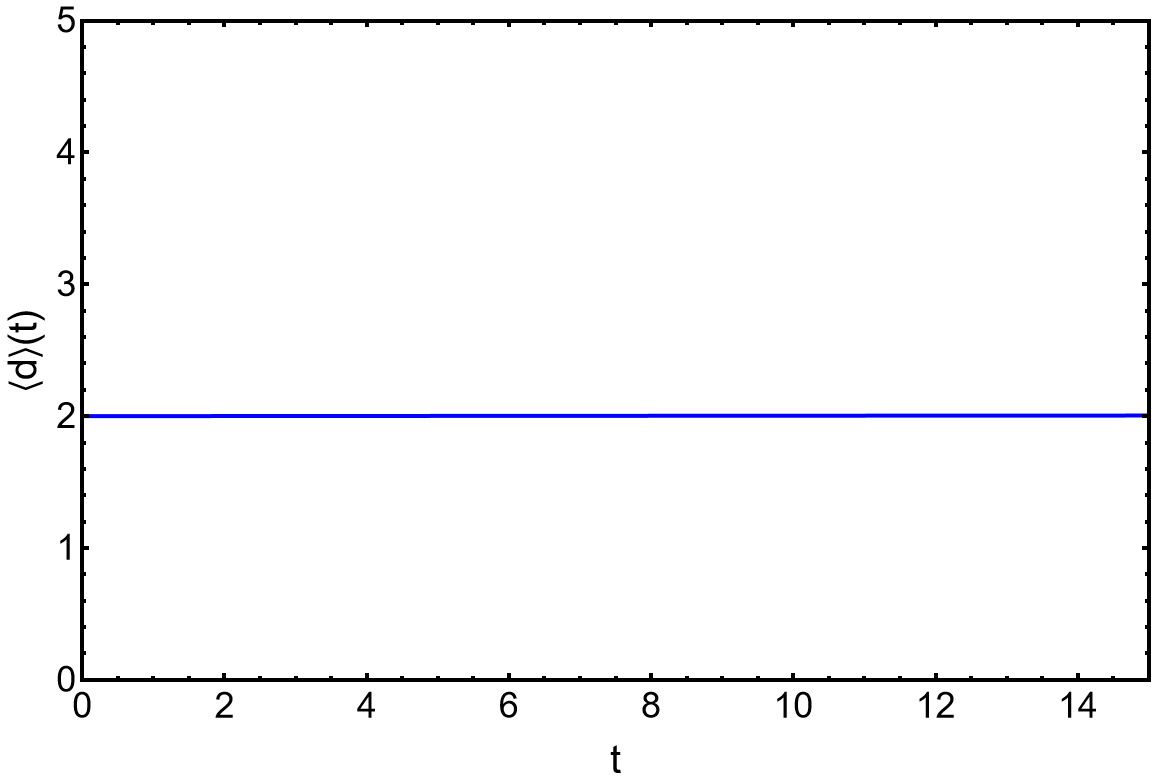} &
    \includegraphics[width=0.18\linewidth]{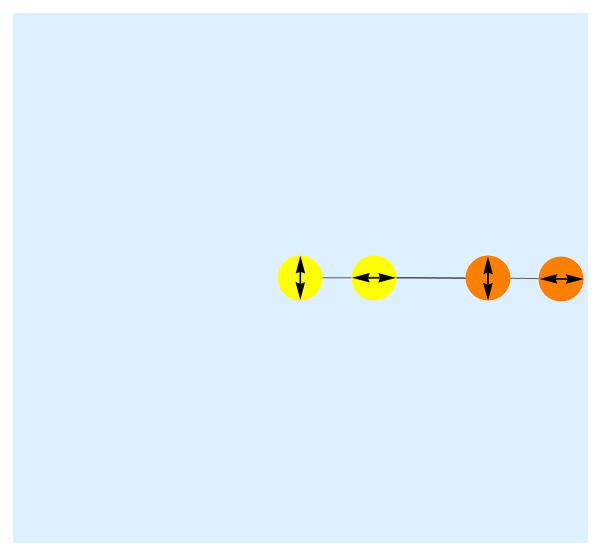} &
    \includegraphics[width=0.18\linewidth]{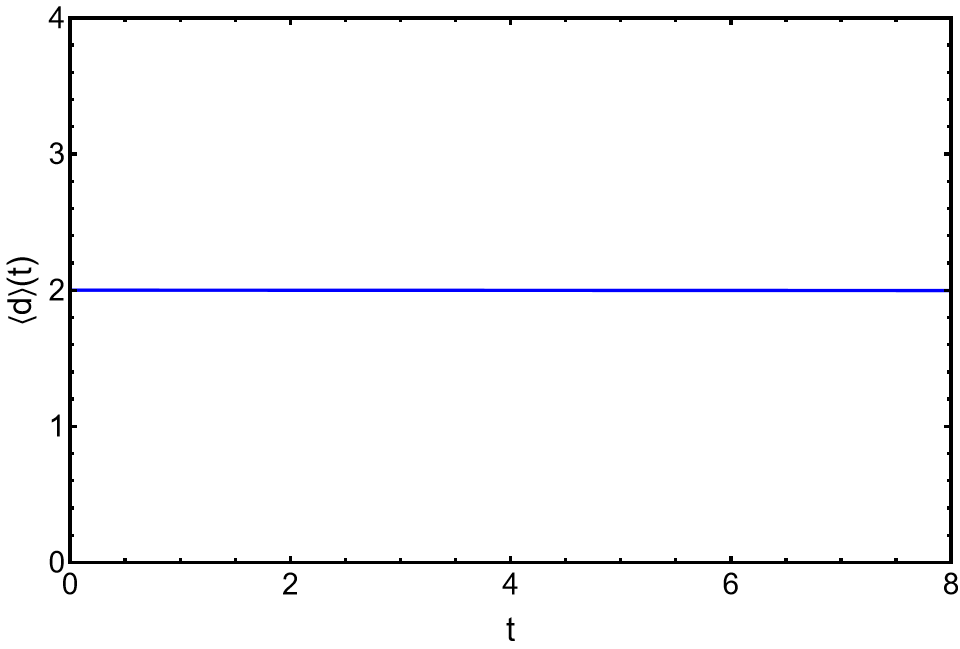} \\[4pt]

    \includegraphics[width=0.18\linewidth]{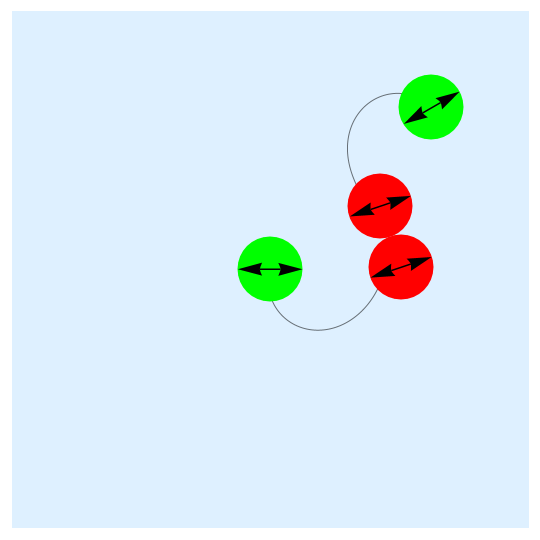} &
    \includegraphics[width=0.18\linewidth]{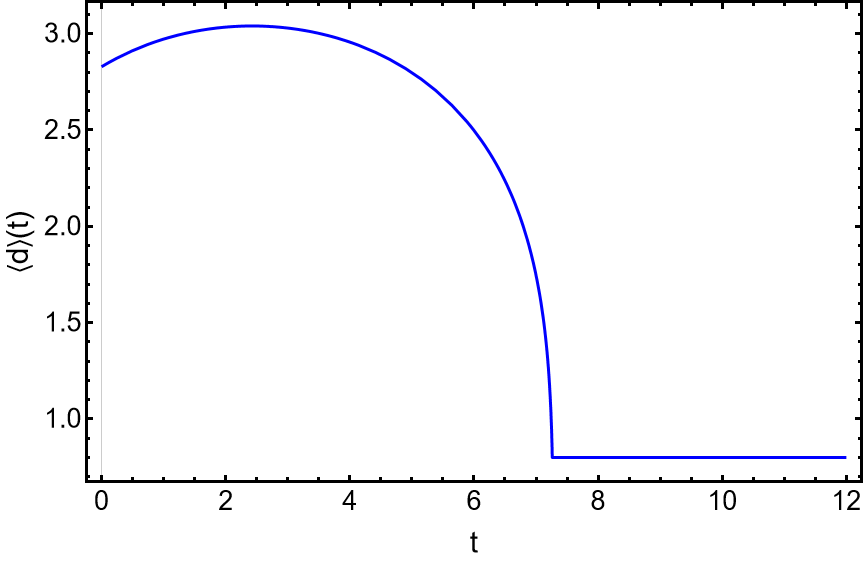} &
    \includegraphics[width=0.18\linewidth]{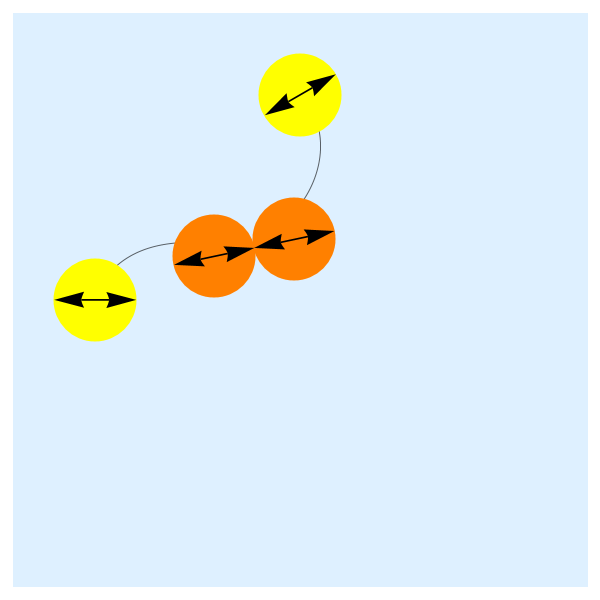} &
    \includegraphics[width=0.18\linewidth]{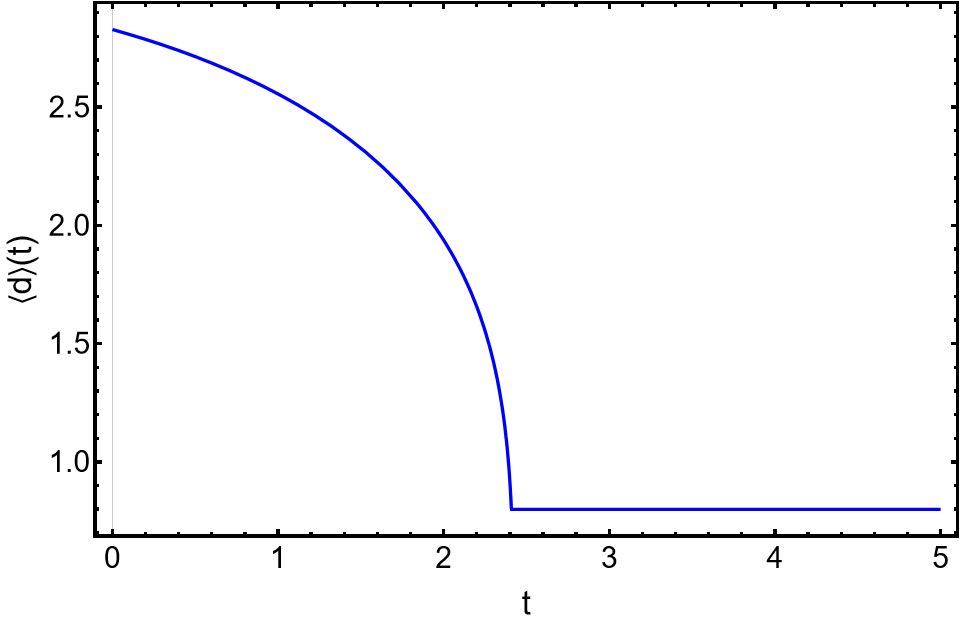} \\[4pt]

    \includegraphics[width=0.18\linewidth]{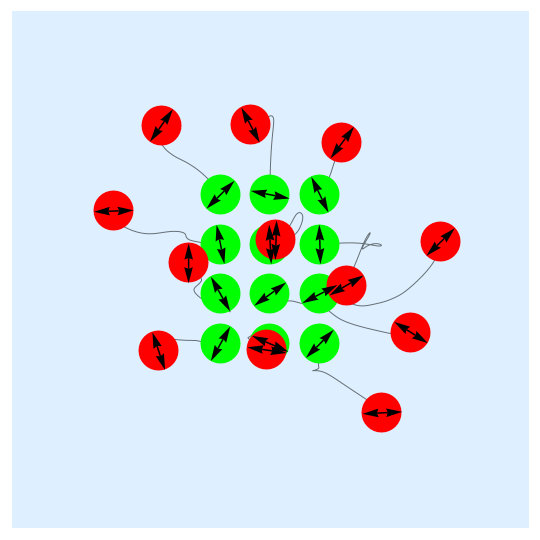} &
    \includegraphics[width=0.18\linewidth]{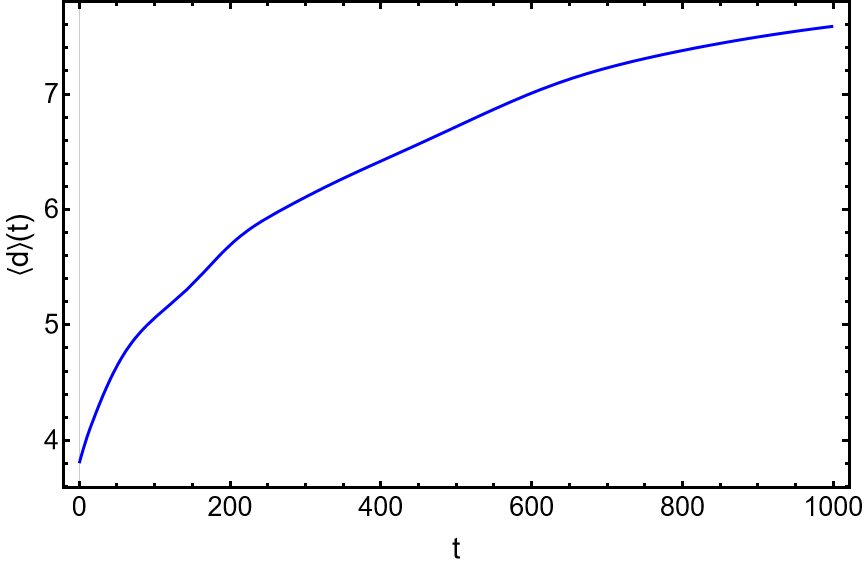} &
    \includegraphics[width=0.18\linewidth]{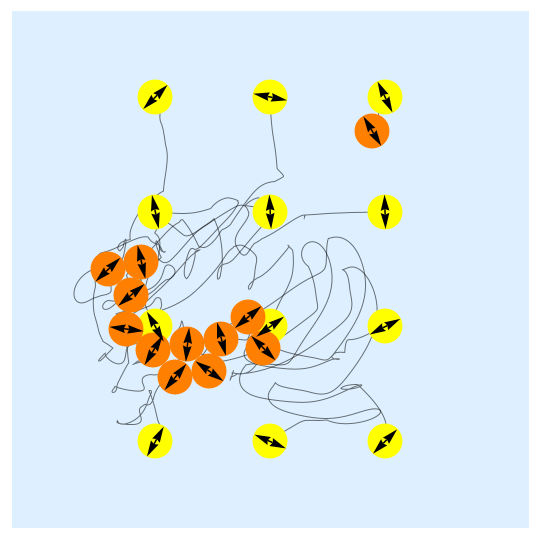} &
    \includegraphics[width=0.18\linewidth]{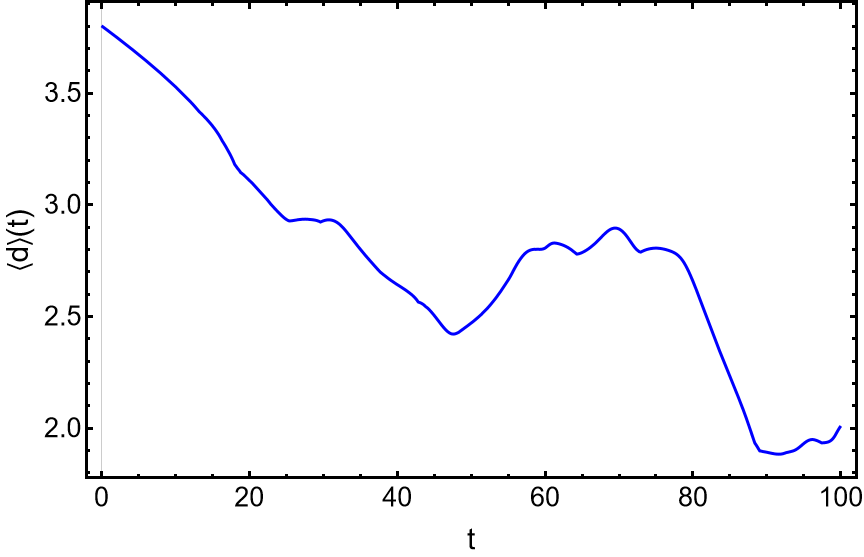} \\
  \end{tabular}

  \caption{
  Far–zone dynamics of pusher and puller dipoles in a compressible supported membrane
  with vanishing odd viscosity and \emph{dynamical} orientations.  
  Rows: axial, side-by-side, perpendicular, random pair, and 12-dipole cluster.  
  Columns: trajectories and mean pair separation for pushers (left) and pullers (right).
  }
  \label{fig:cmp_fz_dyn_all}
\end{figure*}
\begin{figure*}[t]
  \centering
  \begin{tabular}{cccc}
    \multicolumn{4}{c}{\textbf{Compressible membrane (zero odd viscosity) – Near zone – Quenched orientations}} \\[6pt]
    \multicolumn{2}{c}{\textbf{Pusher}} & \multicolumn{2}{c}{\textbf{Puller}} \\
    traj & $\langle d_{ij}\rangle(t)$ & traj & $\langle d_{ij}\rangle(t)$ \\[6pt]

    \includegraphics[width=0.18\linewidth]{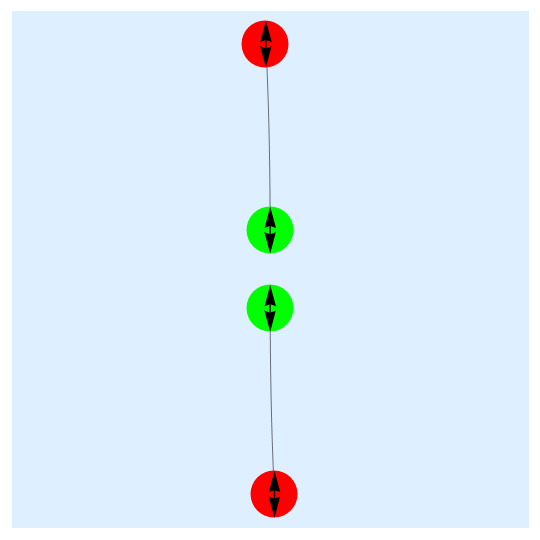} &
    \includegraphics[width=0.18\linewidth]{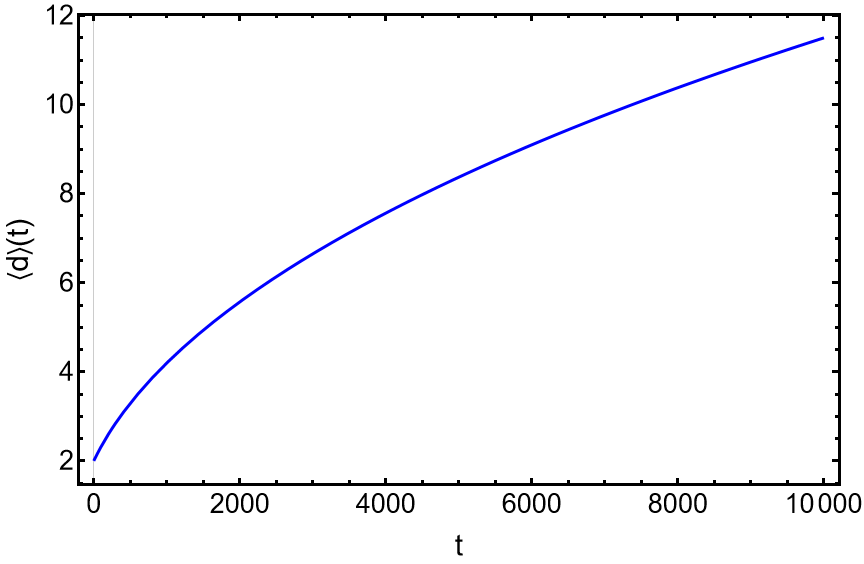} &
    \includegraphics[width=0.18\linewidth]{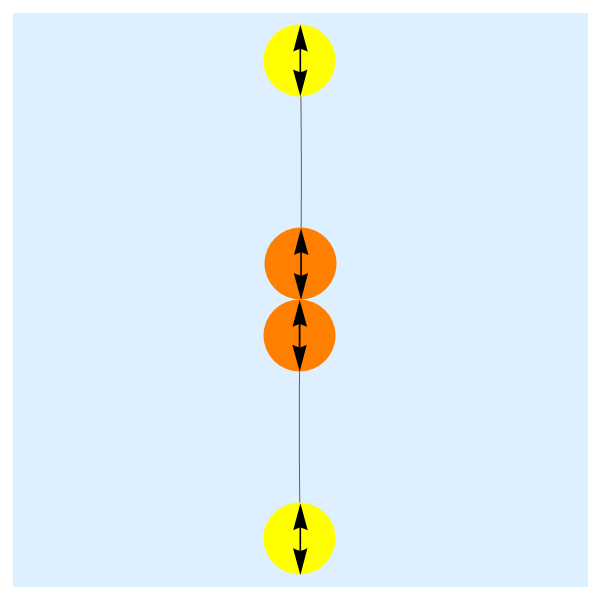} &
    \includegraphics[width=0.18\linewidth]{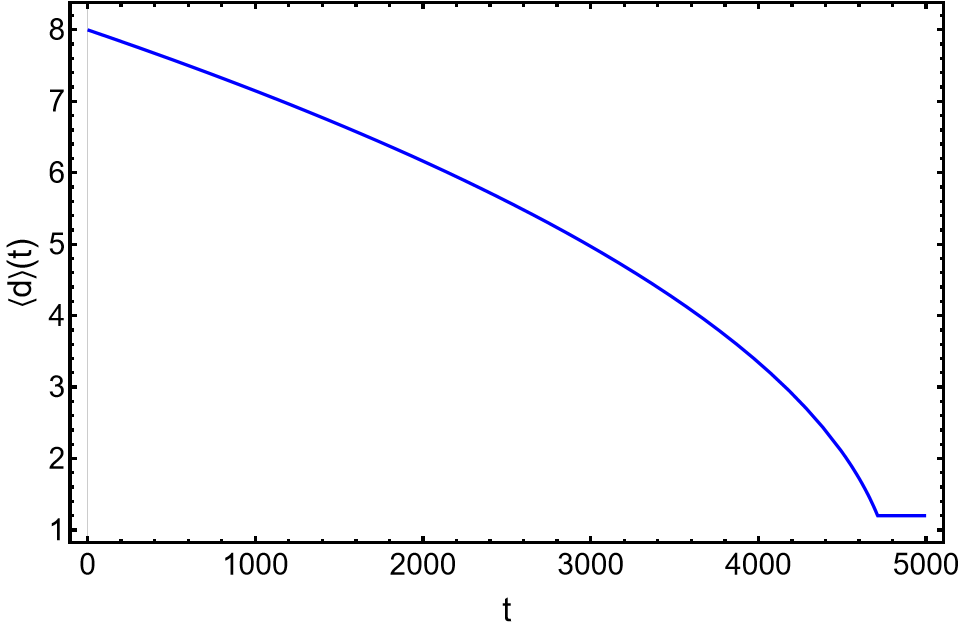} \\[4pt]

    \includegraphics[width=0.18\linewidth]{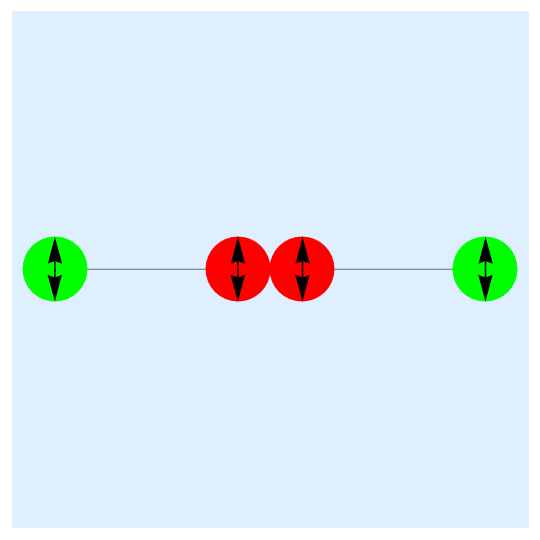} &
    \includegraphics[width=0.18\linewidth]{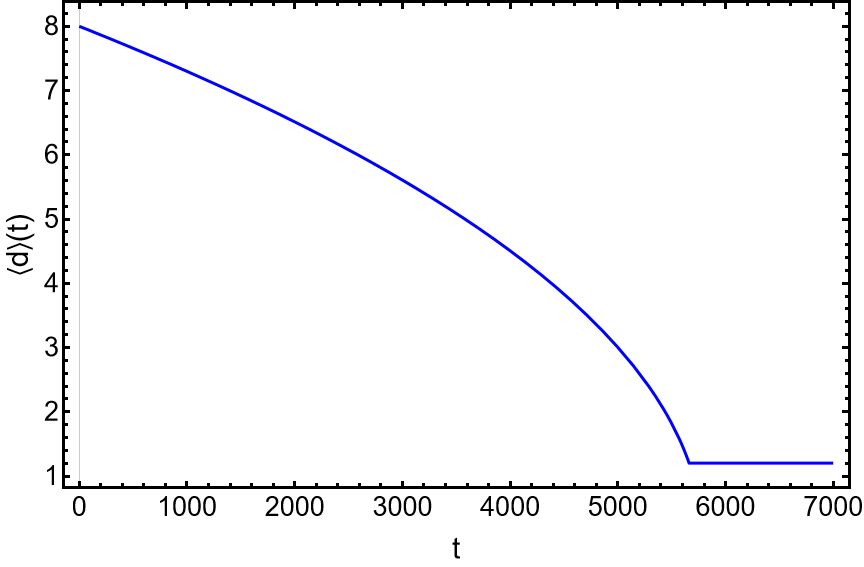} &
    \includegraphics[width=0.18\linewidth]{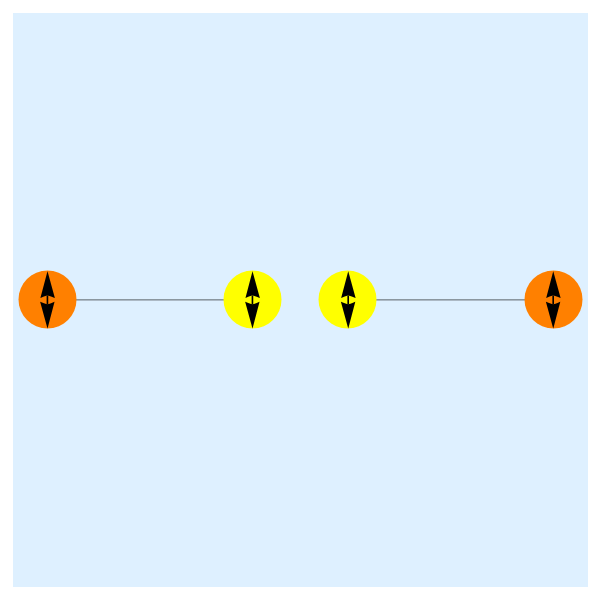} &
    \includegraphics[width=0.18\linewidth]{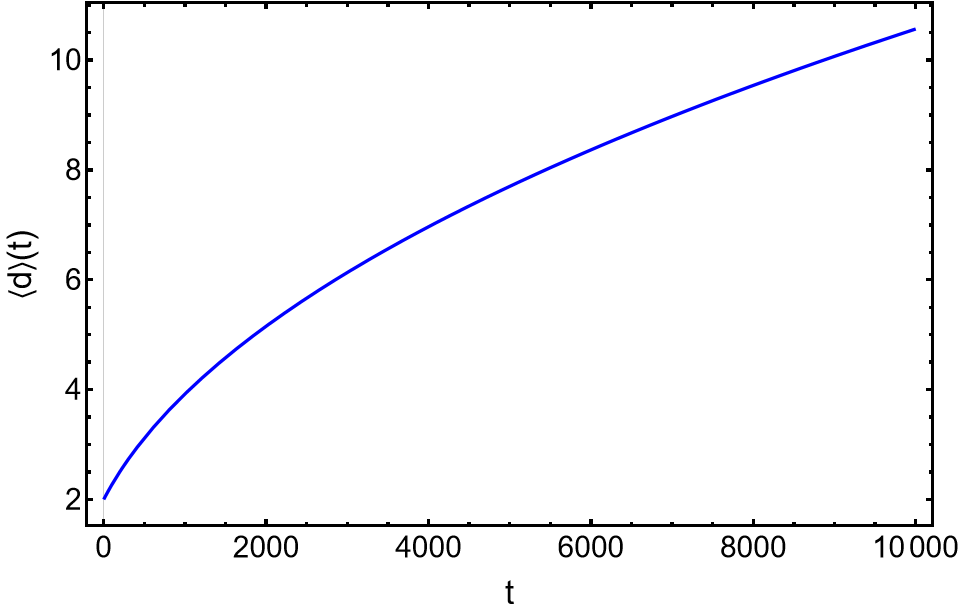} \\[4pt]

    \includegraphics[width=0.18\linewidth]{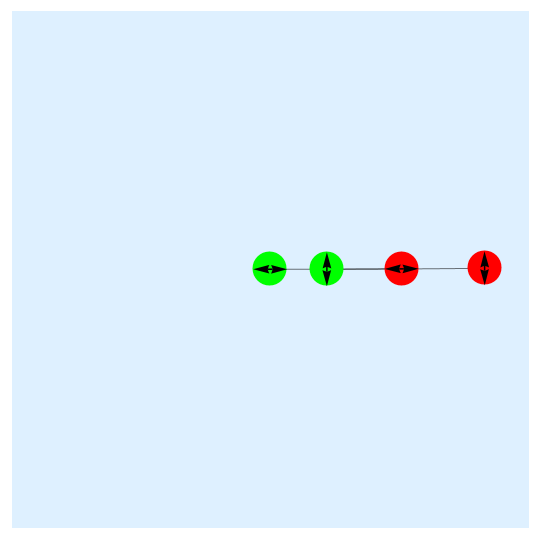} &
    \includegraphics[width=0.18\linewidth]{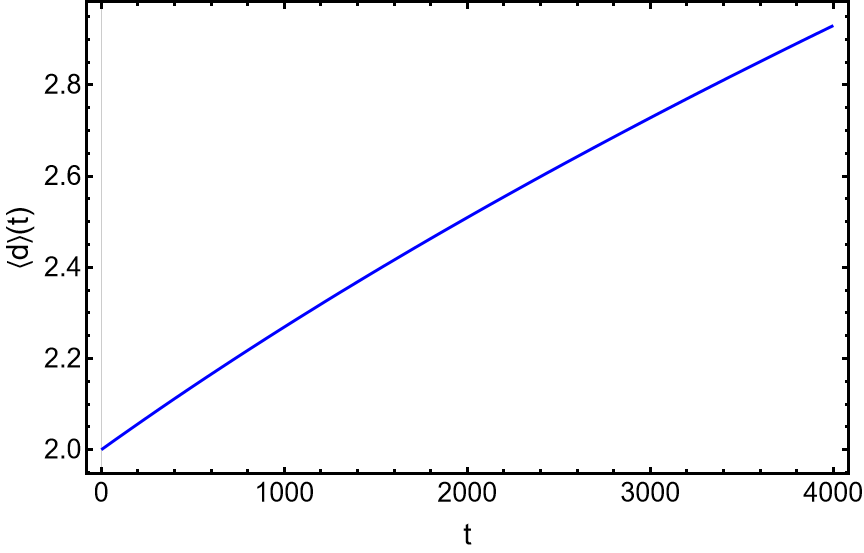} &
    \includegraphics[width=0.18\linewidth]{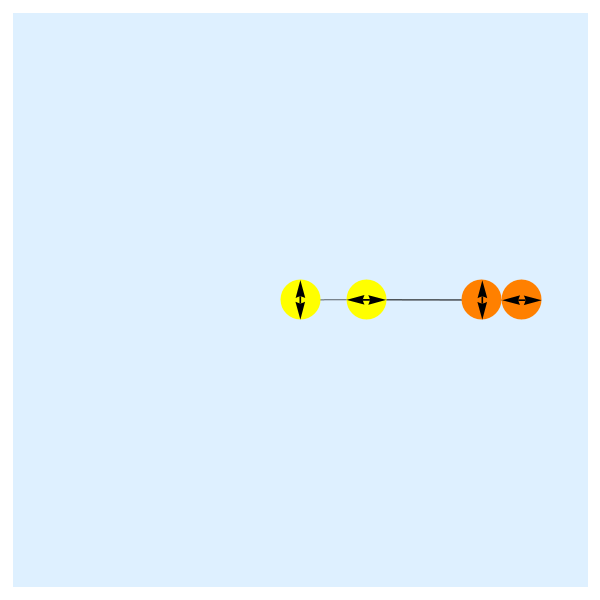} &
    \includegraphics[width=0.18\linewidth]{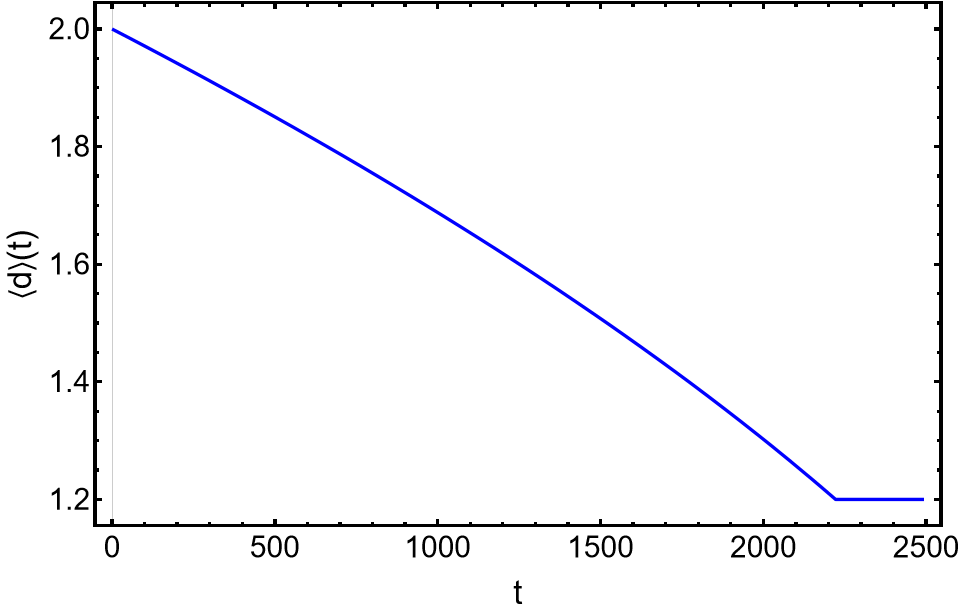} \\[4pt]

    \includegraphics[width=0.18\linewidth]{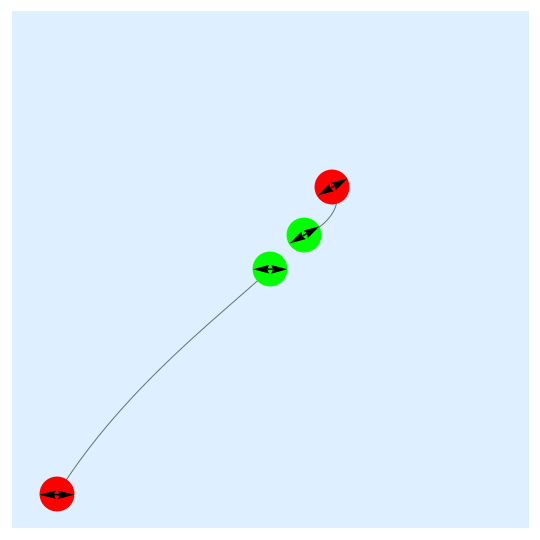} &
    \includegraphics[width=0.18\linewidth]{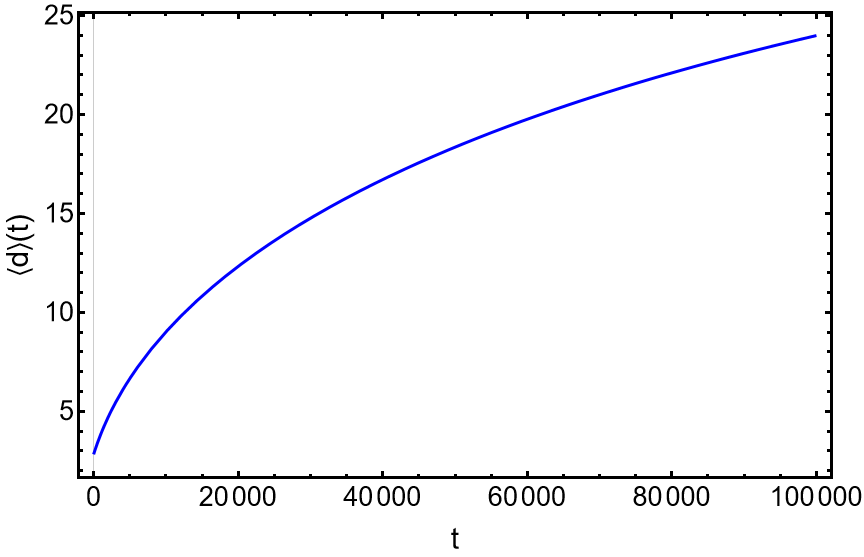} &
    \includegraphics[width=0.18\linewidth]{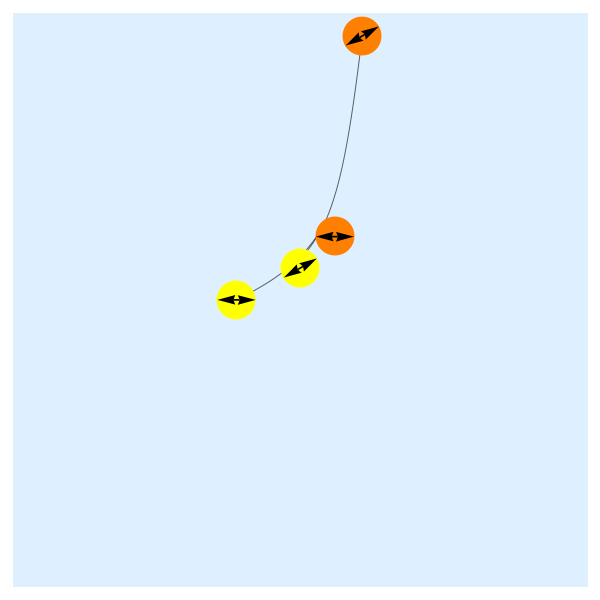} &
    \includegraphics[width=0.18\linewidth]{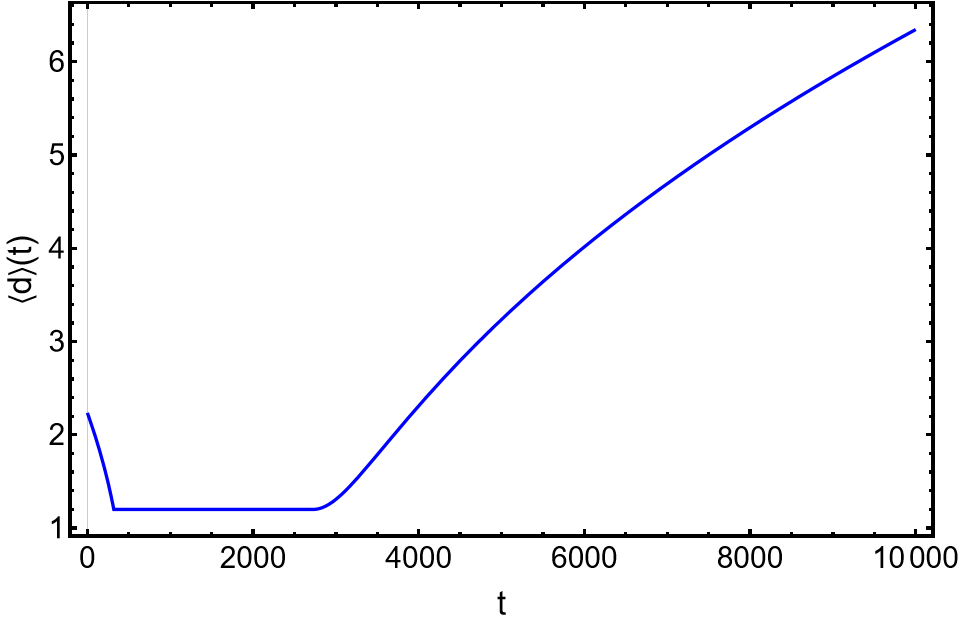} \\[4pt]

    \includegraphics[width=0.18\linewidth]{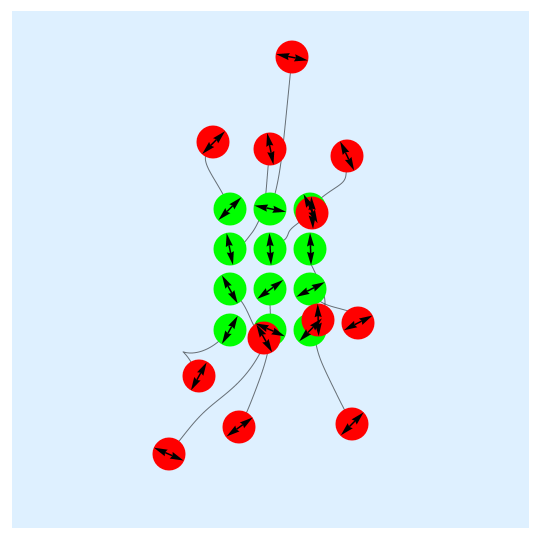} &
    \includegraphics[width=0.18\linewidth]{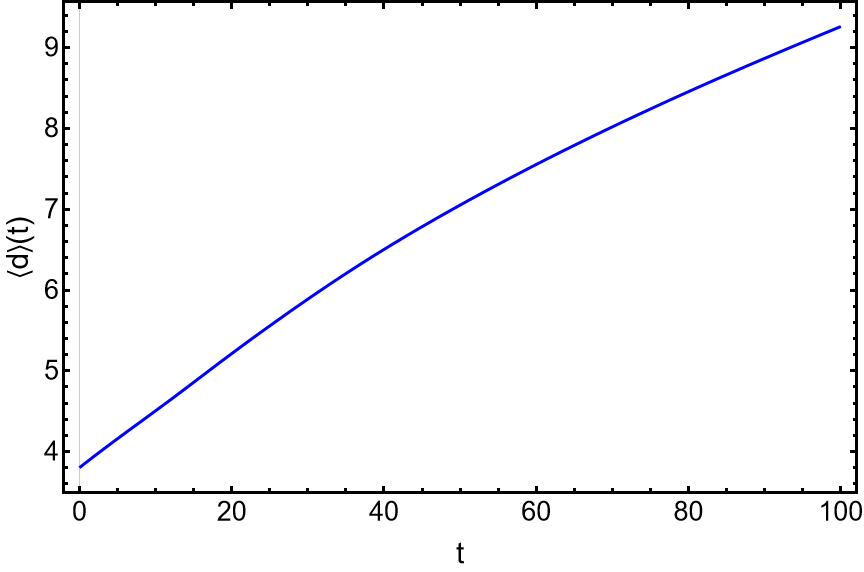} &
    \includegraphics[width=0.18\linewidth]{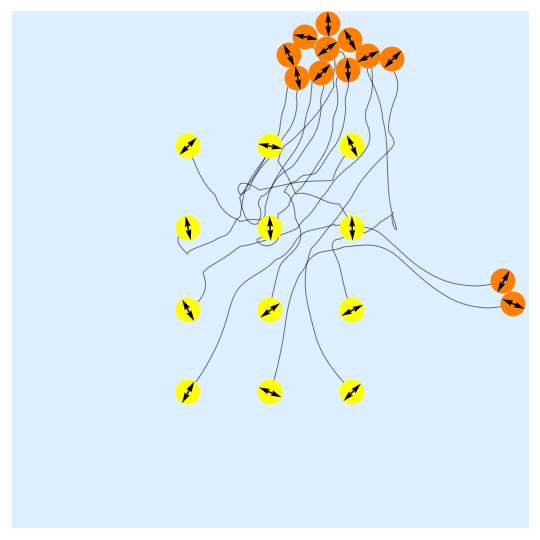} &
    \includegraphics[width=0.18\linewidth]{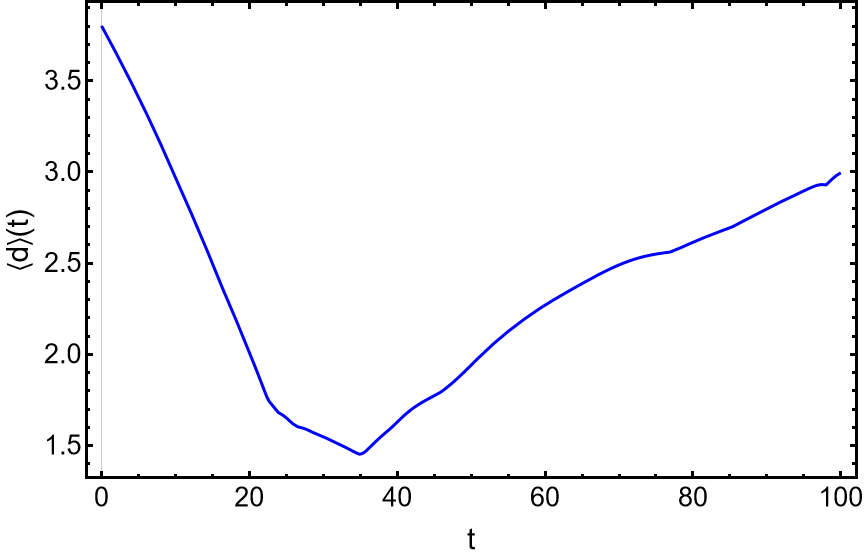} \\
  \end{tabular}

  \caption{
  Near–zone dynamics of pusher and puller dipoles in a compressible supported membrane
  with vanishing odd viscosity and \emph{quenched} orientations.  
  Rows: axial, side-by-side, perpendicular, random pair, and 12-dipole cluster.  
  Columns: trajectories and mean pair separation for pushers (left) and pullers (right).
  }
  \label{fig:cmp_nz_q_all}
\end{figure*}

\begin{figure*}[t]
  \centering
  \begin{tabular}{cccc}
    \multicolumn{4}{c}{\textbf{Compressible membrane (zero odd viscosity) – Far zone – Quenched orientations}} \\[6pt]
    \multicolumn{2}{c}{\textbf{Pusher}} & \multicolumn{2}{c}{\textbf{Puller}} \\
    traj & $\langle d_{ij}\rangle(t)$ & traj & $\langle d_{ij}\rangle(t)$ \\[6pt]

    \includegraphics[width=0.18\linewidth]{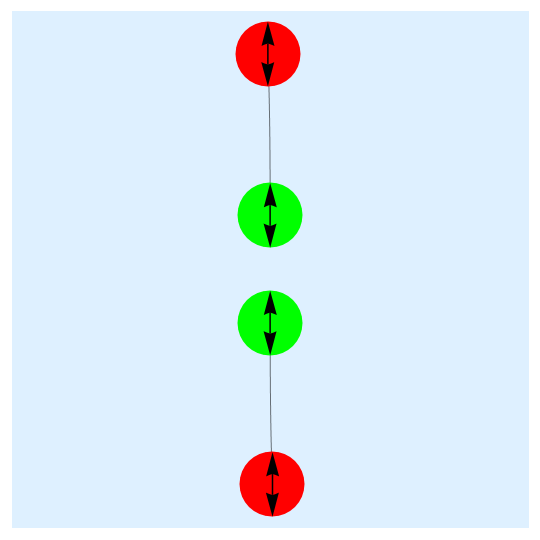} &
    \includegraphics[width=0.18\linewidth]{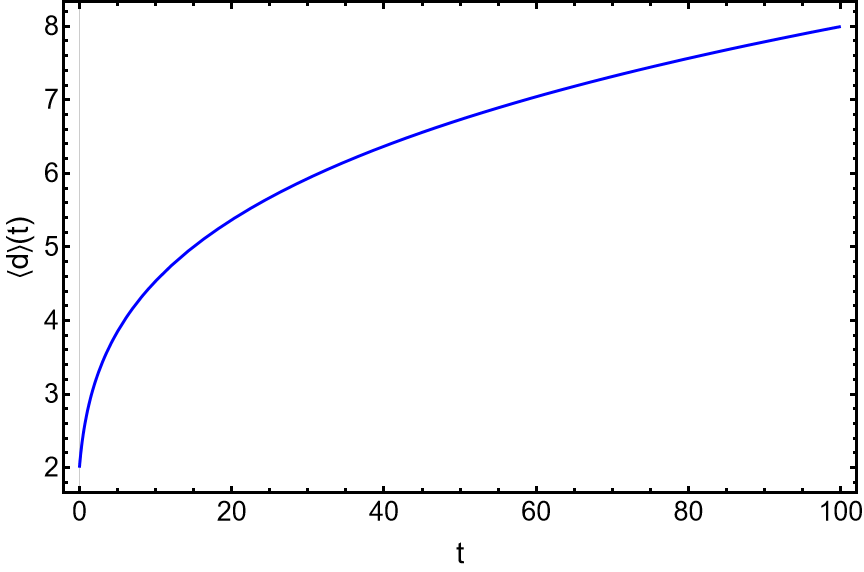} &
    \includegraphics[width=0.18\linewidth]{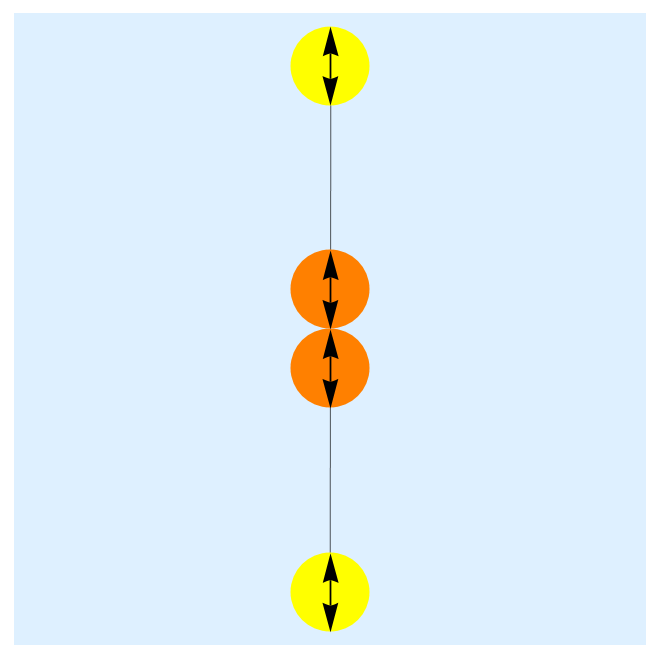} &
    \includegraphics[width=0.18\linewidth]{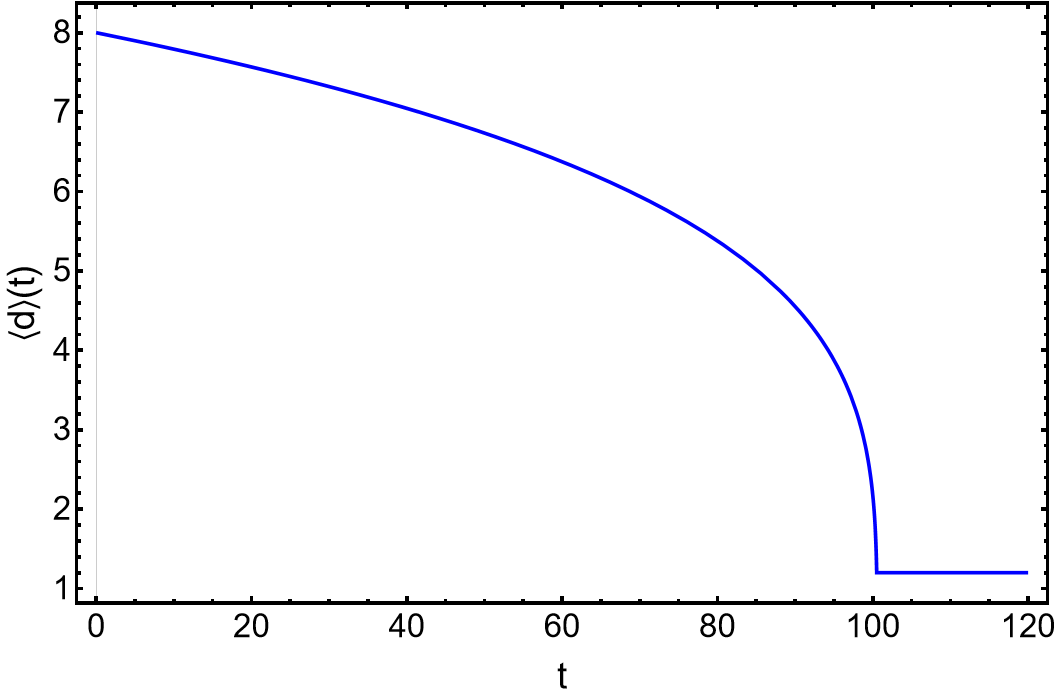} \\[4pt]

    \includegraphics[width=0.18\linewidth]{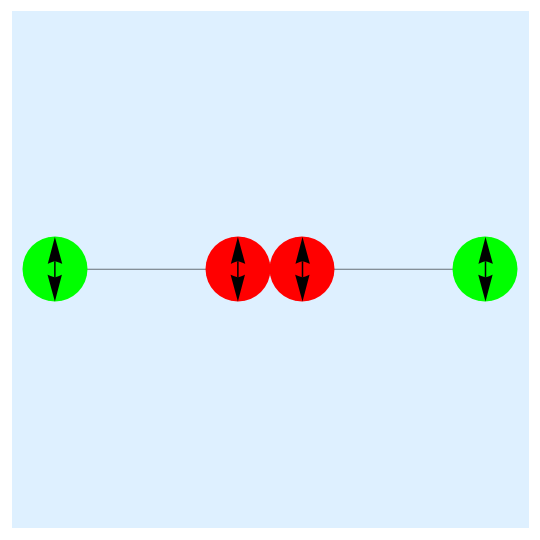} &
    \includegraphics[width=0.18\linewidth]{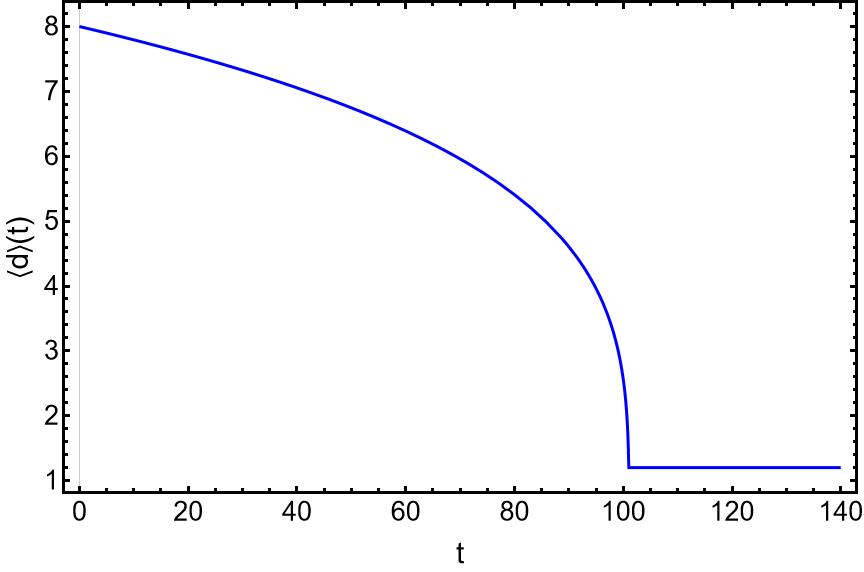} &
    \includegraphics[width=0.18\linewidth]{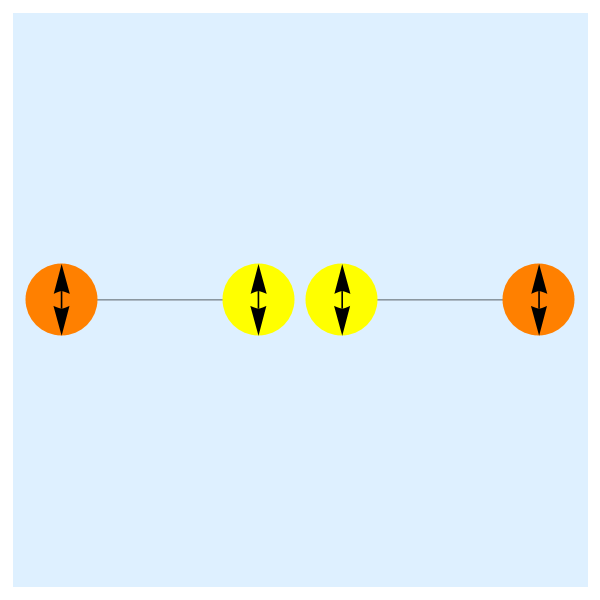} &
    \includegraphics[width=0.18\linewidth]{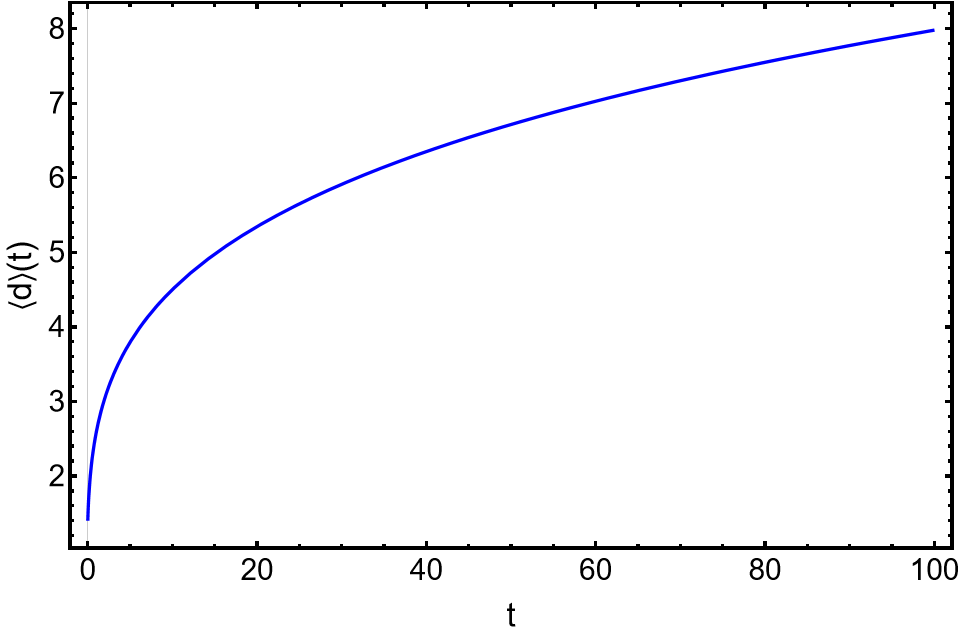} \\[4pt]

    \includegraphics[width=0.18\linewidth]{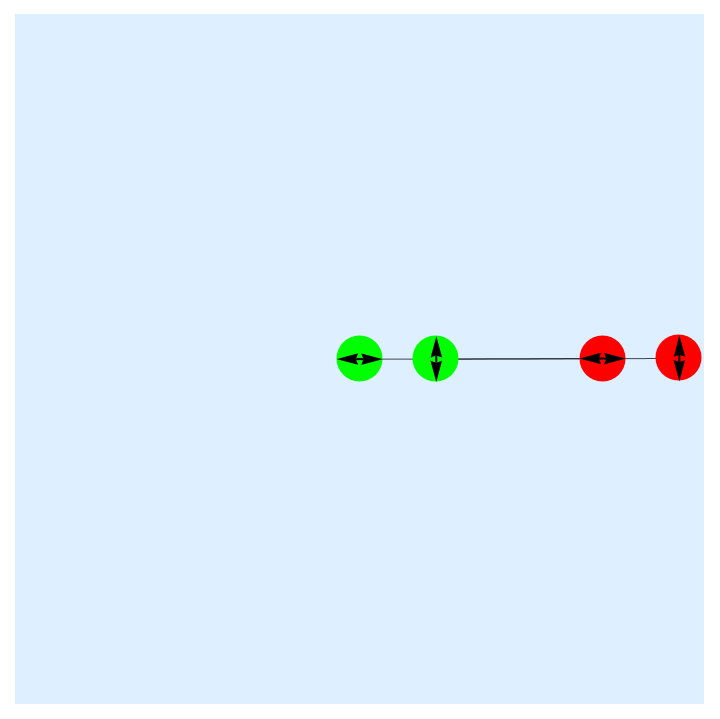} &
    \includegraphics[width=0.18\linewidth]{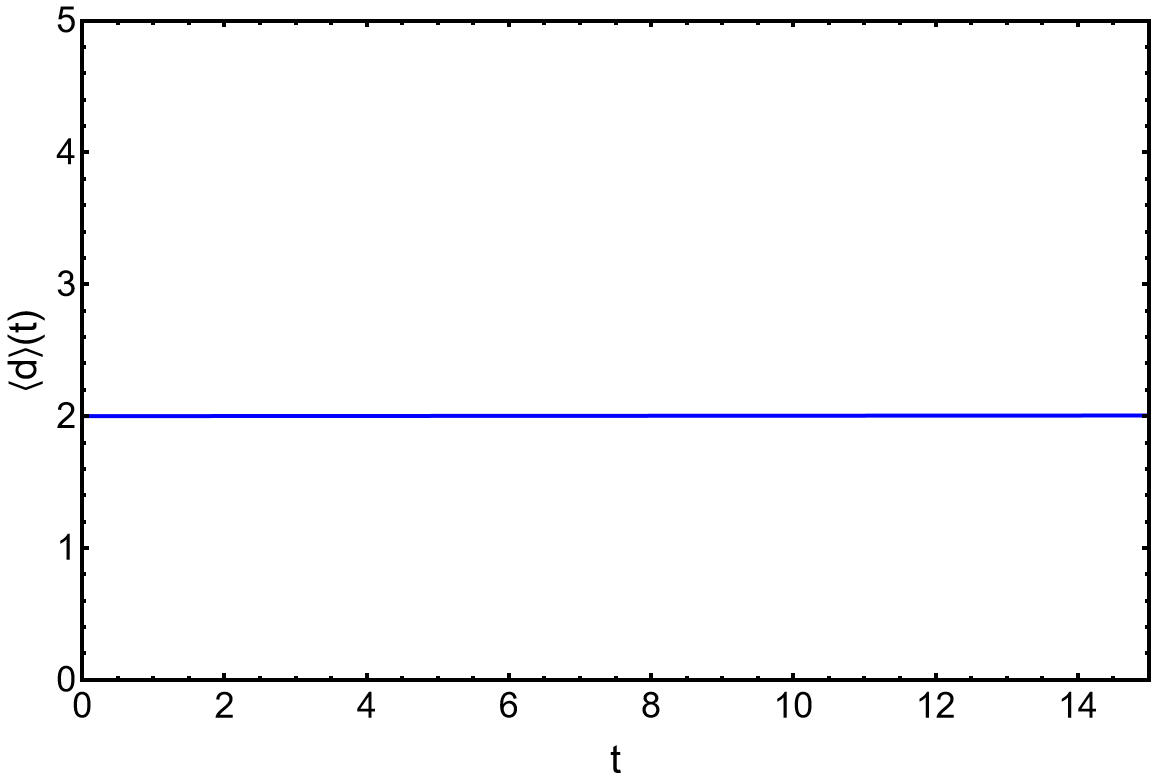} &
    \includegraphics[width=0.18\linewidth]{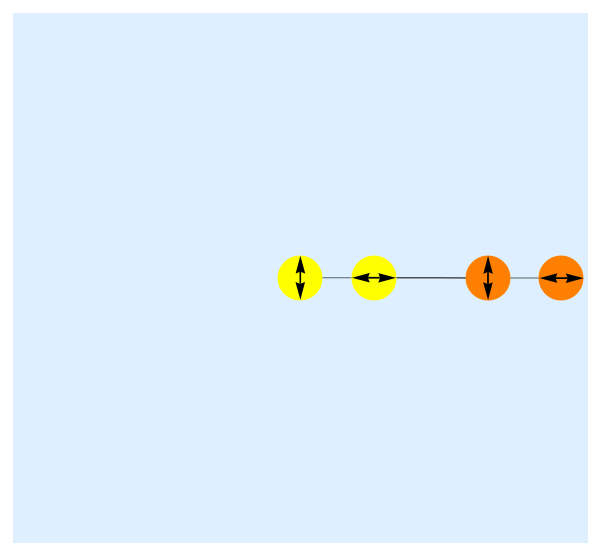} &
    \includegraphics[width=0.18\linewidth]{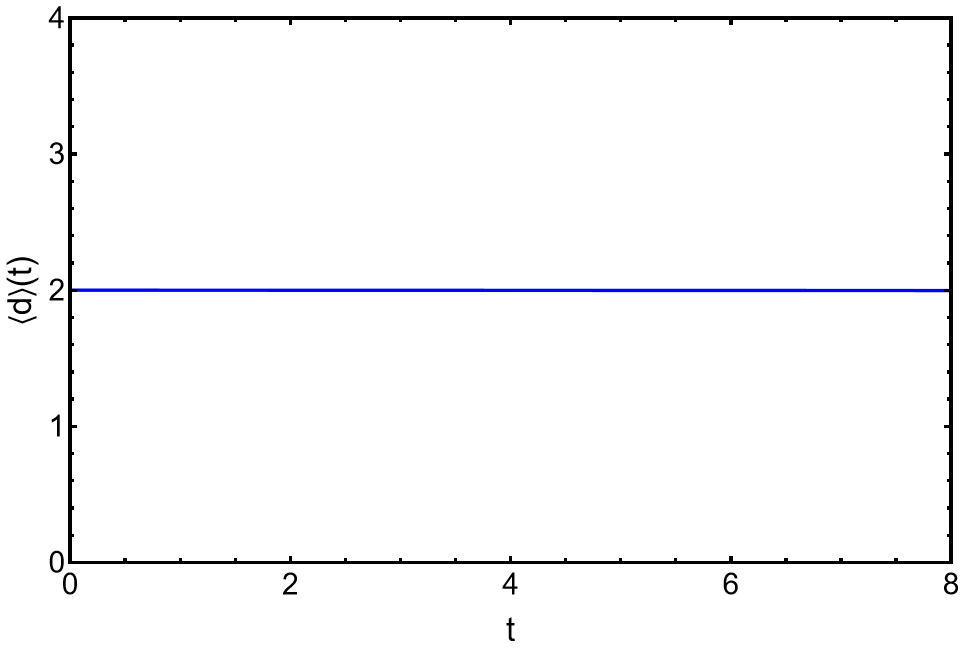} \\[4pt]

    \includegraphics[width=0.18\linewidth]{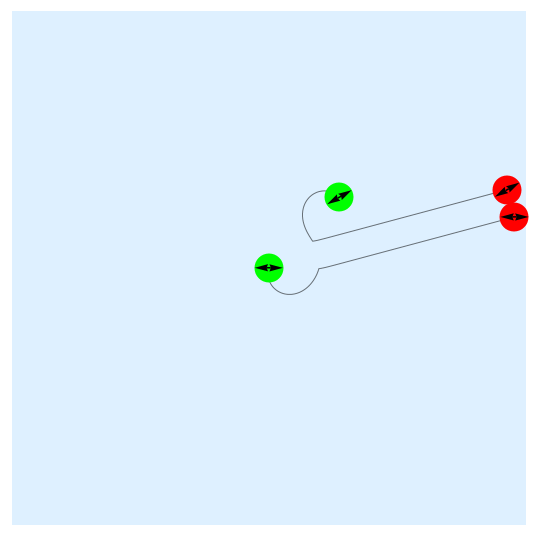} &
    \includegraphics[width=0.18\linewidth]{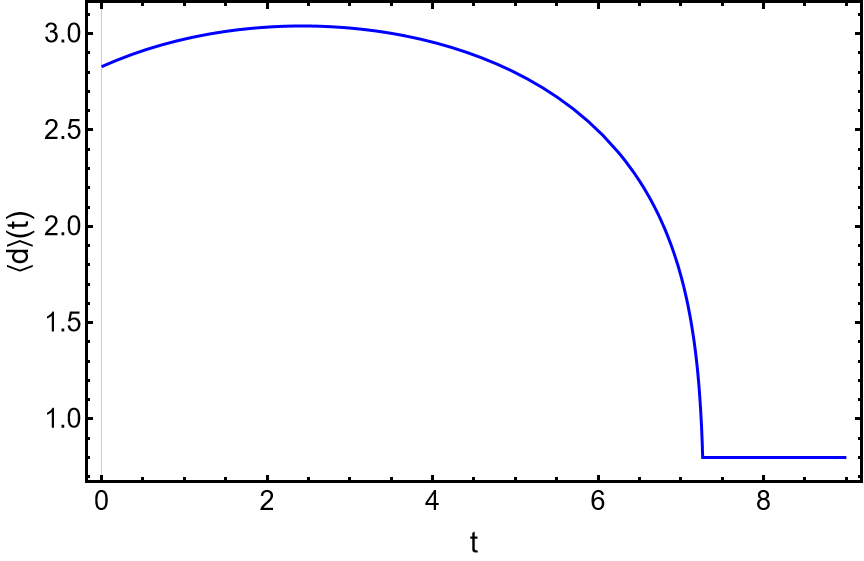} &
    \includegraphics[width=0.18\linewidth]{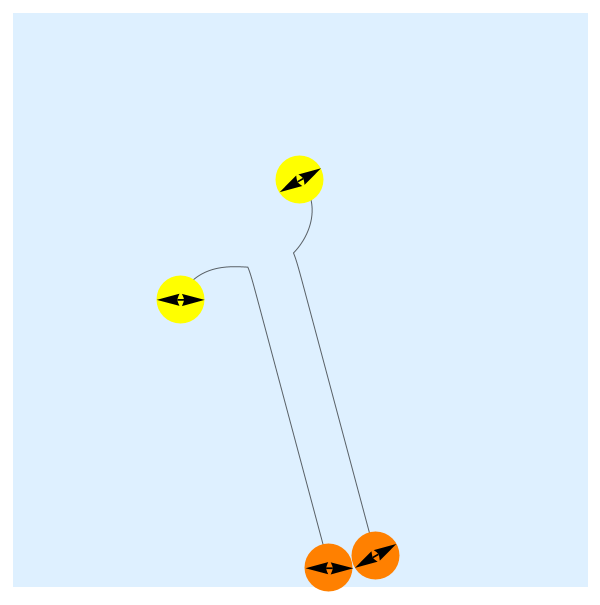} &
    \includegraphics[width=0.18\linewidth]{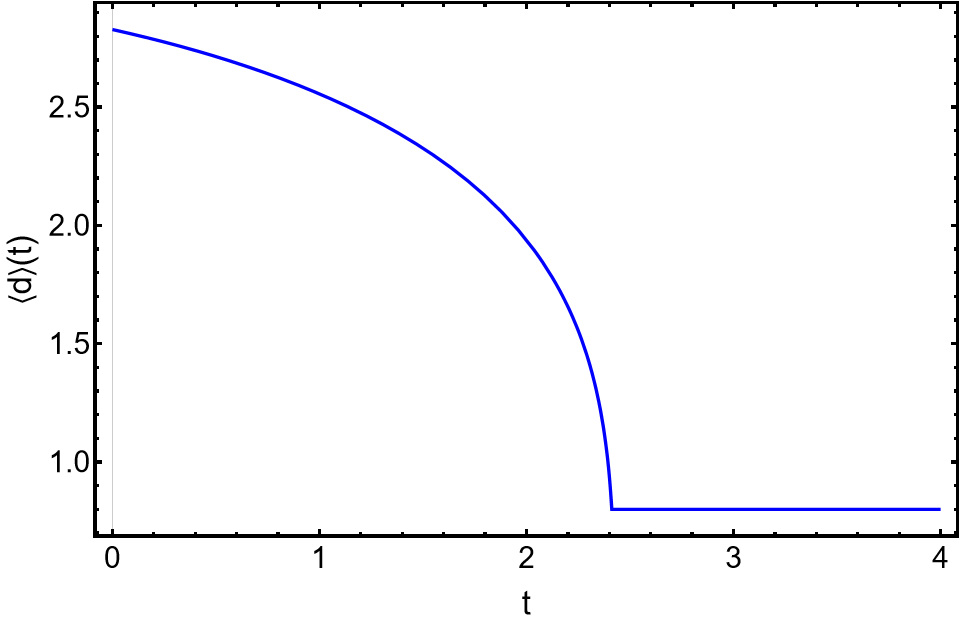} \\[4pt]

    \includegraphics[width=0.18\linewidth]{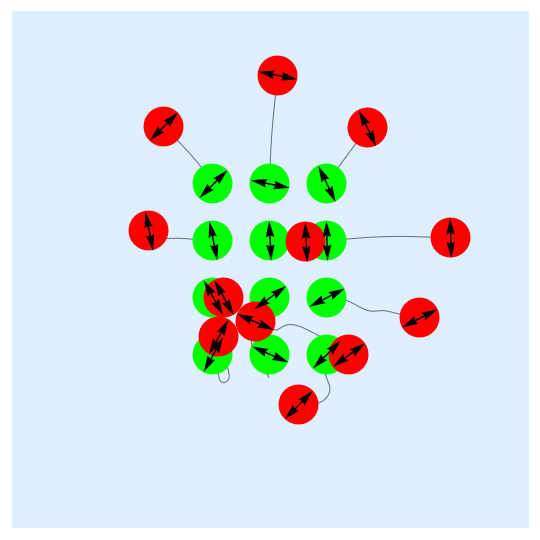} &
    \includegraphics[width=0.18\linewidth]{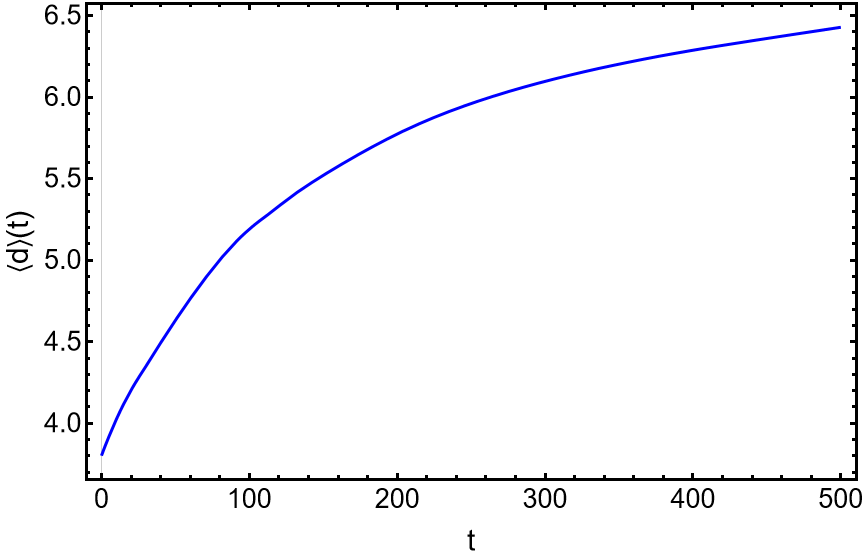} &
    \includegraphics[width=0.18\linewidth]{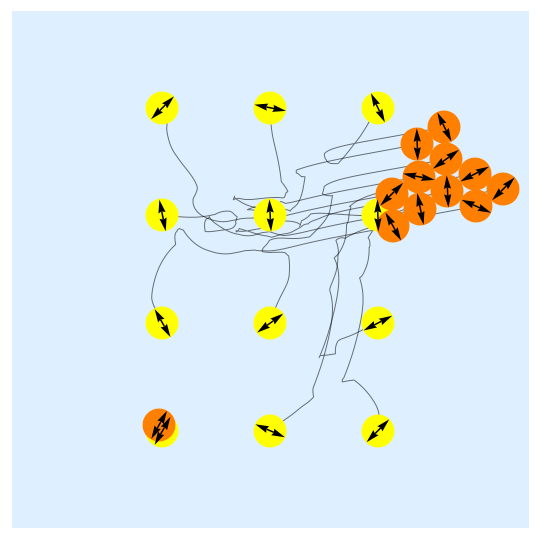} &
    \includegraphics[width=0.18\linewidth]{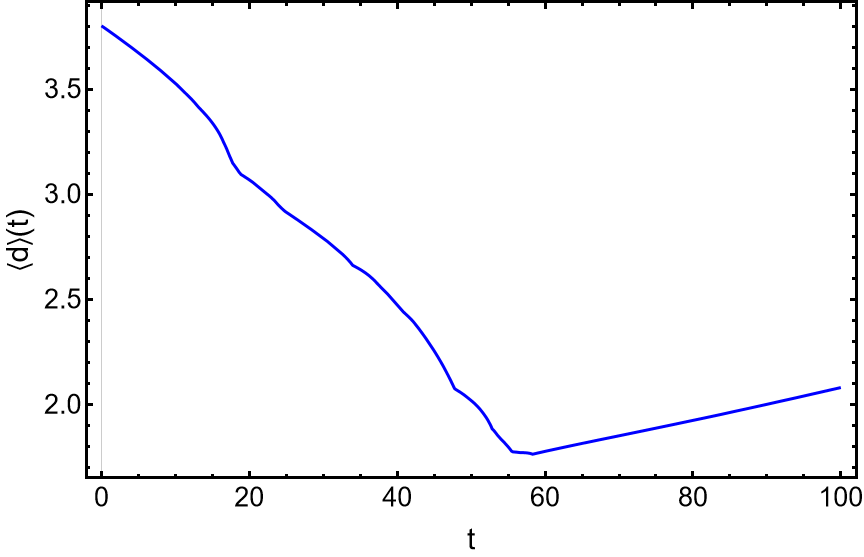} \\
  \end{tabular}

  \caption{
  Far–zone dynamics of pusher and puller dipoles in a compressible supported membrane
  with vanishing odd viscosity and \emph{quenched} orientations.  
  Rows: axial, side-by-side, perpendicular, random pair, and 12-dipole cluster.  
  Columns: trajectories and mean pair separation for pushers (left) and pullers (right).
  }
  \label{fig:cmp_fz_q_all}
\end{figure*}
\begin{figure*}[t]
  \centering
  \begin{tabular}{cccc}
    \multicolumn{4}{c}{\textbf{Compressible membrane with odd viscosity – Near zone – Dynamical orientations}} \\[6pt]
    \multicolumn{2}{c}{\textbf{Pusher}} & \multicolumn{2}{c}{\textbf{Puller}} \\
    traj & $\langle d_{ij}\rangle(t)$ & traj & $\langle d_{ij}\rangle(t)$ \\[6pt]

    \includegraphics[width=0.18\linewidth]{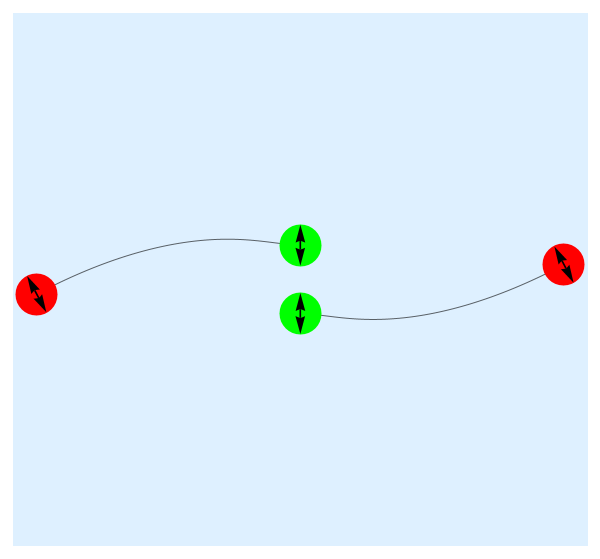} &
    \includegraphics[width=0.18\linewidth]{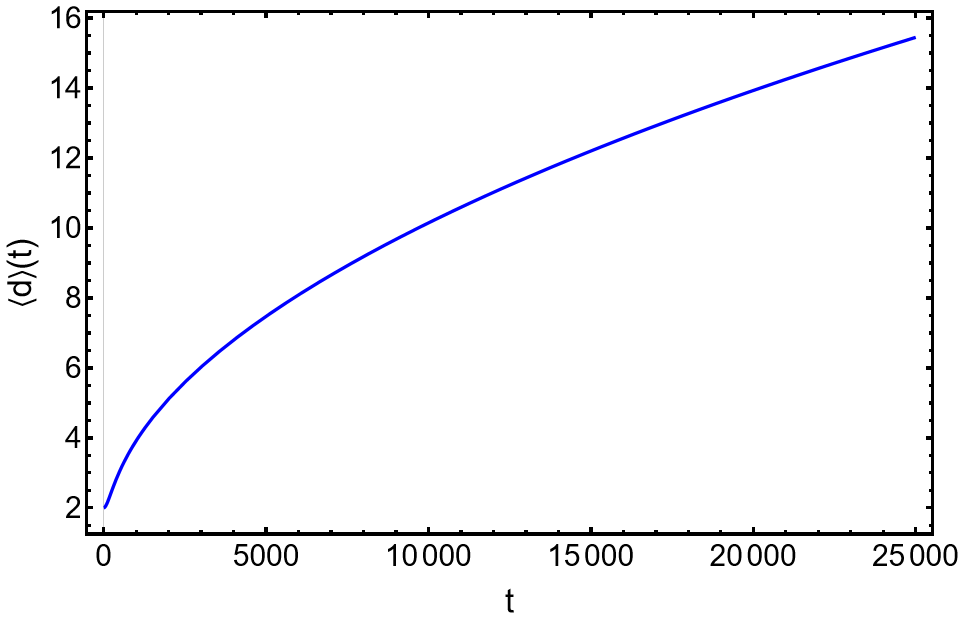} &
    \includegraphics[width=0.18\linewidth]{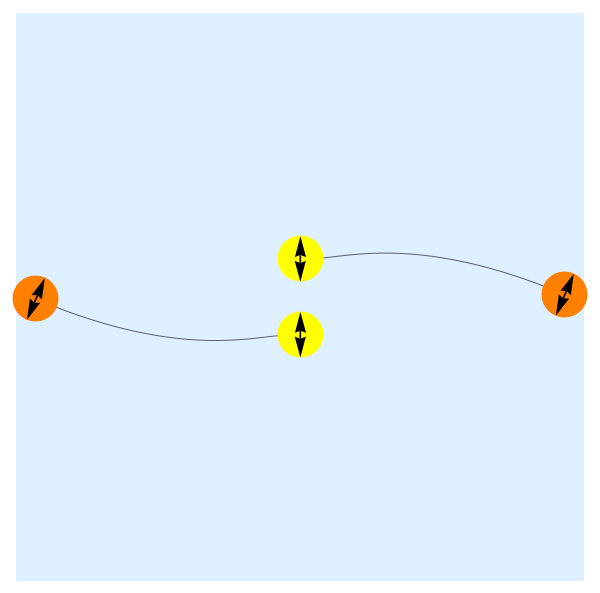} &
    \includegraphics[width=0.18\linewidth]{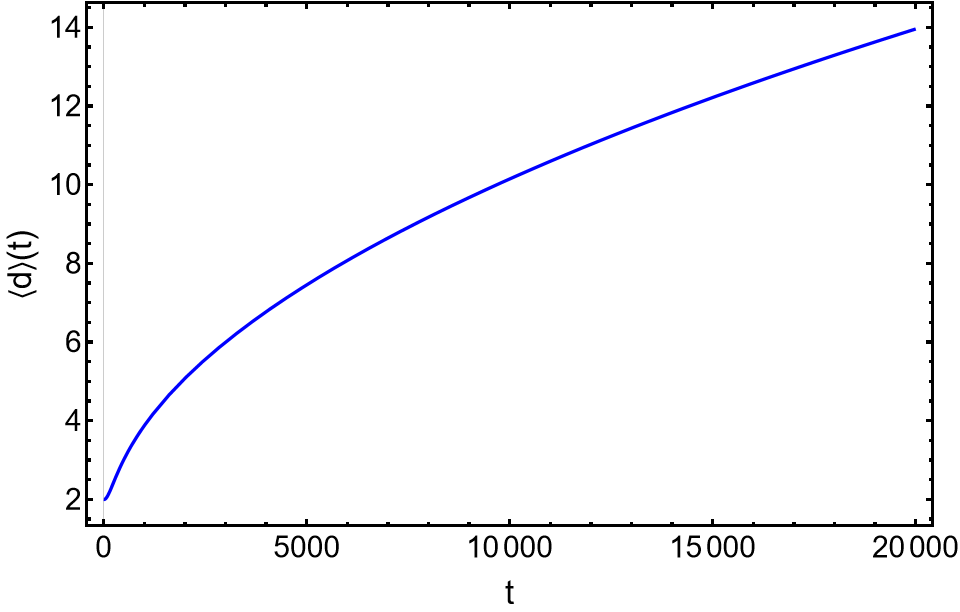} \\[4pt]

    \includegraphics[width=0.18\linewidth]{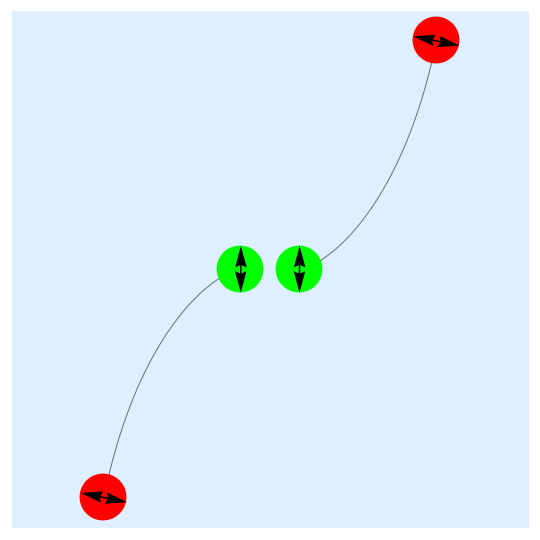} &
    \includegraphics[width=0.18\linewidth]{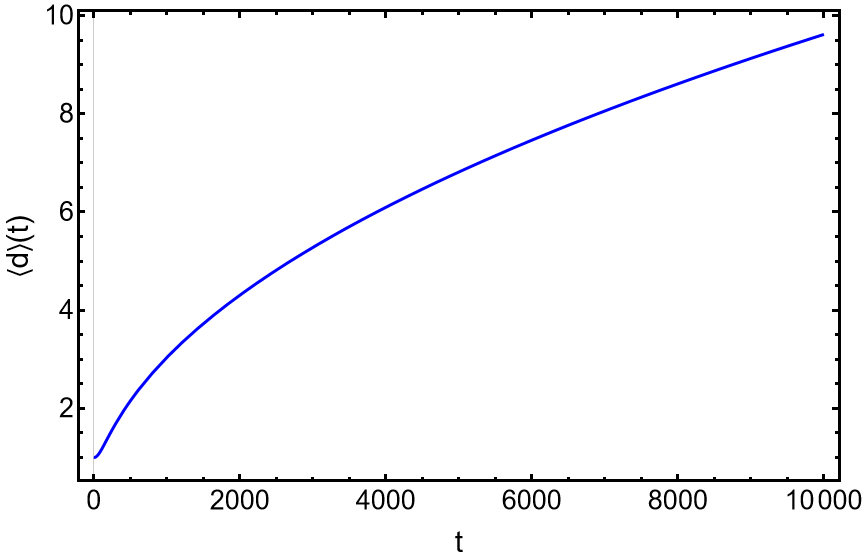} &
    \includegraphics[width=0.18\linewidth]{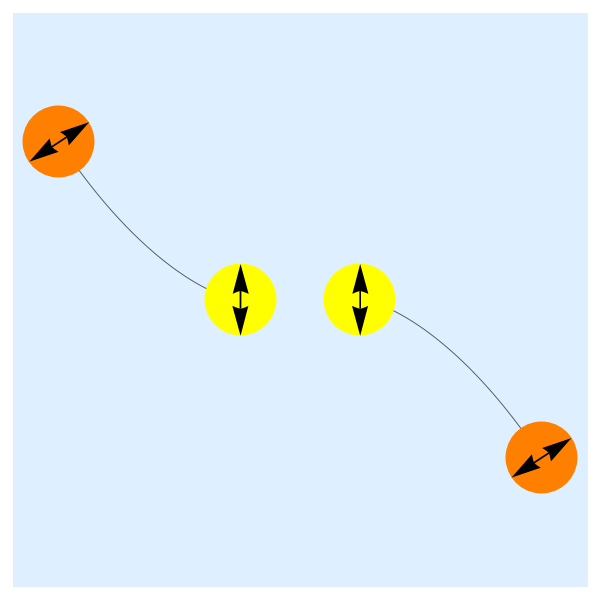} &
    \includegraphics[width=0.18\linewidth]{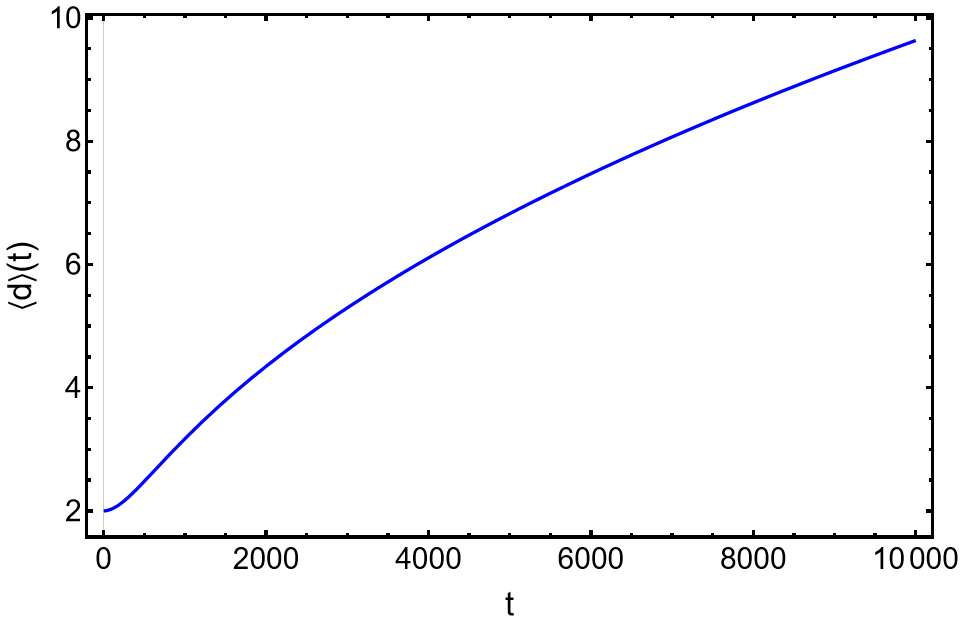} \\[4pt]

    \includegraphics[width=0.18\linewidth]{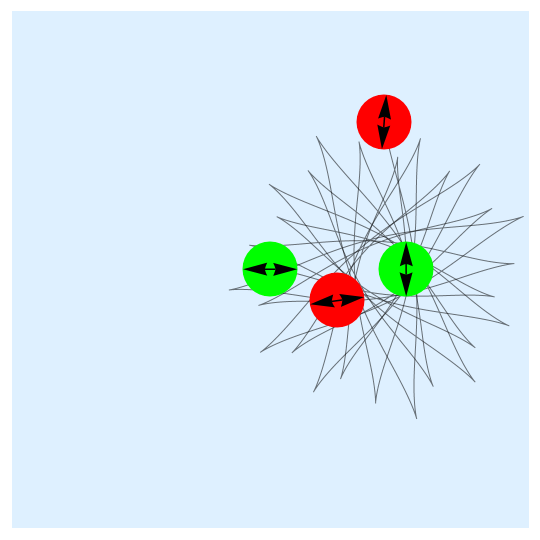} &
    \includegraphics[width=0.18\linewidth]{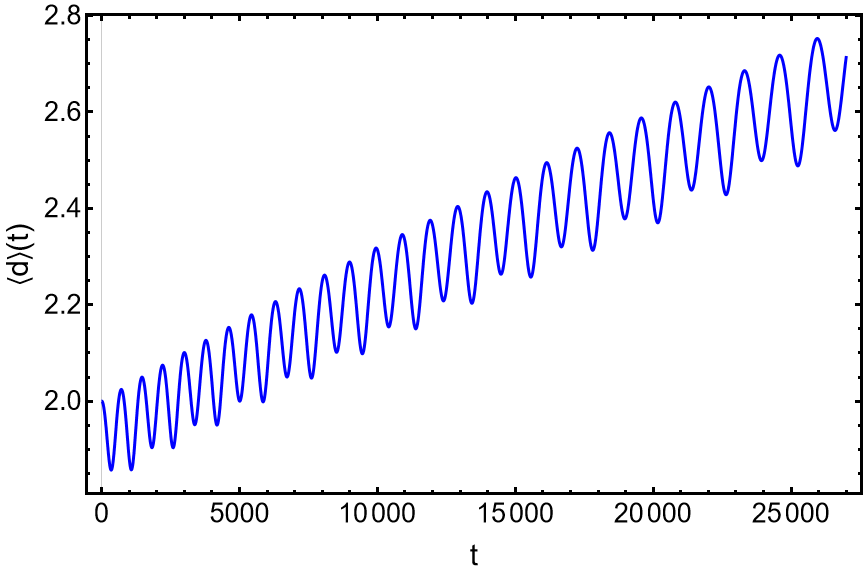} &
    \includegraphics[width=0.18\linewidth]{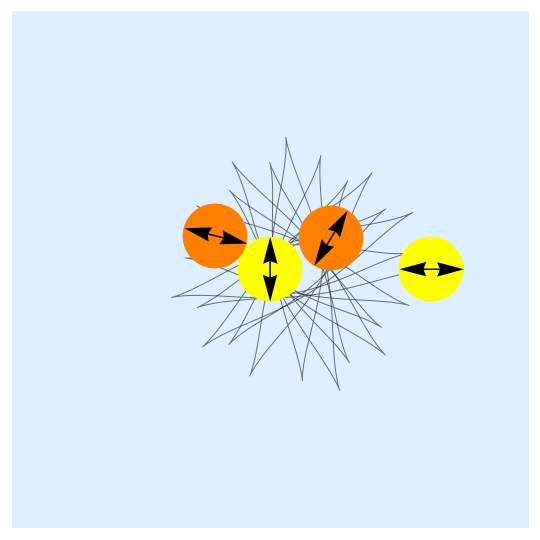} &
    \includegraphics[width=0.18\linewidth]{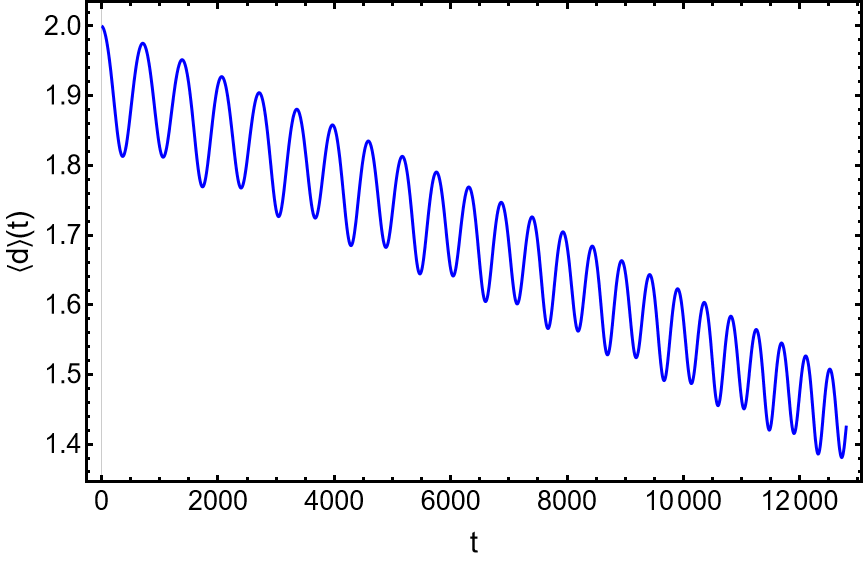} \\[4pt]

    \includegraphics[width=0.18\linewidth]{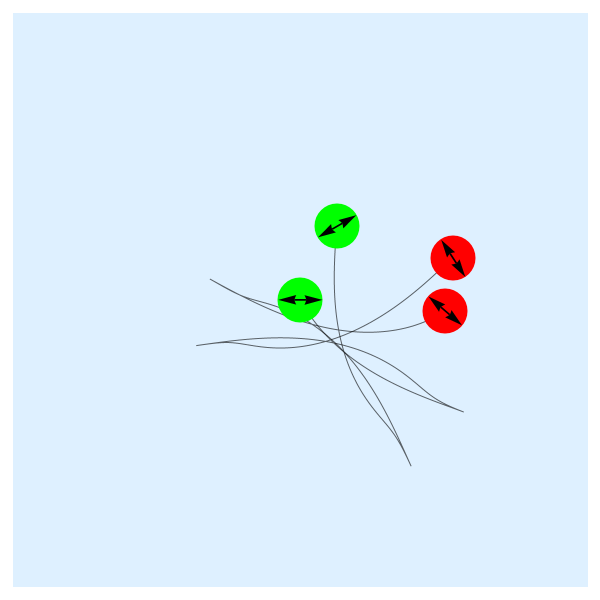} &
    \includegraphics[width=0.18\linewidth]{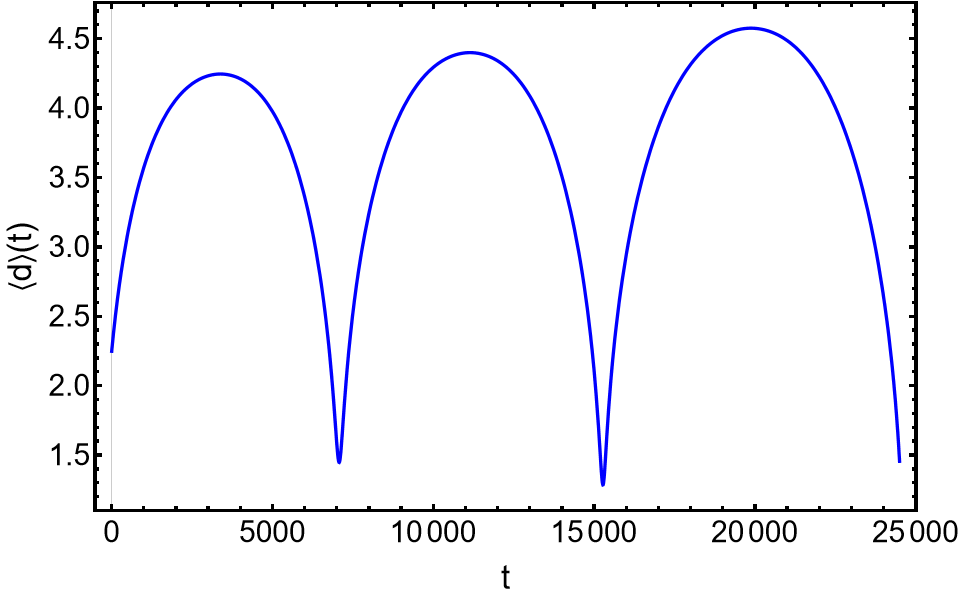} &
    \includegraphics[width=0.18\linewidth]{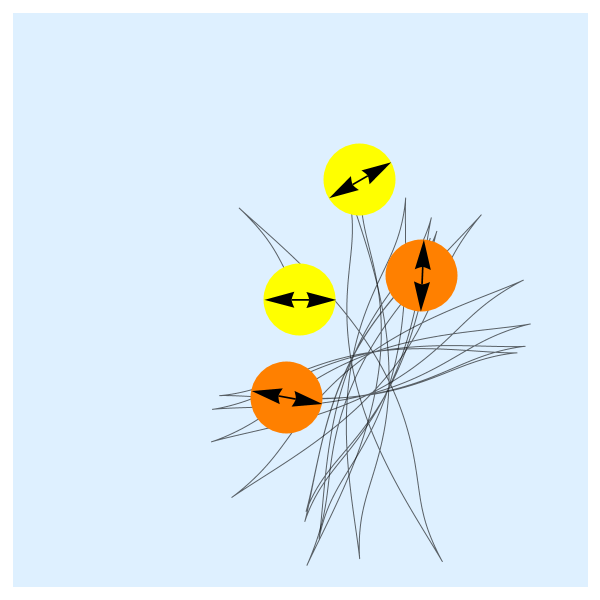} &
    \includegraphics[width=0.18\linewidth]{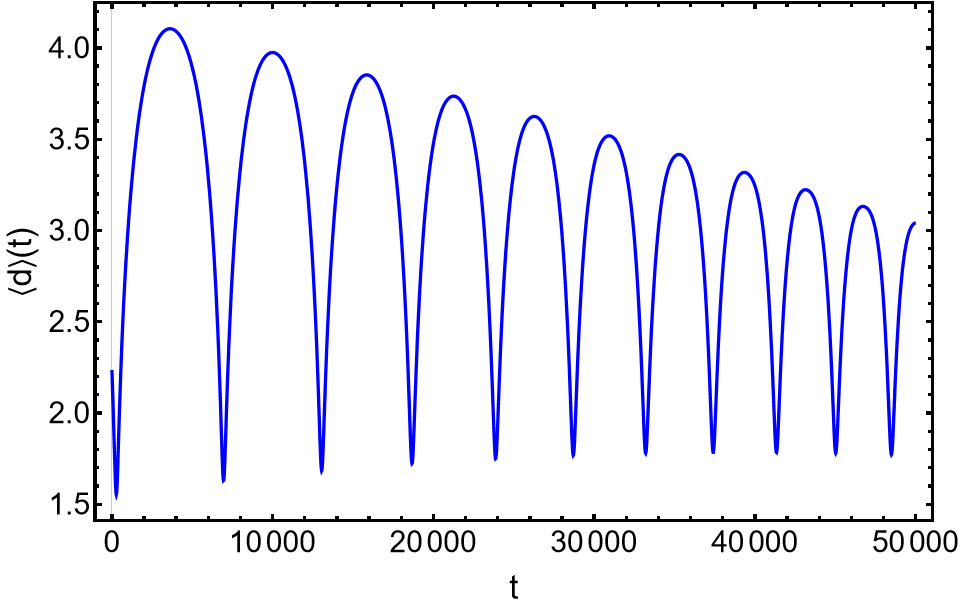} \\[4pt]

    \includegraphics[width=0.18\linewidth]{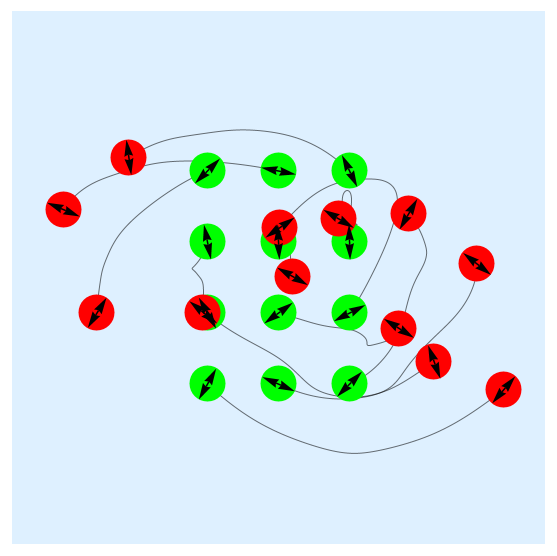} &
    \includegraphics[width=0.18\linewidth]{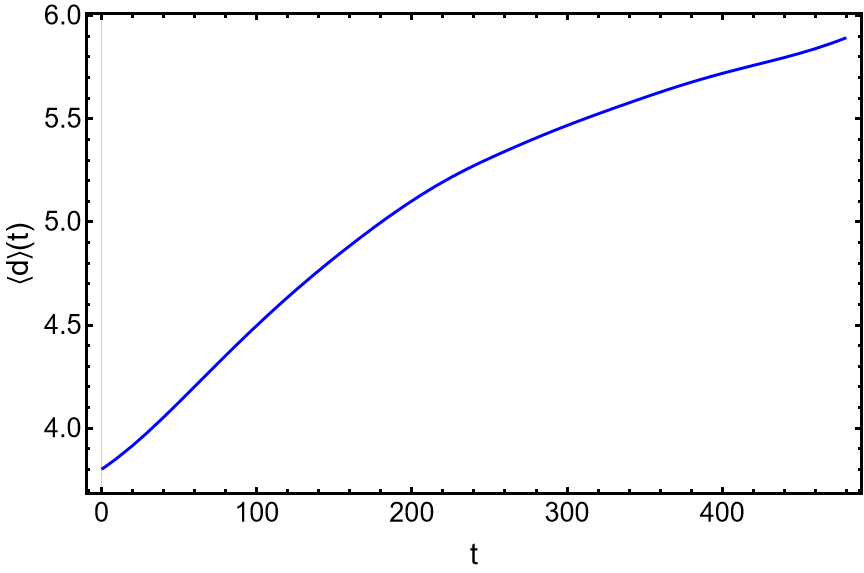} &
    \includegraphics[width=0.18\linewidth]{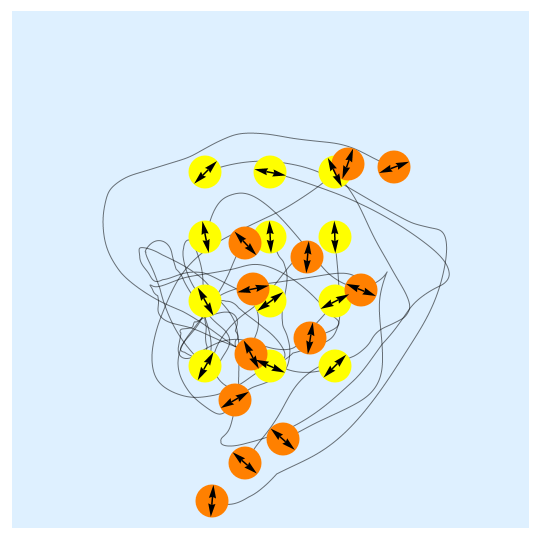} &
    \includegraphics[width=0.18\linewidth]{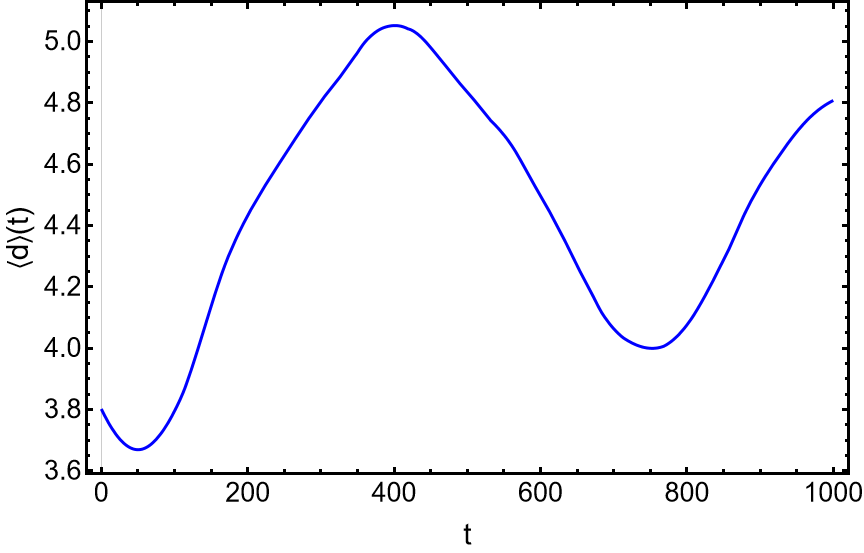} \\
  \end{tabular}

  \caption{
  Near–zone dynamics of pusher and puller dipoles in a compressible supported membrane
  with finite odd viscosity and \emph{dynamical} orientations.  
  Rows: axial, side-by-side, perpendicular, random pair, and 12-dipole cluster.  
  Columns: trajectories and mean pair separation for pushers (left) and pullers (right).
  }
  \label{fig:odd_nz_dyn_all}
\end{figure*}
\begin{figure*}[t]
  \centering
  \begin{tabular}{cccc}
    \multicolumn{4}{c}{\textbf{Compressible membrane with odd viscosity  – Far zone – Dynamical orientations}} \\[6pt]
    \multicolumn{2}{c}{\textbf{Pusher}} & \multicolumn{2}{c}{\textbf{Puller}} \\
    traj & $\langle d_{ij}\rangle(t)$ & traj & $\langle d_{ij}\rangle(t)$ \\[6pt]

    \includegraphics[width=0.18\linewidth]{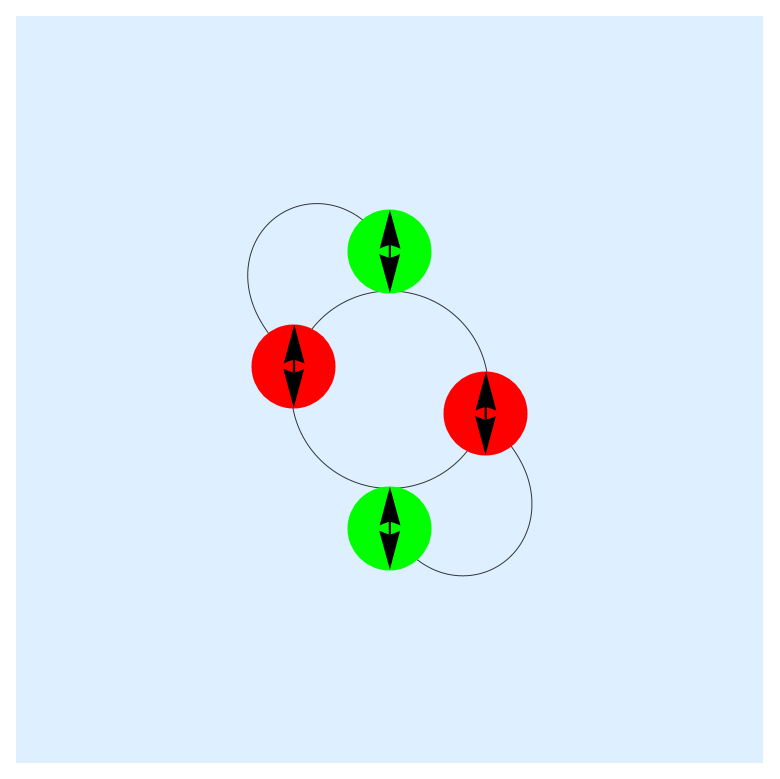} &
    \includegraphics[width=0.18\linewidth]{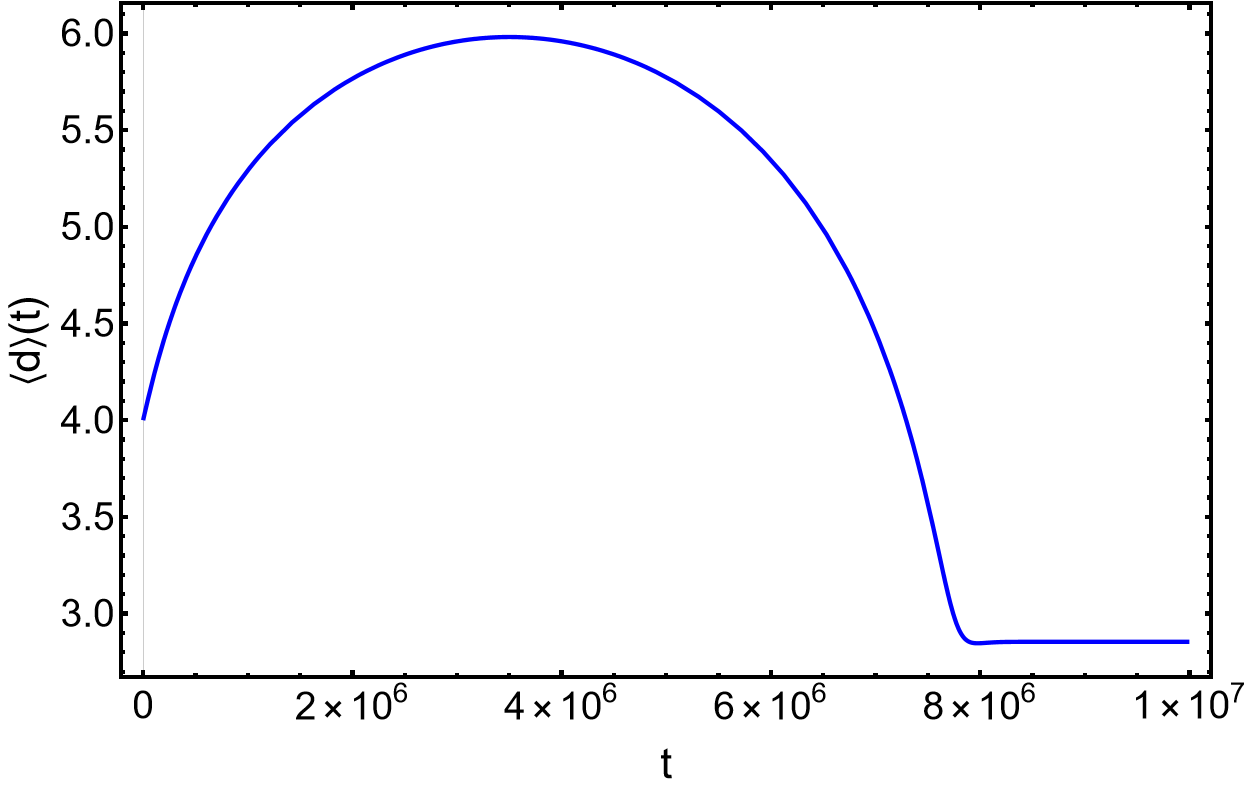} &
    \includegraphics[width=0.18\linewidth]{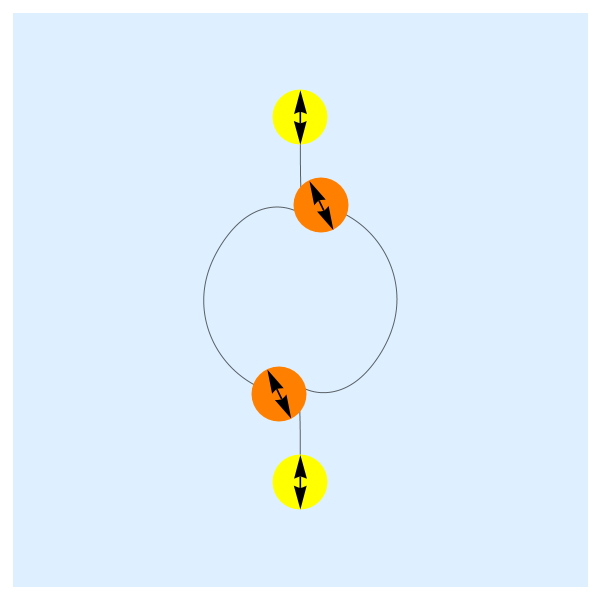} &
    \includegraphics[width=0.18\linewidth]{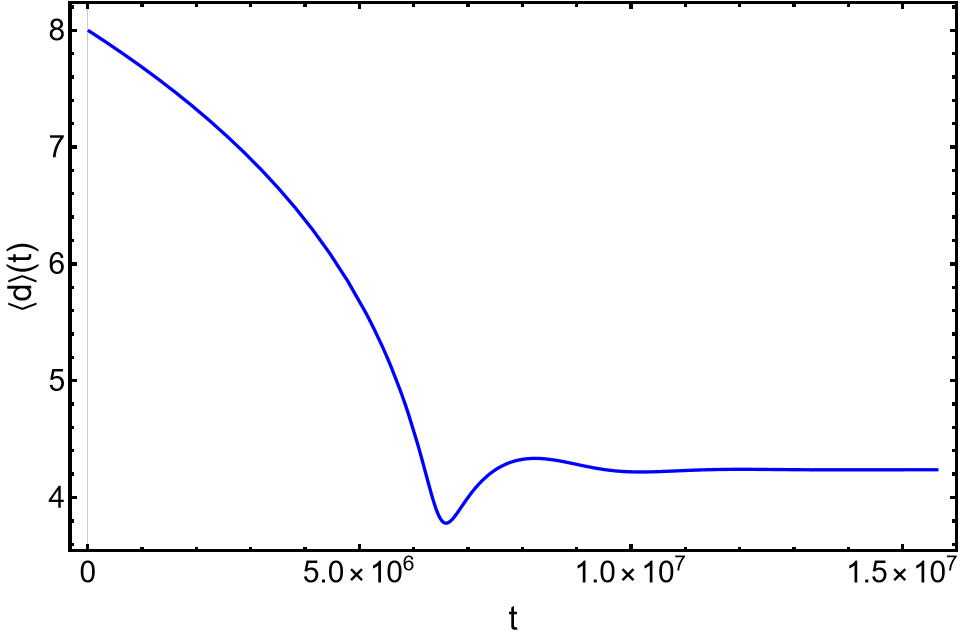} \\[4pt]

    \includegraphics[width=0.18\linewidth]{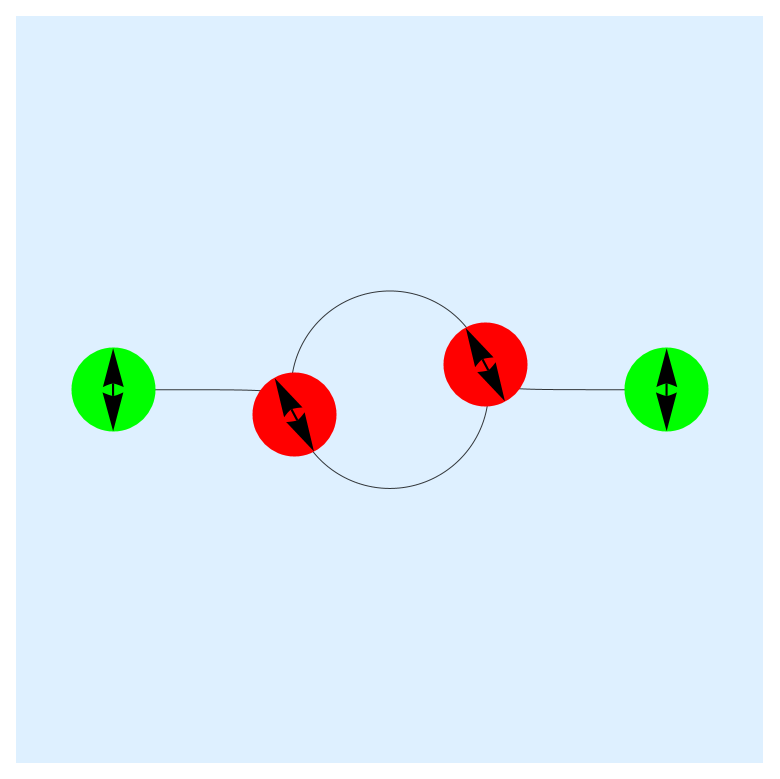} &
    \includegraphics[width=0.18\linewidth]{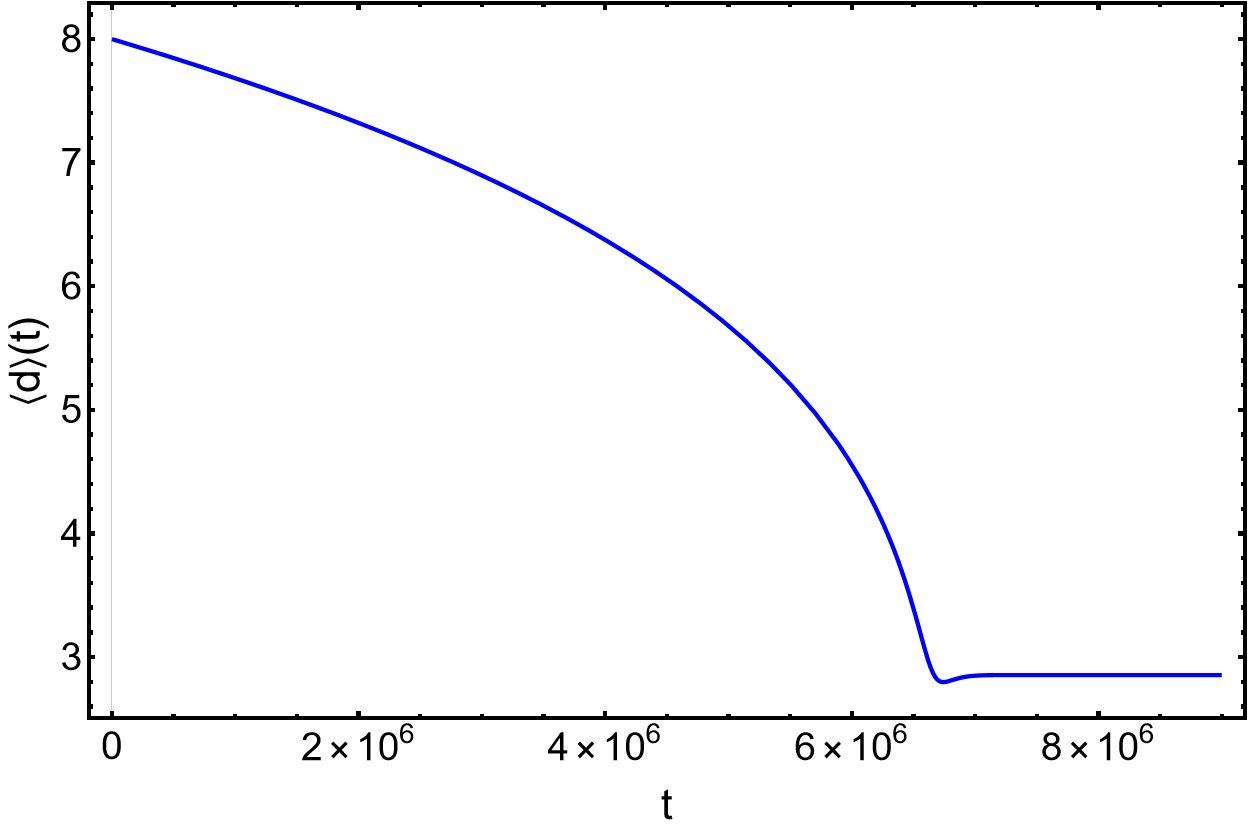} &
    \includegraphics[width=0.18\linewidth]{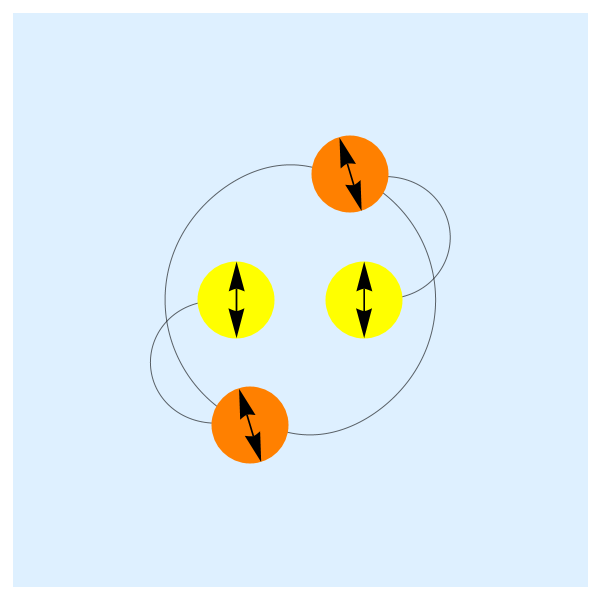} &
    \includegraphics[width=0.18\linewidth]{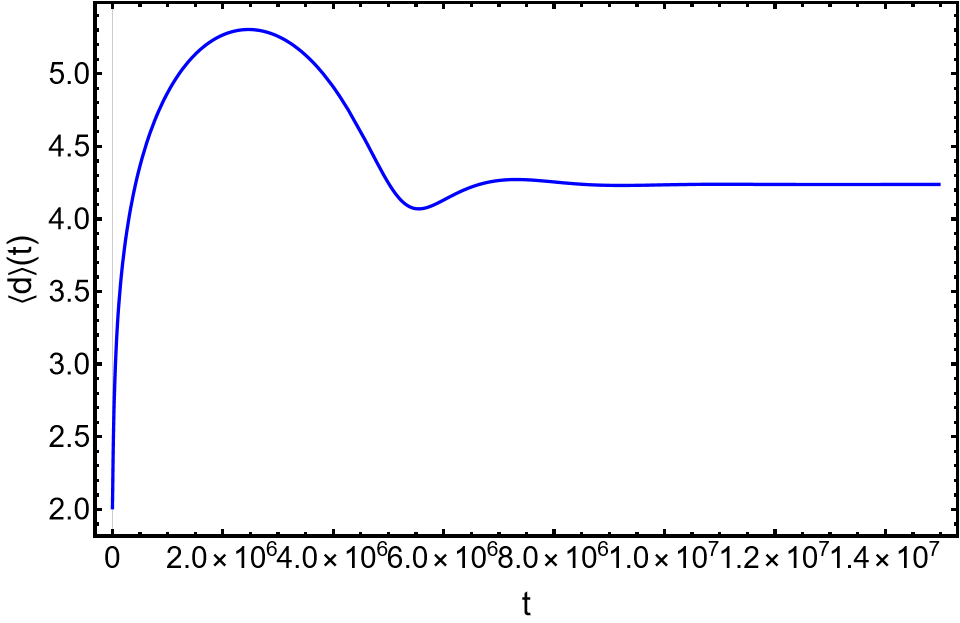} \\[4pt]

    \includegraphics[width=0.18\linewidth]{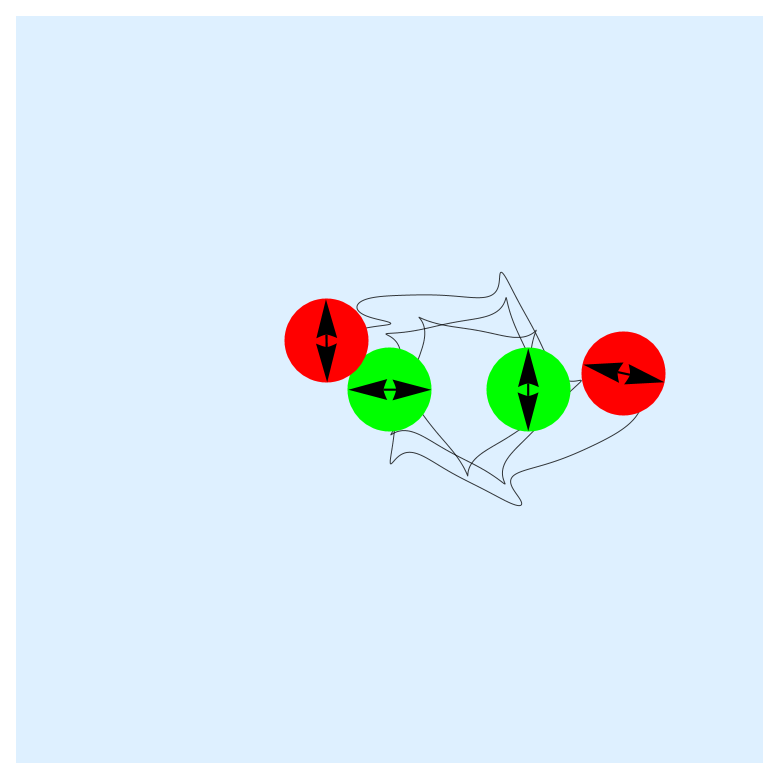} &
    \includegraphics[width=0.18\linewidth]{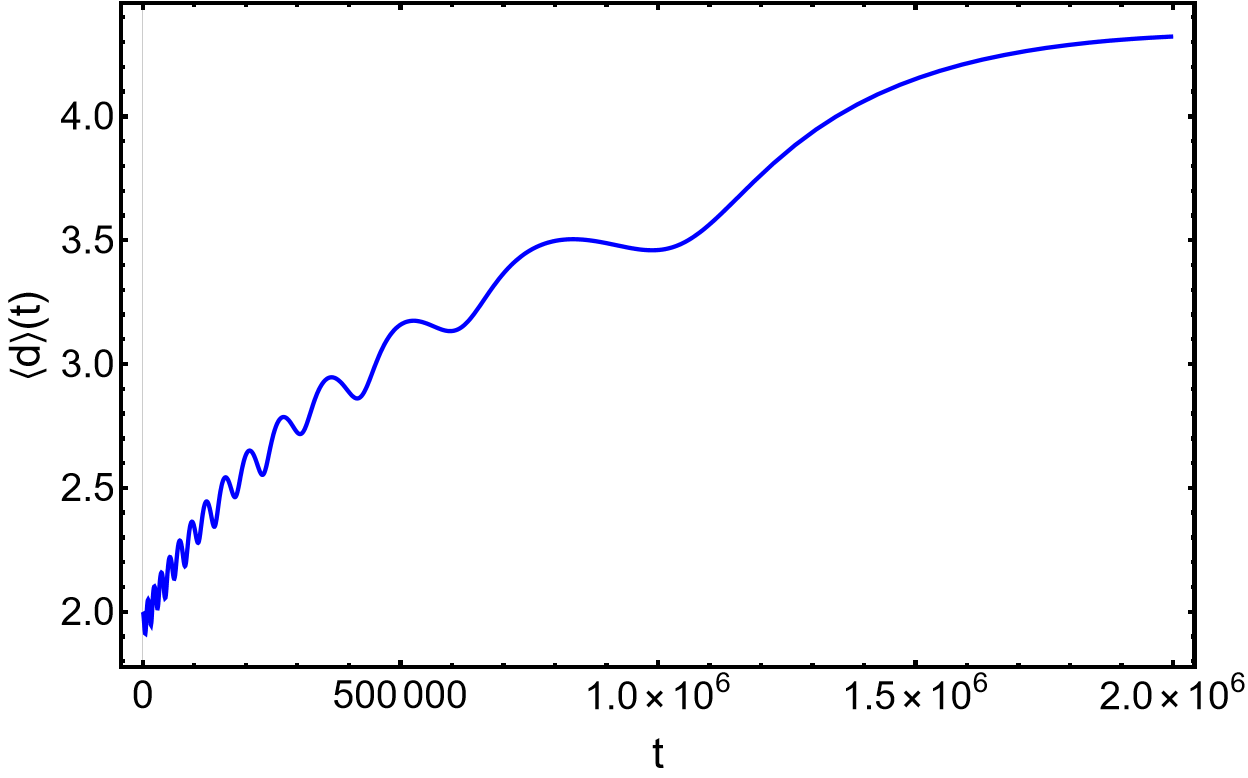} &
    \includegraphics[width=0.18\linewidth]{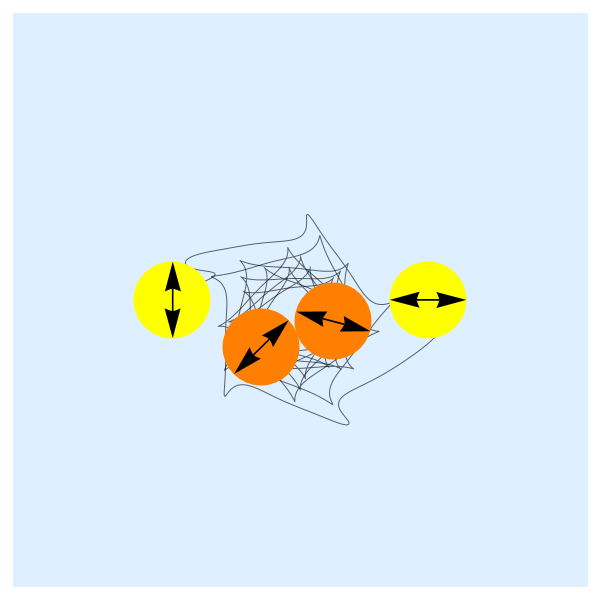} &
    \includegraphics[width=0.18\linewidth]{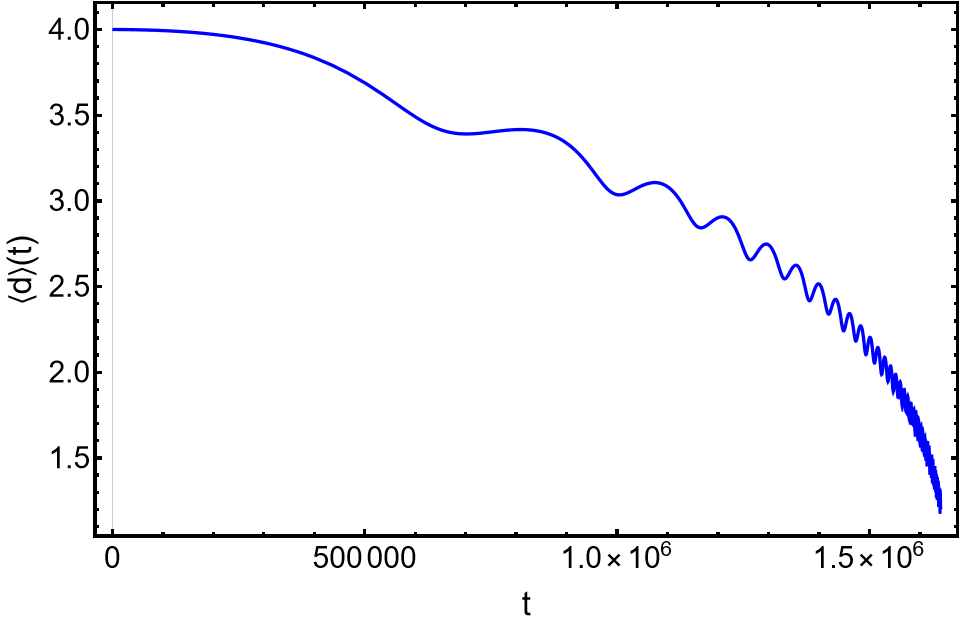} \\[4pt]

    \includegraphics[width=0.18\linewidth]{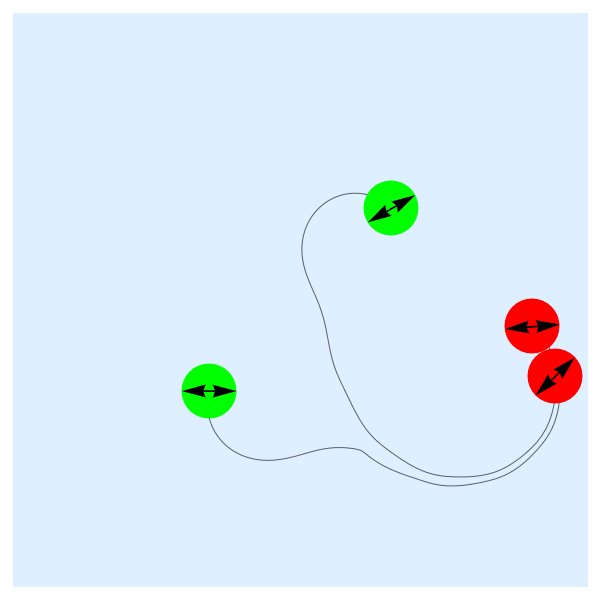} &
    \includegraphics[width=0.18\linewidth]{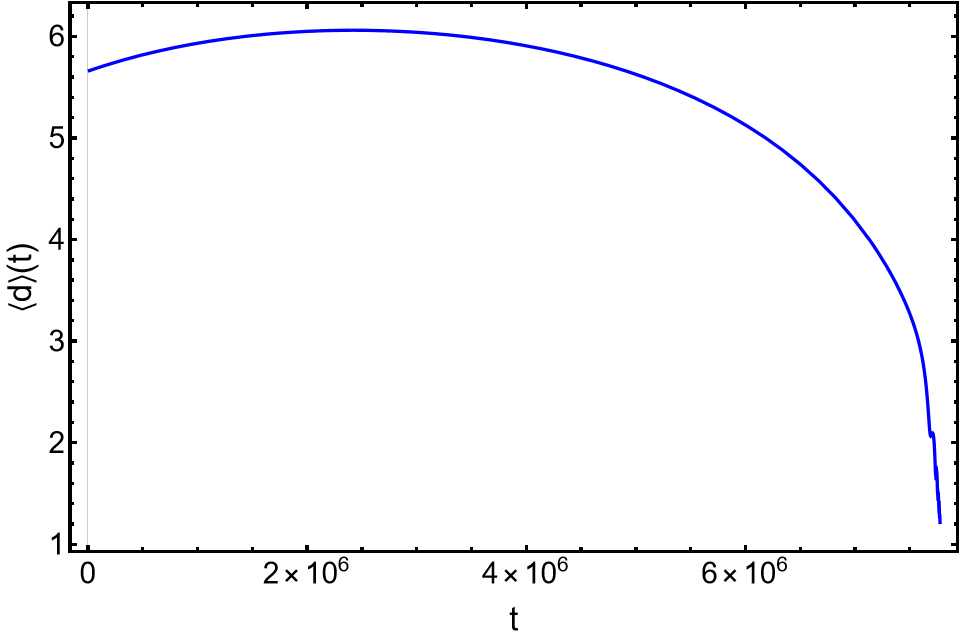} &
    \includegraphics[width=0.18\linewidth]{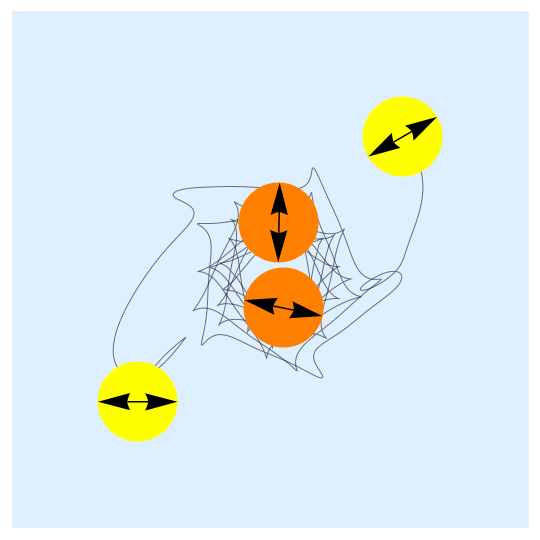} &
    \includegraphics[width=0.18\linewidth]{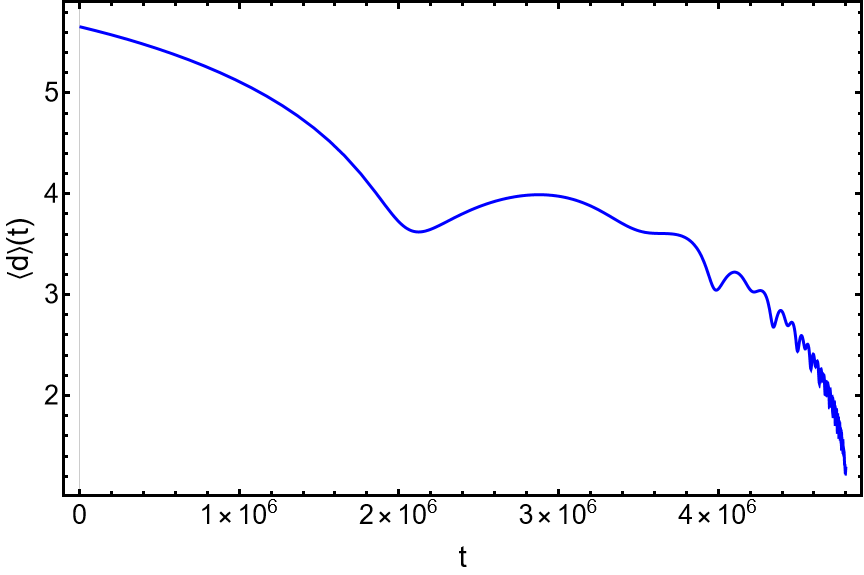} \\[4pt]

    \includegraphics[width=0.18\linewidth]{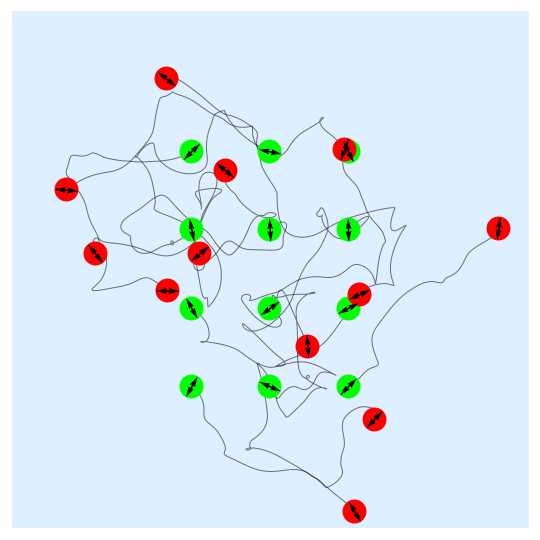} &
    \includegraphics[width=0.18\linewidth]{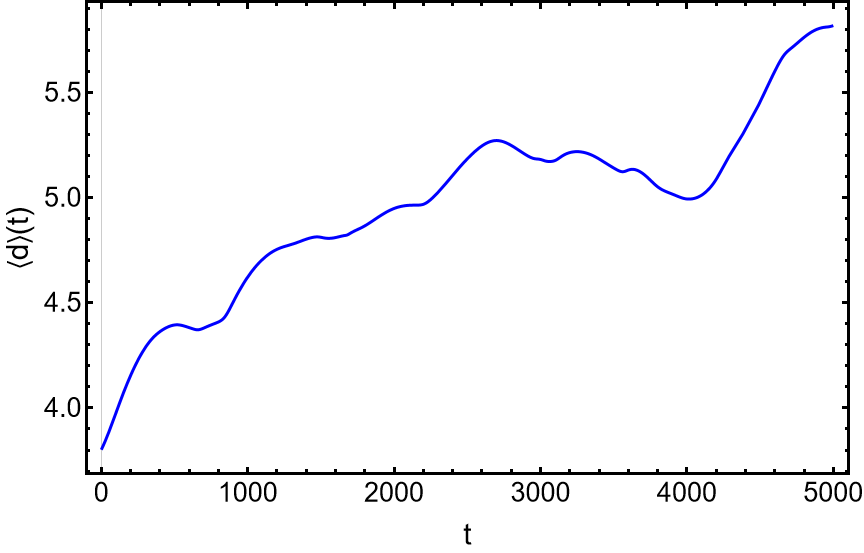} &
    \includegraphics[width=0.18\linewidth]{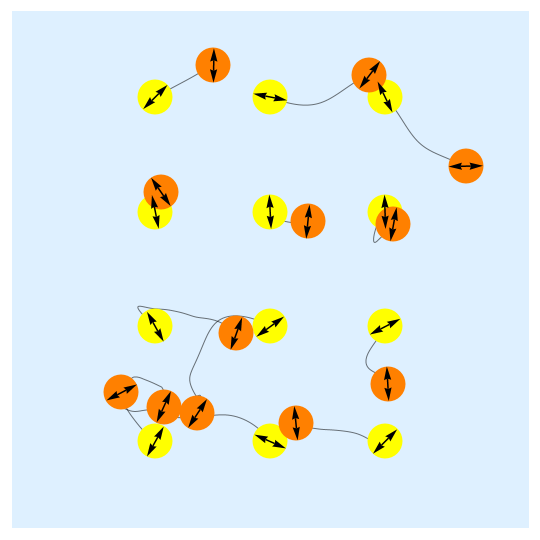} &
    \includegraphics[width=0.18\linewidth]{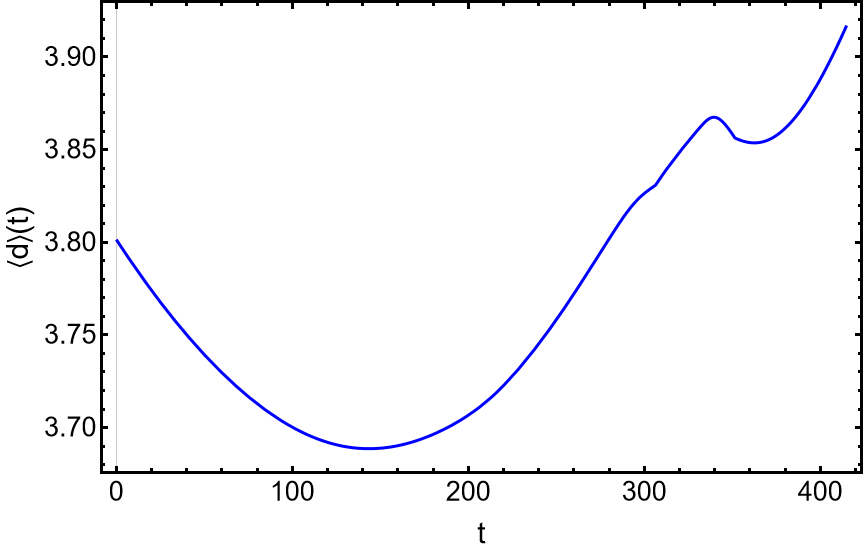} \\
  \end{tabular}

  \caption{
  Far–zone dynamics of pusher and puller dipoles in a compressible supported membrane
  with finite odd viscosity  and \emph{dynamical} orientations.  
  Rows: axial, side-by-side, perpendicular, random pair, and 12-dipole cluster.  
  Columns: trajectories and mean pair separation for pushers (left) and pullers (right).
  }
  \label{fig:odd_fz_dyn_all}
\end{figure*}

\begin{figure*}[t]
  \centering
  \begin{tabular}{cccc}
    \multicolumn{4}{c}{\textbf{Compressible membrane with odd viscosity  – Near zone – Quenched orientations}} \\[6pt]
    \multicolumn{2}{c}{\textbf{Pusher}} & \multicolumn{2}{c}{\textbf{Puller}} \\
    traj & $\langle d_{ij}\rangle(t)$ & traj & $\langle d_{ij}\rangle(t)$ \\[6pt]

    \includegraphics[width=0.18\linewidth]{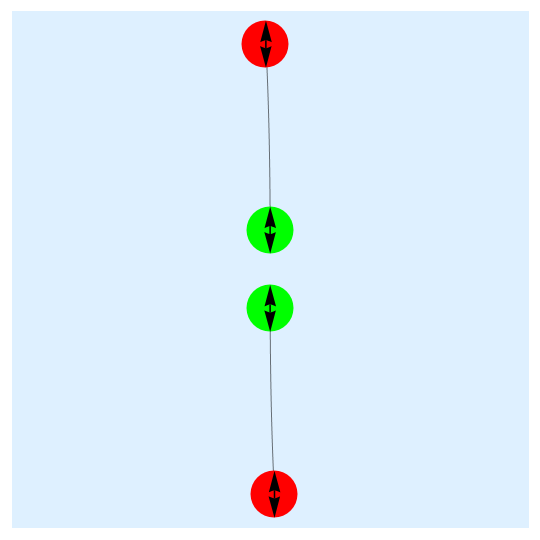} &
    \includegraphics[width=0.18\linewidth]{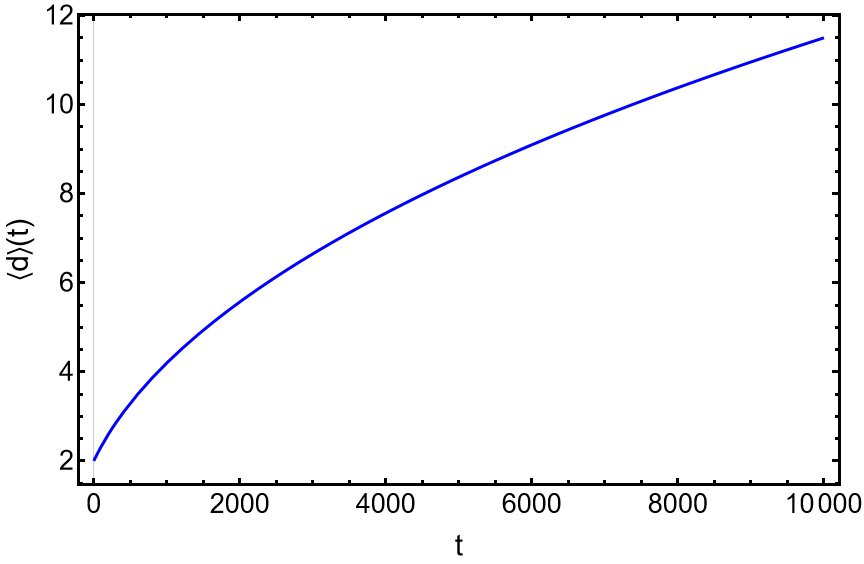} &
    \includegraphics[width=0.18\linewidth]{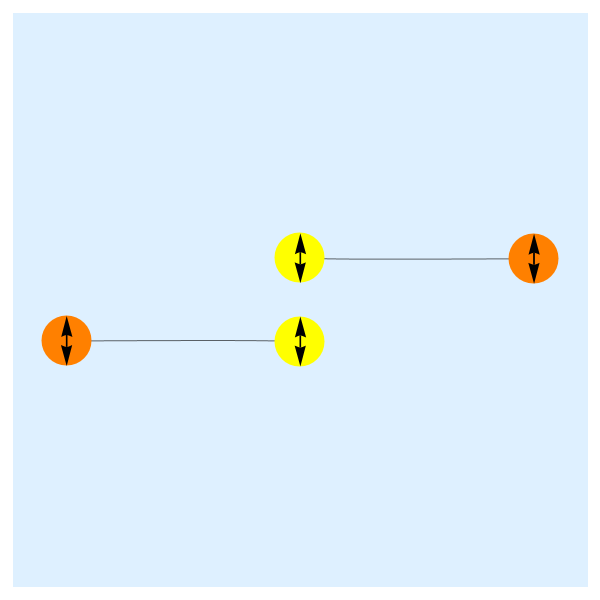} &
    \includegraphics[width=0.18\linewidth]{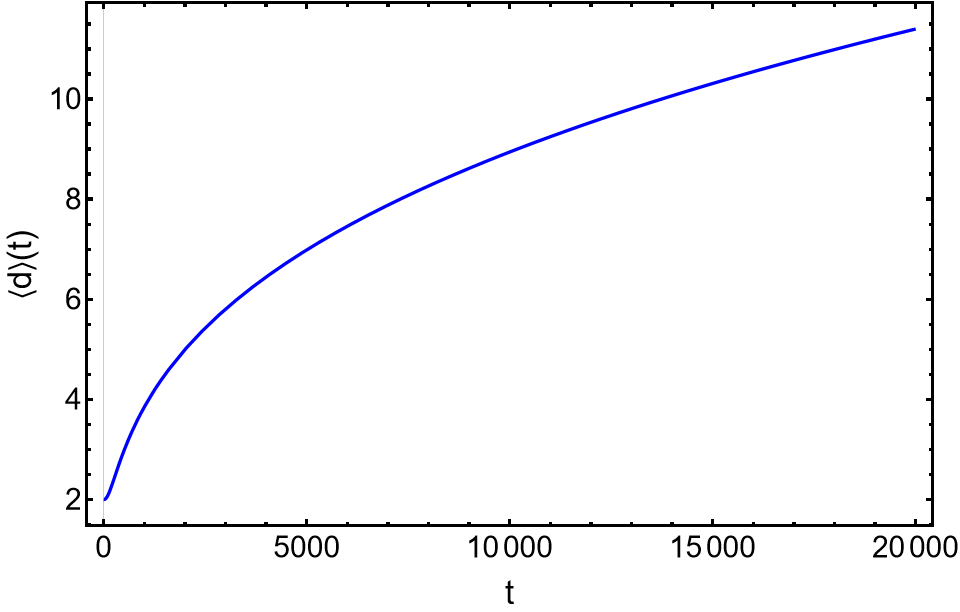} \\[4pt]

    \includegraphics[width=0.18\linewidth]{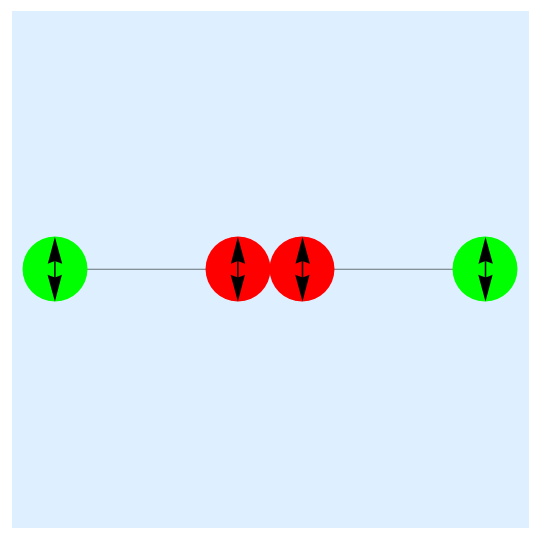} &
    \includegraphics[width=0.18\linewidth]{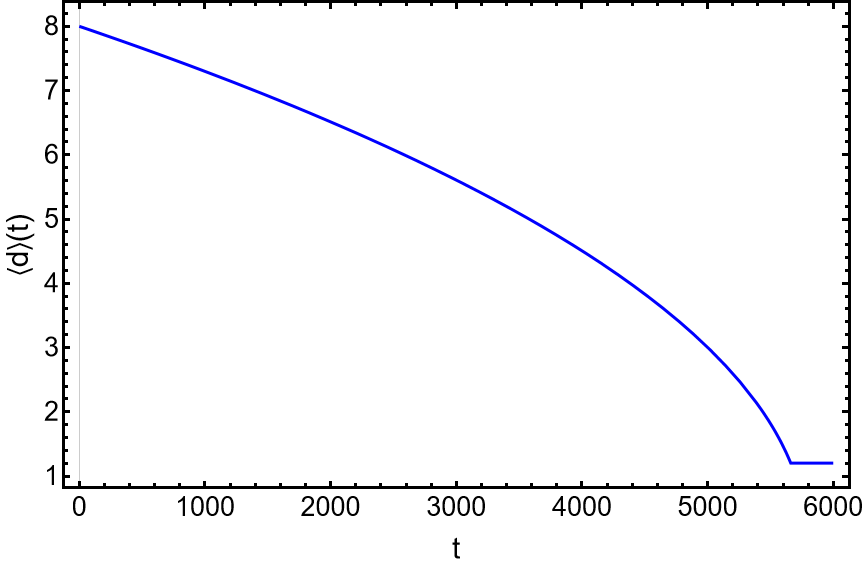} &
    \includegraphics[width=0.18\linewidth]{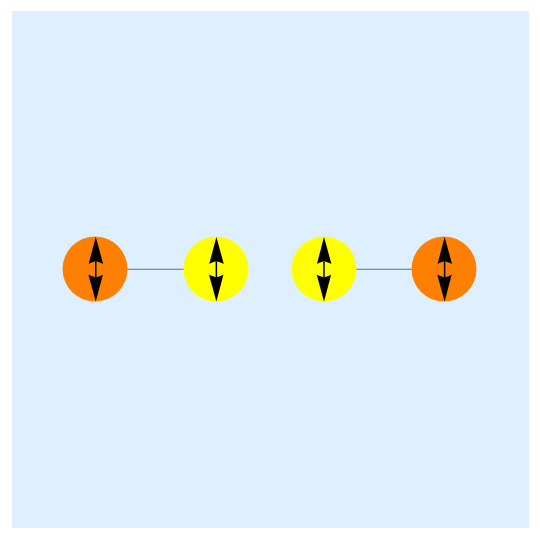} &
    \includegraphics[width=0.18\linewidth]{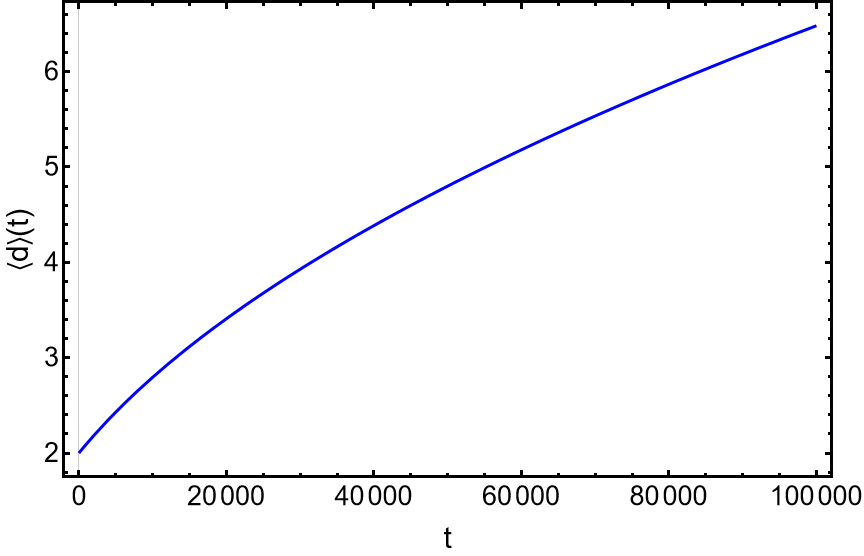} \\[4pt]

    \includegraphics[width=0.18\linewidth]{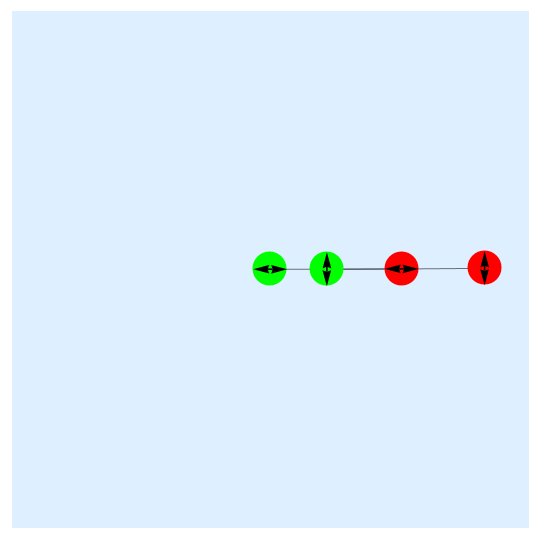} &
    \includegraphics[width=0.18\linewidth]{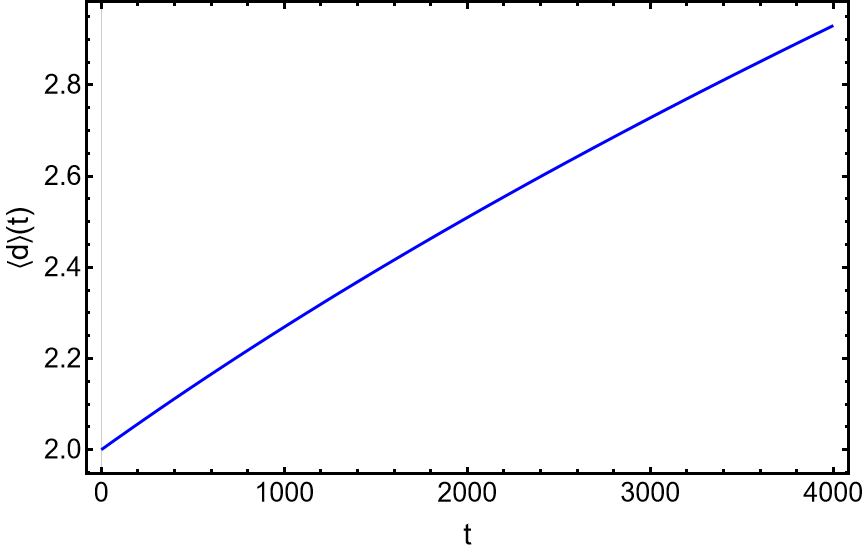} &
    \includegraphics[width=0.18\linewidth]{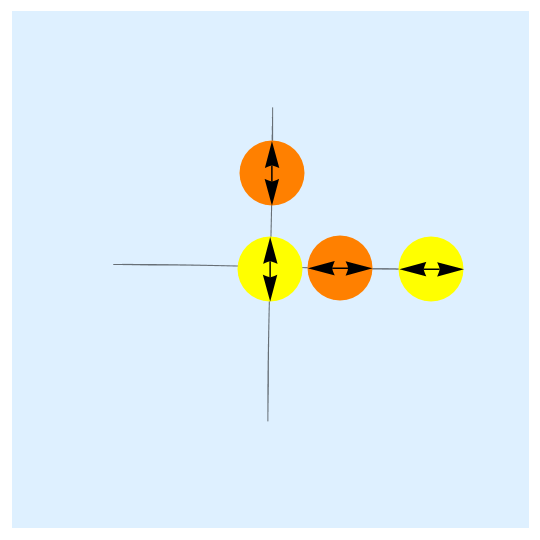} &
    \includegraphics[width=0.18\linewidth]{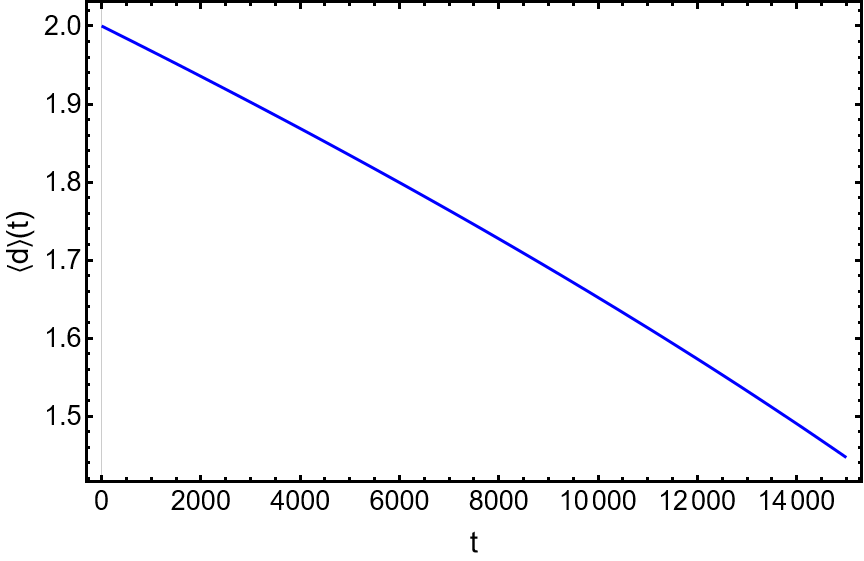} \\[4pt]

    \includegraphics[width=0.18\linewidth]{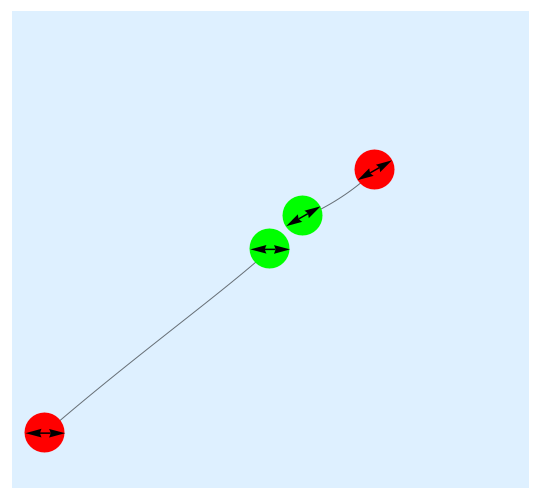} &
    \includegraphics[width=0.18\linewidth]{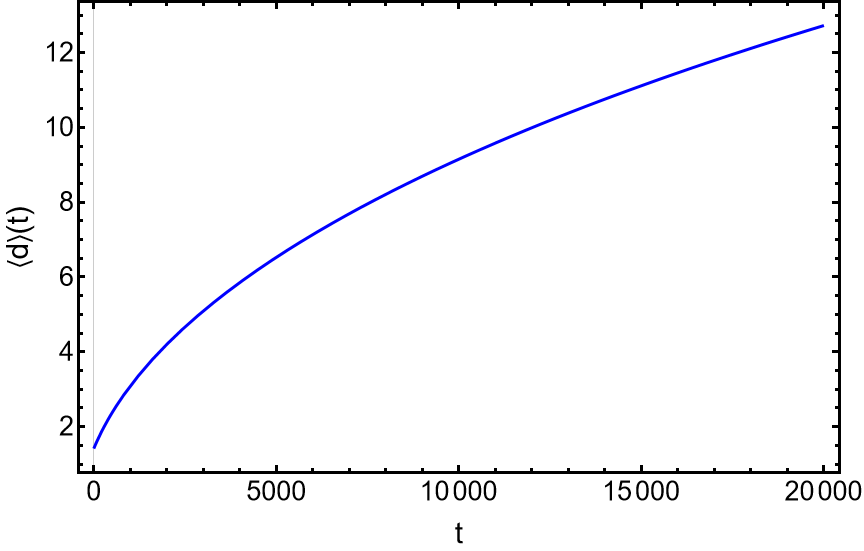} &
    \includegraphics[width=0.18\linewidth]{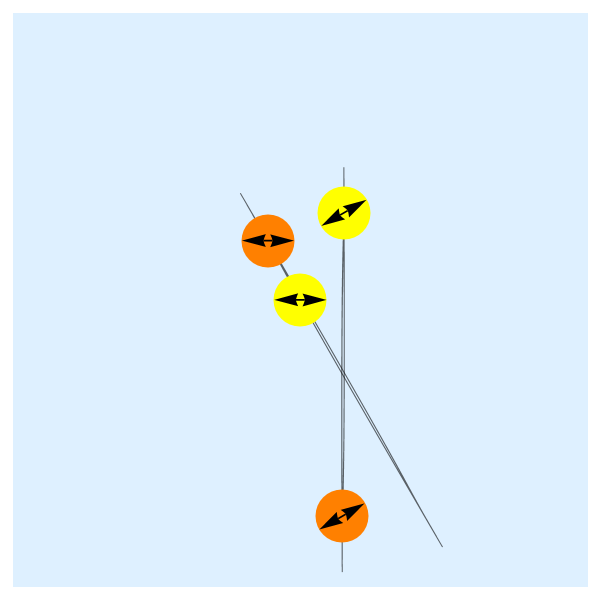} &
    \includegraphics[width=0.18\linewidth]{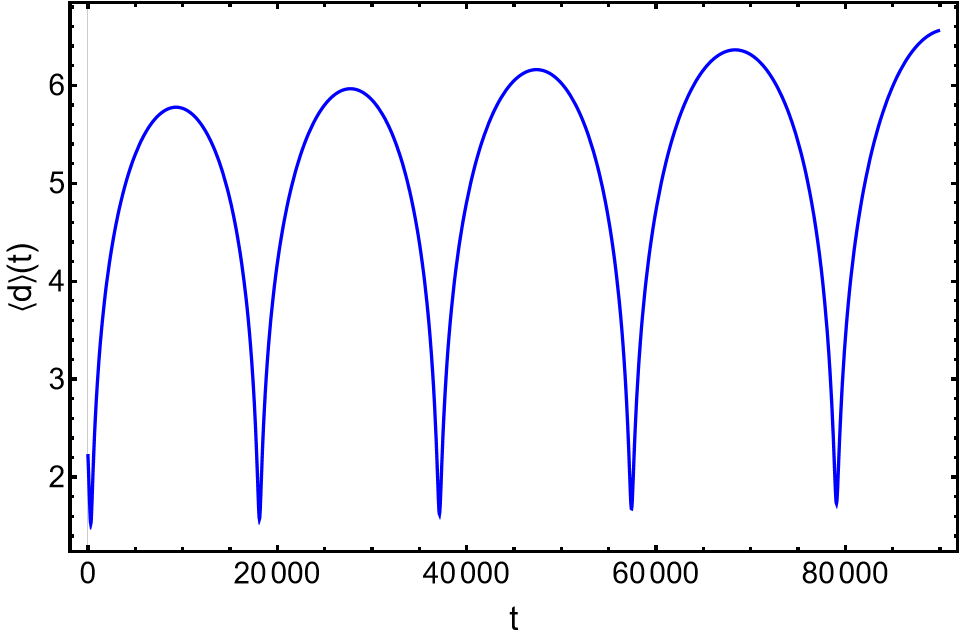} \\[4pt]

    \includegraphics[width=0.18\linewidth]{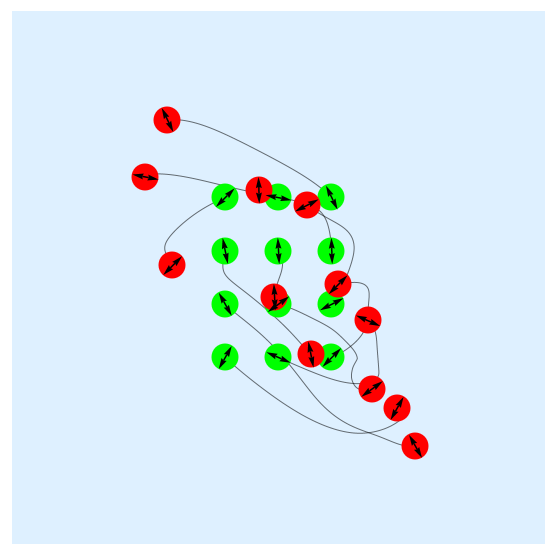} &
    \includegraphics[width=0.18\linewidth]{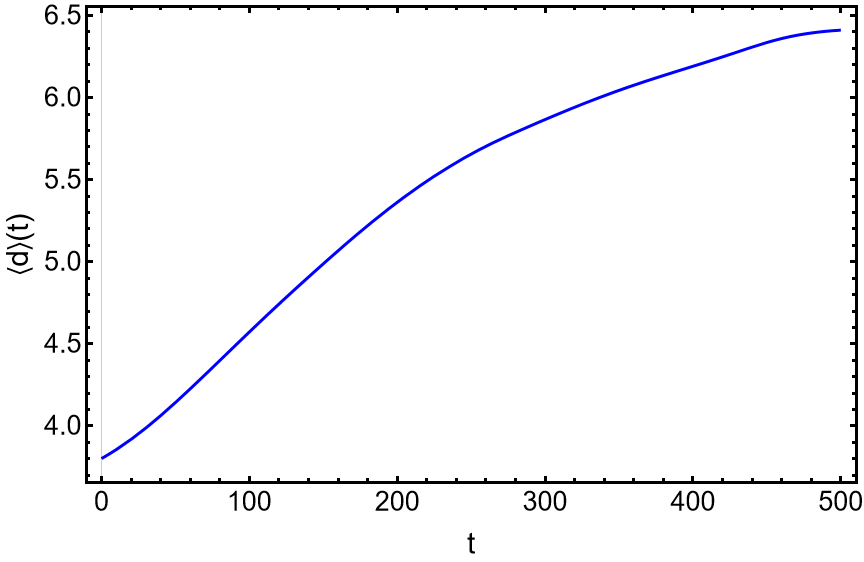} &
    \includegraphics[width=0.18\linewidth]{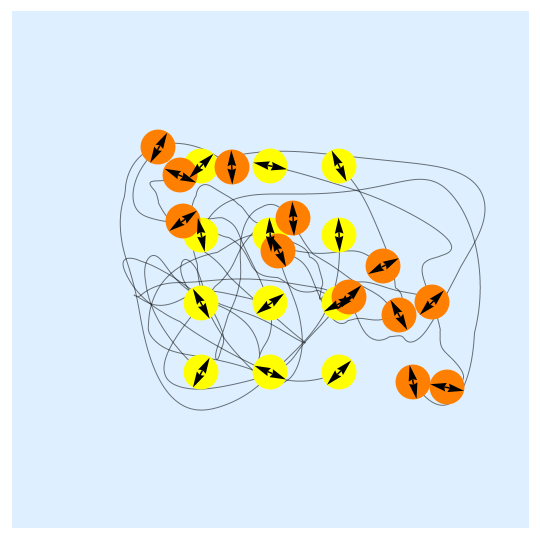} &
    \includegraphics[width=0.18\linewidth]{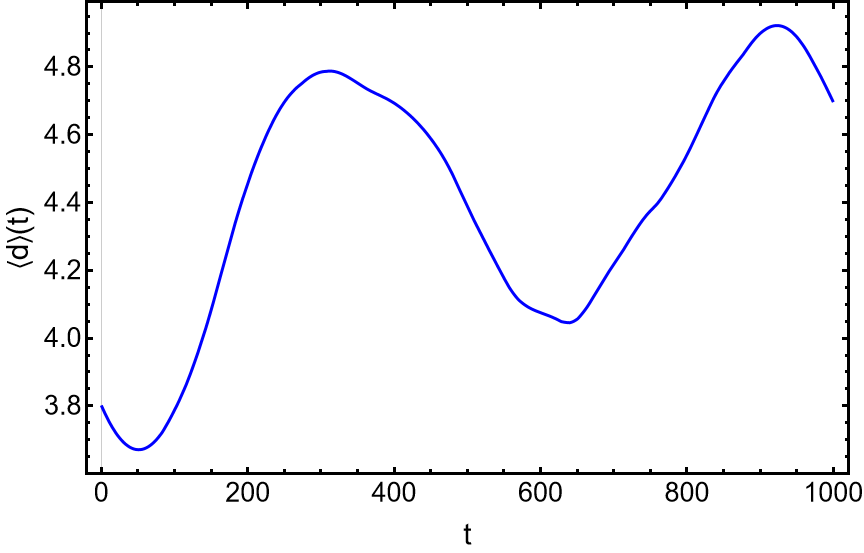} \\
  \end{tabular}

  \caption{
  Near–zone dynamics of pusher and puller dipoles in a compressible supported membrane
  with finite odd viscosity  and \emph{quenched} orientations.  
  Rows: axial, side-by-side, perpendicular, random pair, and 12-dipole cluster.  
  Columns: trajectories and mean pair separation for pushers (left) and pullers (right).
  }
  \label{fig:odd_nz_q_all}
\end{figure*}
\begin{figure*}[t]
  \centering
  \begin{tabular}{cccc}
    \multicolumn{4}{c}{\textbf{Compressible membrane with odd viscosity – Far zone – Quenched orientations}} \\[6pt]
    \multicolumn{2}{c}{\textbf{Pusher}} & \multicolumn{2}{c}{\textbf{Puller}} \\
    traj & $\langle d_{ij}\rangle(t)$ & traj & $\langle d_{ij}\rangle(t)$ \\[6pt]

    \includegraphics[width=0.18\linewidth]{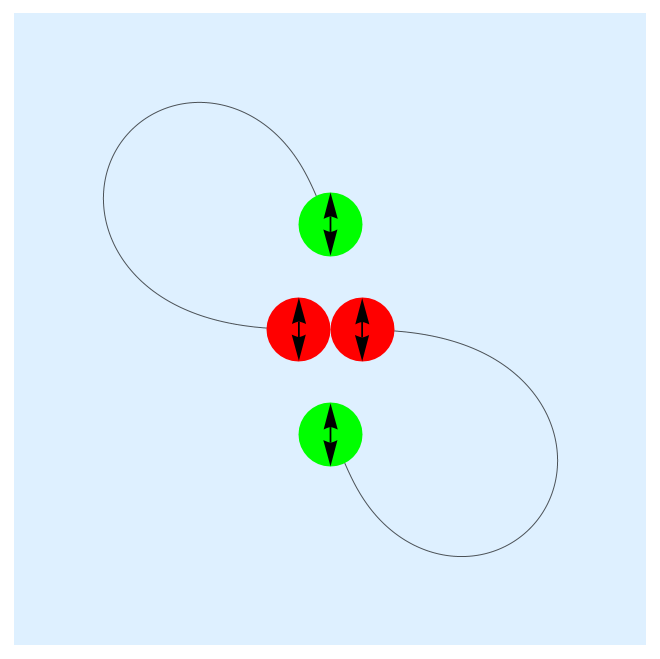} &
    \includegraphics[width=0.18\linewidth]{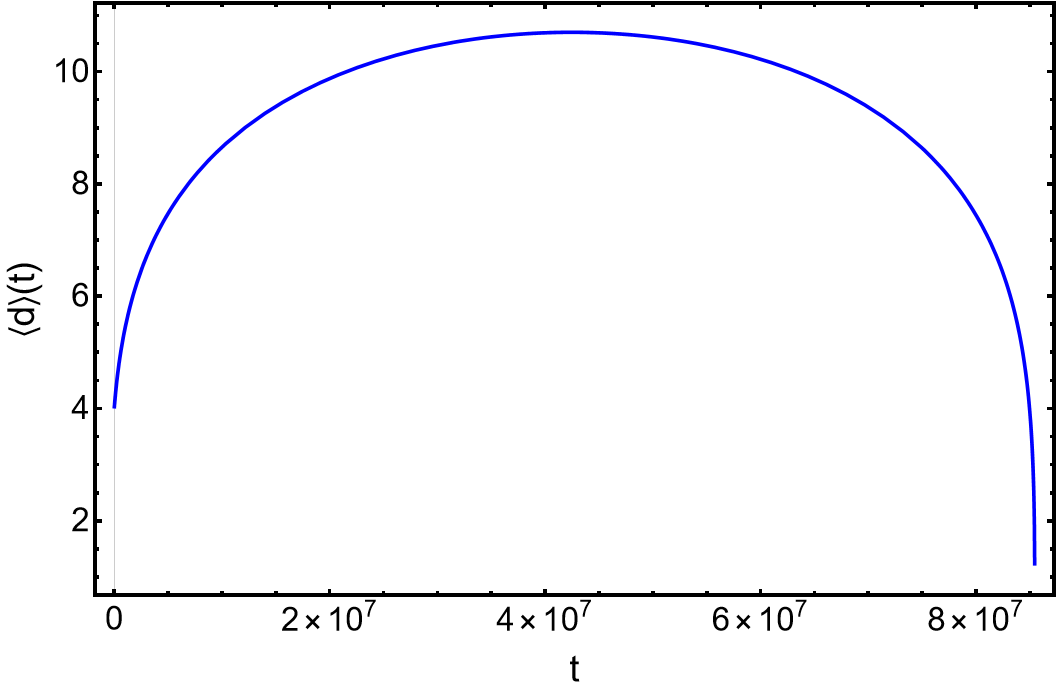} &
    \includegraphics[width=0.18\linewidth]{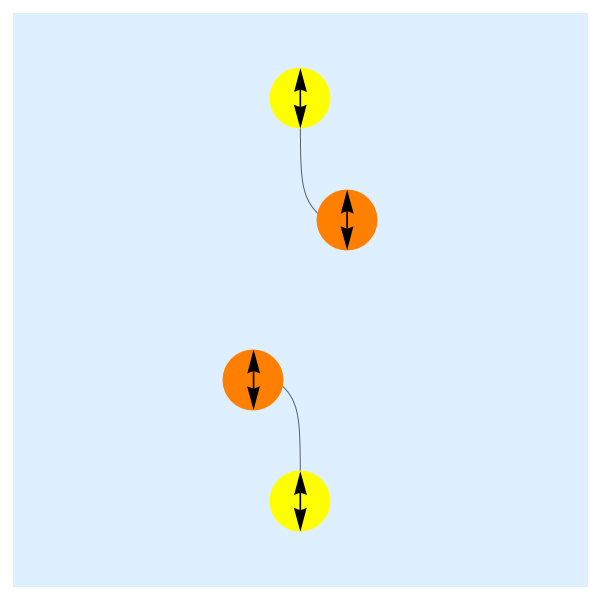} &
    \includegraphics[width=0.18\linewidth]{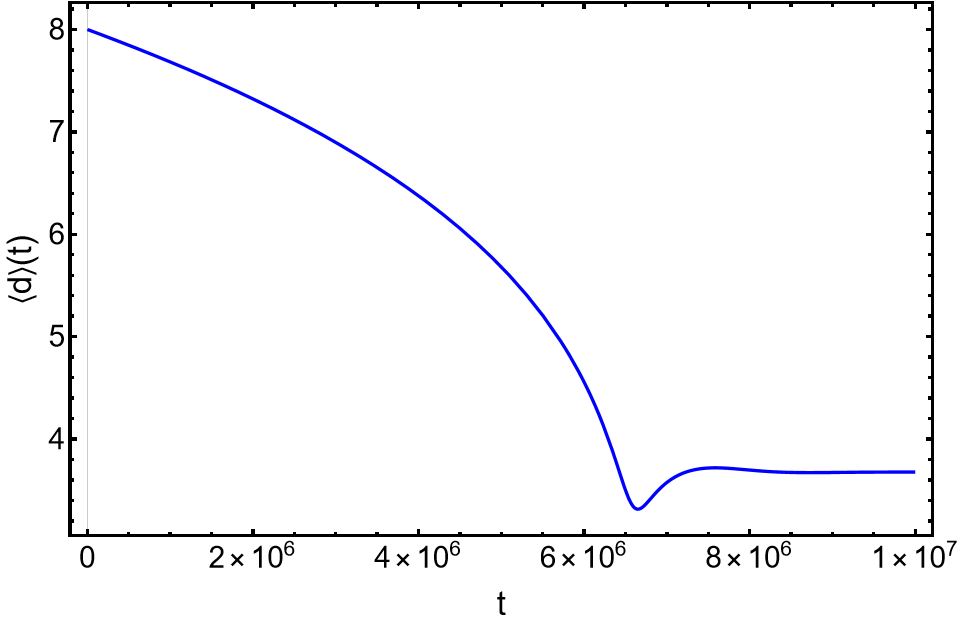} \\[4pt]

    \includegraphics[width=0.18\linewidth]{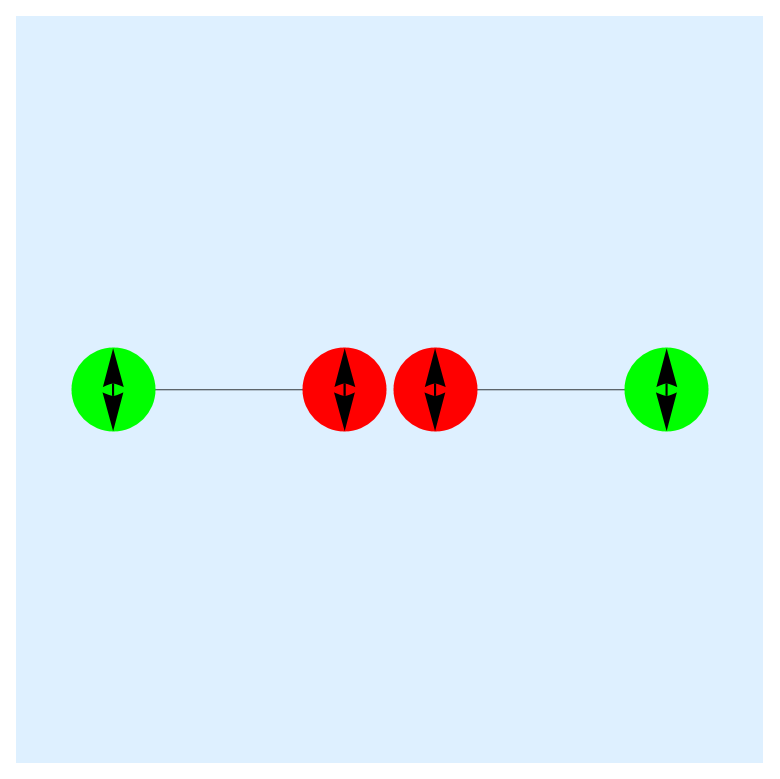} &
    \includegraphics[width=0.18\linewidth]{images/Fcmpoddnqfzpushersbsdist.png} &
    \includegraphics[width=0.18\linewidth]{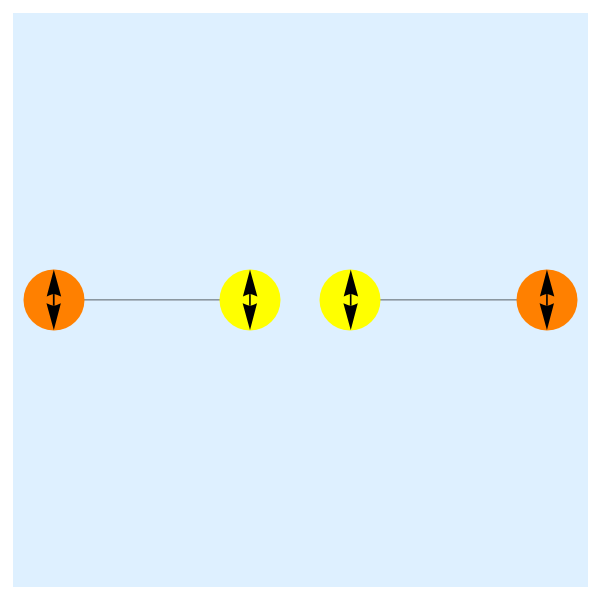} &
    \includegraphics[width=0.18\linewidth]{images/Fcmpoddnqfzpullersbsdist.png} \\[4pt]

    \includegraphics[width=0.18\linewidth]{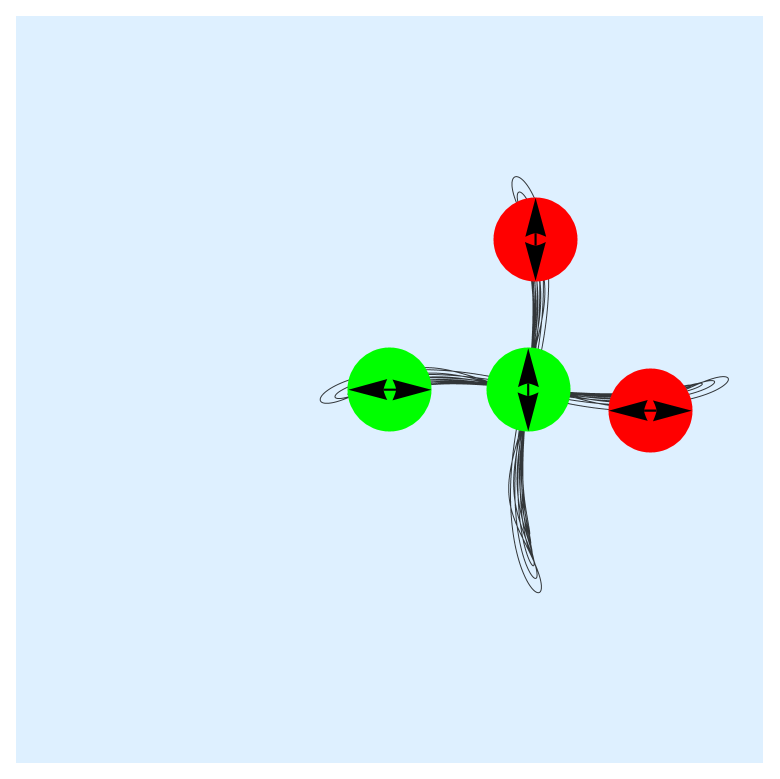} &
    \includegraphics[width=0.18\linewidth]{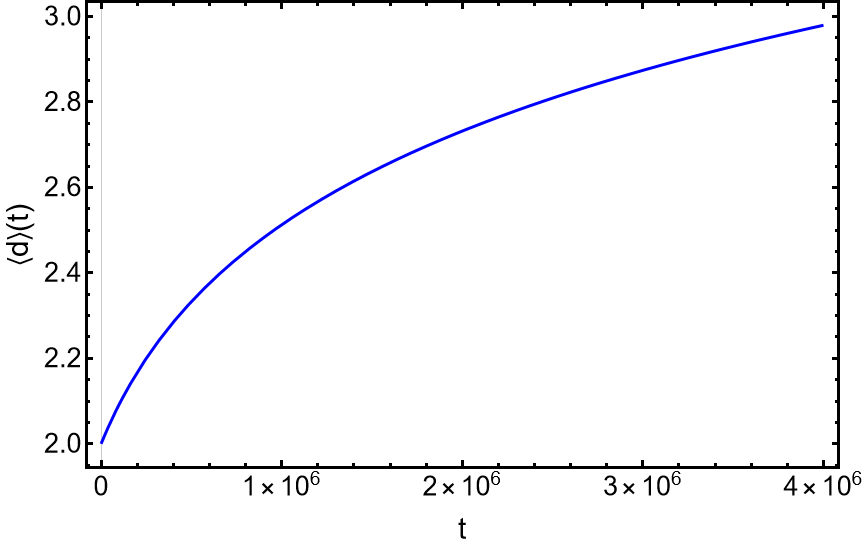} &
    \includegraphics[width=0.18\linewidth]{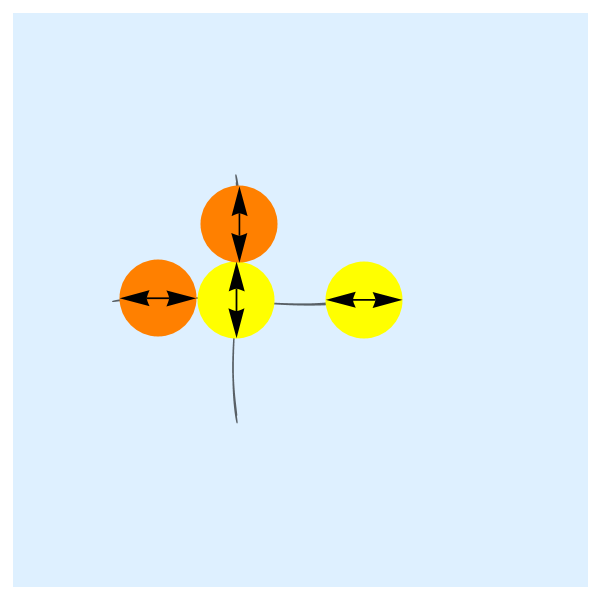} &
    \includegraphics[width=0.18\linewidth]{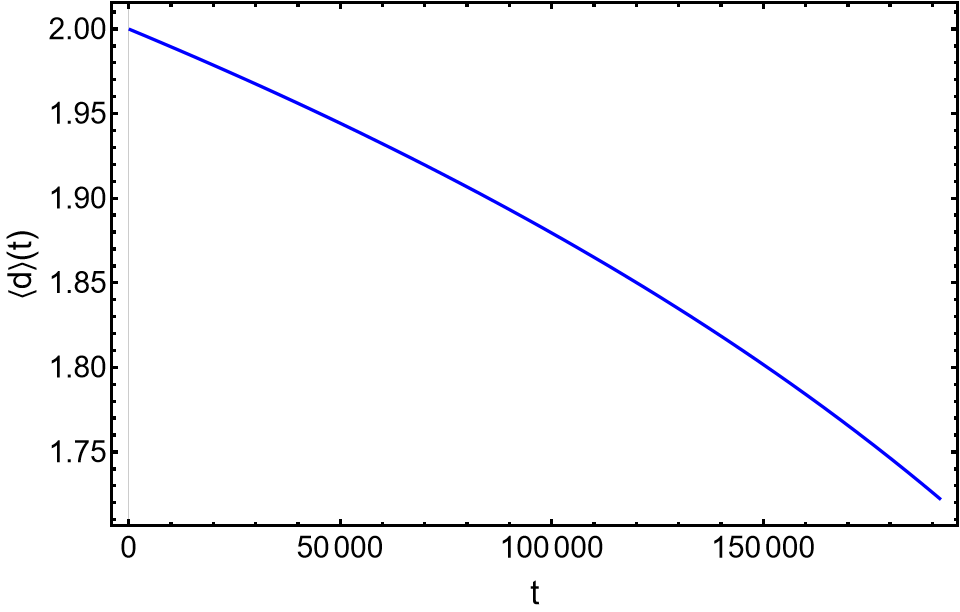} \\[4pt]

    \includegraphics[width=0.18\linewidth]{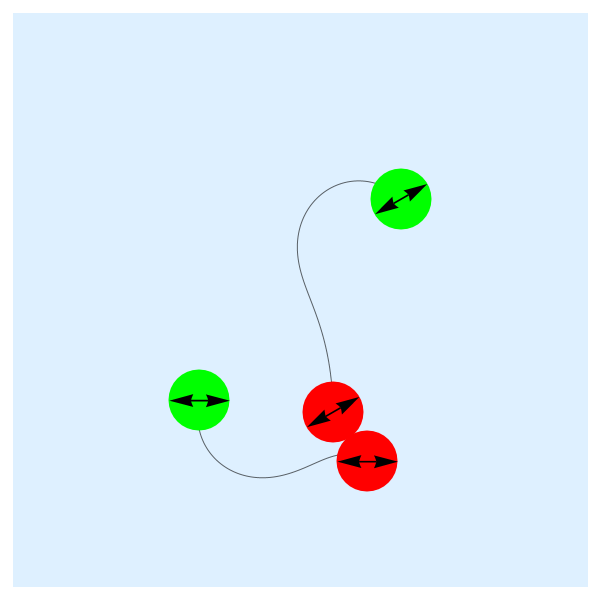} &
    \includegraphics[width=0.18\linewidth]{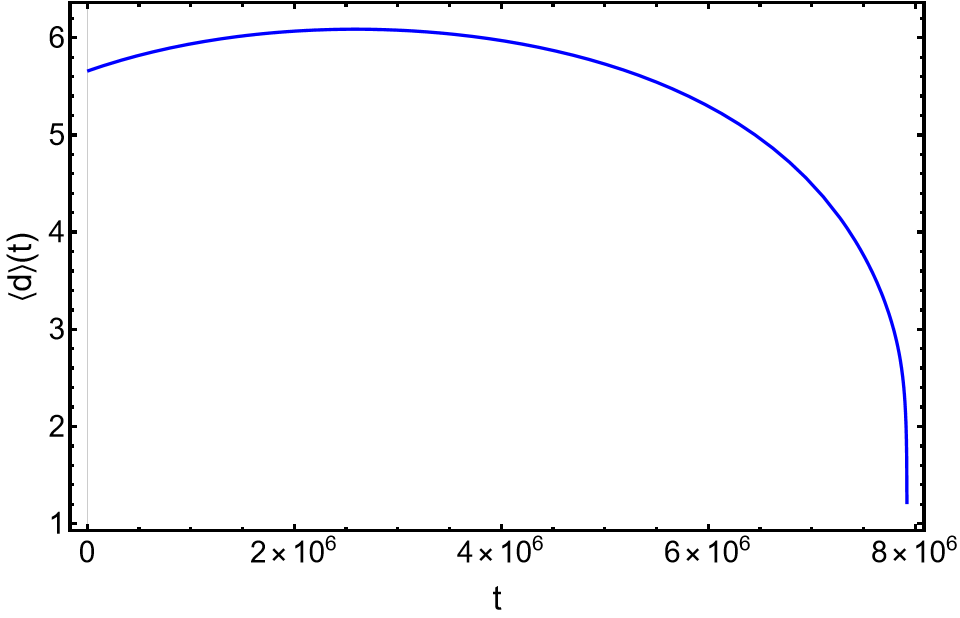} &
    \includegraphics[width=0.18\linewidth]{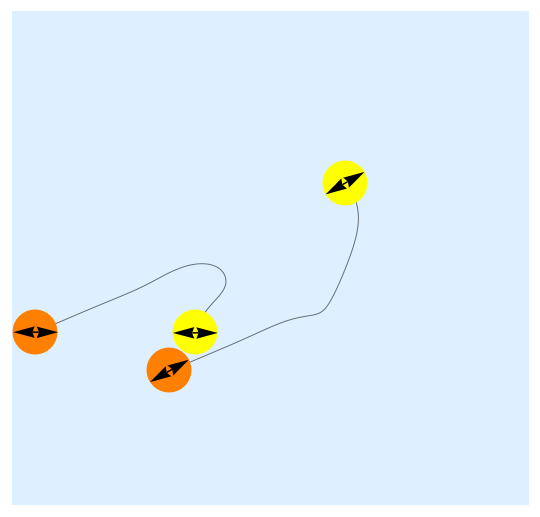} &
    \includegraphics[width=0.18\linewidth]{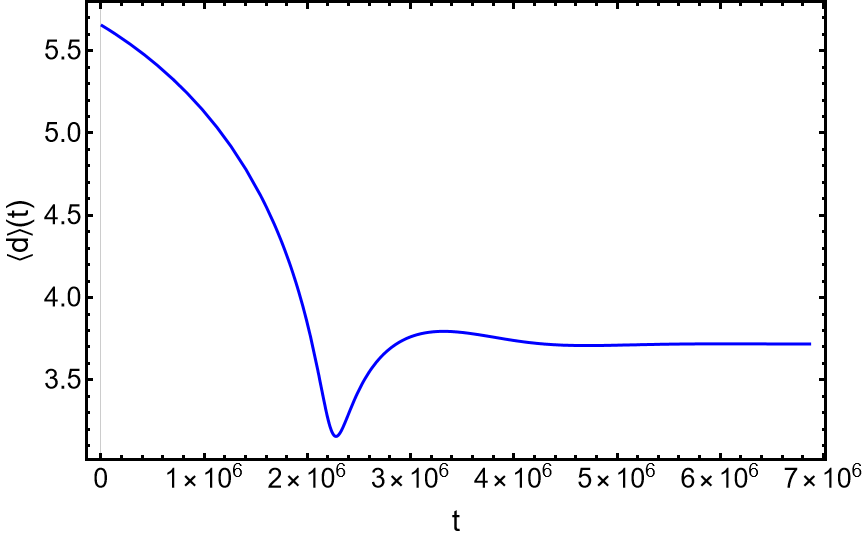} \\[4pt]

    \includegraphics[width=0.18\linewidth]{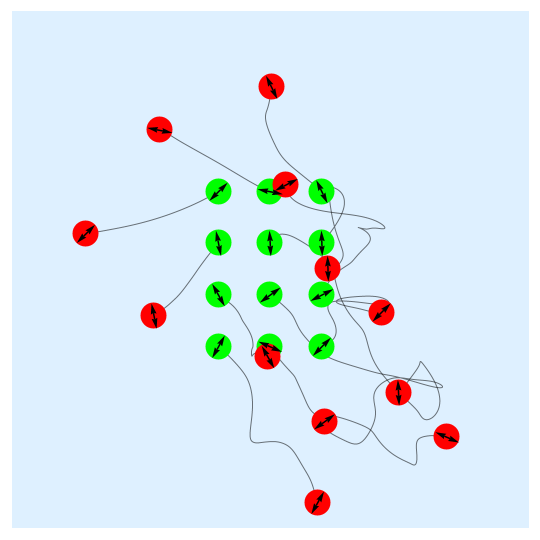} &
    \includegraphics[width=0.18\linewidth]{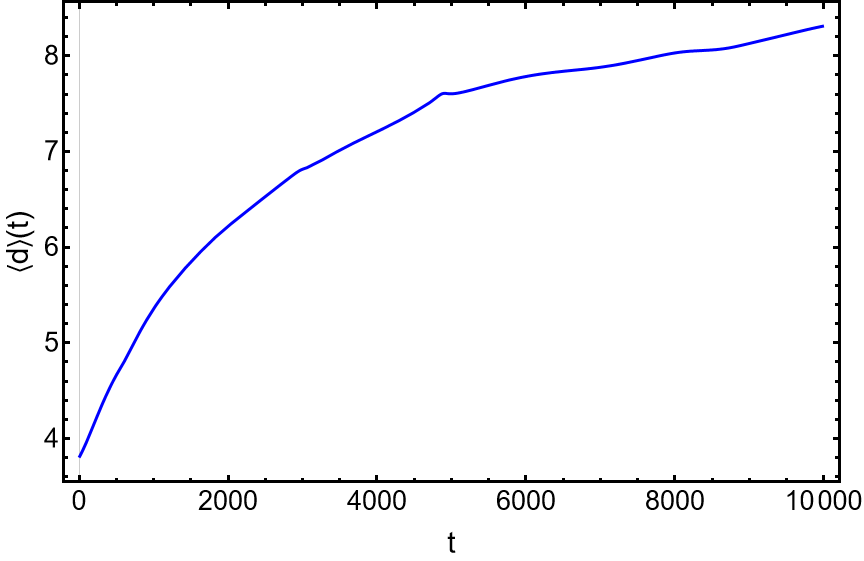} &
    \includegraphics[width=0.18\linewidth]{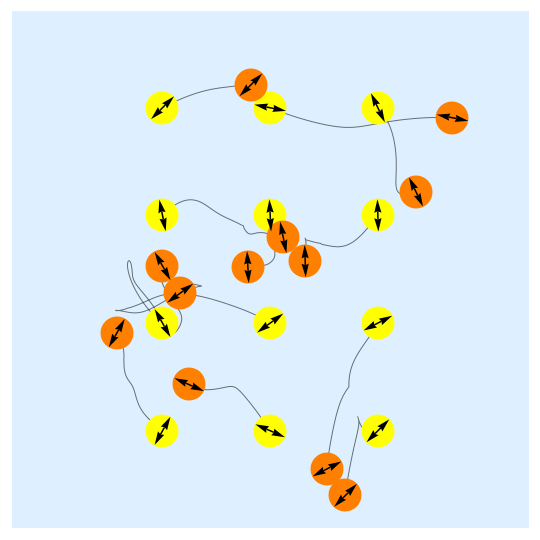} &
    \includegraphics[width=0.18\linewidth]{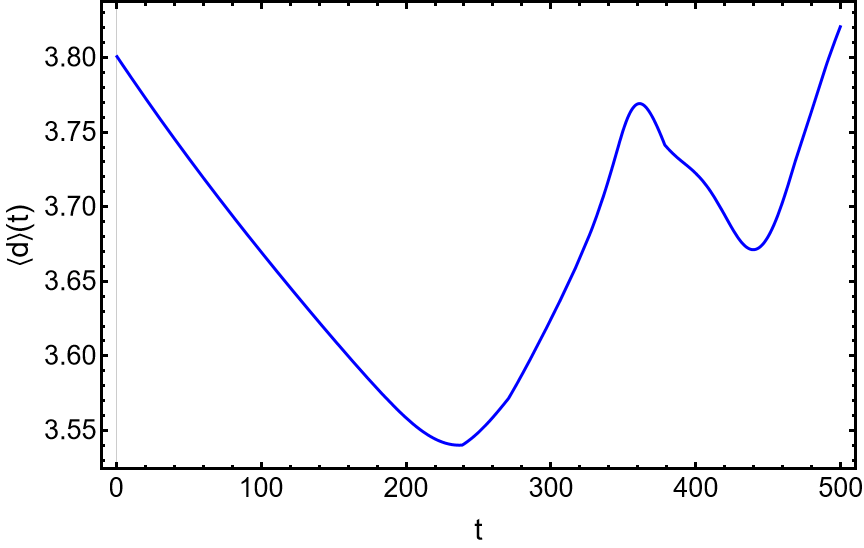} \\
  \end{tabular}

  \caption{
  Far–zone dynamics of pusher and puller dipoles in a compressible supported membrane
  with finite odd viscosity  and \emph{quenched} orientations.  
  Rows: axial, side-by-side, perpendicular, random pair, and 12-dipole cluster.  
  Columns: trajectories and mean pair separation for pushers (left) and pullers (right).
  }
  \label{fig:odd_fz_q_all}
\end{figure*}

\section{Real–space Green's function for a compressible membrane with odd viscosity}
\label{app:greens_real_space}
The starting point of this section is the Fourier–space representation of the Green's tensor
$G_{ij}(\mathbf k)$ obtained by solving the linear Stokes equations (regulated by the Brinkman friction) in momentum
space, as derived for two-dimensional fluid layers with odd viscosity in
Refs.~\cite{HosakaKomuraAndelman2021,Hosaka2023}.
While those works obtain real–space Green's functions in specific limiting
cases, here we carry out the Fourier inversion in full generality for a
compressible supported membrane with finite odd viscosity. Solving the linear Stokes--Brinkman equations for a compressible membrane with
shear, dilatational, and odd viscosities in Fourier space, and eliminating the
pressure using the lubrication--approximated compressibility relation, yields a closed
$2\times2$ linear system for the longitudinal and transverse velocity modes.
Inverting this system gives the mobility tensor
$v_i(\mathbf k)=G_{ij}(\mathbf k)F_j(\mathbf k)$ in terms of three hydrodynamic
screening lengths: the shear screening length
$\kappa^{-1}=\sqrt{\eta_s/\zeta_{\parallel}}$, the compressional screening length
$\lambda^{-1}=\sqrt{h(\eta_s+\eta_d)/(3\eta+h\zeta_{\parallel})}$, and the
odd--viscous length $\nu^{-1}=\sqrt{\eta_o/\zeta_{\perp}}$.
The resulting Green's tensor contains longitudinal, transverse, and
antisymmetric components and reads
\begin{equation}
G_{ij}(\mathbf k)
 = \frac{
  \eta_s (k^2+\kappa^2)\,\hat k_i \hat k_j
 + (\eta_s+\eta_d)(k^2+\lambda^2)\,\bar k_i \bar k_j
 - \eta_o (k^2+\nu^2)\,\varepsilon_{ij}}
 {D(k)},
\label{eq:Gk_def}
\end{equation}
with
\begin{equation}
D(k) =
\eta_s(\eta_s+\eta_d)(k^2+\kappa^2)(k^2+\lambda^2)
+ \eta_o^2 (k^2+\nu^2)^2 .\nn
\end{equation}
Here $\eta_s$ and $\eta_d$ are the shear and dilatational 2D viscosities,
$\kappa^{-1}$, $\lambda^{-1}$ and $\nu^{-1}$ are hydrodynamic screening
lengths defined in main text, and we use the unit vectors
\(
\hat k_i = k_i/k,\;
\bar k_i = -\varepsilon_{ij}\hat k_j
\)
with $\varepsilon_{12}=-\varepsilon_{21}=1$.

The real–space tensor is defined by the inverse Fourier transform
\begin{equation}
G_{ij}(\mathbf r)
 = \int \frac{d^2k}{(2\pi)^2}\;
   e^{i\mathbf k\cdot\mathbf r}\,G_{ij}(\mathbf k),
\qquad
r = |\mathbf r|,\quad
\hat r_i = r_i/r.\nn
\end{equation}
By rotational invariance in the membrane plane the result will take the form
\begin{equation}
G_{ij}(\mathbf r)
 = A_0(r)\,\delta_{ij} + A_1(r)\,\hat r_i \hat r_j + A_2(r)\,\varepsilon_{ij},
\label{eq:Gij_iso_form}
\end{equation}
and then we proceed to obtain explicit expressions for $A_0(r)$, $A_1(r)$ and
$A_2(r)$. It is convenient to decompose the Fourier–space tensor into scalar functions
multiplying isotropic projectors.  Using
\(
\delta_{ij} = \hat k_i \hat k_j + \bar k_i \bar k_j
\)
and
\(
\bar k_i \bar k_j = \delta_{ij} - \hat k_i \hat k_j
\),
the tensor~\eqref{eq:Gk_def} can be rewritten as
\begin{equation}
G_{ij}(\mathbf k)
 = \beta(k)\,\delta_{ij}
 + \bigl[\alpha(k)-\beta(k)\bigr]\hat k_i\hat k_j
 + \gamma(k)\,\varepsilon_{ij},
\label{eq:Gk_alphabeta}
\end{equation}
where
\begin{equation}
\alpha(k) = \frac{\eta_s(k^2+\kappa^2)}{D(k)},\qquad
\beta(k)  = \frac{(\eta_s+\eta_d)(k^2+\lambda^2)}{D(k)},\qquad
\gamma(k) = -\frac{\eta_o(k^2+\nu^2)}{D(k)}.\nn
\label{eq:alphabeta_gamma_def}
\end{equation}

Since all scalar functions depend only on $k=|\mathbf k|$, we can use polar
coordinates $(k,\theta)$ with $\mathbf k\cdot\mathbf r = kr\cos\theta$ and the
identity
\(
\int_0^{2\pi} d\theta\,e^{ikr\cos\theta} = 2\pi J_0(kr)
\)
to reduce the inverse transform to Hankel transforms.  For any isotropic
scalar $F(k)$ we define the Hankel transform
\begin{equation}
\mathcal H\{F\}(r)
 = \int_0^\infty \frac{k\,dk}{2\pi}\,F(k) J_0(kr).
\label{eq:Hankel_def}
\end{equation}
Standard Fourier identities then give
\begin{equation}
G_{ij}(\mathbf r)
 = \delta_{ij}\Psi(r) - \partial_i\partial_j\Phi(r) + \varepsilon_{ij}\chi(r),
\label{eq:Gij_PhiPsiChi}
\end{equation}
with
\begin{equation}
\Psi(r) = \mathcal H\{\beta\}(r),\qquad
\chi(r) = \mathcal H\{\gamma\}(r),\nn
\end{equation}
and
\begin{equation}
\Phi(r)
 = \int_0^\infty \frac{k\,dk}{2\pi}\,
   \frac{\alpha(k)-\beta(k)}{k^2}\,J_0(kr).
\label{eq:Phi_def}
\end{equation}

For a purely radial scalar $\Phi(r)$ we have
\begin{equation}
\partial_i\Phi = \Phi'(r)\,\hat r_i,\qquad
\partial_i\partial_j\Phi
 = \hat r_i\hat r_j\,\Phi''(r)
 + \bigl(\delta_{ij}-\hat r_i\hat r_j\bigr)\frac{\Phi'(r)}{r}.
\label{eq:radial_second_derivative}
\end{equation}
Substituting~\eqref{eq:radial_second_derivative} into
\eqref{eq:Gij_PhiPsiChi} and comparing with the  form
\eqref{eq:Gij_iso_form} yields
\begin{equation}
A_0(r) = \Psi(r) - \frac{\Phi'(r)}{r},\qquad
A_1(r) = -\Phi''(r) + \frac{\Phi'(r)}{r},\qquad
A_2(r) = \chi(r).
\label{eq:A012_from_PhiPsiChi}
\end{equation}
The remainder of the derivation is therefore the evaluation of
$\Psi(r)$, $\chi(r)$, and $\Phi(r)$. Introducing the variable $s=k^2$, so that $D(k)\equiv D(s)$ is a quadratic
polynomial,
\begin{equation}
D(s) = As^2 + Bs + C,
\label{eq:D_quadratic}
\end{equation}
with coefficients
\begin{equation}
A = \eta_s(\eta_s+\eta_d) + \eta_o^2,\qquad
B = \eta_s(\eta_s+\eta_d)(\kappa^2+\lambda^2) + 2\eta_o^2\nu^2,\qquad
C = \eta_s(\eta_s+\eta_d)\kappa^2\lambda^2 + \eta_o^2\nu^4.\nn
\label{eq:ABC_def}
\end{equation}
Let $s_1$ and $s_2$ be the roots of $D(s)$,
\begin{equation}
D(s) = A(s-s_1)(s-s_2),\qquad
s_{1,2} = \frac{-B\pm\sqrt{\Delta}}{2A},\qquad
\Delta = B^2-4AC,
\label{deltadef}
\end{equation}
and let us also define the positive screening masses
\begin{equation}
m_i = \sqrt{-s_i},\qquad i=1,2.\nn
\end{equation}
Each of the scalar functions in~\eqref{eq:alphabeta_gamma_def} is a rational
function of $s$ with denominator $D(s)$ and a linear numerator; therefore it
admits a partial–fraction representation.  For the isotropic sector $\beta$ and
the antisymmetric sector $\gamma$ we write
\begin{equation}
\beta(s) = \frac{(\eta_s+\eta_d)(s+\lambda^2)}{D(s)}
         = \sum_{i=1}^2 \frac{R_i^{(\beta)}}{s-s_i},
\qquad
\gamma(s) = -\frac{\eta_o(s+\nu^2)}{D(s)}
         = \sum_{i=1}^2 \frac{R_i^{(\gamma)}}{s-s_i},
\end{equation}
with residues
\begin{equation}
R_i^{(\beta)} = \frac{(\eta_s+\eta_d)(s_i+\lambda^2)}{A(s_i-s_j)},
\qquad
R_i^{(\gamma)} = -\,\frac{\eta_o(s_i+\nu^2)}{A(s_i-s_j)},
\qquad i\neq j.\nn
\label{eq:Rbeta_Rgamma}
\end{equation}
For the longitudinal sector we consider
\begin{equation}
\frac{\alpha(s)-\beta(s)}{s}
 = \frac{N_\Phi(s)}{sD(s)},
\end{equation}
where
\begin{equation}
N_\Phi(s) = \eta_s(s+\kappa^2) - (\eta_s+\eta_d)(s+\lambda^2)
          = -\eta_d s + \bigl[\eta_s\kappa^2 - (\eta_s+\eta_d)\lambda^2\bigr].\nn
\end{equation}
We decompose this as 
\begin{equation}
\frac{N_\Phi(s)}{sD(s)}
 = \frac{C_0}{s} + \frac{C_1}{s-s_1} + \frac{C_2}{s-s_2},
\label{eq:Phi_partial_fraction}
\end{equation}
with
\begin{equation}
C_0 = \frac{N_\Phi(0)}{D(0)}
    = \frac{\eta_s\kappa^2 - (\eta_s+\eta_d)\lambda^2}{C},\nn
\label{eq:C0_def}
\end{equation}
and
\begin{equation}
C_i = \frac{N_\Phi(s_i)}{s_iD'(s_i)}
    = \frac{-\eta_d s_i + \bigl[\eta_s\kappa^2 - (\eta_s+\eta_d)\lambda^2\bigr]}
           {s_i A(s_i-s_j)},
\qquad i\neq j.\nn
\label{eq:Ci_def}
\end{equation}
The Hankel transform~\eqref{eq:Hankel_def} of each simple pole is given by the
standard identity
\begin{equation}
\int_0^\infty \frac{k\,dk}{2\pi}\,\frac{J_0(kr)}{k^2+m^2}
 = \frac{1}{2\pi}K_0(mr),
\label{eq:Hankel_simple_pole}
\end{equation}
where $K_\nu$ is the modified Bessel function of the second kind.  A logarithmic
kernel arises from the $1/s$ term in~\eqref{eq:Phi_partial_fraction},
\begin{equation}
\int_0^\infty \frac{k\,dk}{2\pi}\,\frac{J_0(kr)}{k^2}
 = \frac{1}{2\pi}\bigl(C_{\rm reg}-\ln r\bigr),\nn
\label{eq:Hankel_log}
\end{equation}
where $C_{\rm reg}$ is a regularization constant that drops out of
the final expressions after differentiation. Using the so-constructed partial–fraction forms we thus obtain
\begin{align}
\Psi(r)
 &= \mathcal H\{\beta\}(r)
  = \frac{1}{2\pi}\sum_{i=1}^2 R_i^{(\beta)} K_0(m_i r),
\\[4pt]
\chi(r)
 &= \mathcal H\{\gamma\}(r)
  = \frac{1}{2\pi}\sum_{i=1}^2 R_i^{(\gamma)} K_0(m_i r),
\\[4pt]
\Phi(r)
 &= \int_0^\infty \frac{k\,dk}{2\pi}\,\frac{\alpha(k)-\beta(k)}{k^2}J_0(kr)
\nonumber\\
 &= -\,\frac{C_0}{2\pi}\ln r
  + \frac{1}{2\pi}\sum_{i=1}^2 C_i K_0(m_i r) + \text{const.}
\label{eq:Psi_chi_Phi_expr}
\end{align}
We differentiate (and use another standard identity  $K_0'(x) = -K_1(x)$ ) to obtain
\begin{align}
\Phi'(r)
 &= -\,\frac{C_0}{2\pi}\frac{1}{r}
    - \frac{1}{2\pi}\sum_{i=1}^2 C_i m_i K_1(m_i r),
\label{eq:Phi_prime}
\\[4pt]
\Phi''(r)
 &= \frac{C_0}{2\pi}\frac{1}{r^2}
    + \frac{1}{2\pi}\sum_{i=1}^2 C_i
       \left[m_i^2 K_0(m_i r) + \frac{m_i}{r}K_1(m_i r)\right].
\label{eq:Phi_2prime}
\end{align}
Substituting Eqs.~\eqref{eq:Psi_chi_Phi_expr}–\eqref{eq:Phi_2prime} into
Eqs.~\eqref{eq:A012_from_PhiPsiChi} yields the explicit coefficients in
Eq.~\eqref{eq:Gij_iso_form}:
\begin{align}
A_0(r)
 &= \frac{1}{2\pi}\sum_{i=1}^2 R_i^{(\beta)} K_0(m_i r)
  + \frac{C_0}{2\pi}\frac{1}{r^2}
  + \frac{1}{2\pi}\sum_{i=1}^2 C_i \frac{m_i}{r} K_1(m_i r),
\\[4pt]
A_1(r)
 &= -\,\frac{1}{2\pi}\sum_{i=1}^2 C_i m_i^2 K_0(m_i r)
    - \frac{1}{\pi}\sum_{i=1}^2 C_i \frac{m_i}{r} K_1(m_i r)
    - \frac{C_0}{\pi}\frac{1}{r^2},
\\[4pt]
A_2(r)
 &= \frac{1}{2\pi}\sum_{i=1}^2 R_i^{(\gamma)} K_0(m_i r),
\end{align}
with the coefficients $A,B,C$, the roots $s_i$, the masses $m_i$, and the
residues $R_i^{(\beta)}$, $R_i^{(\gamma)}$, $C_0$, $C_i$ defined in
Eqs.~\eqref{eq:ABC_def}, \eqref{eq:Rbeta_Rgamma}, \eqref{eq:C0_def},
and~\eqref{eq:Ci_def}.  These expressions coincide with the real–space Green's
function quoted in Eqs.~\eqref{greengen}–\eqref{adefs} of the main text.
\subsection{Structure of the pole contributions}

The spatial structure of the membrane response is governed by the poles
\(m_{1,2}\) of the generalized Stokes operator,
\[
m_{1,2}^2=\frac{B\mp\sqrt{\Delta}}{2A},
\qquad
\Delta=B^2-4AC .
\]
After eliminating \(\nu\) using \(\nu^2=\zeta_\perp/\eta_o\), and treating
\(\zeta_\perp\) as fixed, the coefficients become
\[
A=S+\eta_o^2,\qquad
B=S(\kappa^2+\lambda^2)+2\zeta_\perp\eta_o,\qquad
C=S\kappa^2\lambda^2+\zeta_\perp^2,
\]
where \(S=\eta_s(\eta_s+\eta_d)\).
The discriminant then takes the form
\[
\Delta =
S^2(\kappa^2-\lambda^2)^2
+4S(\kappa^2+\lambda^2)\zeta_\perp\eta_o
-4S\zeta_\perp^2
-4S\kappa^2\lambda^2\eta_o^2 .
\]

Since the coefficient of \(\eta_o^2\) is negative, \(\Delta(\eta_o)\) is a
downward-opening parabola.
Solving \(\Delta=0\) gives the two critical values
\[
\eta_{o,\pm}=
\frac{\zeta_\perp(\kappa^2+\lambda^2)
\pm|\kappa^2-\lambda^2|
\sqrt{\zeta_\perp^2+\kappa^2\eta_s(\eta_s+\eta_d)\lambda^2}}
{2\kappa^2\lambda^2}.
\]

Hence \(\Delta>0\) for \(\eta_{o,-}<\eta_o<\eta_{o,+}\), and the poles
\(m_{1,2}\) are real. In this regime the membrane response is a
superposition of two purely exponentially screened hydrodynamic modes.
Outside this interval, \(\Delta<0\) and the poles form a complex-conjugate
pair \(m_{1,2}=a\pm ib\).

For the positive-parameter branch considered here one has \(A>0\) and
\(B>0\), so the complex poles lie in the right half-plane on the principal
branch and therefore have positive real part. The response thus remains
exponentially screened while acquiring an oscillatory component.
The pole contribution to the Green tensor then behaves asymptotically as
\[
G^{\rm pole}_{ij}(r)
\sim
\frac{e^{-ar}}{\sqrt r}\cos(br-\phi),
\]
up to tensorial prefactors.
When the algebraic coefficient \(C_0\) in the general real-space Green
tensor is nonzero, this oscillatory screened contribution is accompanied
by the algebraic \(1/r^2\) term already present in the full solution.

The three possibilities \(\Delta>0\), \(\Delta=0\), and \(\Delta<0\)
therefore correspond respectively to overdamped-like,
critically damped, and underdamped-like screened hydrodynamic response.

These general results simplify in several useful limits.

In the compressible parity-symmetric limit \(\eta_o\to0\), the
discriminant reduces to
\[
\Delta=[\eta_s(\eta_s+\eta_d)]^2(\kappa^2-\lambda^2)^2\ge0,
\]
so the poles are always real. The screening masses reduce to
\[
m_1^2=\lambda^2,
\qquad
m_2^2=\kappa^2 ,
\]
up to relabeling. The hydrodynamic response therefore decomposes into
independent compressional and shear screening channels.

In the incompressible parity-symmetric limit
\(\eta_o\to0\), \(\eta_d\to\infty\), the longitudinal screening scale
diverges and formally
\[
m_1^2=0,\qquad m_2^2=\kappa^2 .
\]
In this limit the longitudinal sector decouples and its residue vanishes,
so the only nontrivial hydrodynamic screening that remains is the
transverse shear mode with screening length \(\kappa^{-1}\).

Finally, in the identical-screening degenerate limit
\(\kappa=\lambda=\nu=m\), together with
\(\eta_d=3\eta_s\) and
\(\zeta_\perp=(\eta_o/\eta_s)\zeta_\parallel\),
the discriminant vanishes,
\[
\Delta=0,
\qquad
m_1=m_2=m .
\]
The two poles therefore coalesce into a repeated screening mass and the
Green tensor becomes controlled by a single length scale
\(\ell=m^{-1}\).
This limit realizes the coalescence condition \(\Delta=0\) on a special
constrained submanifold of parameter space. In this regime odd viscosity
no longer shifts the pole locations, but instead modifies the residues
and generates the antisymmetric component of the Green tensor.

\section{Parity structure of the Green tensor}

The parity properties of the Green tensor follow directly from the symmetry of the governing hydrodynamic equations. Consider a reflection in the membrane plane described by a matrix $P$ satisfying $P^2=I$ and $\det P=-1$. Under this transformation
\begin{equation}
r_i' = P_{ij} r_j, 
\qquad
v_i'(\mathbf r') = P_{ij} v_j(\mathbf r).
\end{equation}

The isotropic tensor is invariant,
\begin{equation}
P\delta P^T=\delta ,
\end{equation}
whereas the antisymmetric tensor changes sign,
\begin{equation}
P\varepsilon P^T=-\varepsilon .
\end{equation}
Consequently the terms proportional to $\varepsilon_{ij}$ in the hydrodynamic operator change sign under reflection. The hydrodynamic equations are therefore mapped to those of a medium with opposite handedness. Introducing a chirality label $\chi=\pm1$, the Green tensor satisfies the covariance relation
\begin{equation}
G^{(-\chi)}(\mathbf r')
=
P\,G^{(\chi)}(\mathbf r)\,P^T ,
\qquad
\mathbf r' = P\mathbf r .
\label{eq:green_symmetry}
\end{equation}

Rotational isotropy restricts the Green tensor to the form
\begin{equation}
G^{(\chi)}_{ij}(\mathbf r)
=
A_0(r;\chi)\,\delta_{ij}
+
A_1(r;\chi)\,\hat r_i\hat r_j
+
A_2(r;\chi)\,\varepsilon_{ij},
\qquad r=|\mathbf r|.
\end{equation}
Substituting this decomposition into Eq.~\eqref{eq:green_symmetry} and using
\begin{equation}
P(\hat r\hat r)P^T=\widehat{P\mathbf r}\,\widehat{P\mathbf r},
\qquad
P\varepsilon P^T=-\varepsilon ,
\end{equation}
immediately gives
\begin{equation}
A_0(r;\chi)=A_0(r;-\chi), \qquad
A_1(r;\chi)=A_1(r;-\chi), \qquad
A_2(r;\chi)=-A_2(r;-\chi).
\end{equation}

Thus the coefficients multiplying the symmetric tensors $\delta_{ij}$ and $\hat r_i\hat r_j$ are even under chirality reversal, while the coefficient multiplying the antisymmetric tensor $\varepsilon_{ij}$ is odd. This result follows purely from symmetry and does not depend on the detailed analytic form of the Green tensor.

\end{document}